\documentclass{aa}  
\usepackage{float}

\usepackage[dvipsnames]{xcolor}
\usepackage{graphicx}
\usepackage{txfonts}
\usepackage{multirow}
\usepackage{ulem}
\usepackage[hidelinks]{hyperref}
\makeatletter
\renewcommand*\aa@pageof{, page \thepage{} of \pageref*{LastPage}}
\makeatother
\newcommand{\prosp}{{\sc Prospector}}
\newcommand{\cig}{{\sc Cigale}}

\begin{document} 

\title{Polynomial expansion of the star formation history in galaxies}

   \subtitle{}

   \author{D. Jim\'enez-L\'opez\inst{\ref{inst1}} \and P. Corcho-Caballero\inst{\ref{inst1}, \ref{inst2}} \and S. Zamora\inst{\ref{inst1}} \and Y. Ascasibar\inst{\ref{inst1}}}

   \institute{Departamento de F\'{i}sica Te\'orica, Universidad Aut\'onoma de Madrid, Campus de Cantoblanco, Madrid 28049, Spain.\label{inst1}\\\email{daniel.jimenezl@estudiante.uam.es, pablo.corcho@uam.es, sandra.zamora@uam.es, yago.ascasibar@uam.es}
  \and 
  Australian Astronomical Optics, Macquarie University, 105 Delhi Rd, North Ryde, NSW 2113, Australia\label{inst2}
             }
    
   \date{}

 
  \abstract
   {There are typically two different approaches to inferring the mass formation history (MFH) of a given galaxy from its luminosity in different bands. Non-parametric methods are known for their flexibility and accuracy, while parametric models are more computationally efficient.}
   {In this work we propose an alternative, based on a polynomial expansion around the present time, that combines the advantages of both techniques.}
   {In our approach, the MFH is decomposed through an orthonormal basis of N polynomials in lookback time.
   To test the proposed framework, synthetic observations are generated from models based on common analytical approximations (exponential, delayed-$\tau$, and Gaussian star formation histories), as well as cosmological simulations for the Illustris-TNG suite.
   A normalized distance is used to measure the quality of the fit, and the input MFH is compared with the polynomial reconstructions both at the present time and through cosmic evolution.
  Our polynomial expansion is also compared with widely used parametric and non-parametric methods such as \cig\ and \prosp.}
   {The observed luminosities are reproduced with an accuracy of around 10 per cent for a constant star formation rate (N=1) and better for higher-order polynomials.
   Our method provides good results on the reconstruction of the total stellar mass, the star formation rate, and even its first derivative for smooth star formation histories, but it has difficulties in reproducing variations on short timescales and/or star formation histories that peak at the earliest times of the Universe.}
   {The polynomial expansion appears to be a promising alternative to other analytical functions used in parametric methods, combining both speed and flexibility.}

   \maketitle
%
%


\section{Introduction}

One of the fundamental challenges in extragalactic astronomy is the reconstruction of the star formation history (SFH) of a given galaxy from the observed spectral energy distribution (SED).
This is an inverse problem, where the emission at different wavelengths is considered to be the sum over simple stellar populations (SSPs) of different ages and metallicities, with some contribution from the nebular gas.

There are, broadly speaking, two different approaches to carrying out such a reconstruction.
So-called non-parametric methods tackle the problem via a `brute-force' approach, decomposing the SFH as a discrete sum over many individual SSPs with uncorrelated coefficients, with the only constraint that they all have to be positive \citep[e.g.][]{Heavens+00, Cid-Fernandes+05, Ocvirk+06, Tojeiro+07, Koleva+09, MacArthur+09, Sanchez+16, Gomes&Papaderos+17, Cardoso+19}. 

Alternatively, parametric methods describe the SFH in terms of a simple analytical function with only a few free parameters.
A declining exponential is arguably the simplest one, and it has been widely applied to observations of early-type galaxies \citep[e.g.][]{Tinsley+72_Elliptical}.
However, it was soon realized that many objects in the local Universe may display more complex behaviours \citep[e.g.][]{Talbot+71}, and many are indeed better described by a monotonically increasing exponential \citep[e.g.][]{Maraston+10} or a delayed-$\tau$ model of the form $\Psi(t) \equiv \dot M(t) \propto \frac{t}{\tau} e^{-t/\tau}$ \citep[e.g.][]{Lee+10}.
Some authors \citep[e.g.][]{Gladders+13} argue that additional degrees of freedom are necessary in order to capture the diversity of real SFHs and advocate the use of a Gaussian function with two free parameters, whereas other studies are based on a library of exponential SFHs with superimposed star formation bursts \citep[e.g.][]{Kauffmann+03, Walcher+08}, variable galaxy age \citep[e.g.][]{Noll+09}, and/or quenching of the star formation activity at a certain time \citep[e.g.][]{Smethurst+17}.

Non-parametric methods tend to be more flexible and accurate, but they are computationally demanding and subject to numerous degeneracies due to the large number of free variables involved \citep[see e.g.][]{Walcher+11, Conroy+13, Lower+20}.
Parametric methods, on the other hand, would be advantageous in this respect, but their simplicity makes it difficult to explore the whole range of possible SFHs.

In the present work we propose a new kind of parametric model that attempts to combine the advantages of both approaches.
We model the stellar mass buildup of a galaxy in terms of the mass formation history (MFH),
\begin{equation}
M(t) = \int_0^{t} \Psi(t')\ {\rm d} t',
\end{equation}
where $\Psi(t)$ represents the usual SFH in terms of the instantaneous star formation rate (SFR).
We note that $M(t)$ is the total mass of stars formed up to time $t$ rather than the actual stellar mass, $M_*(t)$, which takes stellar evolution into account.
Our main idea is to expand the MFH around the present time as a polynomial series based on a normalized lookback time,
\begin{equation}
    \hat t \equiv \frac{t_0-t}{t_0}
,\end{equation}
that varies from $\hat t = 0$ at the present time, $t=t_0$, to $\hat t=1$ at the Big Bang, $t=0$.
This provides a simple model where the free parameters have a straightforward physical interpretation (the total stellar mass formed, the current SFR, and its time derivatives) and whose number can be arbitrarily set by the user according to the amount and quality of the available data.

The details of our polynomial expansion are thoroughly presented in Sect.~\ref{sec:formalism}, and the method is calibrated against a set of synthetic observations based on analytical and numerical SFHs described in Sect.~\ref{sec:test}.
Section~\ref{sec:results} evaluates the accuracy of the reconstruction, and the main caveats and potential improvements are discussed in Sect.~\ref{sec:discussion}.
Our main conclusions are briefly summarized in Sect.~\ref{sec:conclusions}.


\section{Mathematical formalism}
\label{sec:formalism}

In terms of $\hat t$, the MFH can be written as
\begin{equation}
M(\hat t) = \int_0^{(1-\hat t)t_0} \Psi(t)\ {\rm d} t.
\end{equation}
We made the simplifying assumption that galaxies do not lose mass: neglecting stellar lifetimes as well as tidal stripping, $M(\hat t)$ must be a monotonically decreasing function of $\hat t$.
We expanded this function around the present time as a polynomial series in the normalized lookback time, $\hat t$.
Imposing that $M(1)=0$ at the beginning of the Universe, we considered $N$ polynomials
\begin{equation}
B_{i}^{(N)}(\hat t)=\sum_{n=0}^{i} \beta_{n, i}^{(N)}\, \mathcal{M}_n^{(N)}(\hat t)
\label{eq:pol_basis}
\end{equation}
that are a linear combination of primordial terms of the form
\begin{equation}
\mathcal{M}_n^{(N)}(\hat t) = \hat t^n-\hat t^N,
\label{eq_primordial_MFH}
\end{equation}
where $i$ varies from 0 to $N-1$, and the coefficients $\beta_{n, i}$ are set by ensuring that the elements $B_i$ form an orthonormal basis that satisfies
\begin{equation}
B_i^{(N)}(\hat t) \cdot B_j^{(N)}(\hat t) = \delta_{ij}
\end{equation}
according to a suitable definition of the scalar product.

\begin{table*}
\caption{Specific luminosities on the $ugriz$ photometric system (in erg/s/M$_\odot$) associated with the primordial polynomial MFH $\{\mathcal{M}_n^{(N)}(\hat t)\}_{n=0,N-1}$ up to $N=5$ for the {\sc Popstar} SSPs.}
\centering

\begin{tabular}{ccrrrrr}
 \hline
 \hline
 $N$ & $\mathcal{M}_n^{(N)}(\hat t)$ &  $\mathcal{L}_n^{(N)}(u)$~~ &  $\mathcal{L}_n^{(N)}(g)$~~ &  $\mathcal{L}_n^{(N)}(r)$~~ &  $\mathcal{L}_n^{(N)}(i)$~~ &  $\mathcal{L}_n^{(N)}(z)$~~ \\
 \hline
 1 & $1-\hat t$ & $1.22 \cdot10^{30}$& $1.08 \cdot10^{32}$& $1.13 \cdot10^{32}$& $7.33 \cdot10^{31}$& $1.28 \cdot10^{31}$\\
 \hline
 2 & $1-\hat t^2$ & $1.73 \cdot10^{30}$& $3.15 \cdot10^{31}$& $4.89 \cdot10^{31}$& $3.63 \cdot10^{31}$& $7.01 \cdot10^{30}$\\
 & $\hat t-\hat t^2$ & $-1.04 \cdot10^{31}$& $-7.60 \cdot10^{31}$& $-6.39 \cdot10^{31}$& $-3.70 \cdot10^{31}$& $-5.78 \cdot10^{30}$\\
 \hline
 3 & $1-\hat t^3$ & $1.18 \cdot10^{30}$& $2.40 \cdot10^{31}$& $3.96 \cdot10^{31}$& $3.01 \cdot10^{31}$& $5.94 \cdot10^{30}$\\
 & $\hat t-\hat t^3$ & $-1.10 \cdot10^{31}$& $-8.36 \cdot10^{31}$& $-7.31 \cdot10^{31}$& $-4.31 \cdot10^{31}$& $-6.85 \cdot10^{30}$\\
 & $\hat t^2-\hat t^3$ & $-5.55 \cdot10^{29}$& $-7.54 \cdot10^{31}$& $-9.20 \cdot10^{30}$& $-6.15 \cdot10^{30}$& $-1.07 \cdot10^{30}$\\
 \hline
 4 & $1-\hat t^4$ & $1.02 \cdot10^{30}$& $2.16 \cdot10^{31}$& $3.63 \cdot10^{31}$& $2.78 \cdot10^{31}$& $5.51 \cdot10^{30}$\\
 & $\hat t-\hat t^4$ & $-1.12 \cdot10^{31}$& $-8.59 \cdot10^{31}$& $-7.64 \cdot10^{31}$& $-4.54 \cdot10^{31}$& $-7.27 \cdot10^{30}$\\
 & $\hat t^2-\hat t^4$ & $-7.13 \cdot10^{29}$& $-9.93 \cdot10^{30}$& $-1.25 \cdot10^{31}$& $-8.49 \cdot10^{30}$& $-1.49 \cdot10^{30}$\\
 & $\hat t^3-\hat t^4$ & $-1.58 \cdot10^{29}$& $-2.38 \cdot10^{30}$& $-3.31 \cdot10^{30}$& $-2.34 \cdot10^{30}$& $-4.25 \cdot10^{29}$\\
 \hline
 5 & $1-\hat t^5$ & $9.43 \cdot10^{29}$& $2.04 \cdot10^{31}$& $3.46 \cdot10^{31}$& $2.66 \cdot10^{31}$& $5.30 \cdot10^{30}$\\
 & $\hat t-\hat t^5$ & $-1.12 \cdot10^{31}$& $-8.71 \cdot10^{31}$& $-7.81 \cdot10^{31}$& $-4.67 \cdot10^{31}$& $-7.50 \cdot10^{30}$\\
 & $\hat t^2-\hat t^5$ & $-7.89 \cdot10^{29}$& $-1.11 \cdot10^{31}$& $-1.42 \cdot10^{31}$& $-9.70 \cdot10^{30}$& $-1.72 \cdot10^{30}$\\
 & $\hat t^3-\hat t^5$ & $-2.34 \cdot10^{29}$& $-3.55 \cdot10^{30}$& $-5.00 \cdot10^{30}$& $-3.55 \cdot10^{30}$& $-6.48 \cdot10^{29}$\\
 & $\hat t^4-\hat t^5$ & $-7.54 \cdot10^{28}$& $-1.16 \cdot10^{30}$& $-1.68 \cdot10^{30}$& $-1.21 \cdot10^{30}$& $-2.23 \cdot10^{29}$\\
\hline
\end{tabular}
\label{tab:Lprimordial}
\end{table*}

\begin{table*}
\caption{Coefficients of basis polynomials $\{B_i^{(N)}(\hat t)\}_{i=0,N-1}$ up to $N=5$.}
\centering
\begin{tabular}{rrrrrr}
\hline
\hline
 N & $\beta_{0,i}^{(N)}$~~ ~~& $\beta_{1,i}^{(N)}$~~ ~~& $\beta_{2,i}^{(N)}$~~ ~~& $\beta_{3,i}^{(N)}$~~ ~~& $\beta_{4,i}^{(N)}$~~ ~~\\
\hline
1 & $5.78 \cdot10^{-33}$ \\
\hline
2 & $1.45 \cdot10^{-32}$ \\
  & $4.04 \cdot10^{-32}$ & $2.78 \cdot10^{-32}$ \\
\hline
3 & $1.80 \cdot10^{-32}$ \\
  & $4.90 \cdot10^{-32}$ & $2.42 \cdot10^{-32}$ \\
  & $5.28 \cdot10^{-31}$ & $-2.06 \cdot10^{-31}$ & $3.96 \cdot10^{-30}$ \\
\hline
4 & $1.96 \cdot10^{-32}$ \\
  & $5.38 \cdot10^{-32}$ & $2.34 \cdot10^{-32}$ \\
  & $6.45 \cdot10^{-31}$ & $-2.00 \cdot10^{-31}$ & $3.13 \cdot10^{-30}$ \\
  & $2.01 \cdot10^{-30}$ & $1.24 \cdot10^{-31}$ & $-1.45 \cdot10^{-29}$ & $7.41 \cdot10^{-29}$ \\
\hline
5 & $2.06 \cdot10^{-32}$ \\
  & $5.66 \cdot10^{-32}$ & $2.31 \cdot10^{-32}$ \\
  & $7.17 \cdot10^{-31}$ & $-1.98 \cdot10^{-31}$ & $2.87 \cdot10^{-30}$ \\
  & $2.20 \cdot10^{-30}$ & $1.10 \cdot10^{-31}$ & $-1.30 \cdot10^{-29}$ & $5.05 \cdot10^{-29}$ \\
  & $1.05 \cdot10^{-29}$ & $-9.68 \cdot10^{-31}$ & $8.24 \cdot10^{-29}$ & $-1.19 \cdot10^{-27}$ & $3.10 \cdot10^{-27}$ \\
\hline
\end{tabular}
\label{tab:basis_coefficients}
\end{table*}

In the end, it is our aim to find an MFH that matches a given set of observables.
More precisely, we considered $N_\nu$ independent luminosity measurements, $L_{\rm obs}(\nu)$, in different frequency bands, $\nu$.
Therefore, we needed to compute the luminosities, $\mathcal{L}_n^{(N)}(\nu)$, of the $N$ primordial polynomials,  $\mathcal{M}_n^{(N)}(\hat t)$, as well as those of the basis elements,
\begin{equation}
L_{i}^{(N)}(\nu) = \sum_{n=0}^{i} \beta_{n, i}^{(N)}\, \mathcal{L}_n^{(N)}(\nu).
\end{equation}
We then parameterized the MFH of a given galaxy as a linear combination,
\begin{equation}
M^{(N)}(\hat t) = \sum_i \delta_i^{(N)}\, B_i^{(N)}(\hat t)
\label{eq:mass_reconstruction}
,\end{equation}
of our basis elements and then minimized the distance,
\begin{equation}
d^2 = \sum_\nu \left[ L_M(\nu) - \ L_{\rm obs}(\nu) \right]^2
,\end{equation}
between the predicted luminosities,
\begin{equation}
L_M(\nu) = \sum_i \delta_i^{(N)}\, L_i^{(N)}(\nu)
,\end{equation}
and the observed values, $L_{\rm obs}(\nu)$.
Minimizing with respect to the coefficients $\delta_i$,
\begin{equation}
\frac{\partial d^2}{\partial \delta_i^{(N)}} = 0
,\end{equation}
implies
\begin{equation}
\delta_i^{(N)} =  \sum_\nu L_{\rm obs}(\nu)\ L_i^{(N)}(\nu)
\end{equation}
if we define the scalar product as
\begin{equation}
B_i^{(N)} \cdot B_j^{(N)} \equiv \sum_\nu L_i^{(N)}(\nu)\ L_j^{(N)}(\nu).
\label{eq_scalar_product}
\end{equation}

We note that the number, $N$, of basis vectors must be equal to or smaller than the number, $N_\nu$, of observables for the system to be determined.
If $N > N_\nu$, infinitely many degenerate solutions are possible, and thus the proposed method cannot be applied.
When $N$ equals $N_\nu$, the basis will uniquely cover the whole observable space, and any data will be exactly reproduced ($d=0$) by one and only one model.
If $N<N_\nu$, only a subspace of all possible observables will be accessible as a linear combination of the basis elements, and the normalized distance to the observational measurements will be, in general, larger than zero.
As long as $N \le N_\nu$, the scalar product will yield the coefficients of the polynomial MFH whose luminosities are as close as possible to the observed values.
However, there is no guarantee that such an optimal fit will be physically meaningful, especially if the actual MFH cannot be well approximated by a polynomial.

In particular, it is perfectly possible that the optimal reconstruction (in terms of best reproducing the observed luminosities) yields negative star formation for arbitrary periods of time.
For that reason, we looked for the zeroes of $\Psi(\hat t)$ and used them to divide the MFH into separate intervals with a constant sign.
Then, we computed their individual luminosities (using the same polynomial coefficients, but only considering star formation within the selected time interval) and rescaled their MFHs by a constant factor to provide the best possible fit.
This, of course, ensures positive star formation, but the quality of the fit necessarily worsens compared to the original result obtained for $0 \le \hat t \le 1$.
We repeated this process for all polynomial degrees $N \le N_\nu$ and identified the values of $\hat t_{\rm min}$, $\hat t_{\rm max}$, and $N$ that provide the best match to the observed data.

\section{Synthetic observations}
\label{sec:test}

To test the ability of our formalism to reconstruct a realistic non-polynomial MFH, we simulated the observed luminosities for a set of synthetic galaxies.
We first considered three different analytical SFHs that are fairly representative of real galaxies and have often been used to fit observational data: (i) a simple exponential, $\Psi(t) \propto e^{-t/\tau}$, with characteristic time $\tau$; (ii) a delayed-$\tau$ model of the form $\Psi(t) \propto \frac{t}{\tau} e^{-t/\tau}$; and (iii) a Gaussian, $\Psi(t) \propto e^{-\frac{(t_0-t-\alpha)^2}{2~\sigma^2}}$, where $\alpha$ denotes the peak age of the SFH (i.e. $\Psi$ peaks at a lookback time $t_0-\alpha$) and $\sigma = \frac{\rm FWHM}{\sqrt{8\ln(2)}}$ sets its full-width half maximum (FWHM). 

We varied all our timescales from 0.1 to 13.5~Gyr.
Short values of $\tau$ represent very old galaxies that formed the vast majority of their stars in the early Universe, while a Gaussian with a FWHM much shorter than $t_0$ would describe a star formation burst of age $\alpha$.
On the other hand, large values of $\tau$ or FWHM yield smooth SFHs that can be well approximated by a low-order polynomial.

In addition, we also applied our algorithm to the SFHs of $22155$ galaxies selected from the IllustrisTNG-100-1 simulation, part of the IllustrisTNG magnetohydrodynamic cosmological simulations suite\footnote{\url{https://www.tng-project.org/}} \citep[][]{Illustris17, Illustris2017a, Illustris2017b, Illustris2018, Illustris2018a}.
They were run with the moving-mesh {\sc Arepo} code \citep{Springel10} and include a vast range of sub-grid physical processes that may play a critical role in galaxy evolution, such as gas cooling, star formation, supermassive black hole or supernova growth, and active galactic nucleus feedback \citep[see][for details]{Weinberger+18, Pillepich+18a}.
The TNG100-1 run consists of a cubic volume with box length $\sim110$ Mpc at z=0 and dark mater and baryonic mass resolutions of $7.5\times10^6~M_\odot$ and $1.4\times10^6~M_\odot$, respectively, and we selected all galaxies with stellar masses in the range $10^9<M_*/M_\odot<10^{12}$.

For every synthetic galaxy, we computed the luminosities, $L_{\rm syn}(\nu)$, in
the $ugriz$ photometric system ($N_\nu=5$) of the Sloan Digital Sky Survey \citep[SDSS;][]{York+00} as an integral,
\begin{equation}
L(\nu) = \int_0^{t_0} \Psi(t)\ \mathcal{L}_{\rm SSP}(\nu, t_0-t)\ {\rm d}\hat t
\label{eq_popsynth}
,\end{equation}
over SSPs whose specific luminosities per unit stellar mass, $\mathcal{L}_{\rm SSP}$, have been obtained from the {\sc Popstar} evolutionary synthesis models \citep{Popstar+09, Popstar+12, Popstar+13, Millan-Irigoyen+21}.
These models provide a grid of 106 SSP ages, ranging from $10^5$~yr to $\sim 16$~Gyr, and we used the results appropriate for solar metallicity (Z = 0.02) and a \citet{Kroupa+01} initial mass function (IMF).

The $ugriz$ luminosities of our primordial MFHs are quoted in Table~\ref{tab:Lprimordial}.
We note that the values for $n>0$ are all negative because our primordial terms are not monotonically decreasing functions of $\hat t$, and thus the instantaneous SFRs, $\Psi(\hat t) = \frac{N\,\hat t^{N-1} - n\,\hat t^{n-1}}{t_0}$, may become negative.
As mentioned above, this is an important feature of our models: in order to take advantage of the simple and computationally efficient scalar product, one cannot enforce positivity of the SFR at all times, which is a major source of unphysical results.
The precise coefficients that yield orthonormal bases $\{B_i(\hat t)\}_{i=0,N-1}$ of degree $N$ up to five in this particular photometric system are quoted in Table~\ref{tab:basis_coefficients}.

\section{Results}
\label{sec:results}

\begin{figure}
        \centering
        \includegraphics[width=\linewidth]{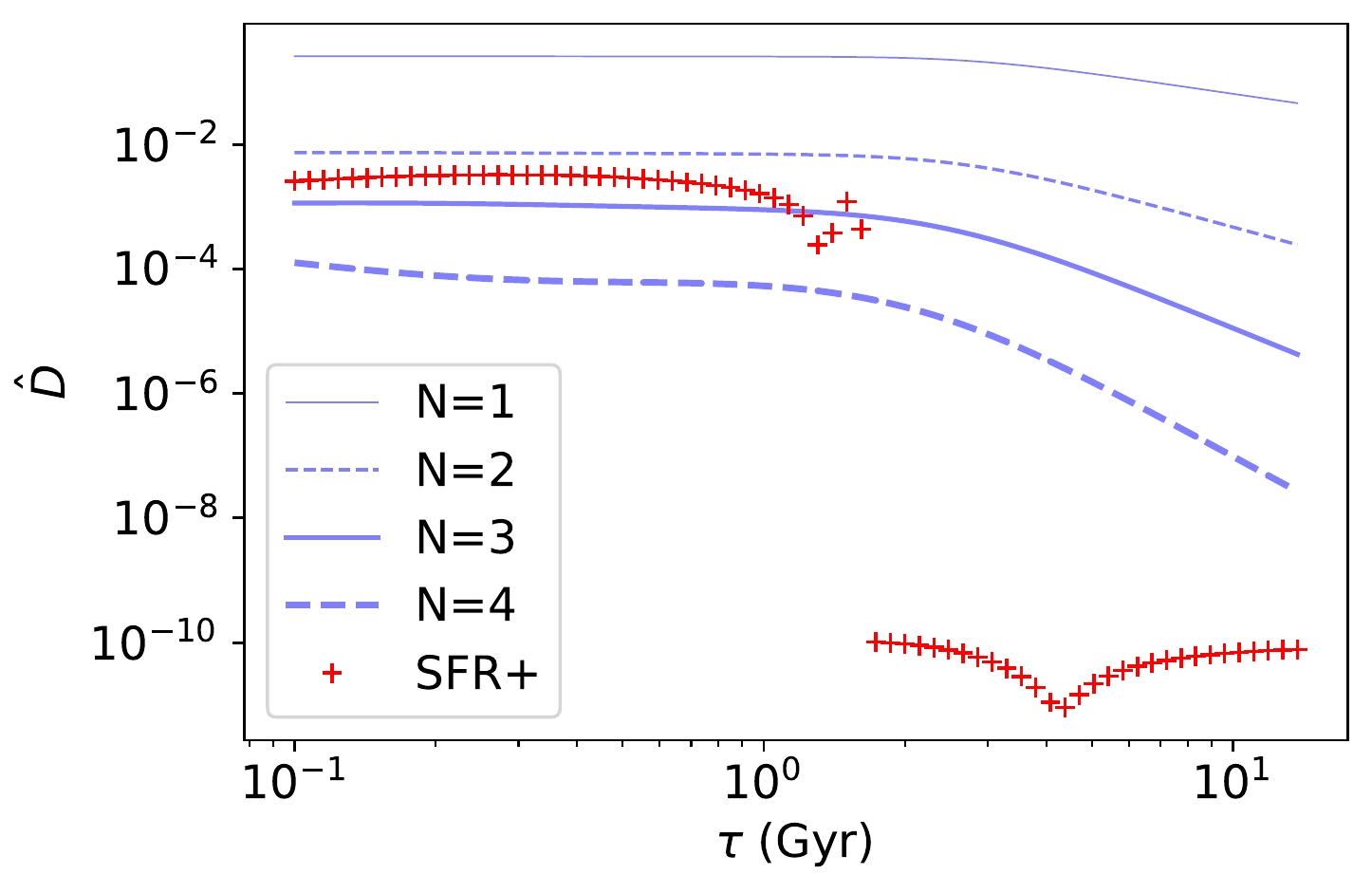}
        \includegraphics[width=\linewidth]{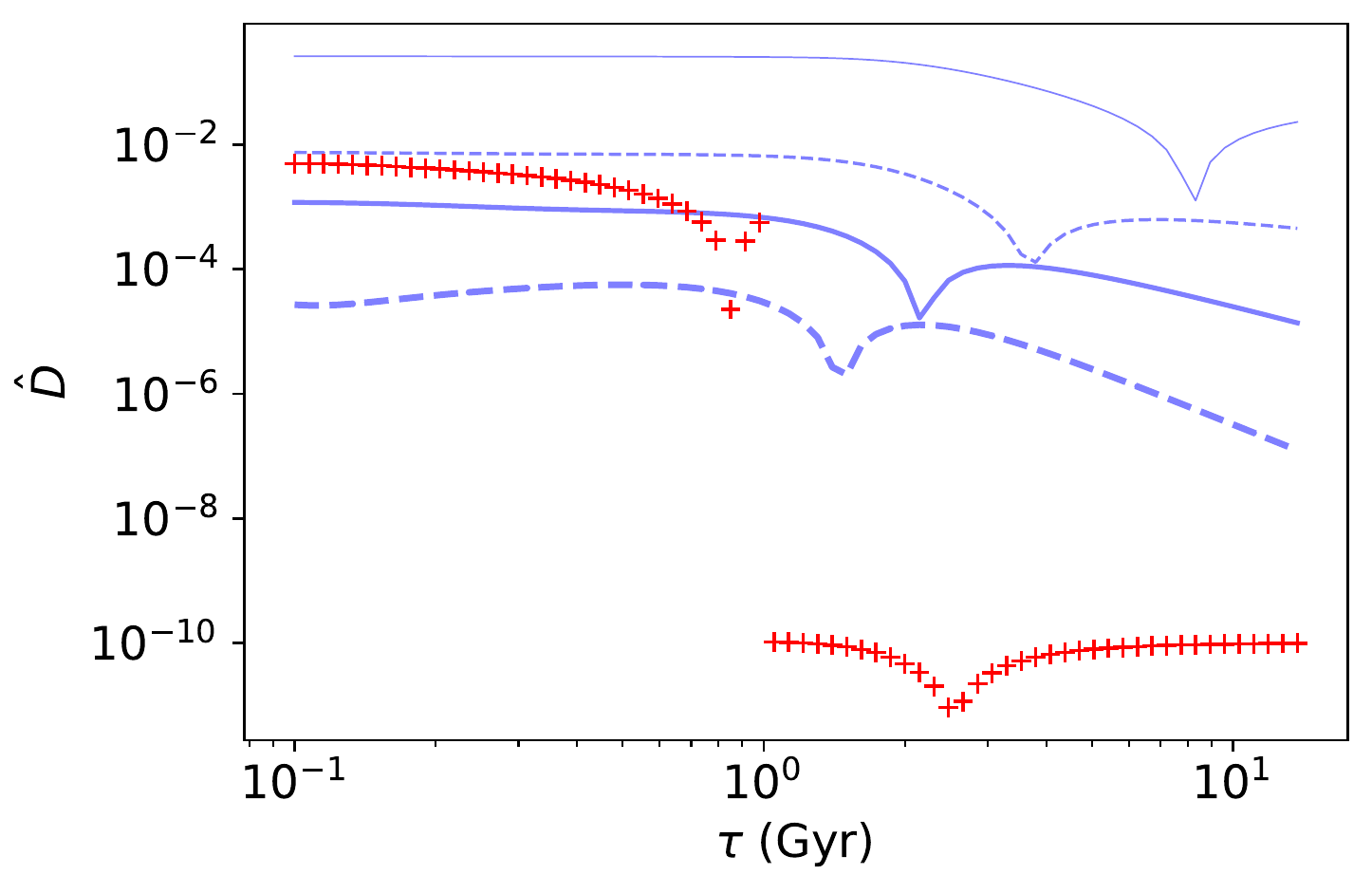}
        \caption{
Best fitting normalized distances associated with each polynomial degree, $N$ (blue lines) and the best positive-SFR fit (red crosses) for different characteristic times of the exponential (upper panel) and delayed-$\tau$ (bottom panel) analytical models.}
        \label{fig:distances_1}
\end{figure}

We first evaluated the ability of our polynomial description to reproduce the luminosities and colours of realistic galaxies by comparing the synthetic observations with the best possible fits that can be achieved with polynomials of different order.
Our ultimate goal is, however, to recover the true underlying MFH, which will be addressed separately in Sect.~\ref{sec:reconstruction} and compared with the results obtained for the Illustris simulations by the \cig\ and \prosp\ inference codes in Sect.~\ref{sec:comparison}.

\subsection{Fit quality}
\label{sec:fit_quality}

To quantify the agreement between the synthetic measurement and the best polynomial fit, we computed the normalized distance, $\hat D$, as
\begin{equation}
    \hat D^2 = \frac{ d^2 }{ \sum_\nu L_{obs}^2(\nu) }.
    \label{eq:norm_d}
\end{equation}
By definition, if the actual MFH were indeed a polynomial of degree $N_{\rm true}$, a perfect fit with $d=0$ would be obtained by our procedure as long as $N \ge N_{\rm true}$.
Otherwise, since we only have five independent observables, $N=5$ will always be able to obtain a perfect fit, although this does not necessarily imply that the polynomial that best fits the observation provides a good description of the actual MFH.
The polynomial basis with $N<5$ will only cover a subset (hyperplane) of the whole five-dimensional space of all possible observations.
If the true MFH is close to a polynomial, its luminosities will be close to this hyperplane.
On the other hand, if the MFH is very different from a polynomial, such as a short star formation burst, the observed luminosities may or may not be far away from the hyperplane covered by the linear combinations of the basis polynomials.

Figure~\ref{fig:distances_1} shows the normalized distance~\eqref{eq:norm_d} as a function of the characteristic timescales of exponential and delayed-$\tau$ MFHs (the distances associated with the fifth degree polynomial are compatible with zero up to truncation errors).
In general terms, higher-order polynomials always achieve an equal or better fit (lower distance) compared to previous orders.
In addition, we also observe that the fit is usually better for longer timescales (i.e. a smoother MFH) than for those with narrower SFR periods.
For cases where the highest polynomial degrees give rise to negative values of the SFR (typically narrow star formation bursts with small $\tau$), the best positive-SFR fit will have distances comparable to $N=3$.
For smoother histories, with long values of $\tau$, there are no periods of negative SFR, and the best fit is assigned to the $N=5$ expansion.

\begin{figure}
        \centering
        \includegraphics[width=.69\linewidth]{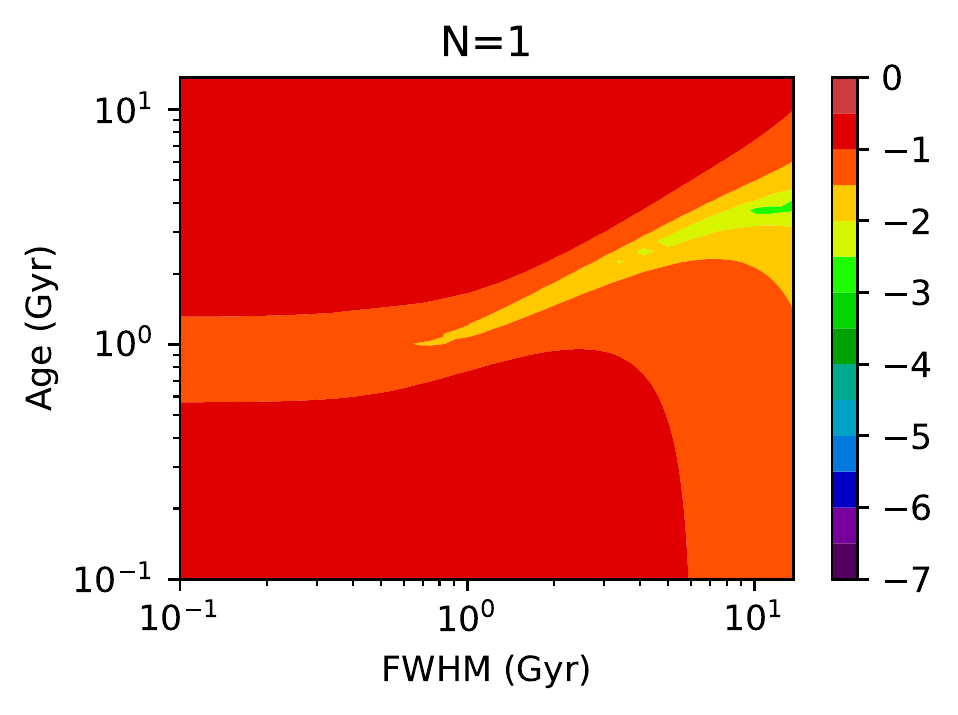}
        \includegraphics[width=.69\linewidth]{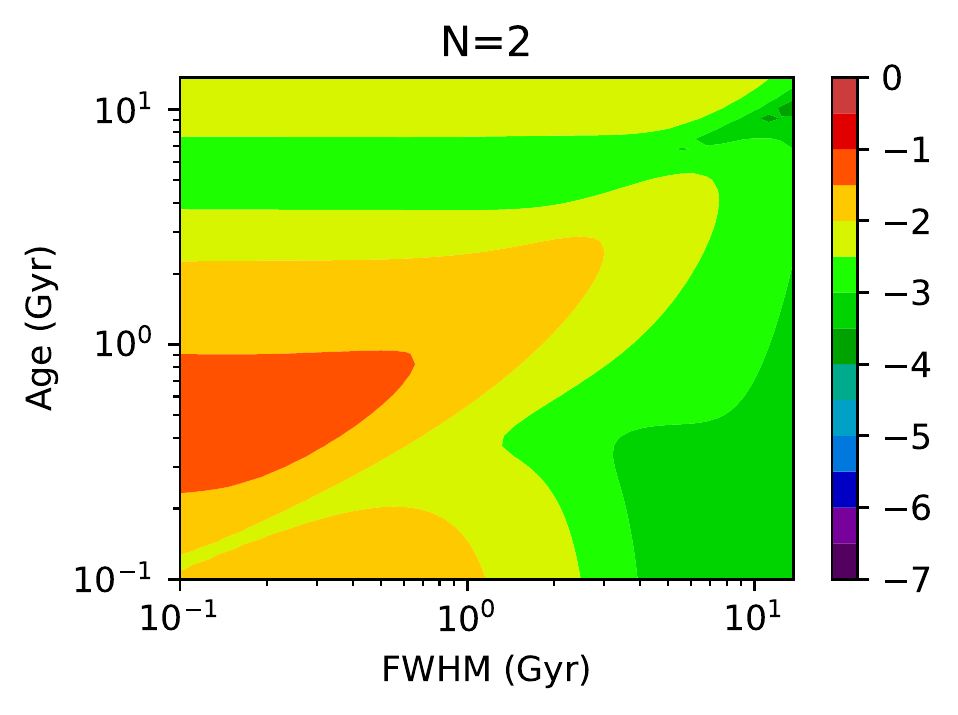}
        \includegraphics[width=.69\linewidth]{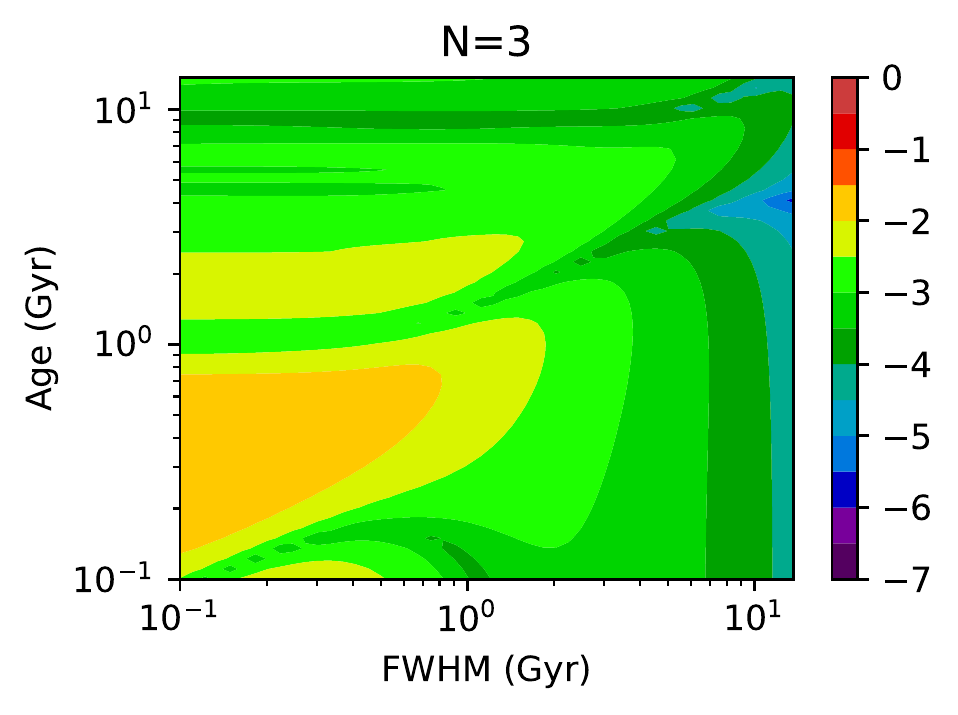}
        \includegraphics[width=.69\linewidth]{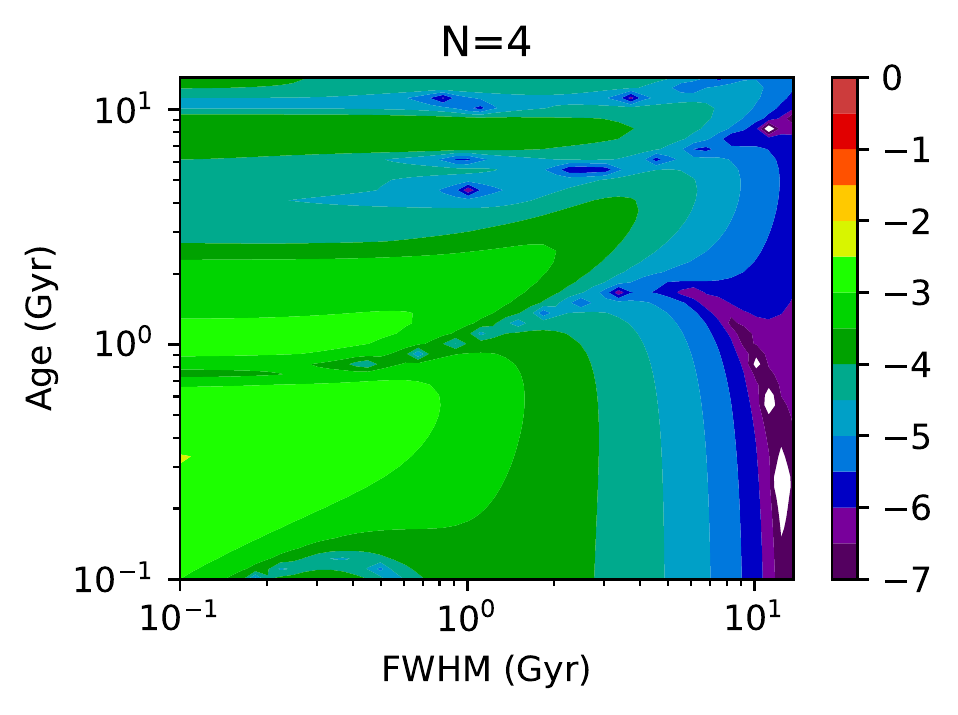}
        \includegraphics[width=.69\linewidth]{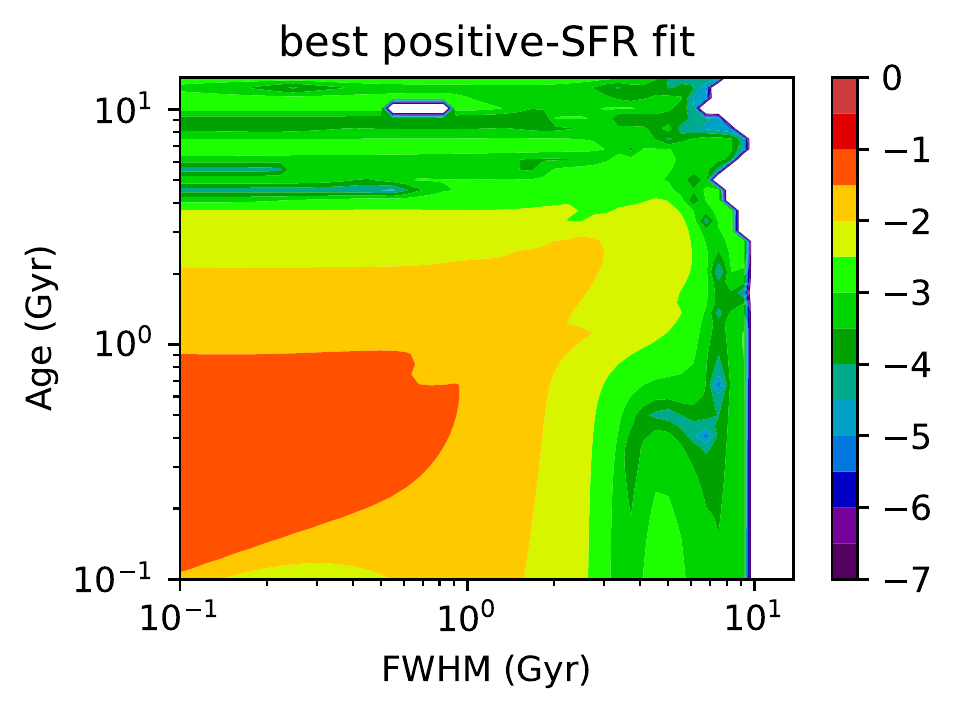}
        \caption{
        Best fitting normalized distances associated with each polynomial degree, $N$, and the best positive-SFR fit for different ages and FWHMs of the Gaussian synthetic SFHs.
    }
        \label{fig:distances_gauss}
\end{figure}

\begin{figure}
        \centering
        \includegraphics[width=\linewidth]{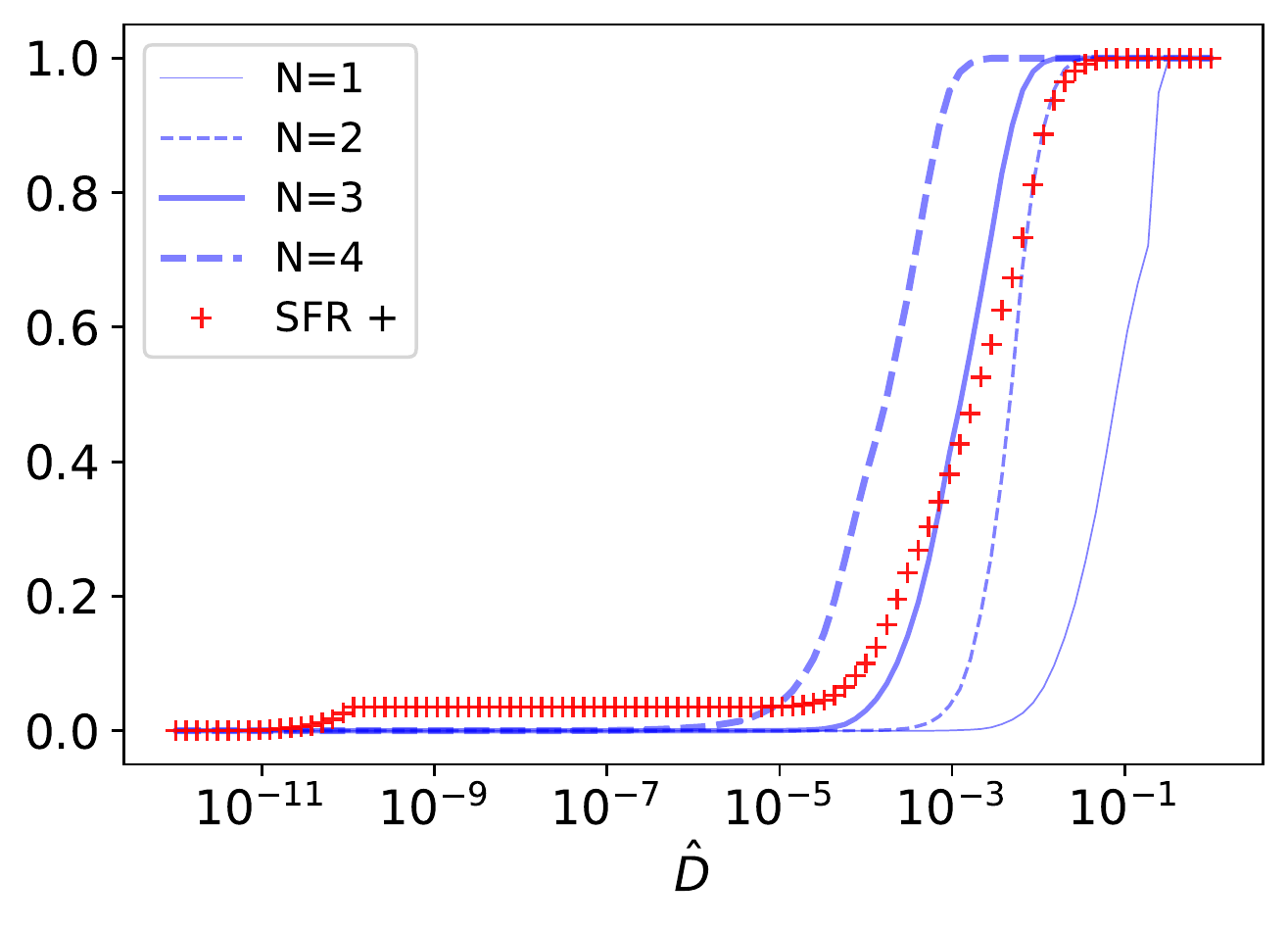}

\caption{
Fraction of galaxies with distances under a certain value, $\hat D$, associated with each polynomial degree, $N$ (blue lines) and the best positive-SFR fit (red crosses) for the simulations from the Illustris sample.}
        \label{fig:distances_Illustris}
\end{figure}

A similar behaviour is shown for the Gaussian SFHs in Fig.~\ref{fig:distances_gauss}, where we now represent the distances as a function of the two model parameters: the burst age and the FWHM. Smaller distances are found for higher-order polynomials and MFHs with wider FWHMs (smoother histories). White sections on the map represent values under $\hat D = 10^{-7}$ (zero up to truncation errors). Once again, the best positive-SFR fit is similar to $N\sim 2-3$ for SFHs that vary on short timescales and are equal to $N=5$ for smooth functions that can be accurately described by a polynomial.

\begin{figure}
        \centering
        \includegraphics[width=0.49\linewidth]{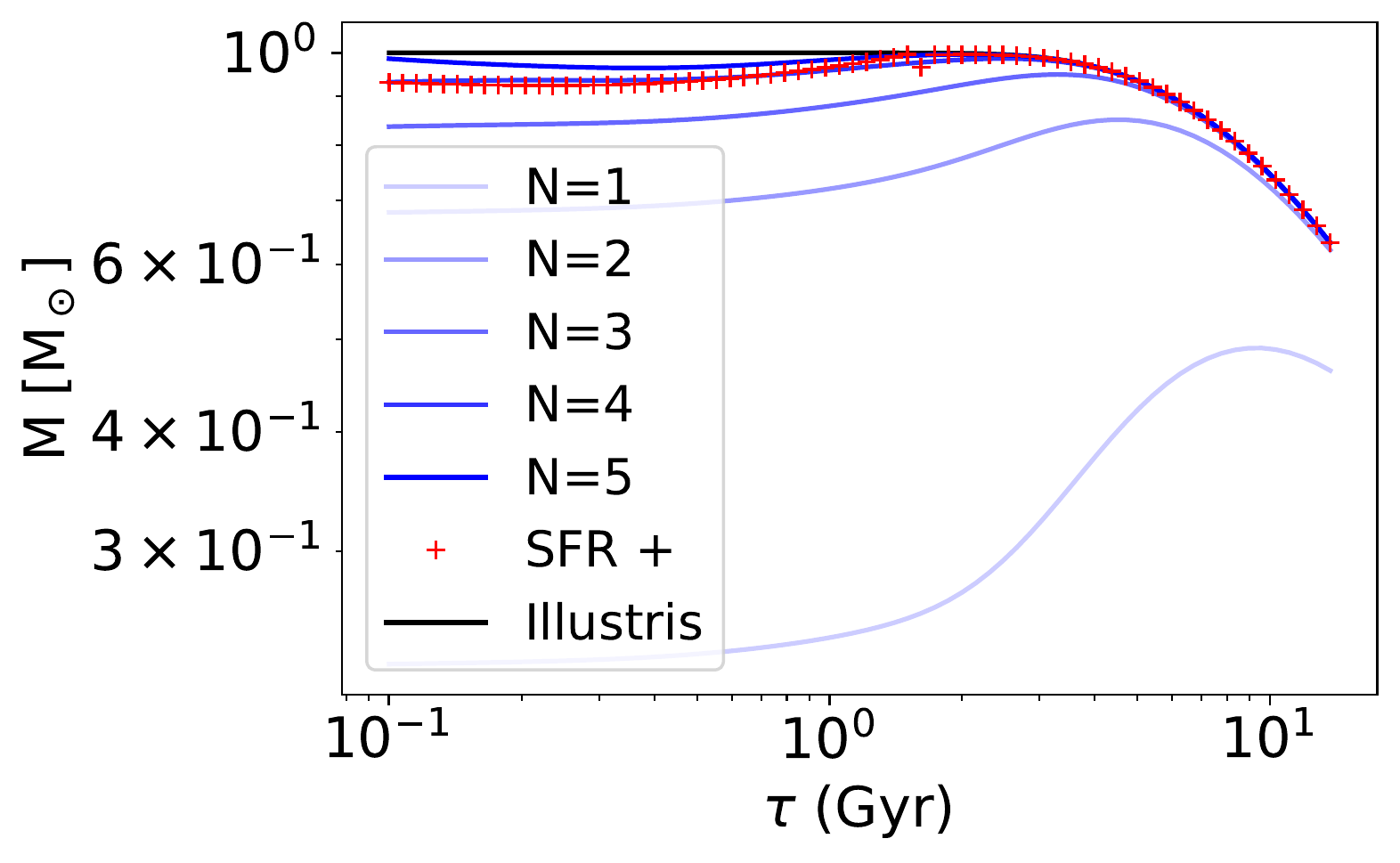}
        \includegraphics[width=0.49\linewidth]{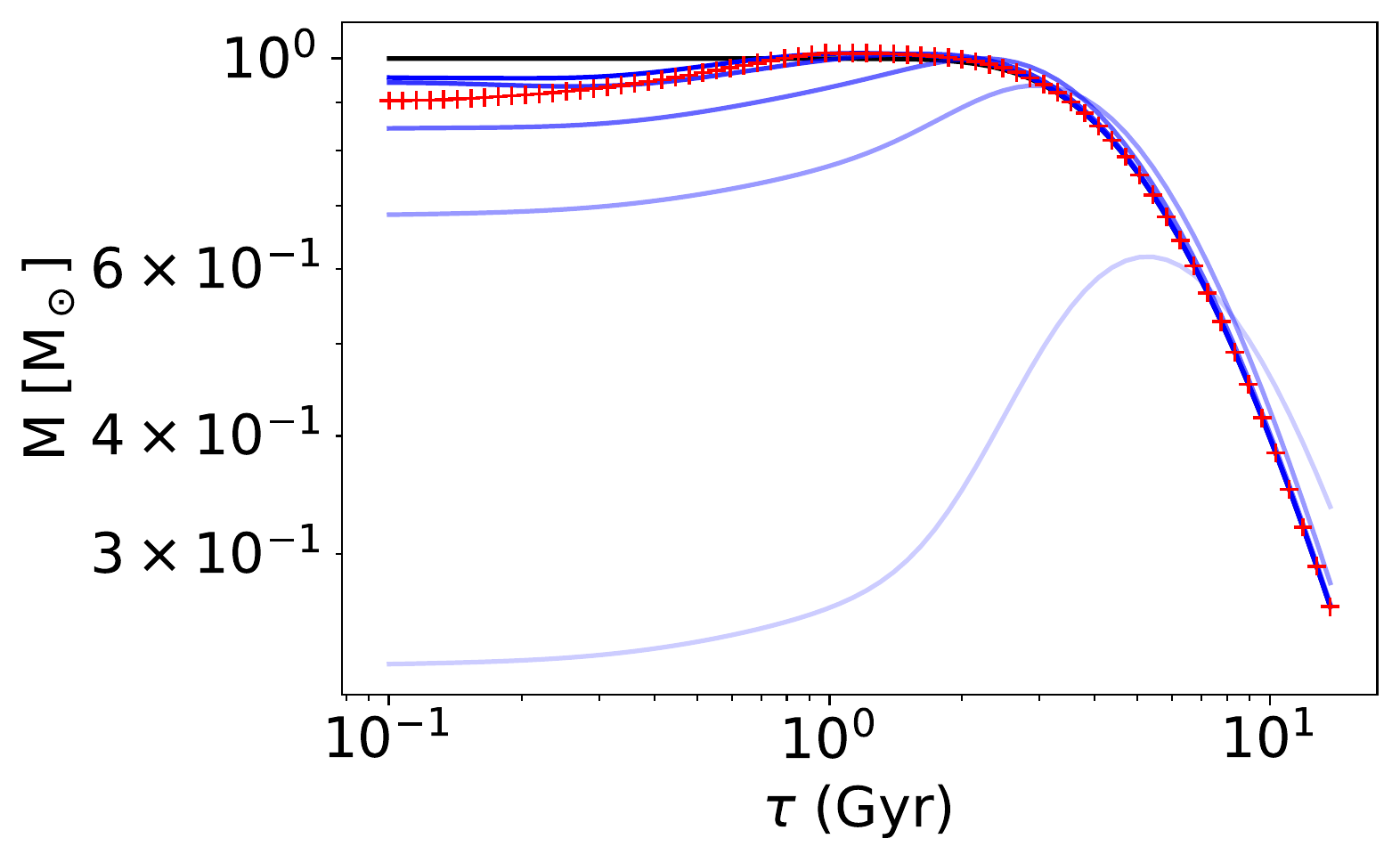}\\

        \includegraphics[width=0.49\linewidth]{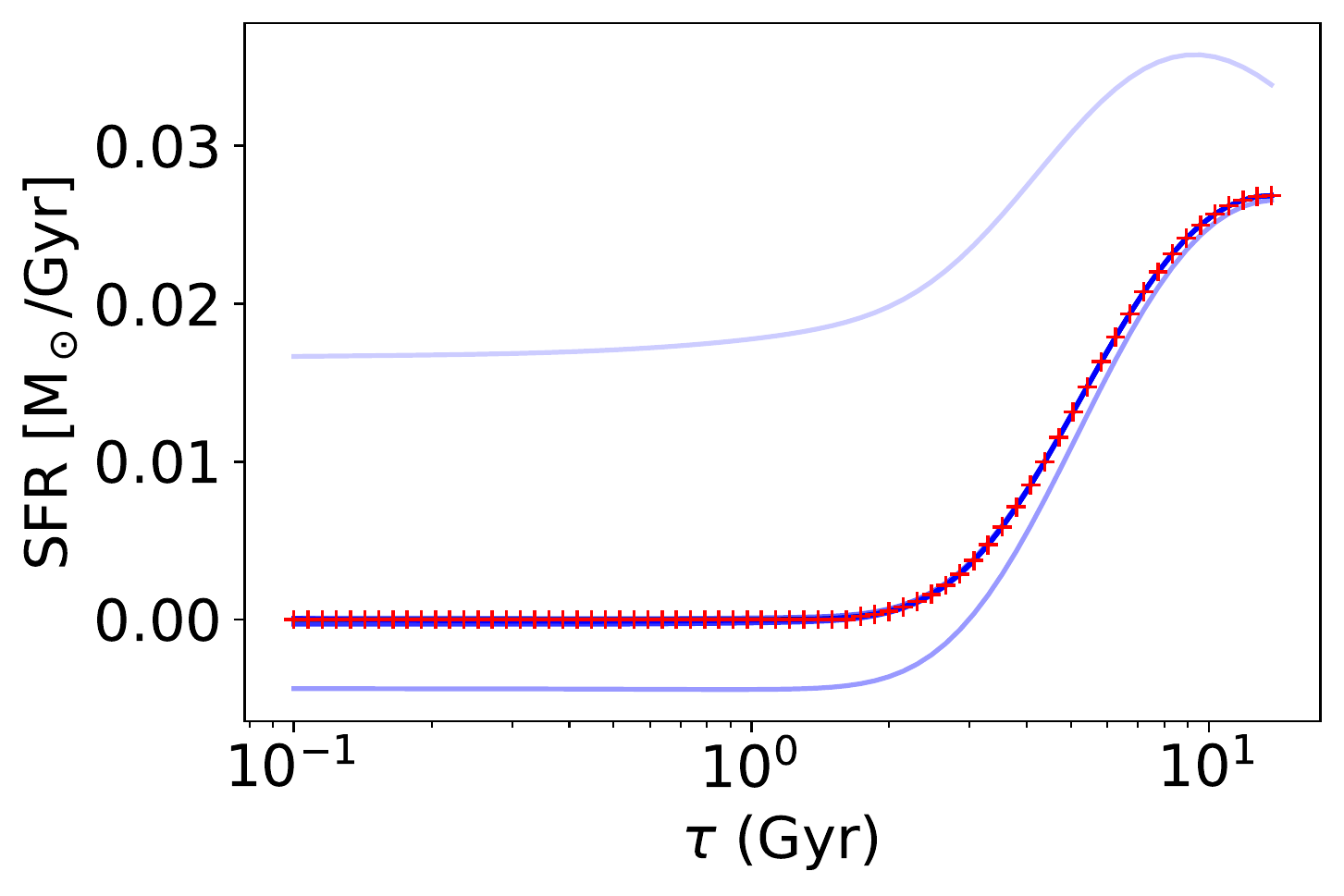}
        \includegraphics[width=0.49\linewidth]{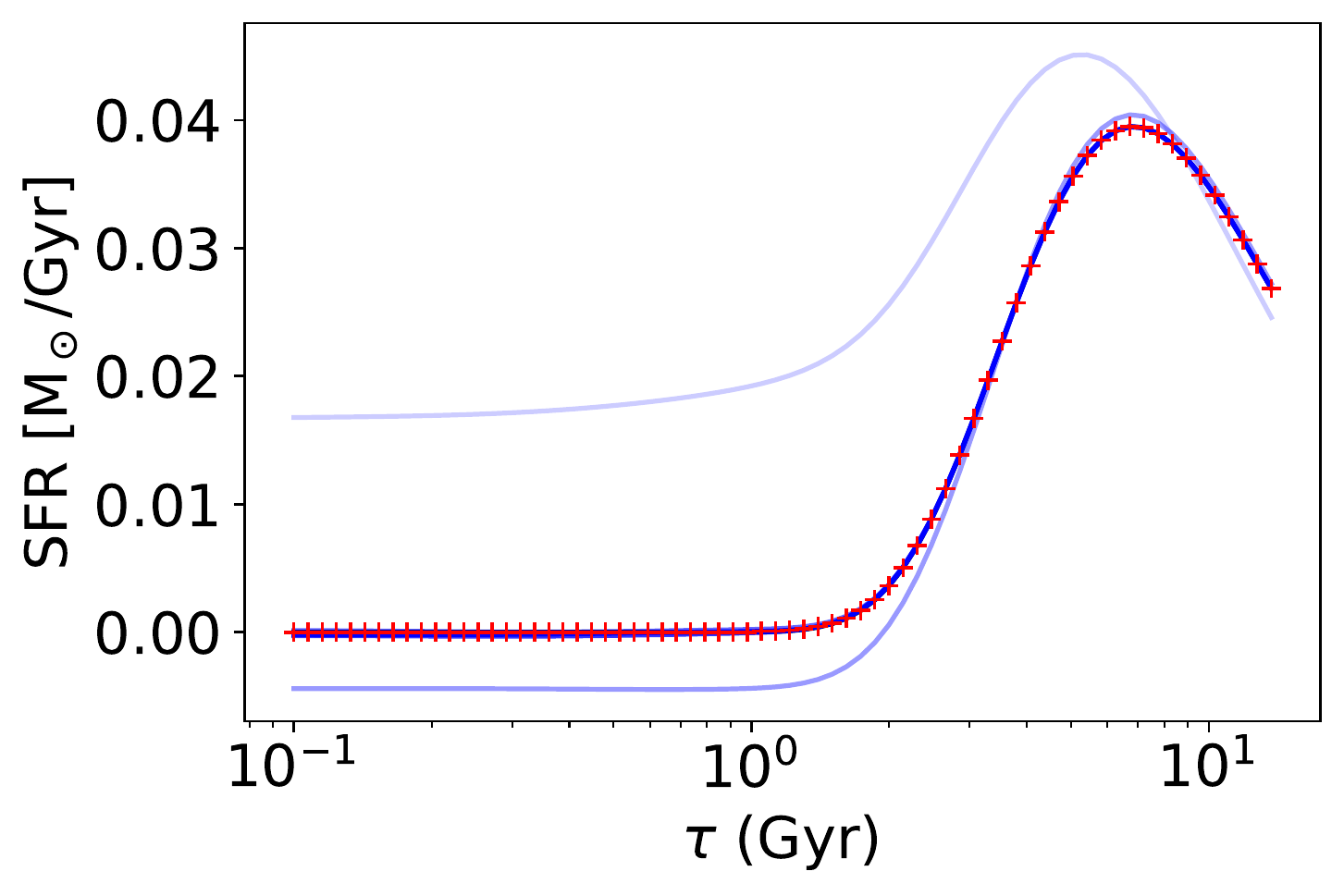}\\

        \includegraphics[width=0.49\linewidth]{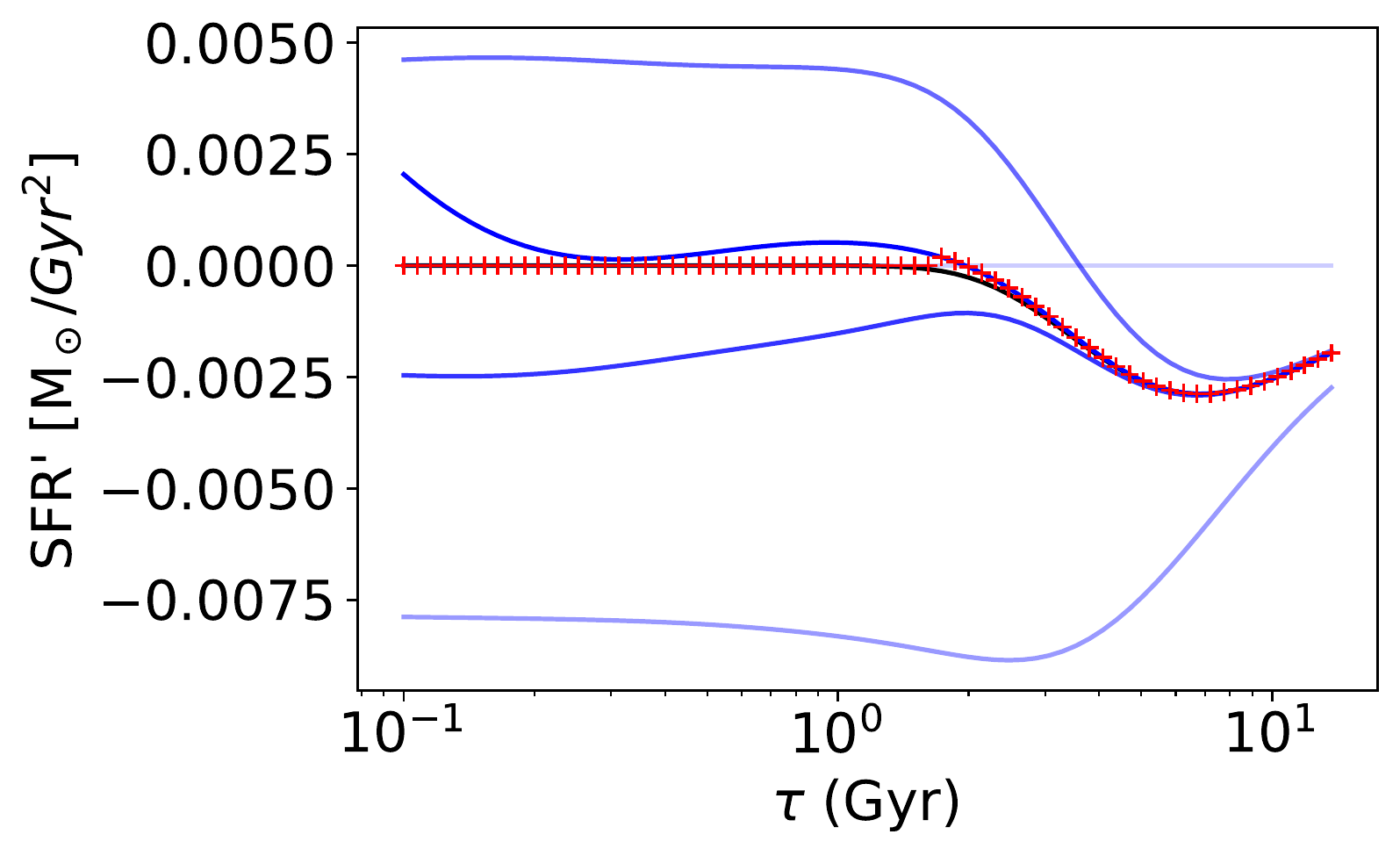}
        \includegraphics[width=0.49\linewidth]{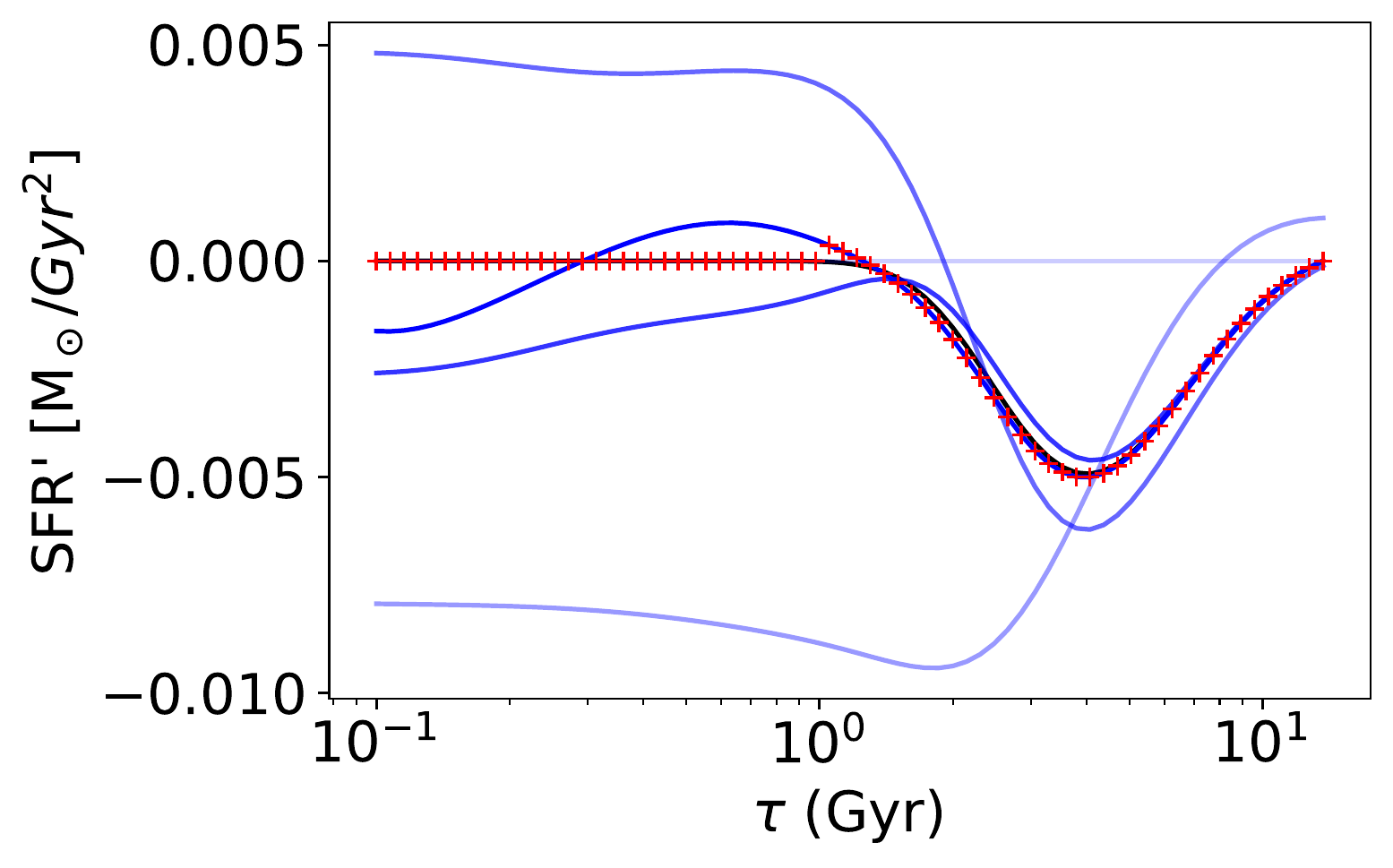}\\

        \includegraphics[width=0.49\linewidth]{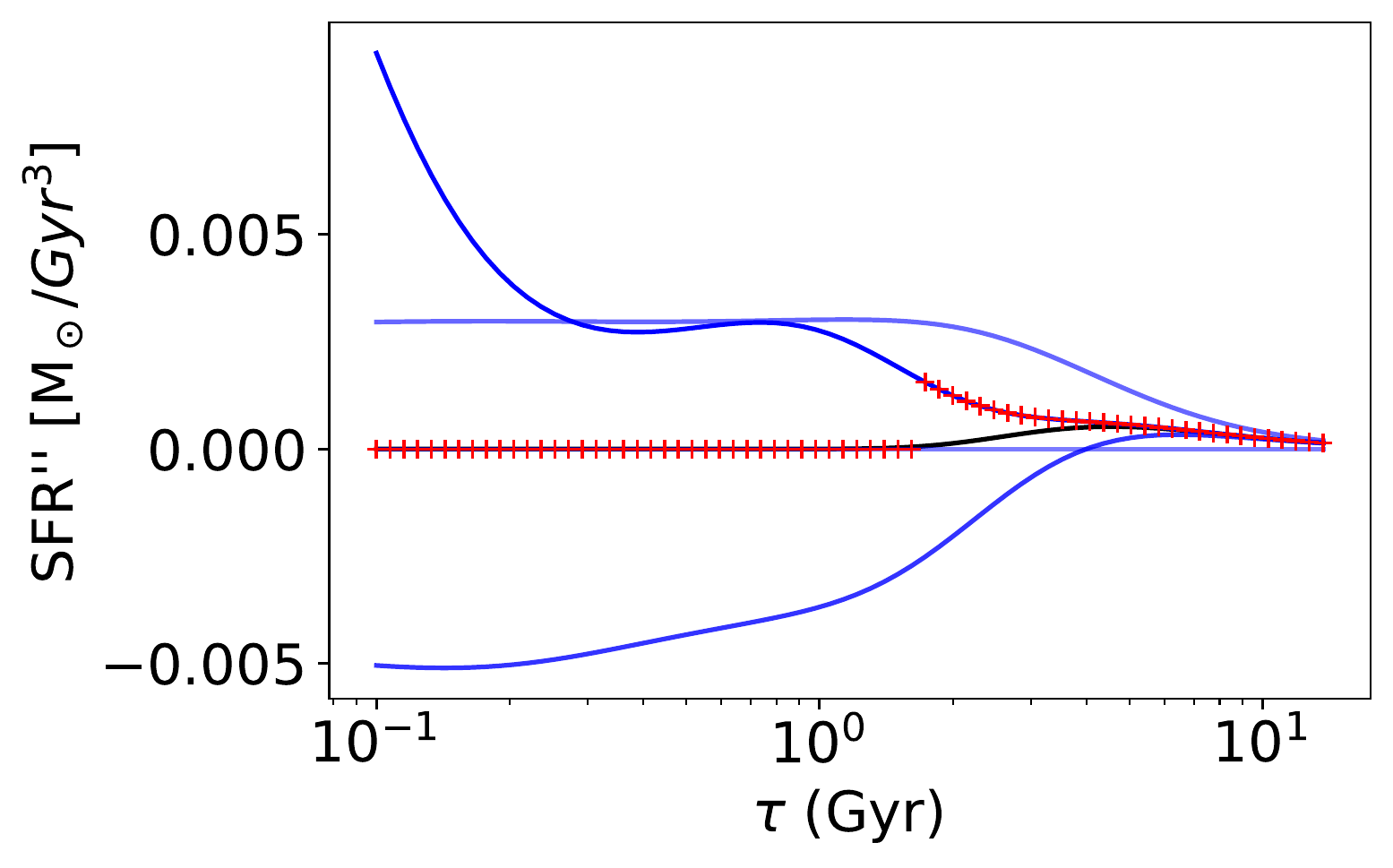}
        \includegraphics[width=0.49\linewidth]{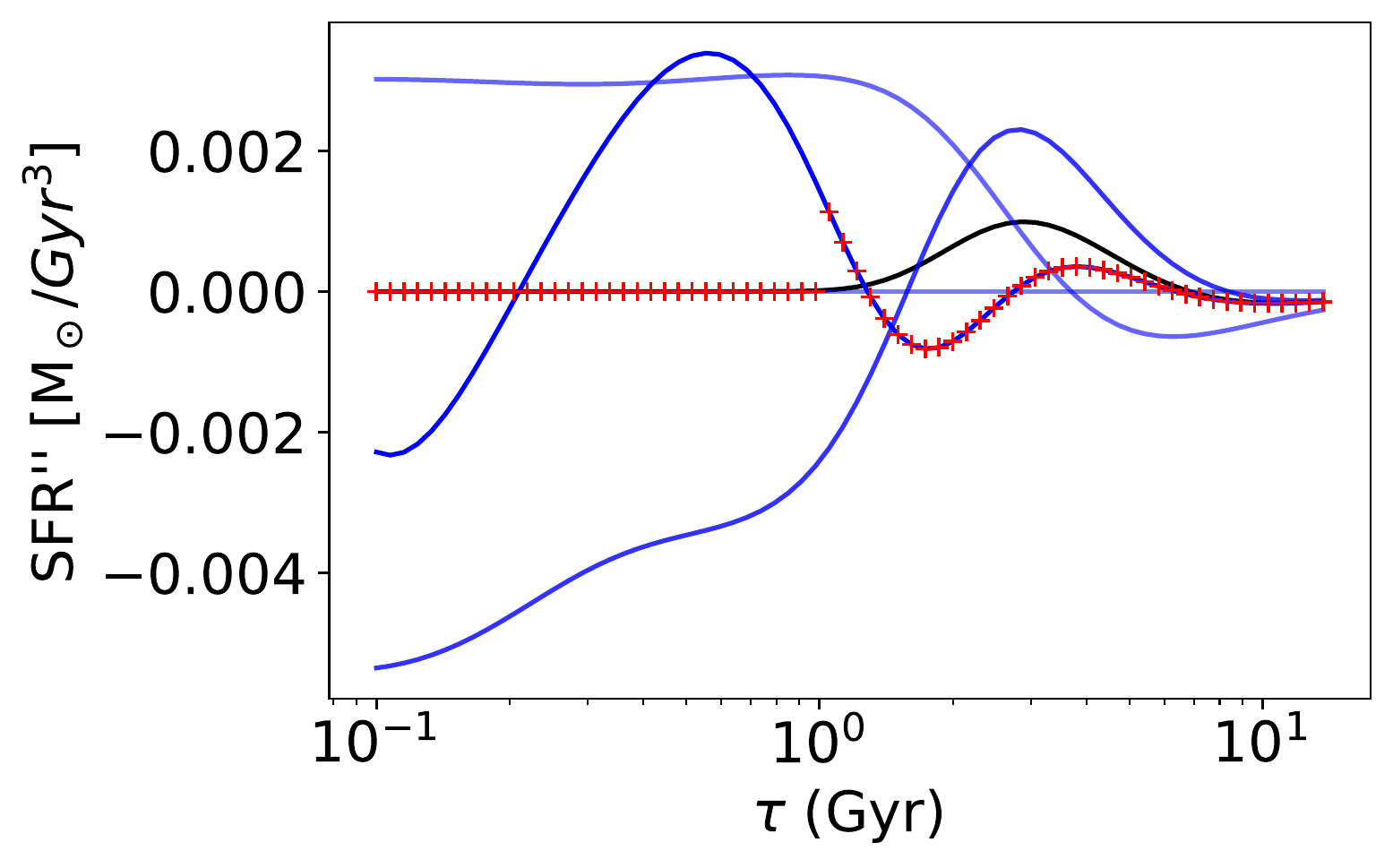}\\
        
        \caption{Estimates, from top to bottom, of the total stellar mass formed, the SFR, and its first and second time derivatives, reconstructed at the present time ($\hat t$=0) for the exponential (left) and delayed-$\tau$ (right) synthetic SFHs with different timescales, $\tau$, compared with the correct values (solid black lines). Red crosses correspond to the best positive-SFR fit.
}
        \label{fig:coefficients_tau}
\end{figure}

\begin{figure*}
\centering
  \begin{tabular}{@{}cccc@{}}
  Mass & SFR & SFR' & SFR'' \\
  \turnbox{90}{\hspace{13 mm} N=1}
    \includegraphics[width=.23\textwidth]{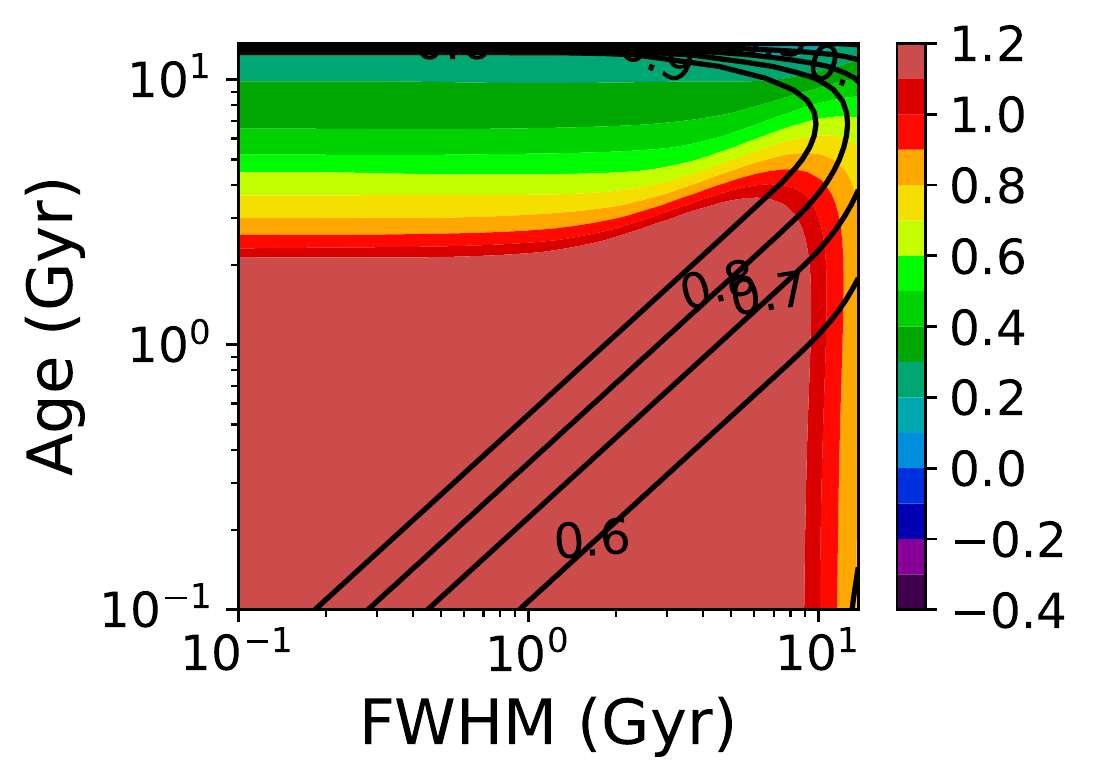} &
    \includegraphics[width=.23\textwidth]{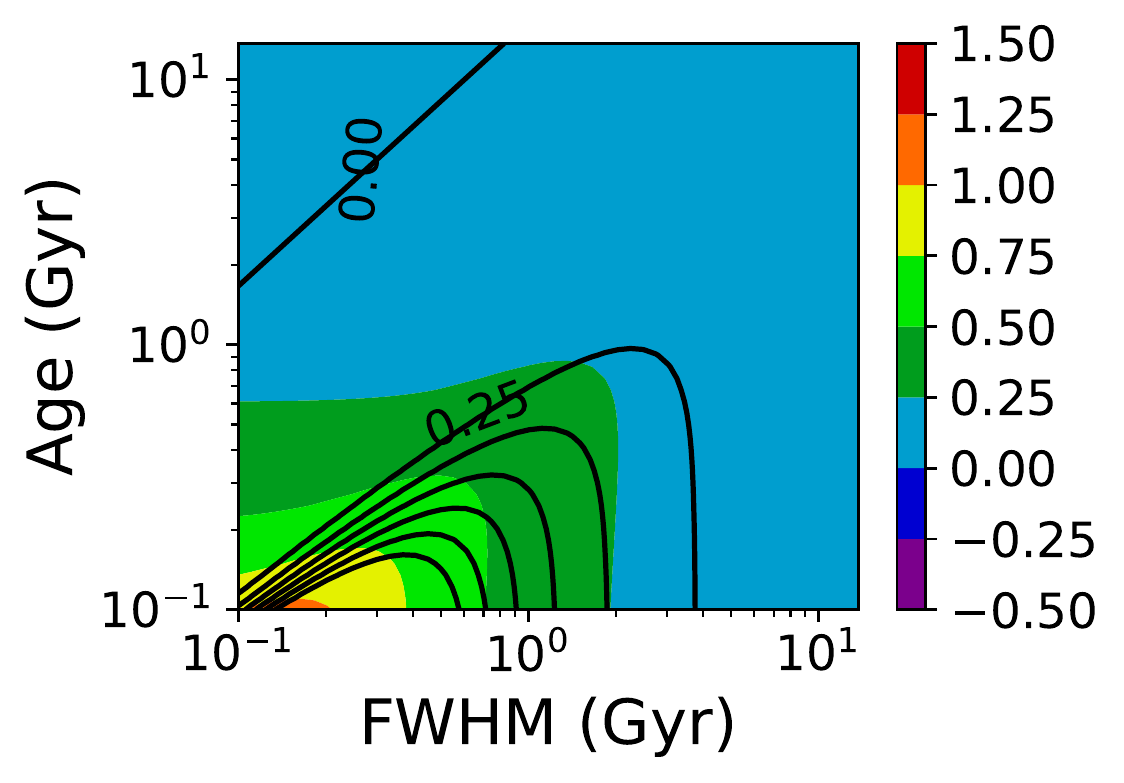} &
    &
    \\
\turnbox{90}{\hspace{13 mm} N=2}
    \includegraphics[width=.23\textwidth]{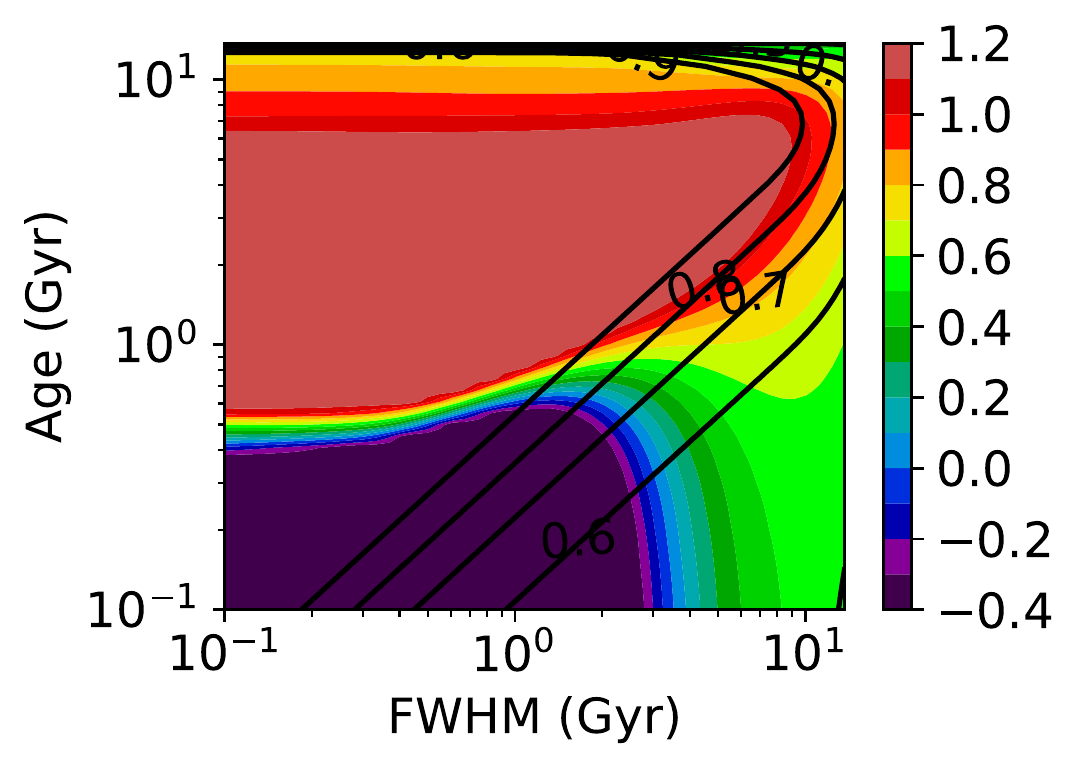} &
    \includegraphics[width=.23\textwidth]{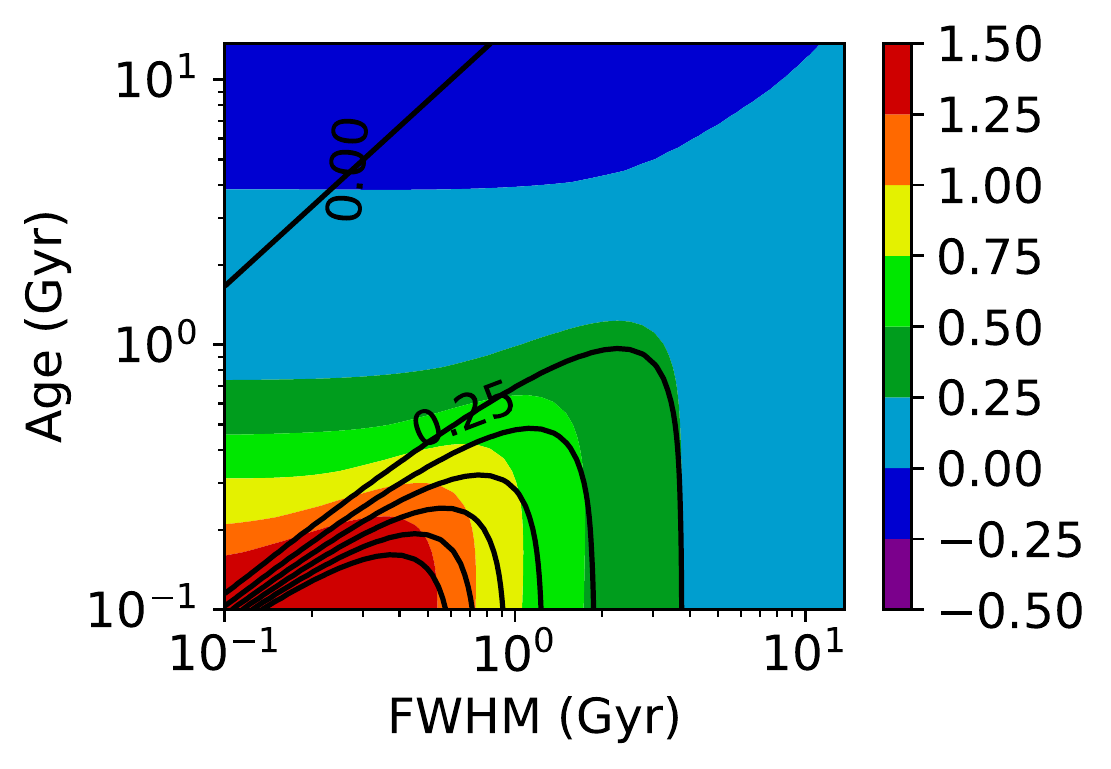} &
    \includegraphics[width=.23\textwidth]{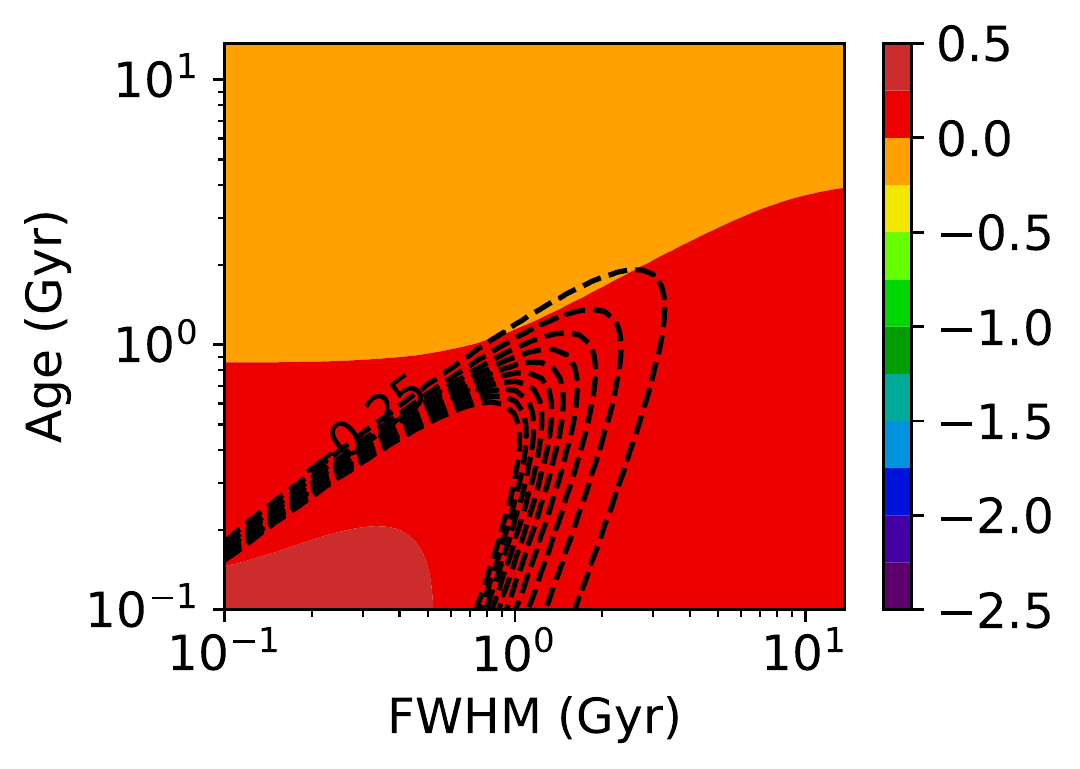} &
    \\
\turnbox{90}{\hspace{13 mm} N=3}
    \includegraphics[width=.23\textwidth]{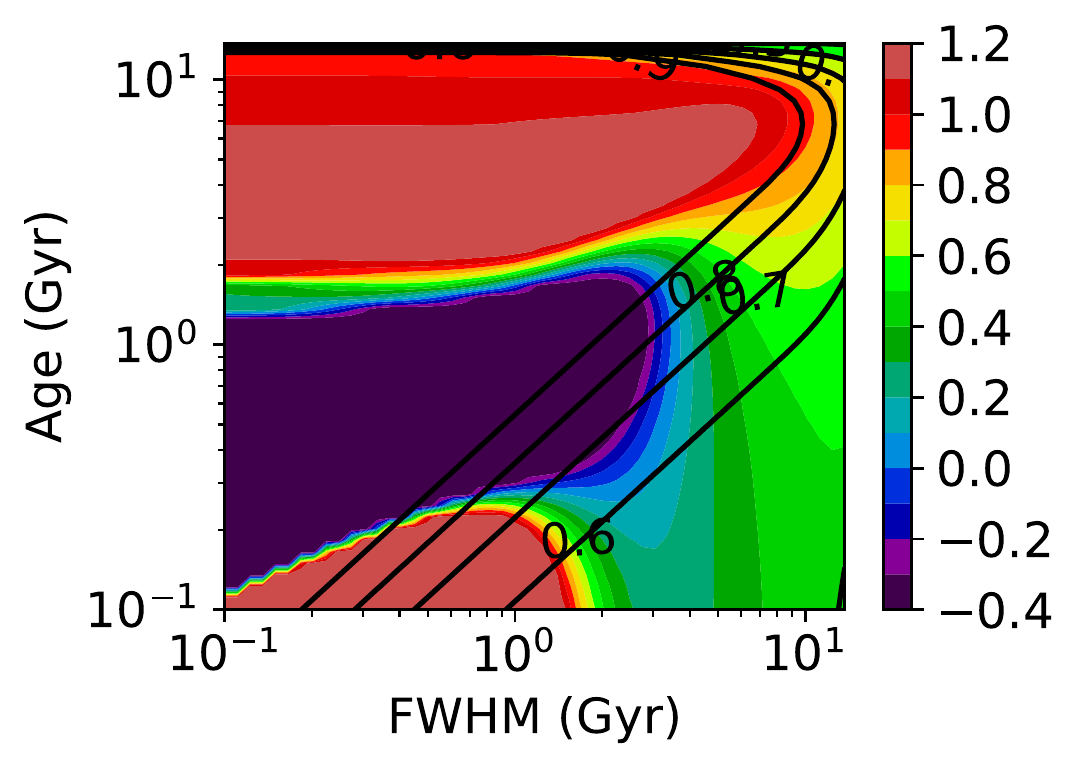} &
    \includegraphics[width=.23\textwidth]{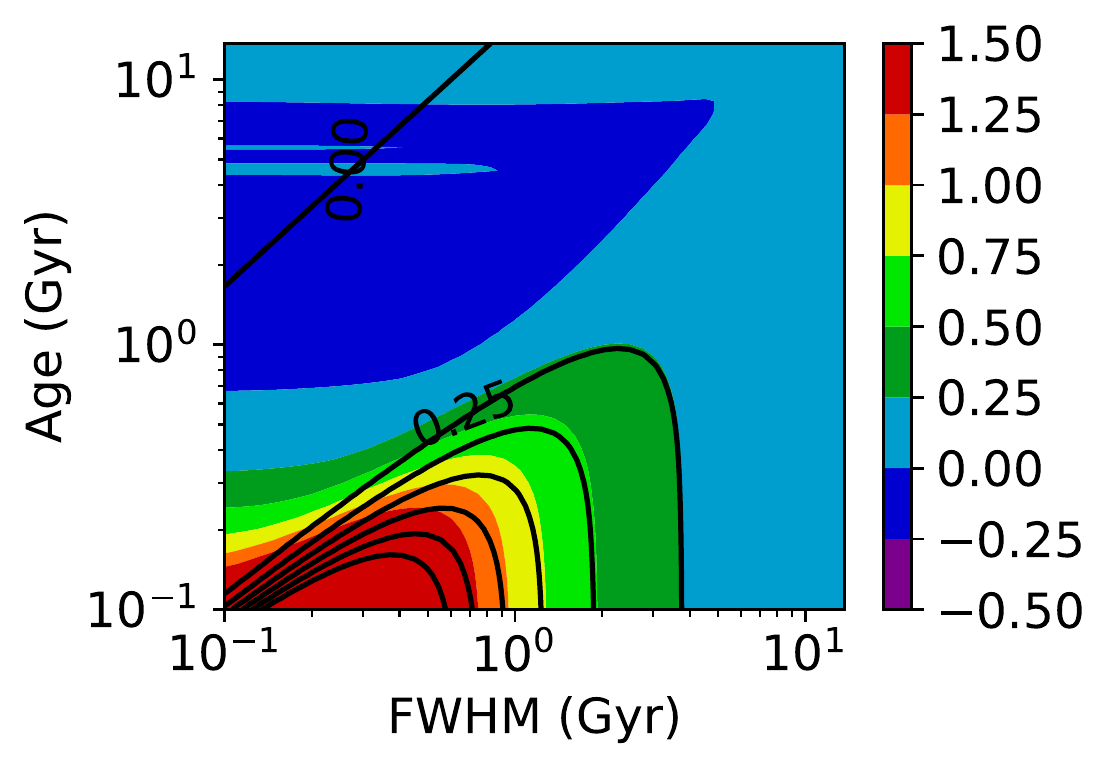} &
    \includegraphics[width=.23\textwidth]{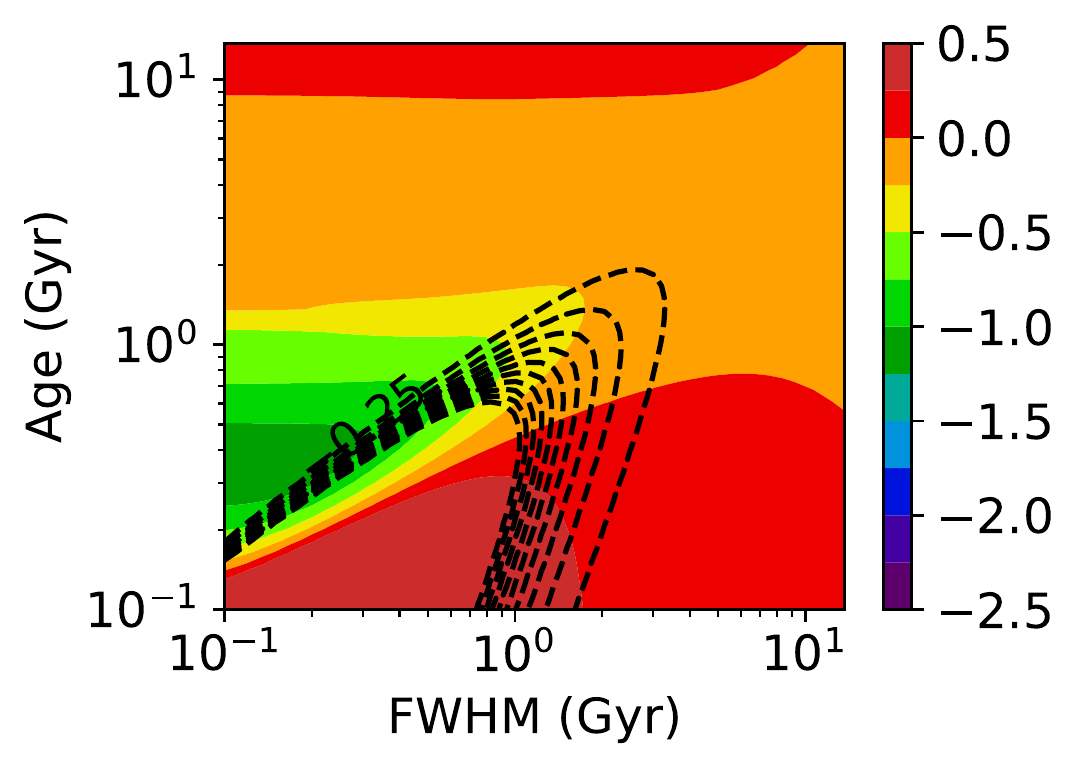} &
    \includegraphics[width=.23\textwidth]{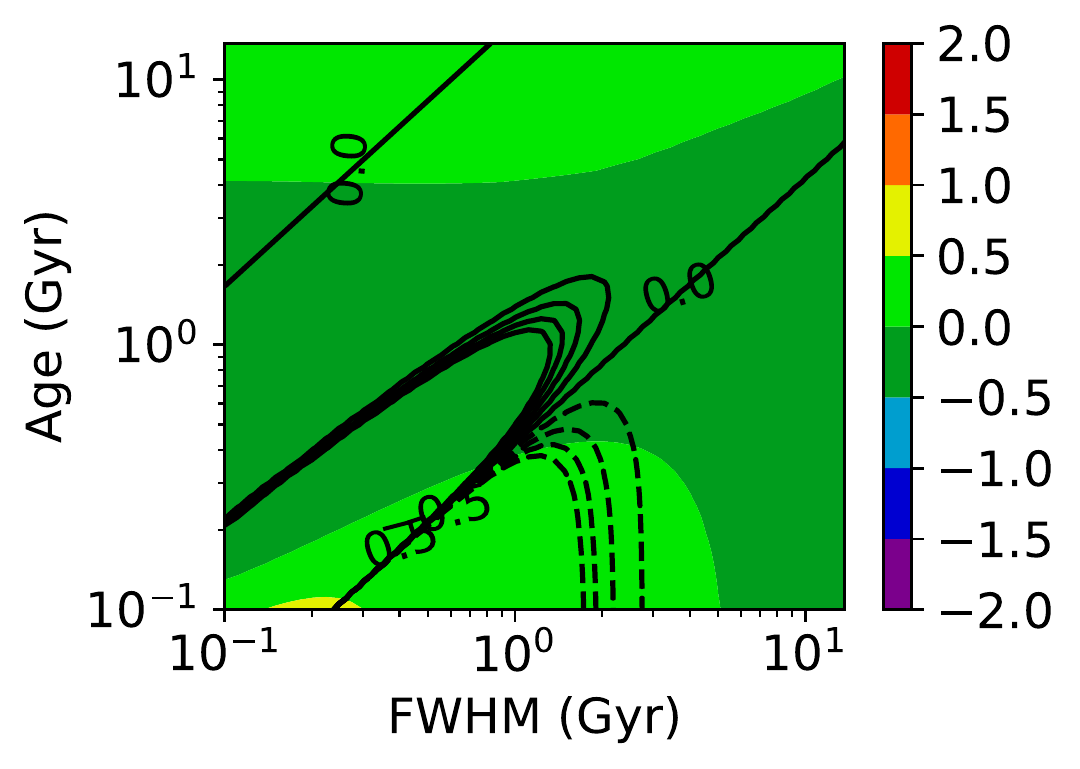} \\
\turnbox{90}{\hspace{13 mm} N=4}
    \includegraphics[width=.23\textwidth]{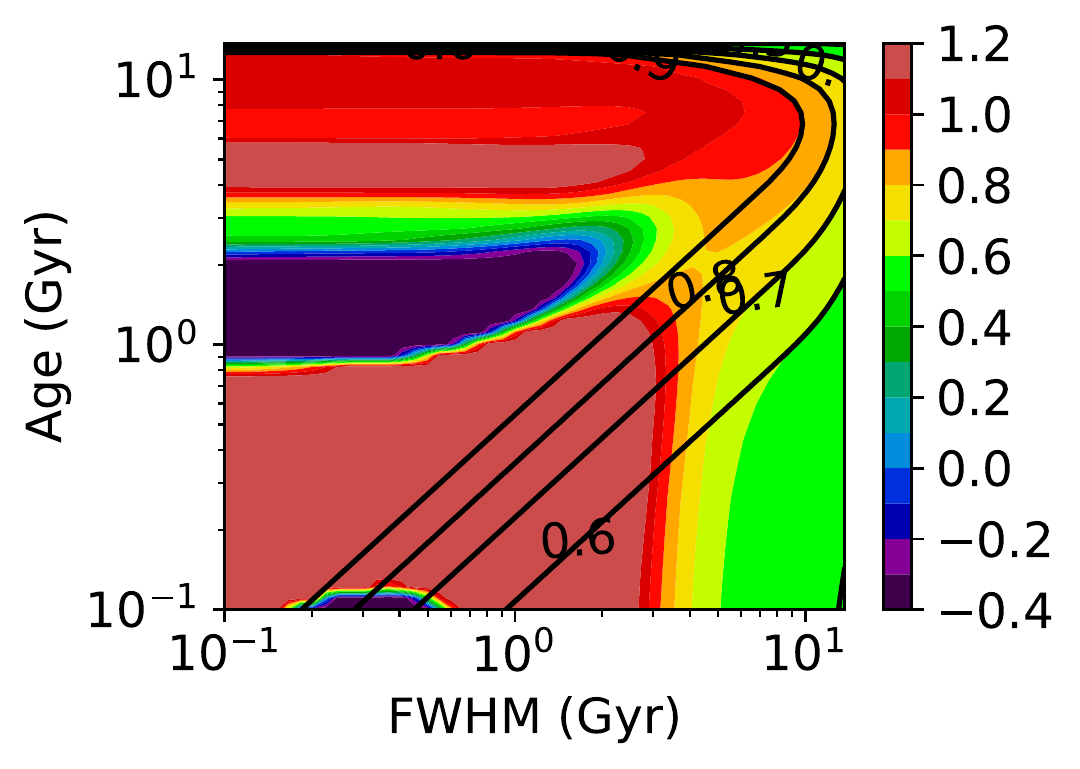} &
    \includegraphics[width=.23\textwidth]{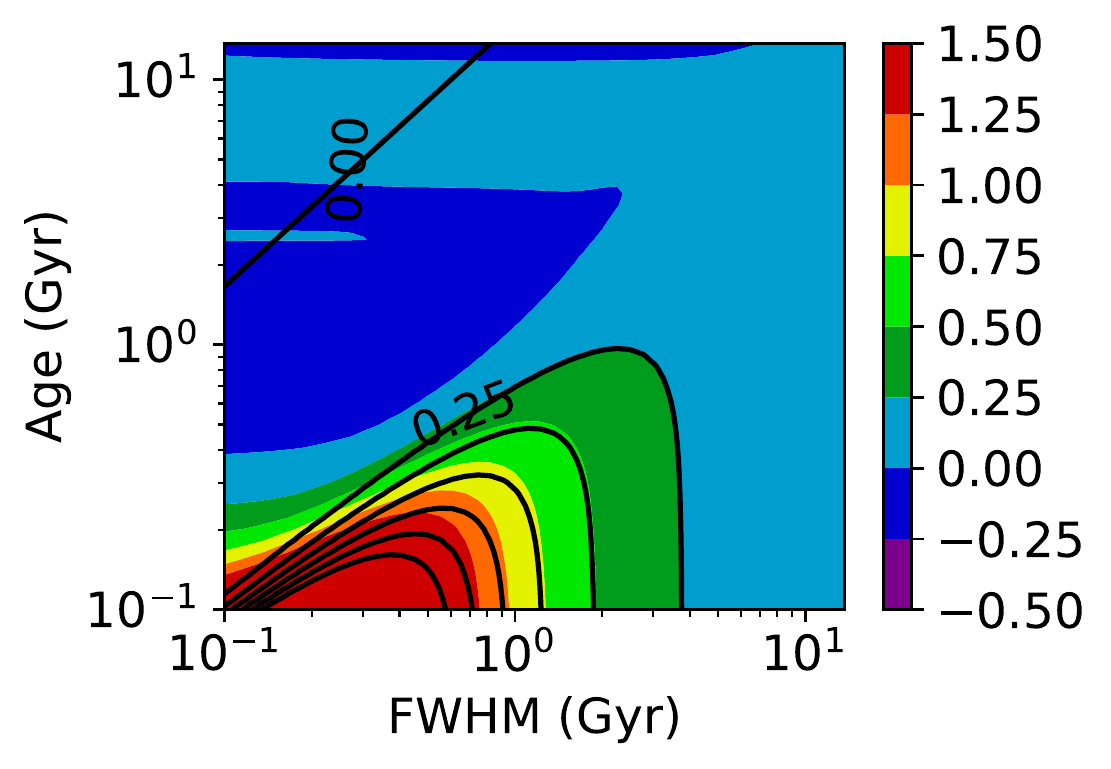} &
    \includegraphics[width=.23\textwidth]{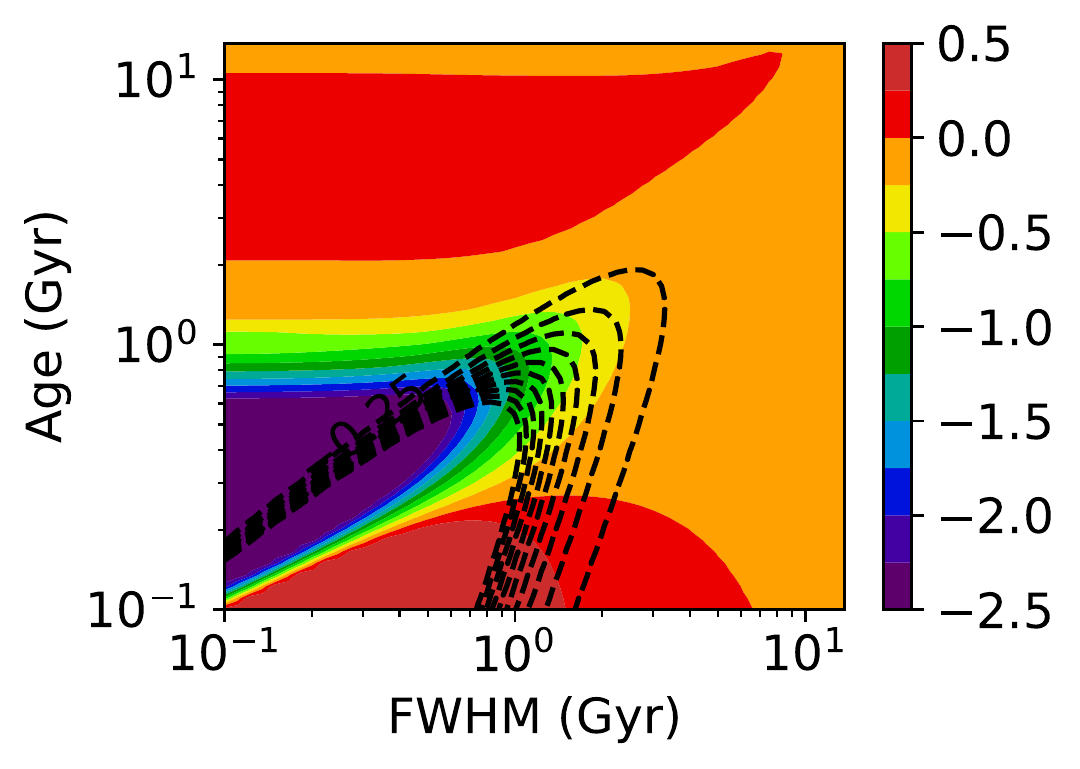} &
    \includegraphics[width=.23\textwidth]{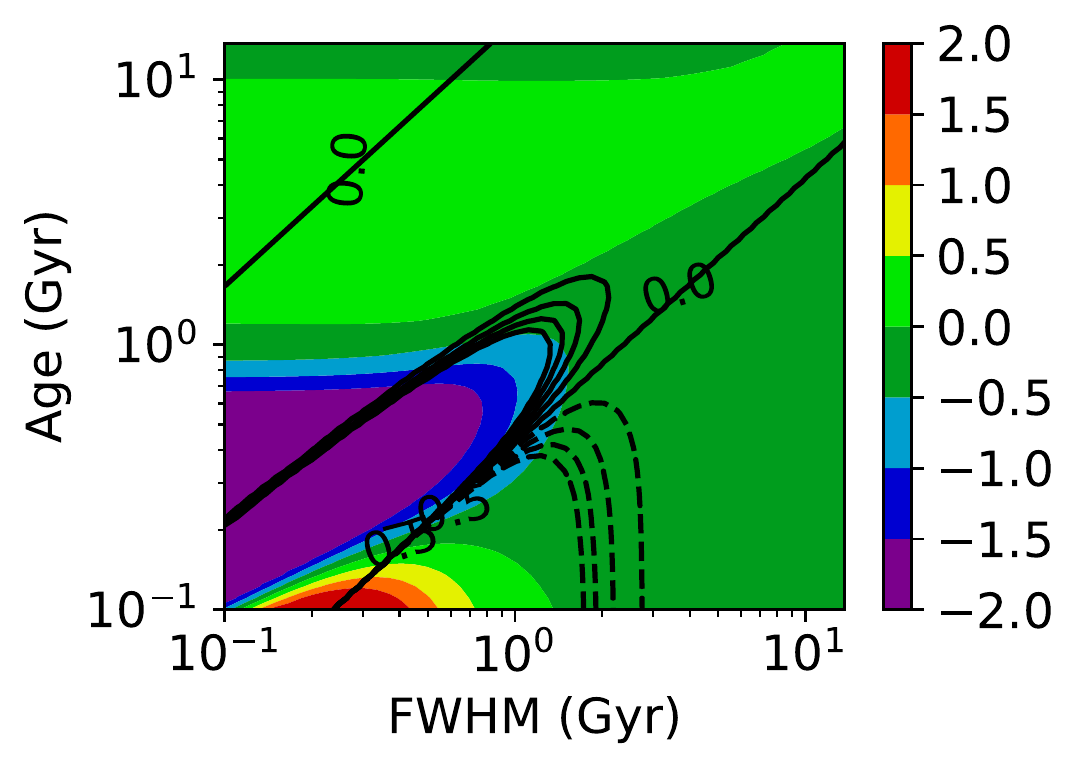} \\
\turnbox{90}{\hspace{13 mm} N=5}
    \includegraphics[width=.23\textwidth]{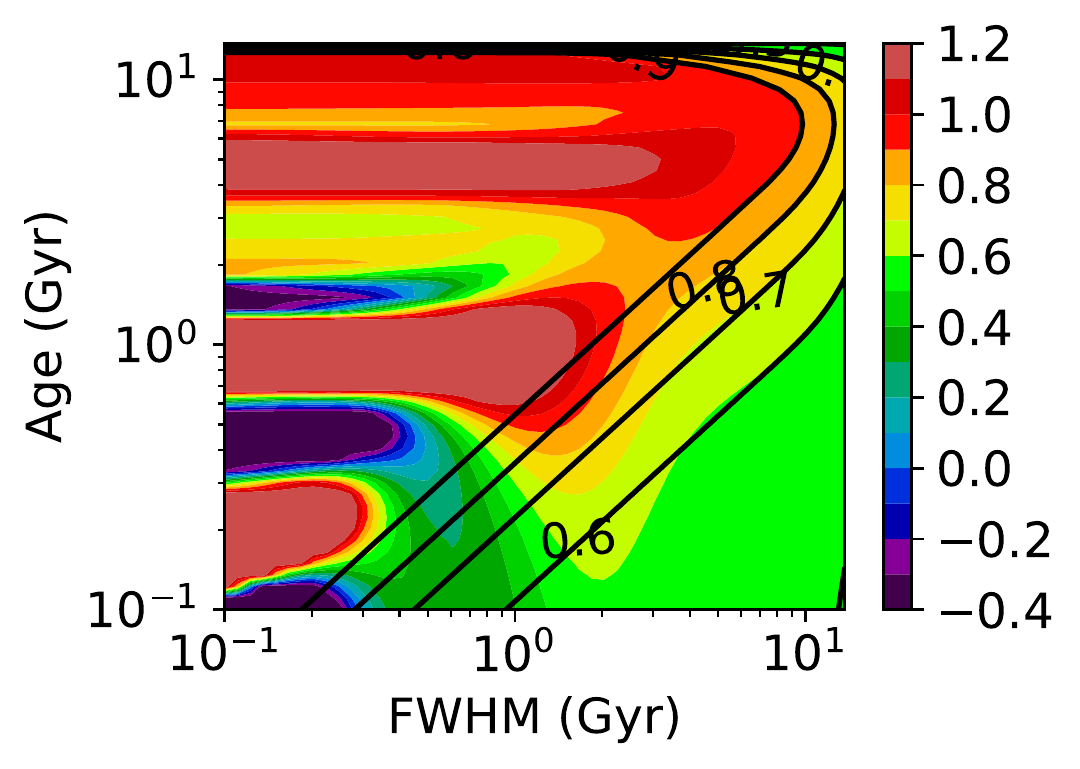} &
    \includegraphics[width=.23\textwidth]{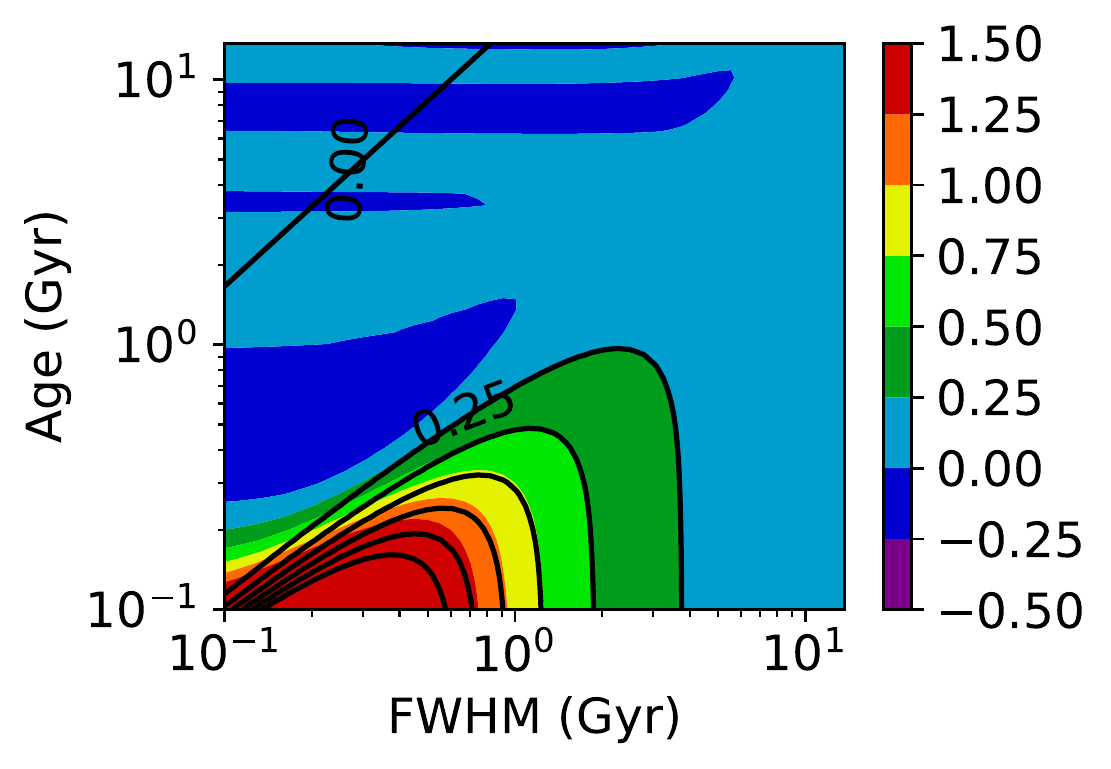} &
    \includegraphics[width=.23\textwidth]{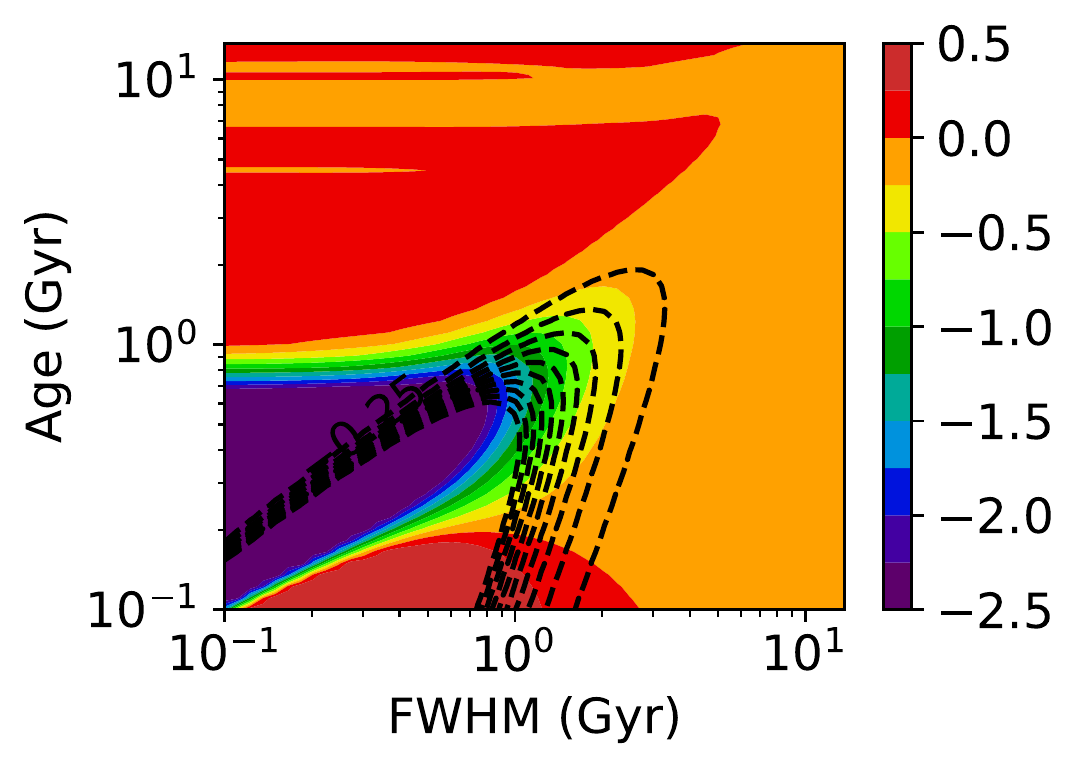} &
    \includegraphics[width=.23\textwidth]{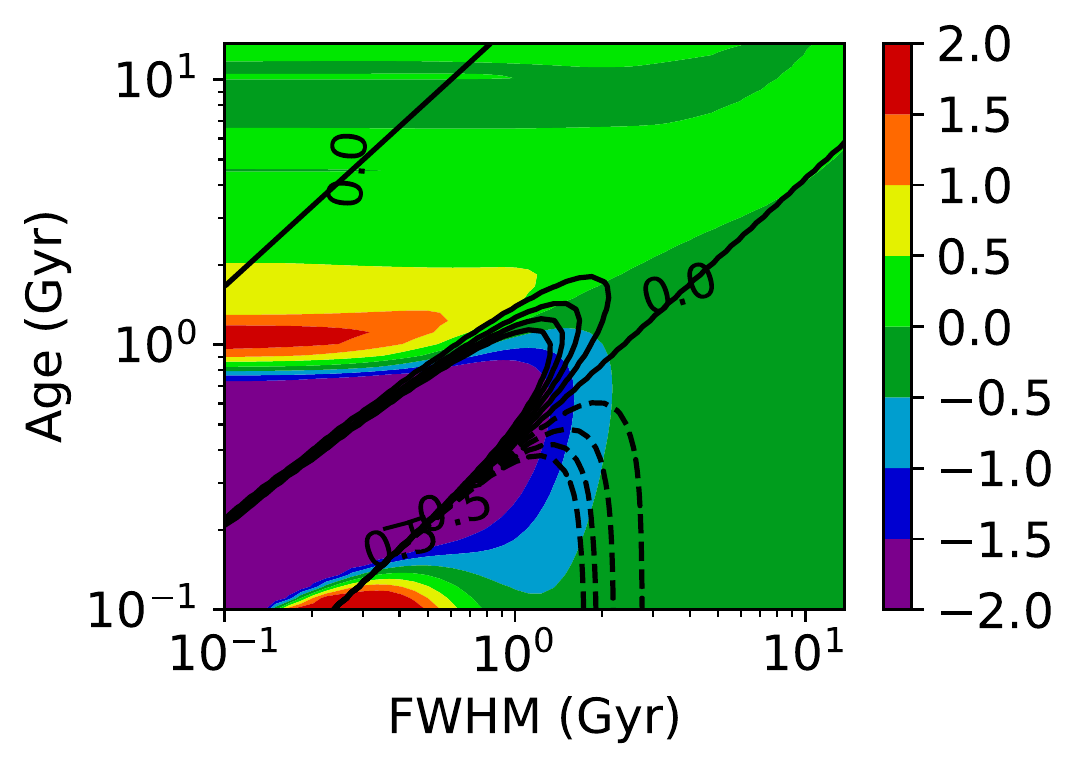} \\
\turnbox{90}{\hspace{12 mm} SFR +}
    \includegraphics[width=.23\textwidth]{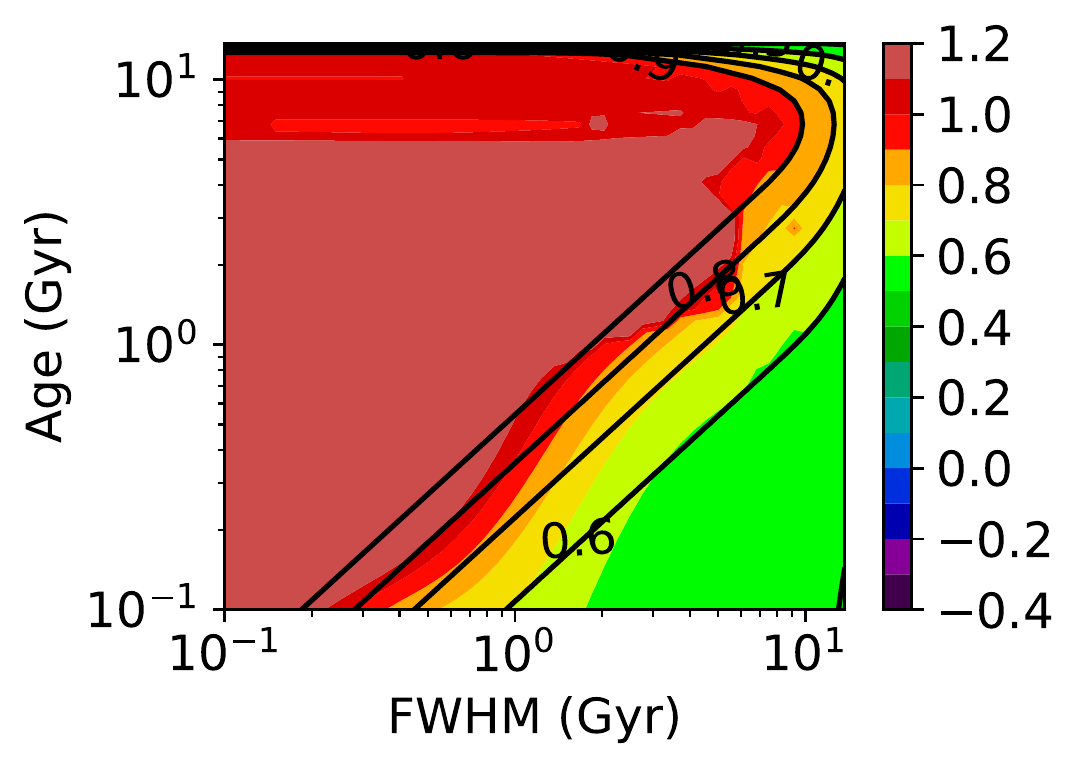} &
    \includegraphics[width=.23\textwidth]{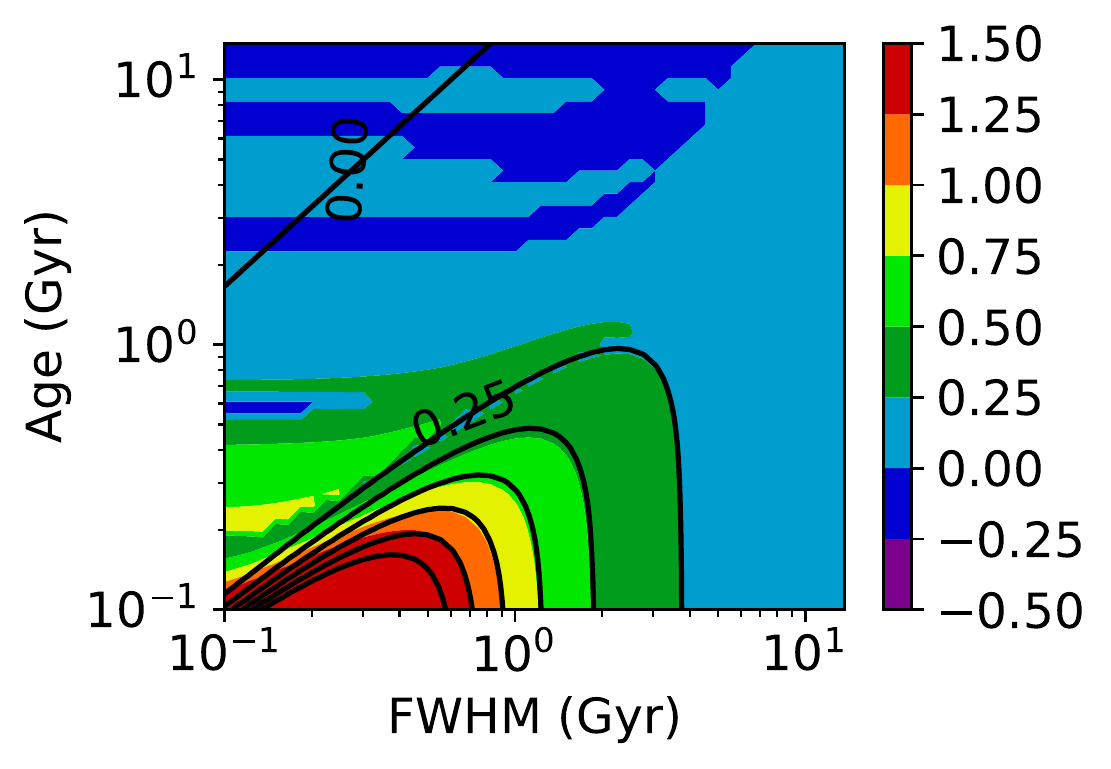} &
    \includegraphics[width=.23\textwidth]{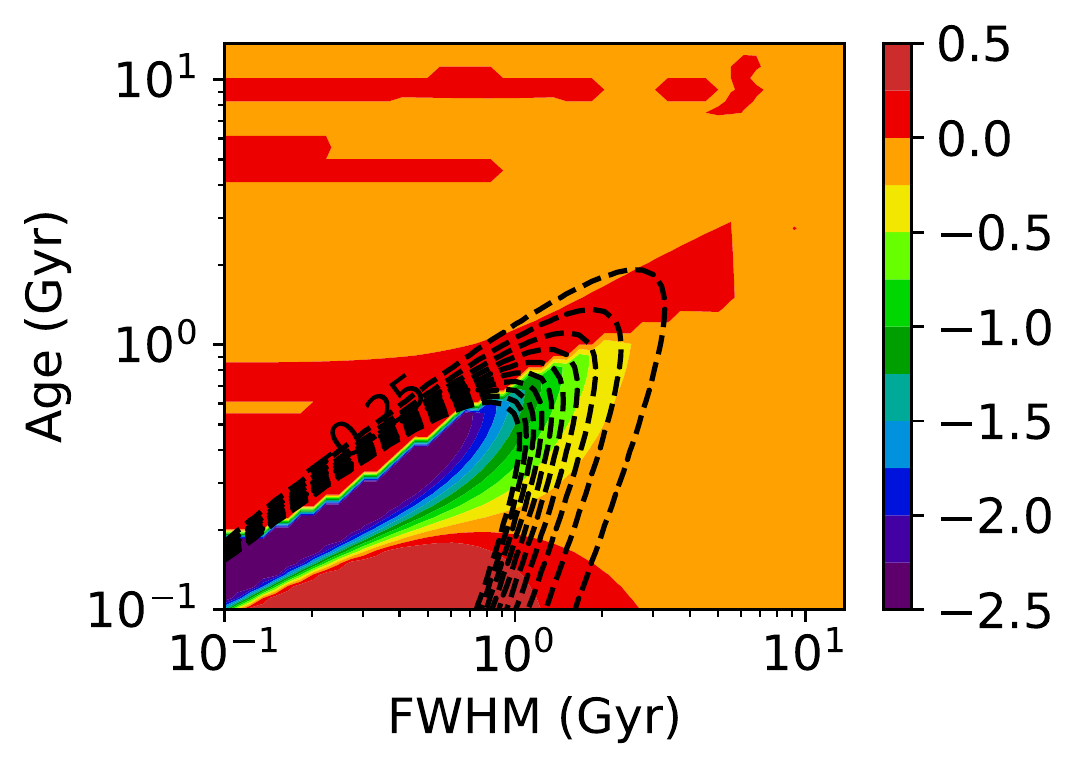} &
    \includegraphics[width=.23\textwidth]{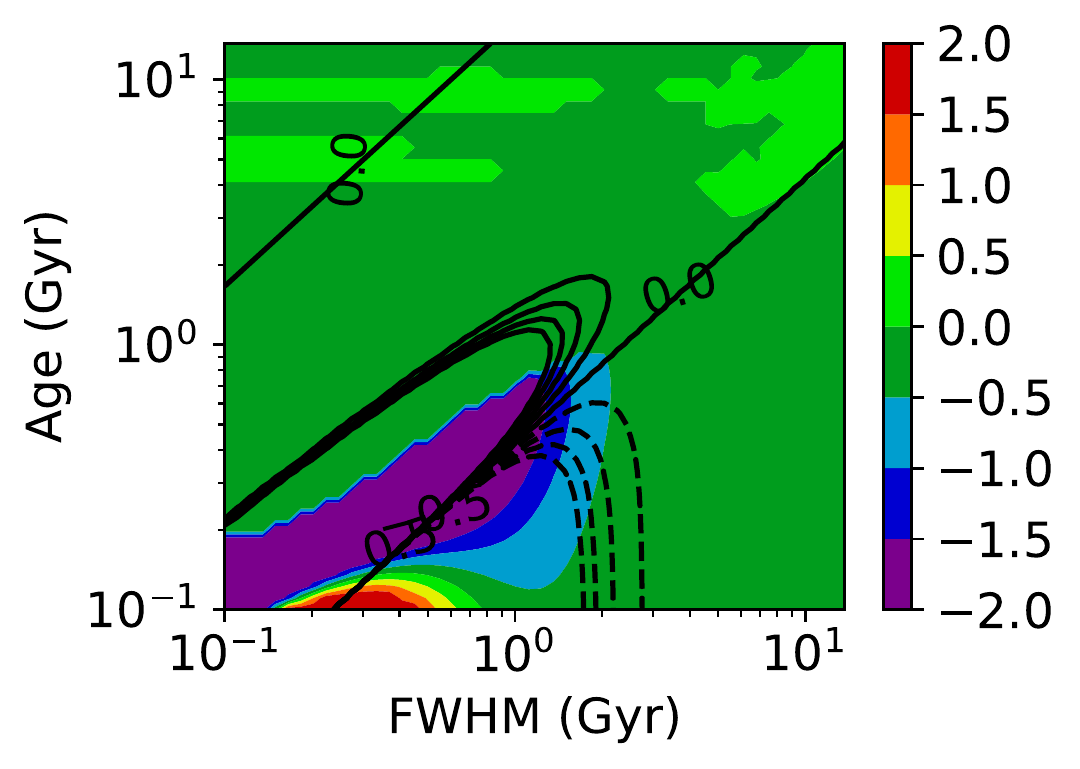} \\
    
  \end{tabular}
  \caption{
  Estimates, from left to right, of the total stellar mass formed, the SFR, and its first and second time derivatives, reconstructed at the present time ($\hat t$=0) for the Gaussian synthetic SFHs with different ages and FWHMs, compared with the correct values (black contours). $N$ increases from top to bottom, where the last row (SFR+) corresponds to the best positive-SFR polynomial fit.}
  \label{fig:coefficients_gauss}
\end{figure*}

We note that, for some particular combinations of the input parameters, the fits obtained with $N$ and $N+1$ polynomials are basically identical for a certain $N$, which yields the sharp troughs observed for the delayed-$\tau$ model in Fig.~\ref{fig:distances_1} and the darker stripes on the Gaussian colour maps in Fig.~\ref{fig:distances_gauss}.
From the definition in Eq.~\eqref{eq:mass_reconstruction}, one can readily see that the MFH reconstructed by the $N$-th and $(N+1)$-th degree polynomials are identical when the sum of the coefficients associated with $\hat t^{N+1}$ vanishes (see Figs.~\ref{fig:coefficients_tau} and~\ref{fig:coefficients_gauss}), something that never occurs for the exponential law.

In order to illustrate the distances in luminosity space assigned to the results from the Illustris simulations, we plot the cumulative fraction of galaxies under a certain distance, $\hat D$, in Fig.~\ref{fig:distances_Illustris}.
We observe that only 3.5\% of the sample has purely positive SFRs for $N=5$.
The vast majority of galaxies have the best positive-SFR fit with an assigned distance close to the values expected for $N=2-3$.

In general, we always find that $\hat D \la 10^{-2}$ for $N \ge 2$.
In other words, a linear SFH (a quadratic MFH) is able to reproduce the observed luminosities with an accuracy of the order of one per cent or even better.
While it is obvious that this would indeed be, in most cases, an overly simplistic reconstruction of the underlying analytical model, the accuracy of the measured luminosity should be higher if we want to discriminate the difference with respect to a higher-order reconstruction.

\subsection{Quality of the reconstruction}
\label{sec:reconstruction}

As stated above, our polynomial MFHs do not cover all possible histories, but just a subset that would be most appropriate for smooth functions.
Then, one can imagine that SFHs that vary on short timescales may be harder to reproduce than galaxies that build up their mass more uniformly over cosmic time.

The reconstruction of the stellar mass, the SFR, and its first and second derivatives at the present time, $\hat t = 0$ (i.e. the physical quantities upon which the polynomial expansion is based), are shown in Fig.~\ref{fig:coefficients_tau} for different values of the characteristic timescale, $\tau$, for the exponential and delayed-tau models.
The total mass is well reproduced (always better than 20 per cent) for $N > 2$, regardless of the value of $\tau$.
For timescales above a few gigayears, the recovery is also good for $N=2$; only the first-order polynomial (i.e. a constant SFR) differs from the true mass by more than a few per cent.
We note, however, that, according to Fig.~\ref{fig:distances_1}, one may rule out $N=1$ in favour of $N=2$ as long as the relative uncertainty of the observational measurements remains below $\sim 10$ per cent.
The current instantaneous SFR may also be accurately estimated for large $\tau$ even with the simplest polynomials.
For $\tau < 1$~Gyr, $N=1$ and $N=2$ fail to yield an appropriate model, but $N \ge 3$ is still able to qualitatively reproduce the mass and SFR of the input model at the present time, although the low SFRs are impossible to recover accurately as they approach zero.
For short timescales, the results of our polynomial expansion should not be taken at face value.
On the other hand, when $\tau$ becomes comparable to the age of the Universe, it is notable that $N=4$ and $N=5$ can even provide a reasonable estimate of the time derivative of the SFR, although their description of the second derivative is, at best, only a very crude approximation. The best positive-SFR fit sets the SFR (and therefore its derivatives) to zero for $\hat t < \hat t_{\rm min}$, and thus it actually helps recover a better estimation of the derivatives at present time for low values of the characteristic timescale, $\tau$, than any polynomial degree.

These trends are also observed for the Gaussian models in Fig.~\ref{fig:coefficients_gauss}, where the results are shown for a range of values for the age and FWHM of the stellar burst. The polynomial description is able to capture the mass and the SFR, and it even crudely estimates its derivatives at the present time for large FWHM values.
Unfortunately, when this parameter drops below a few gigayears, the MFH becomes increasingly similar to a short star formation burst, not amenable to a description in terms of smooth polynomials, and the estimates provided by our unconstrained polynomials cannot be trusted, even if the observed luminosities are accurately reproduced.
The best positive-SFR fit automatically selects low-order polynomials ($N \sim 3$) for short FWHMs, and it always provides meaningful results, although it struggles to reproduce the SFHs with FWHM$<1$~Gyr.
The mass and SFR returned by our method are fairly accurate, but the first derivative of the SFR is always difficult to recover and must be interpreted, at best, in a qualitative sense.
The second time derivative is never accurately recovered.

\begin{figure}
        \centering
        \includegraphics[width=.73\linewidth]{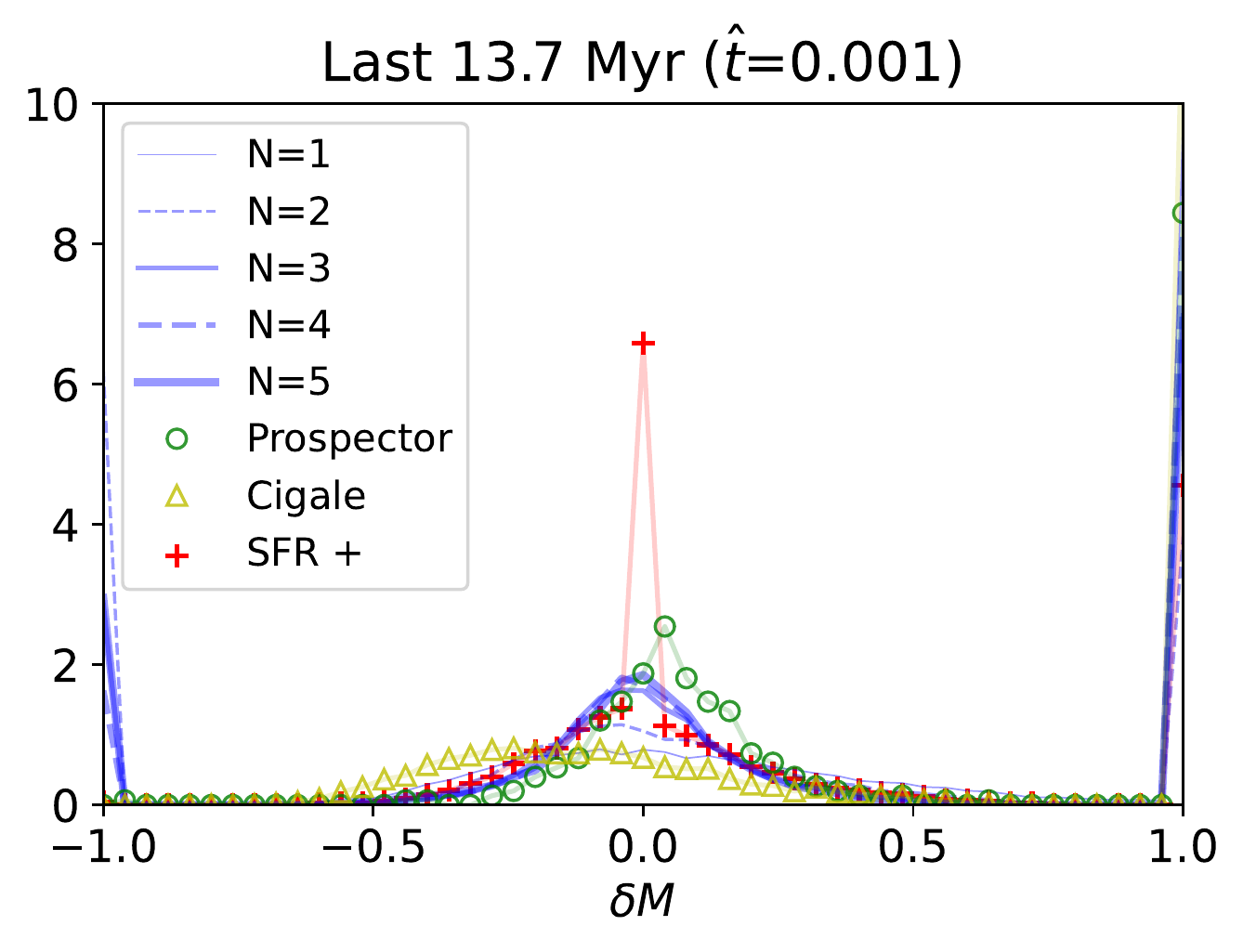}
        \includegraphics[width=.73\linewidth]{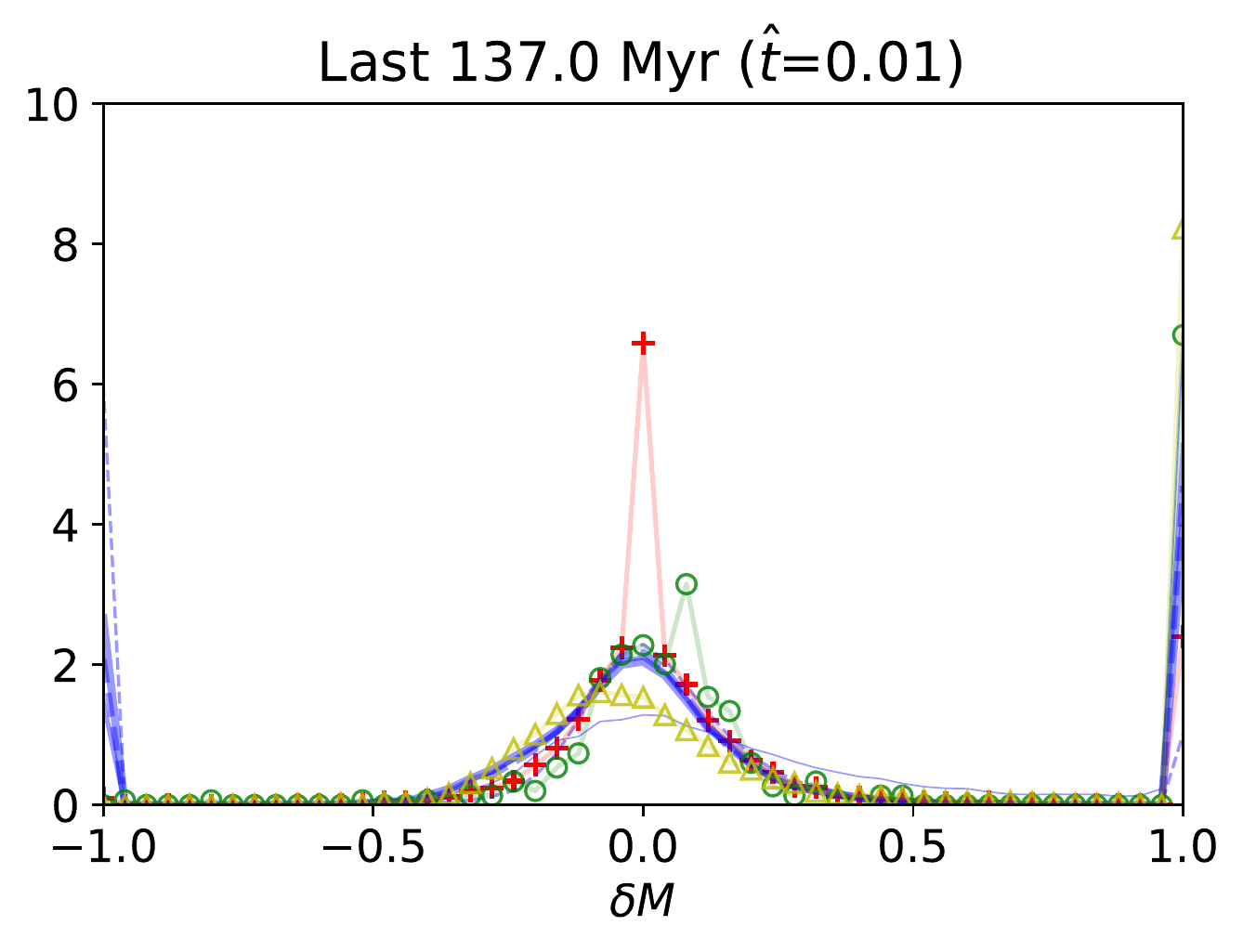}
    \includegraphics[width=.73\linewidth]{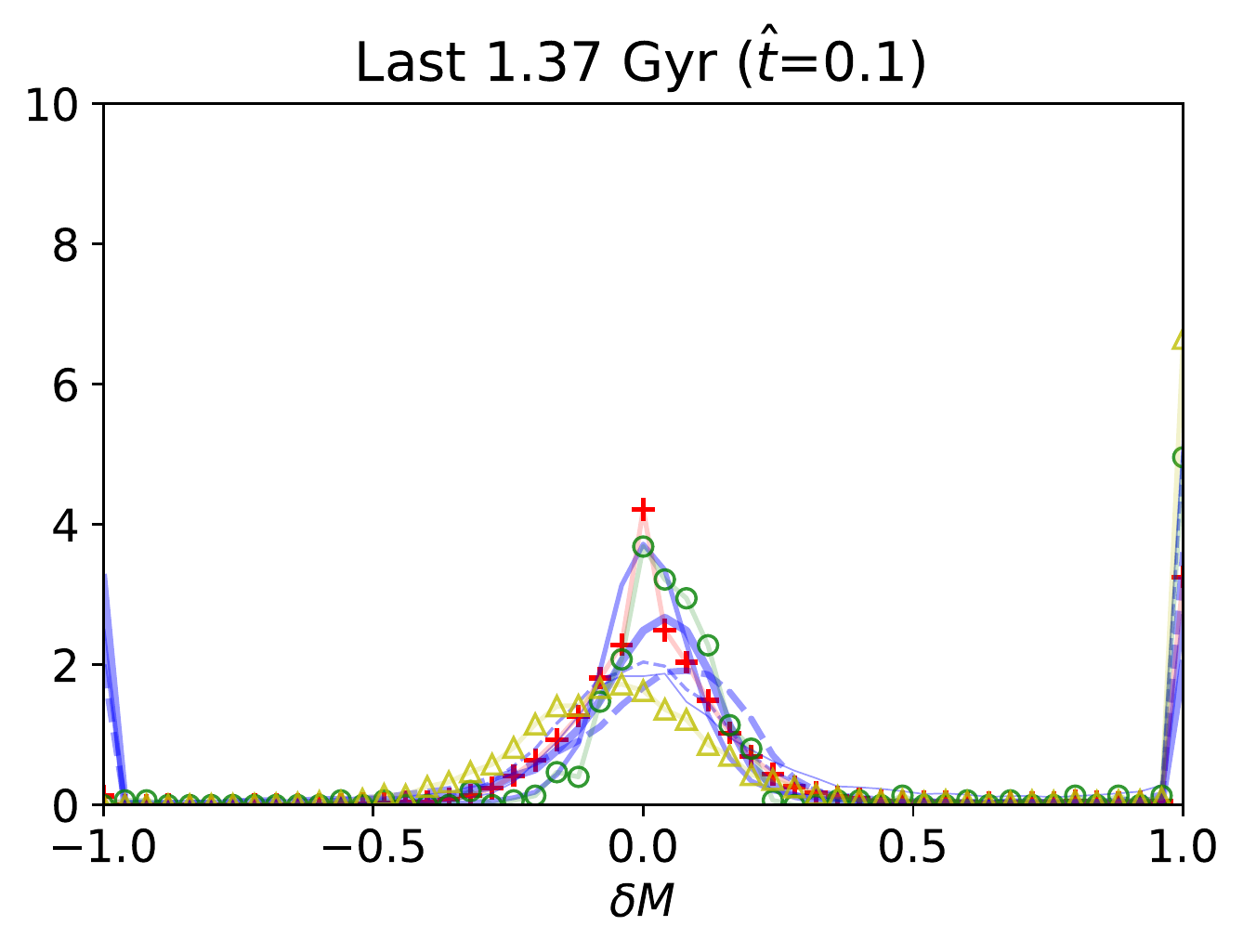}
        \includegraphics[width=.73\linewidth]{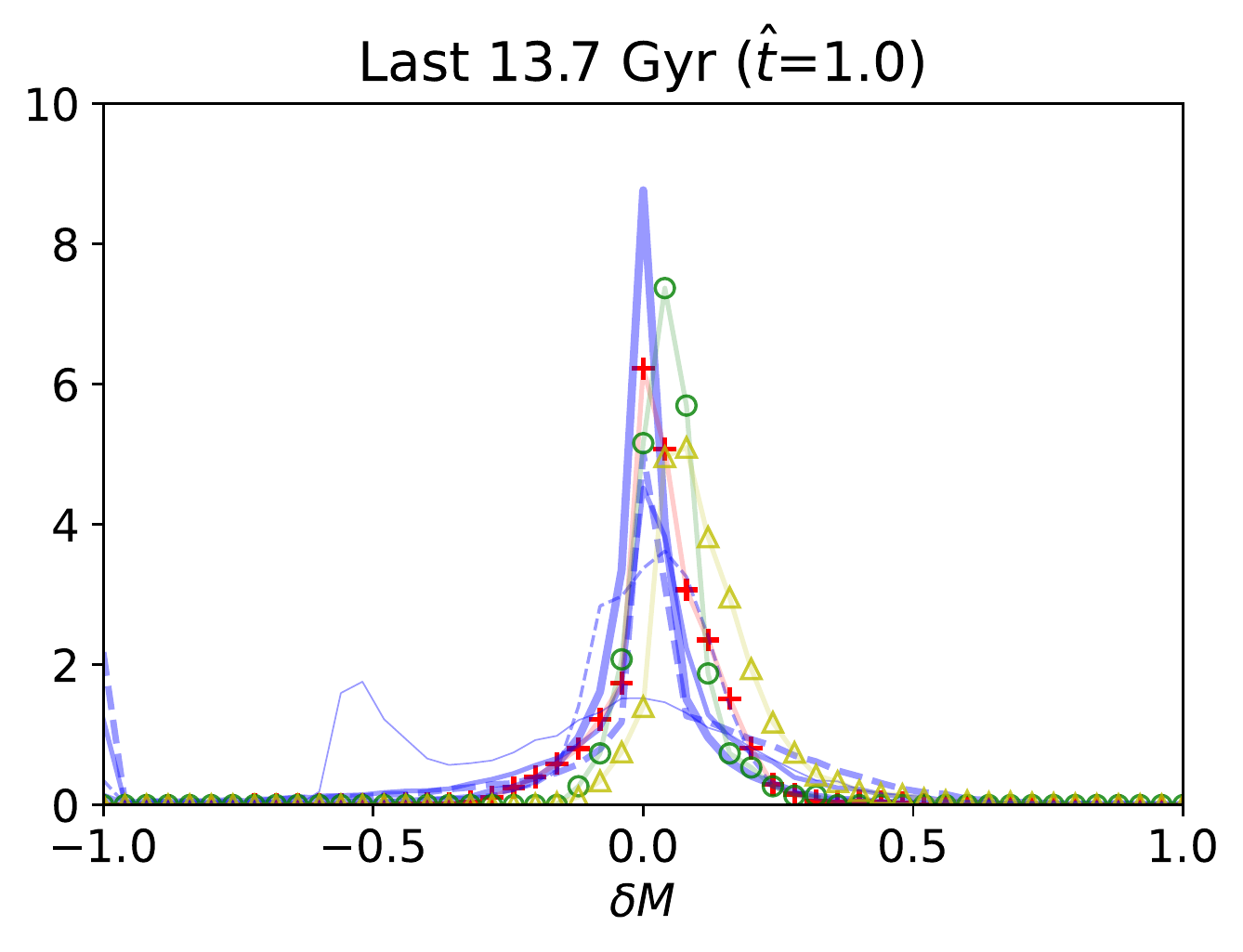}
        
        \caption{Probability densities for different values of $\hat t$ of the $\delta M$ associated with the different polynomial methods (blue lines), the best positive-SFR fit (red crosses), the \prosp\ model (green circles), and the \cig\ fit (yellow triangles).
}
        \label{fig:delta_M}
\end{figure}

In contrast to the analytical models, the SFHs of the simulated galaxies are inherently stochastic due to the finite mass of the stellar particles, and therefore the instantaneous SFR (let alone its derivatives) are not trivially defined and are severely affected by Poisson noise.
Thus, we assessed the quality of our reconstruction in terms of the relative difference
\begin{equation}
\color{black}\delta M(\hat t) = \frac{\Delta M_{\rm poly}-\Delta M_{\rm sim}}{|\Delta M_{\rm poly}|+|\Delta M_{\rm sim}|}
\label{eq:delta_M}
\end{equation}
between the mass $\Delta M(\hat t) = M(0) - M(\hat t)$) formed since a certain $\hat t$ in our model and the simulation. We note that $-1 \le \delta M \le 1$, where $\delta M = 0$ indicates a perfect reconstruction, while $\delta M = -1$ for $\Delta M_{\rm poly} \le 0$ and $\delta M = 1$ for $\Delta M_{\rm sim}=0$ and $\Delta M_{\rm sim} > 0$.
When our best positive-SFR fit correctly identifies that no mass was formed in the simulation during the requested interval, we set $\delta M = 0$.

The probability distribution of $\delta M$ at $\hat t =\{10^{-3},\ 10^{-2},\ 0.1, 1\}$ is represented in Fig.~\ref{fig:delta_M}.
These time intervals trace stellar populations of very different ages, and we think they capture most of the relevant information about the SFH of the galaxy.
We find that the mass formed in each interval is recovered with an accuracy of the order of 0.2~dex, although our unconstrained polynomials yield negative masses for a sizeable fraction of the objects.
By construction, the best positive-SFR fit completely eliminates the peak at $\delta M = -1$, which is associated with these null or negative masses, and it helps detect quenched galaxies, bringing them from $\delta M = 1$ to $\delta M = 0$.

One can directly extrapolate our reconstruction towards the past in an attempt to infer the full time evolution of the MFH from the observed luminosities.
From Figs.~\ref{fig:rec_exp} and~\ref{fig:rec_dt}, we find that the polynomial expansion provides a good description for exponential and delayed-$\tau$ models with high values of the characteristic timescale, $\tau$ (above several gigayears), whereas the fit is not as good for values much shorter than the age of the Universe, even for the highest $N$ degrees.

We can also appreciate a similar behaviour for Gaussian SFHs.
Figures~\ref{fig:Gauss_rec_mass}, \ref{fig:Gauss_rec_SFR}, and~\ref{fig:Gauss_rec_SFRP} show the MFH, the SFR, and its first derivative as a function of $\hat t$ for different values of the FWHM and the peak age of the Gaussian.
In general, the accuracy of the polynomial reconstructions is very sensitive to these parameters.
Very narrow SFHs can never be well reproduced, not even by the highest degree polynomials.
At best, one can recover qualitative information, such as an approximate estimate of the height and width of the star formation peak.
The polynomial fit becomes a better description as the FWHM increases above a few gigayears, and the SFH is fairly well reproduced, especially for the higher-order polynomials (perhaps excluding the extrapolation towards the earliest moments of the Universe, although this problem is significantly alleviated for the best positive-SFR fit).

For a small sample of randomly selected Illustris galaxies covering a broad mass range, Fig.~\ref{fig:Illustris_reconstruction_mass} exemplifies how a constant SFR ($N=1$) tends to underestimate the mass-to-light ratio and thus the reconstructed mass.
The overall buildup of stellar mass in recent times is nicely recovered for $N > 1$; however, extrapolating back to the first gigayear of cosmic evolution is always problematic, and our simple polynomials often feature unphysical oscillations.
The best positive-SFR prescription fully corrects for this undesirable behaviour and yields MFHs that faithfully reproduce the simulation results at all times.
The instantaneous SFR (Fig.~\ref{fig:Illustris_reconstruction_sfr}) is also accurately recovered, even for early cosmic times.
At recent epochs, the smooth polynomial solution is representative of the average discrete SFR of the simulated galaxies, and our best positive-SFR fit is not only able to successfully identify quenched galaxies, but also to provide an order-of-magnitude estimate of their death time \citep{Corcho-Caballero+21}.

\subsection{Comparison with other methods}
\label{sec:comparison}

In order to put our results into context, we compared them with two other algorithms representative of state-of-the-art parametric and non-parametric algorithms.
More precisely, we fitted our Illustris galaxies with the publicly available codes \cig\ and \prosp, which are widely used in the literature.

\cig\ \citep[Code Investigating GALaxy Emission;][]{2019A&A...622A.103B} is an energy-balance code developed for modelling the observed galaxy SEDs from X-rays to radio wavelengths. This package applies a Bayesian strategy through the analysis of the likelihood distribution that takes into account the uncertainties on the observations and the effect of intrinsic degeneracies between physical parameters. It is characterized by a series of modules that (i) build stellar population models, (ii) consider the dust absorbing and re-emitting in the infrared wavelengths, (iii) add non-thermal sources of dust emission, (iv) include the interstellar lines, (v) take redshift effects  into account, (vi) calculate best-fit models, and (vii) estimate weighted galaxy properties such as the SFR, attenuation, dust luminosity, stellar mass, and many other physical quantities.

We first generated synthetic photometry in the ugriz bands from the simulated SFHs using the \citet{bc03} stellar basis with a \citet{Chabrier2003} IMF and metallicity Z = 0.02.
Then, the synthetic luminosities were fit to a grid of models with delayed-$\tau$ SFHs, typically used to reconstruct the SFHs of large observational galaxy samples \citep[e.g.][]{Noll+09, Ciesla+15}.
The characteristic timescale was varied from $0.1$ to $7$ Gyr in linear steps of 0.1 Gyr (a total of 70 models), and the age of the stellar population (i.e. the onset of star formation) was 13.7 Gyr.

The code spent about $0.1-0.6$ seconds fitting each galaxy on a standard computer using four cores. Most of that time was actually invested in input and output operations. The fit itself is negligible in comparison, and in this sense there in no practical difference in terms of computational time with respect to our method, both methods being virtually instantaneous for the simple setup considered in the present work.
In terms of accuracy, one can clearly see in Fig.~\ref{fig:delta_M} that results obtained with the delayed-$\tau$ model are comparable to our polynomial fits for the total stellar mass, although the dispersion around $\delta M =0$ appears to increase for the mass formed on shorter timescales.

While parametric models may indeed provide a reasonable description of the SFHs, their limited number of degrees of freedom prevents them from fully capturing the diversity found in individual galaxies \citep[see e.g.][and references therein]{Lower+20}.
On the other hand, non-parametric models enable a more thorough exploration of the possible SFHs at the expense of a significantly higher computation time.
Here, we calibrated the performance of our algorithm against the Bayesian inference code \prosp~\citep{ben_johnson_17, ascl19, 2021ApJS}, which performs a Monte Carlo sampling to compute the posterior parameter distribution of relatively complex models.
We used the Flexible Stellar Population Synthesis ({\sc FSPS}) stellar population synthesis models \citep[][]{Conroy_2009, Conroy_2010} to compute the synthetic luminosities of the Illustris galaxies, setting a default signal-to-noise ratio for each photometric band of $S/N=100$.
These data were then fit with the default continuity non-parametric model, which estimates the difference $\Delta\log(\rm SFR)$ between adjacent time bins, weighted by a Student's-t distribution to prevent sharp changes \citep[see][for details]{Leja+19a}.
We chose $N=7$ adaptive time bins and the same default parameters as discussed in \citet{Leja+19a}.
Dust extinction and metallicity were fixed to $A_V=0$ and $Z=Z_\odot$, respectively.

\prosp\ spends of the order of $\sim 20\pm{10}$ minutes fitting an individual galaxy.
Running the code on the full sample would have been computationally unfeasible, and therefore we randomly selected 373 objects with the intent of sampling the whole mass ($9\leq\log_{10}(M_*/{\rm M}_\odot)\leq11$) and specific star formation rate ($\rm sSFR$) range ($-12 \leq \log_{10}(sSFR/\rm{yr}^{-1})<-9$) as uniformly as possible, verifying by visual inspection that continuous SFHs, as well as suddenly quenched galaxies with different death times, are duly represented.
One can see in Figs. \ref{fig:Illustris_reconstruction_mass} and \ref{fig:Illustris_reconstruction_sfr} that, in general, \prosp\ provides a good fit to the MFH, and it has sufficient flexibility to accurately recover variations in the instantaneous SFR in recent epochs.
As shown in Fig.~\ref{fig:delta_M}, the differences $\delta M$ with respect to the mass formed in different time intervals is comparable to the results obtained by our polynomial reconstruction.
If anything, the best positive-SFR fit seems to be slightly more successful in correctly identifying recently quenched galaxies.

\section{Discussion}
\label{sec:discussion}

Our results show that the proposed polynomial expansion is able to successfully reproduce representative smooth SFHs from photometric observables, as long as most of the stellar mass is formed over long characteristic timescales ($\tau$ or FWHM comparable to the age of the Universe).
On the other hand, the method has difficulties when the MFH features abruptly changes or peaks in the early Universe.
In any case, it offers a compromise between the stability, computational efficiency, and straightforward physical interpretation of parametric methods and the flexibility of the non-parametric approach.
Technically, the fit only involves a scalar product, providing a significant advantage with respect to non-parametric methods in terms of computing time. 
The coefficients of the expansion are trivially related to the current mass, the SFR, and its derivatives with respect to time at the moment the object was observed.

Extrapolating backwards, the fit provides a fairly accurate reconstruction of the SFH when it is indeed smooth, although the unconstrained polynomials may provide unphysical negative values, especially at higher orders.
For that reason, we also computed the best positive-SFR fit, where a finite time interval $0 \le \hat t_{\rm min} < \hat t < \hat t_{\rm max} \le 1$ is identified where star formation is positive and definite.
No stars are formed for $t < t_{\rm min}$ nor $t < t_{\rm max}$.
Our results show that this model is capable of reproducing not only  simple analytical SFHs, but also the complex MFH found in cosmological numerical simulations.
The accuracy of the best positive-SFR fit rivals, or even surpasses, that of state-of-the-art parametric and non-parametric algorithms such as \prosp\ and \cig, involving a negligible computational burden.
A particularly relevant aspect where this prescription has proven to excel is the identification of quenched galaxies from the synthetic photometry.
As discussed by \citet{Corcho-Caballero+21}, many of the simulated galaxies have not formed any stars over the last few gigayears, while it is at present unclear whether such `quenching' of the star formation activity is observed in the real Universe \citep[cf.][]{Ciesla+18,Aufort+20,Corcho-Caballero+20}.
Based on the distribution of the relative difference $\delta M$ shown in Fig.~\ref{fig:delta_M}, we argue that the best positive-SFR fit provides a promising alternative for tackling this particular problem.

Nevertheless, there are also important caveats and potential improvements to the scheme proposed here.
First and foremost, observational uncertainties must be taken into account in order to provide error estimates on the recovered quantities.
Typical errors in the SDSS fluxes are of the order of 2\% in the $u$ band and 1\% in $g$, $r$, $i$, and $z$.
As noted above, this is larger than the values of $\hat D$ obtained for $N \ge 2$.
Therefore, it is unlikely that meaningful results will be obtained by increasing the polynomial degree, $N$, further than this limit.
This, in turn, suggests that constraining even the first derivative of the SFR will be difficult in practice.
A more quantitative assessment will be carried out in future work.

In addition, our tests do not consider the effects of metallicity and dust extinction.
Treating these non-linear parameters represents a major improvement to the proposed framework, but it would be necessary to apply it to real observational data.
A Bayesian approach is arguably the most appropriate way of quantifying the posterior probability distribution of the model parameters, at the expense of computing time.
A significant advantage would be that the priors would make it possible to supplement our best positive-SFR fit with additional constraints, such as assigning less (or zero) weight to the individual polynomial terms that provide a significant amount of negative mass formation.
More elaborate priors based on chemical evolution constraints could also be explored, and the initial and final times,  $t_{\rm min}$ and $t_{\rm max}$, could be left to vary as two additional free parameters.
These would again behave non-linearly, but it is straightforward to include them within a Bayesian framework.

Another trivial, albeit important, extension would be to consider multi-wavelength photometric observations in other bands (e.g. infrared and/or ultraviolet) that contain valuable information, as well as spectroscopic data.
Including additional observables, appropriately weighted to yield a combined likelihood, would make it possible to increase $\hat D$ and discriminate between models that are very similar in the optical colours considered in the present work.
This is arguably the most promising avenue for increasing the maximum degree, $N$, of the polynomial basis in a practical application of the algorithm.

\section{Conclusions}
\label{sec:conclusions}

Here we propose an analytical description of the MFH of a given galaxy in terms of a polynomial expansion around the present time.
Our tests against a set of synthetic observations in the SDSS $ugriz$ photometric system of exponential, delayed-$\tau$, and Gaussian SFHs, as well as a set of galaxies from the Illustris-TNG suite of cosmological simulations, show that:
\begin{enumerate}
\item The polynomial reconstruction of order $N=1$ (constant SFR) is able to reproduce the synthetic luminosities with an accuracy of the order of 10 per cent or better.
The accuracy improves with $N$, with higher-order polynomials reaching values that are always below 1 per cent.
\item For smooth SFHs that vary on timescales comparable to the age of the Universe, the polynomial expansion does indeed provide a faithful reconstruction of the actual SFH, especially at recent times. On the other hand, our approach is not well suited to describe an SFH that varies on short timescales or peaks in the very early Universe, yielding negative stellar masses and/or SFRs.
\item Such unphysical results can be avoided by selecting the best positive-SFR fit, where star formation is always positive within the time interval $0 \le \hat t_{\rm min} < \hat t < \hat t_{\rm max} \le 1$ and vanishes outside this range.
This prescription helps polynomial models recover a more accurate estimation of the mass formed at the most recent times and identify quenched galaxies in the numerical simulations.
\item The results of the best positive-SFR fit are highly competitive with state-of-the-art parametric and non-parametric methods, both in terms of accuracy and computational speed.
\end{enumerate}

We conclude that the proposed polynomial expansion represents a promising alternative to the analytical functions that have been traditionally used in parametric methods.
The polynomial description is more flexible than the exponential or delayed-$\tau$ models, and the number of free parameters can be trivially adapted to the amount and quality of the available data.
Its physical interpretation in terms of the current total mass, SFR, and higher-order time derivatives is straightforward, and its linear nature is amenable to an extremely fast computation, which could in principle be applied to large photometric surveys on a pixel-by-pixel basis.

Nonetheless, further work is still necessary before our approach can be applied to real measurements.
Most importantly, the effects of metallicity, dust extinction, and observational uncertainties must be taken into account, and we will explore a Bayesian extension of the present formalism in the near future.
This powerful statistical framework will enable us to test the ability of the polynomial expansion to tackle the problem under realistic conditions.

\begin{acknowledgements}
We thank the anonymous referee for a helpful, constructive report, that has encouraged us to develop the best positive-SFR fit, add numerical simulations to our test suite, and compare our results with other codes in the literature. 
This work has been funded by the Spanish Research Agency (grant PID2019-107408GB-C42 / AEI / 10.13039/501100011033).
S.Z. also acknowledges support from contract BES-2017-080509 associated to grant AYA2016-79724-C4-1-P from the Spanish Ministry of Economy, Industry and Competitiveness through the MINECO-FEDER research.

\end{acknowledgements}



\bibliographystyle{aa}
\bibliography{references.bib} 
\begin{appendix}

\section{Reconstructed histories}

\begin{figure*}
\centering
  \begin{tabular}{@{}cccc@{}}
  $\tau=0.5$~Gyr & $\tau=3.5$~Gyr & $\tau=7.5$~Gyr & $\tau=13.5$~Gyr \\
    \includegraphics[width=.23\textwidth]{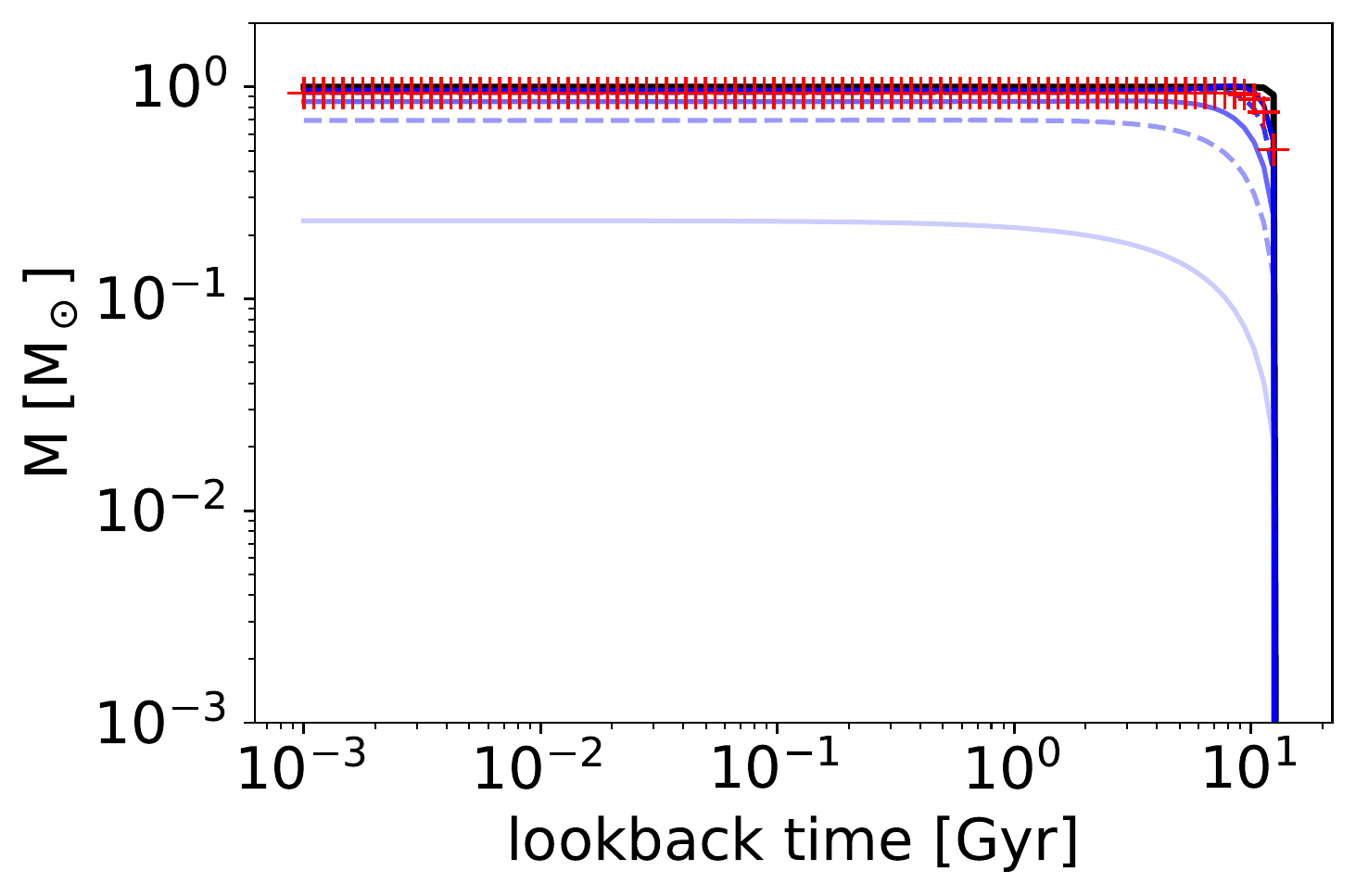} &
    \includegraphics[width=.23\textwidth]{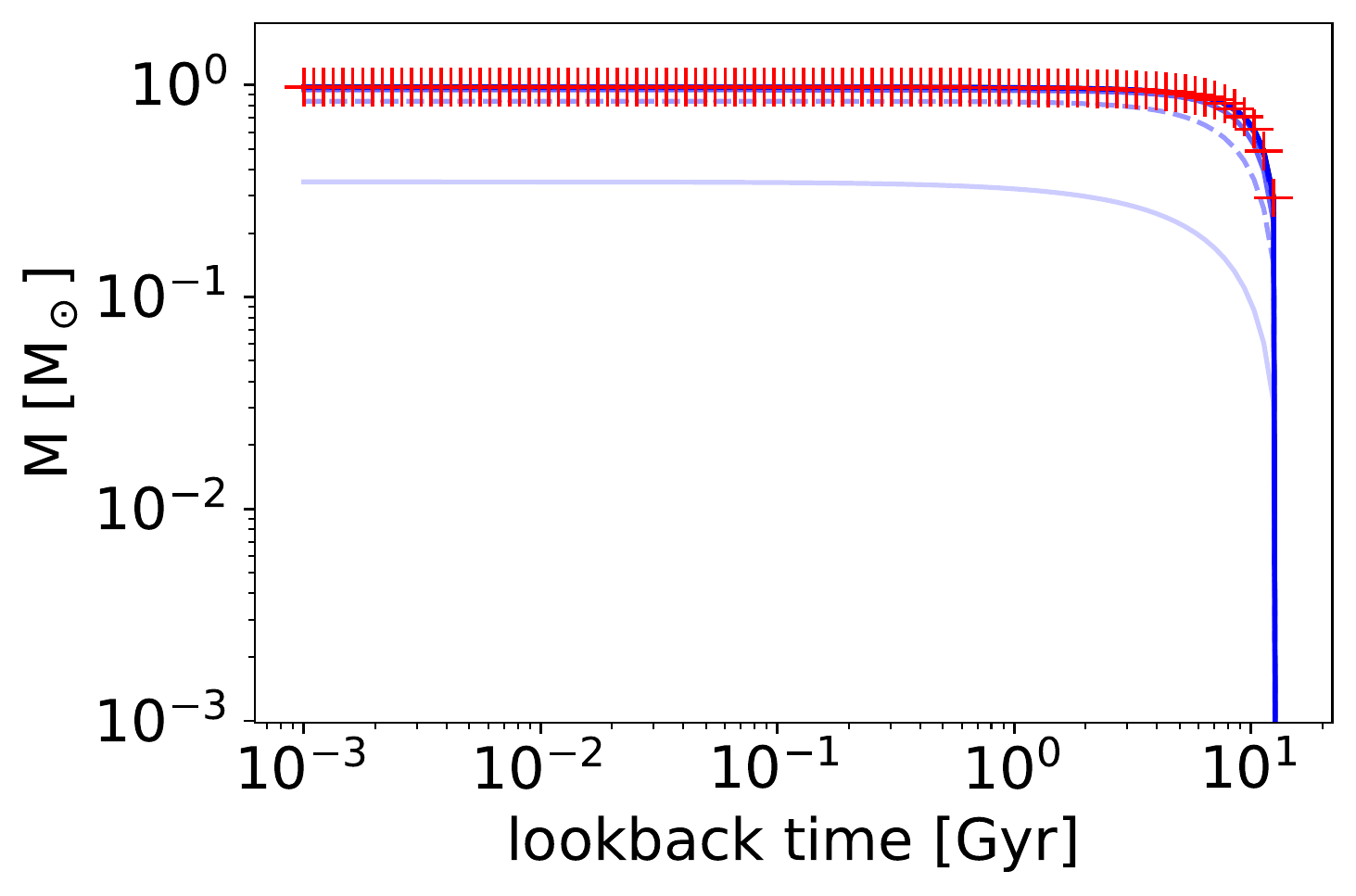} &
    \includegraphics[width=.23\textwidth]{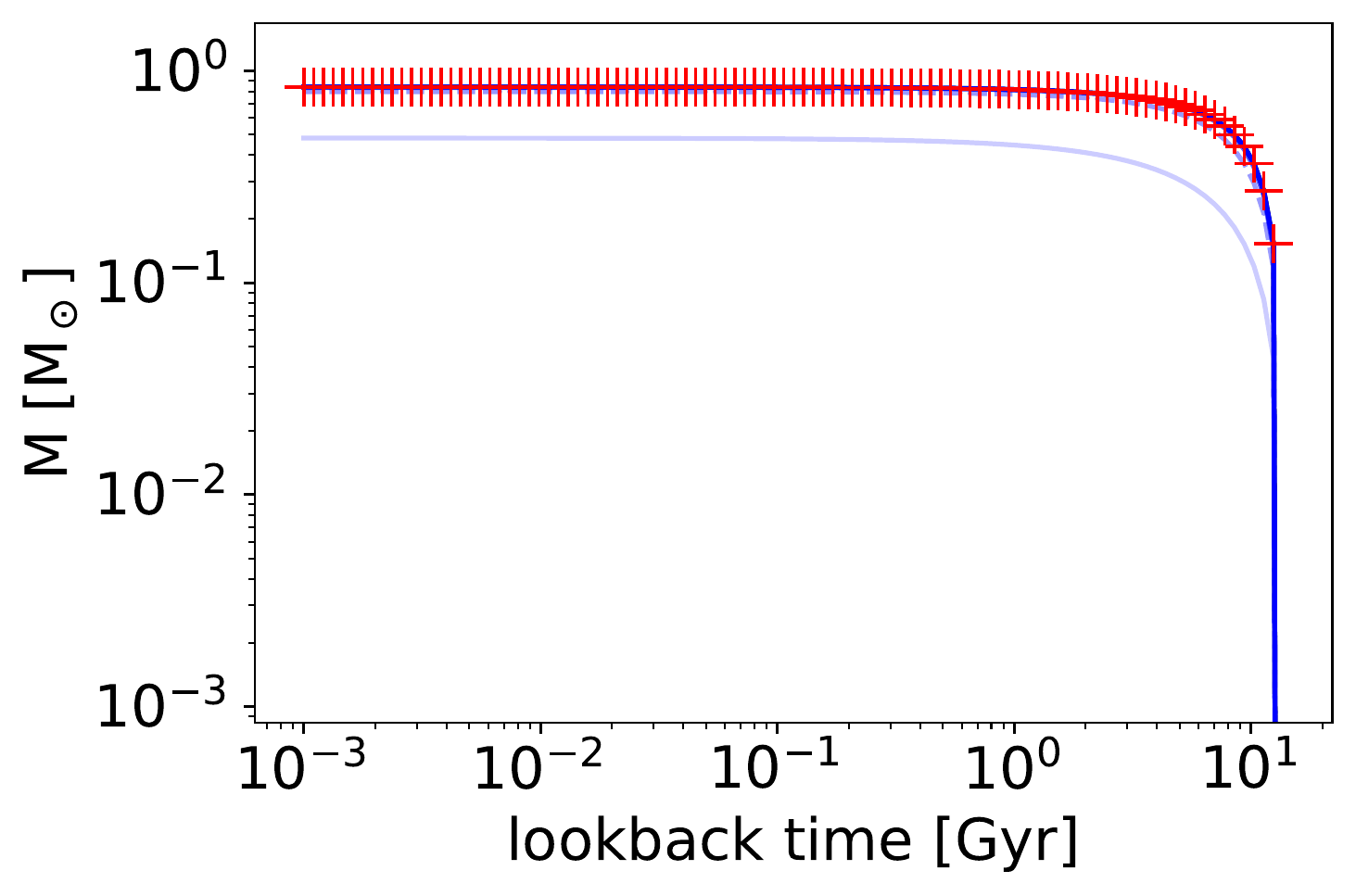} &
    \includegraphics[width=.23\textwidth]{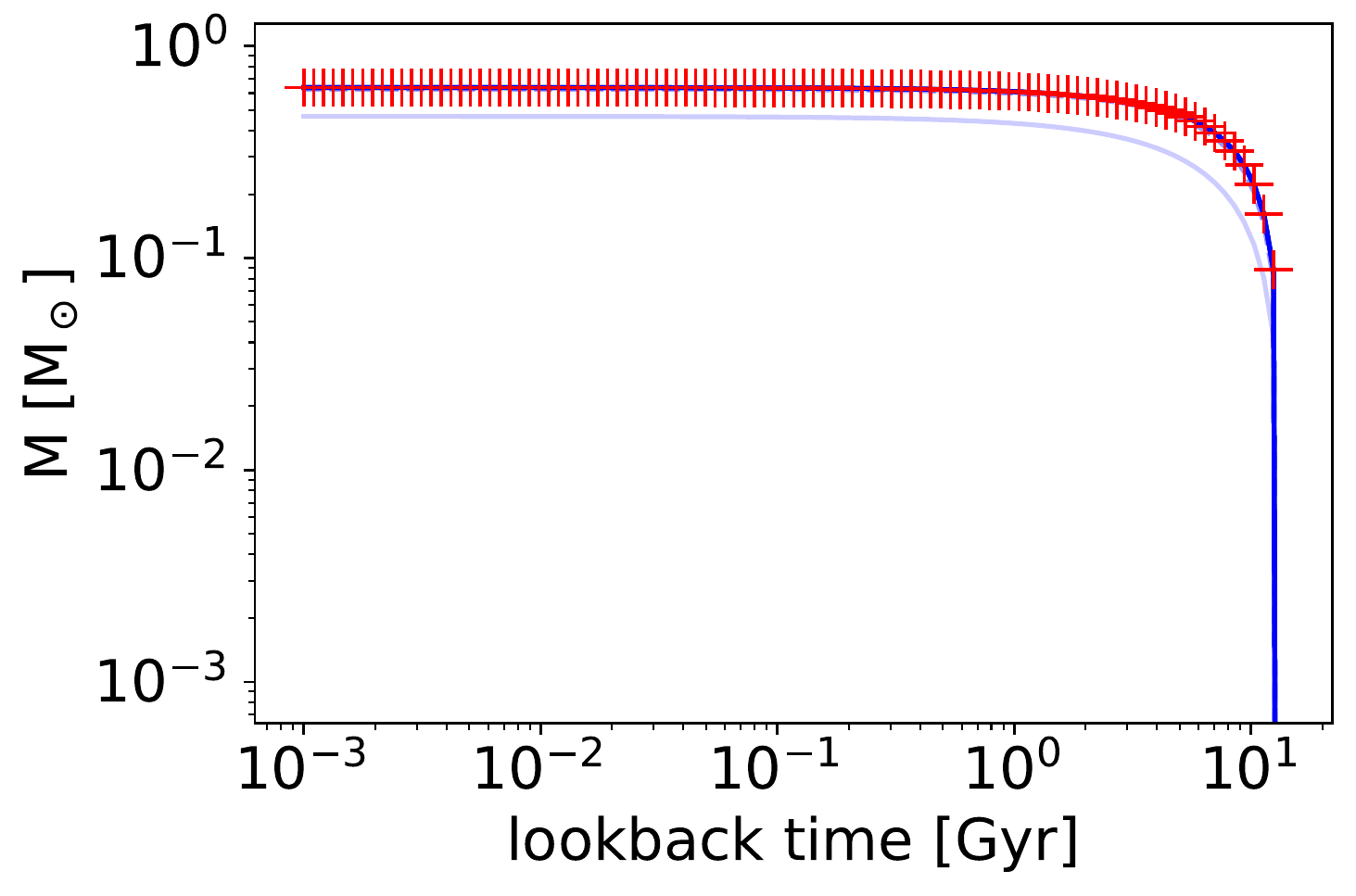}   \\
    \includegraphics[width=.23\textwidth]{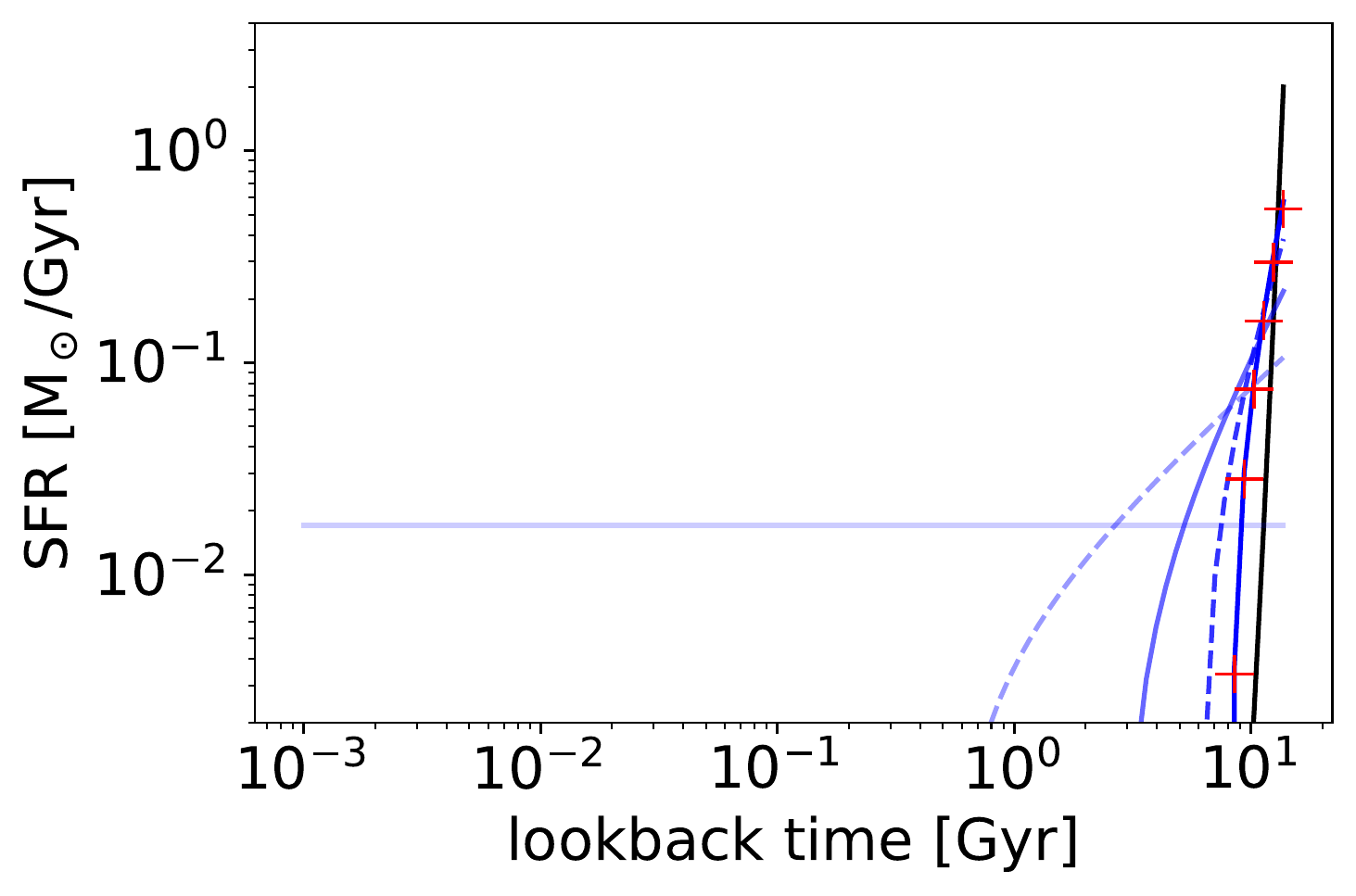} &
    \includegraphics[width=.23\textwidth]{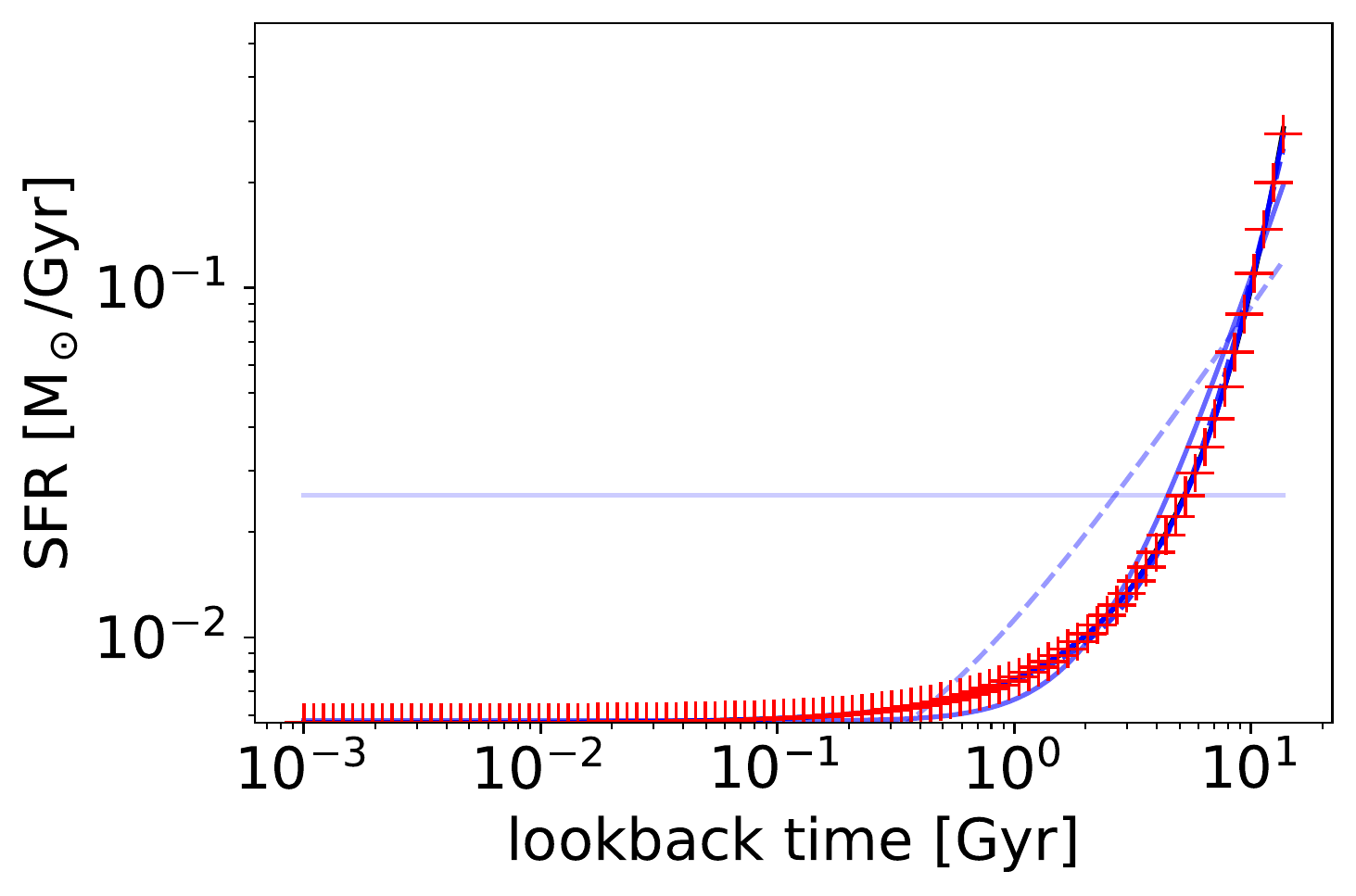} &
    \includegraphics[width=.23\textwidth]{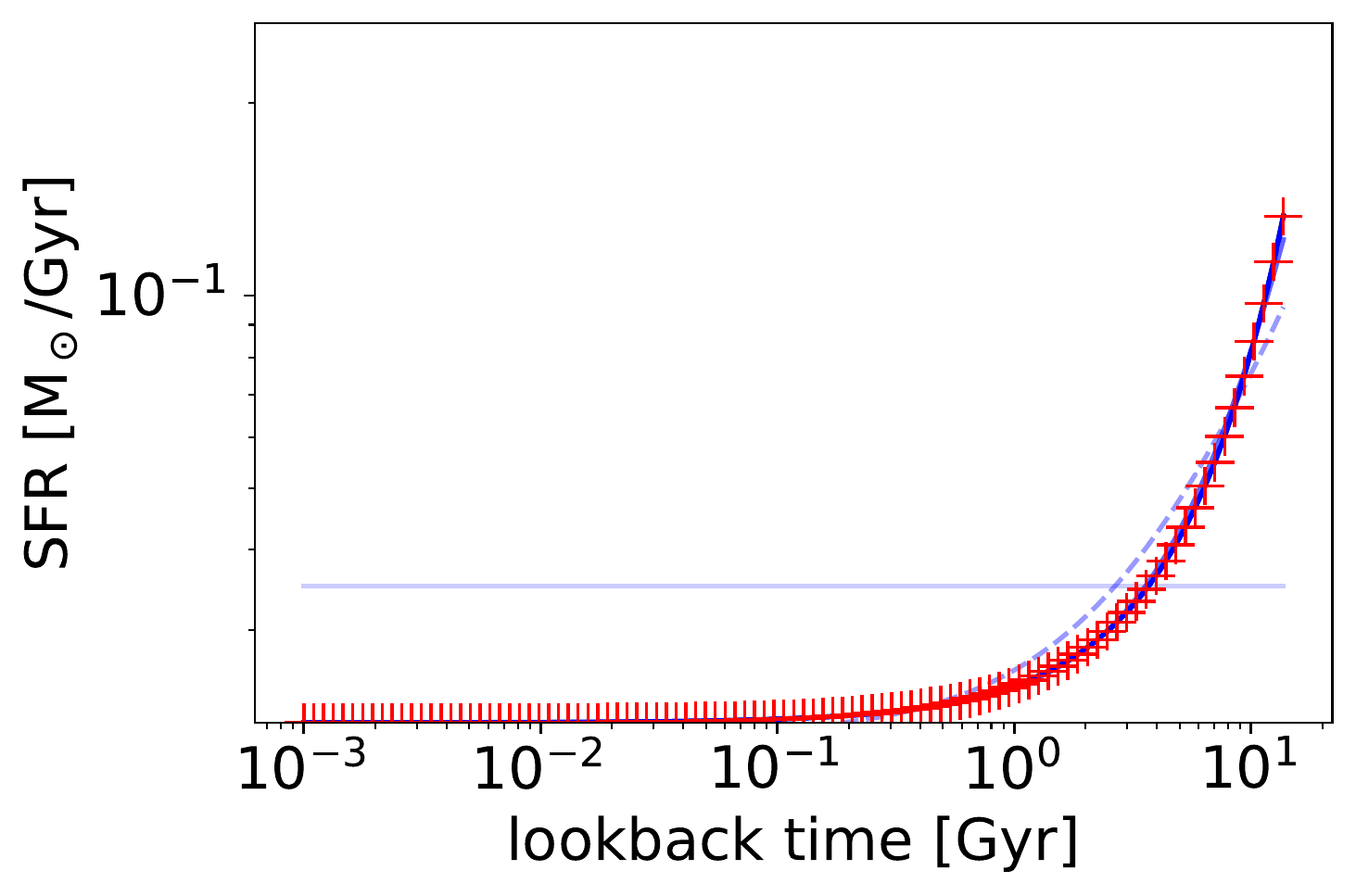} &
    \includegraphics[width=.23\textwidth]{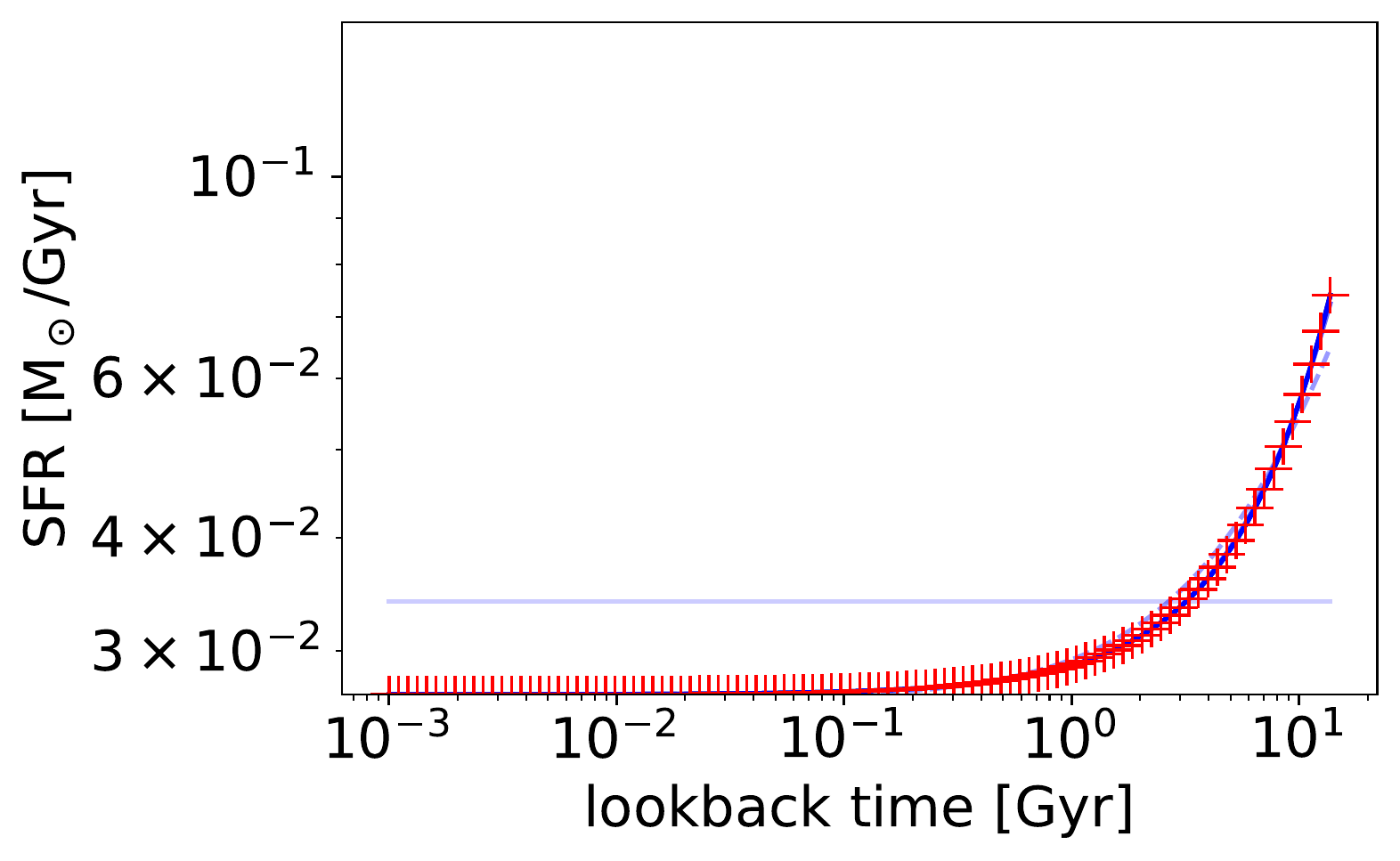}   \\
    \includegraphics[width=.23\textwidth]{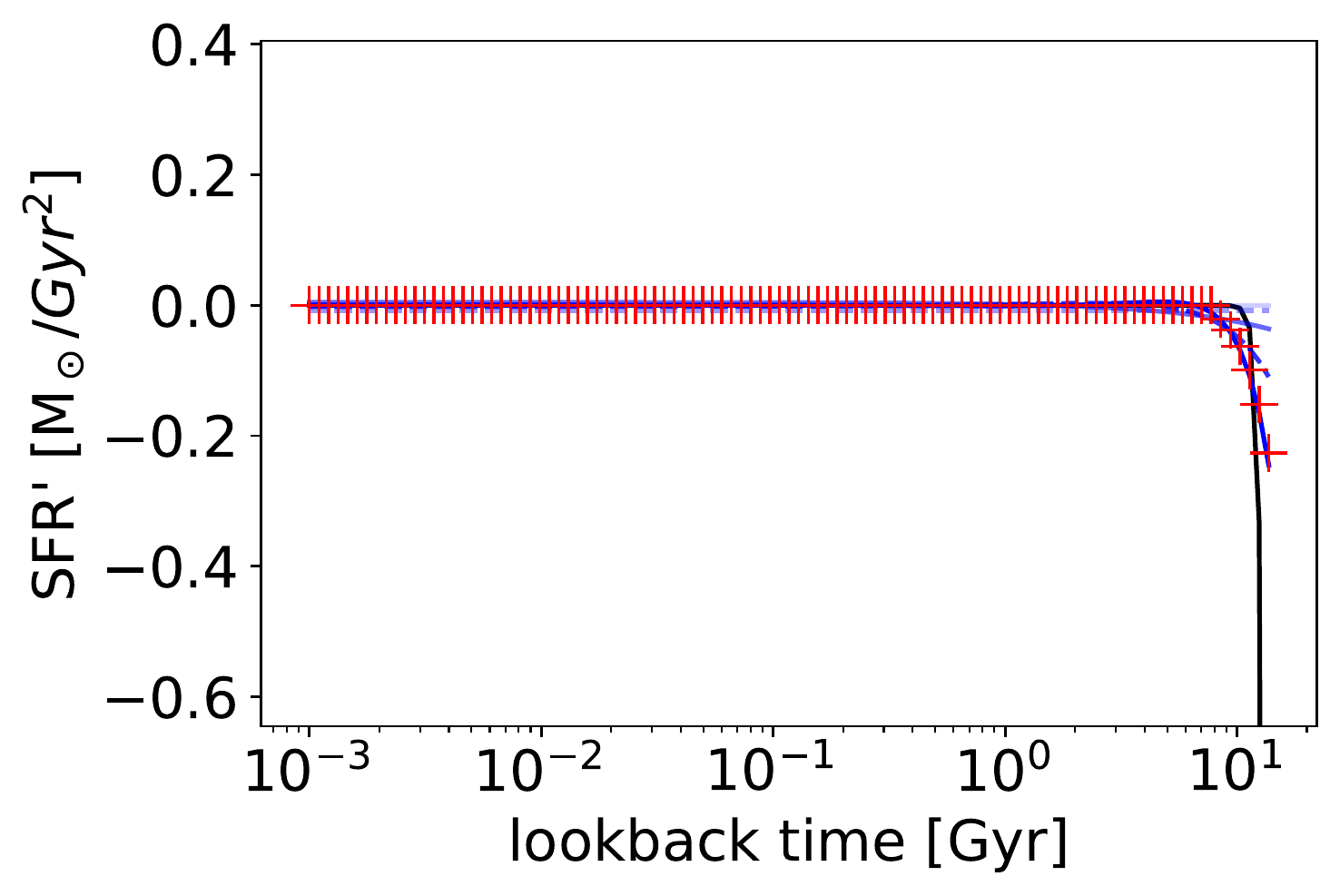} &
    \includegraphics[width=.23\textwidth]{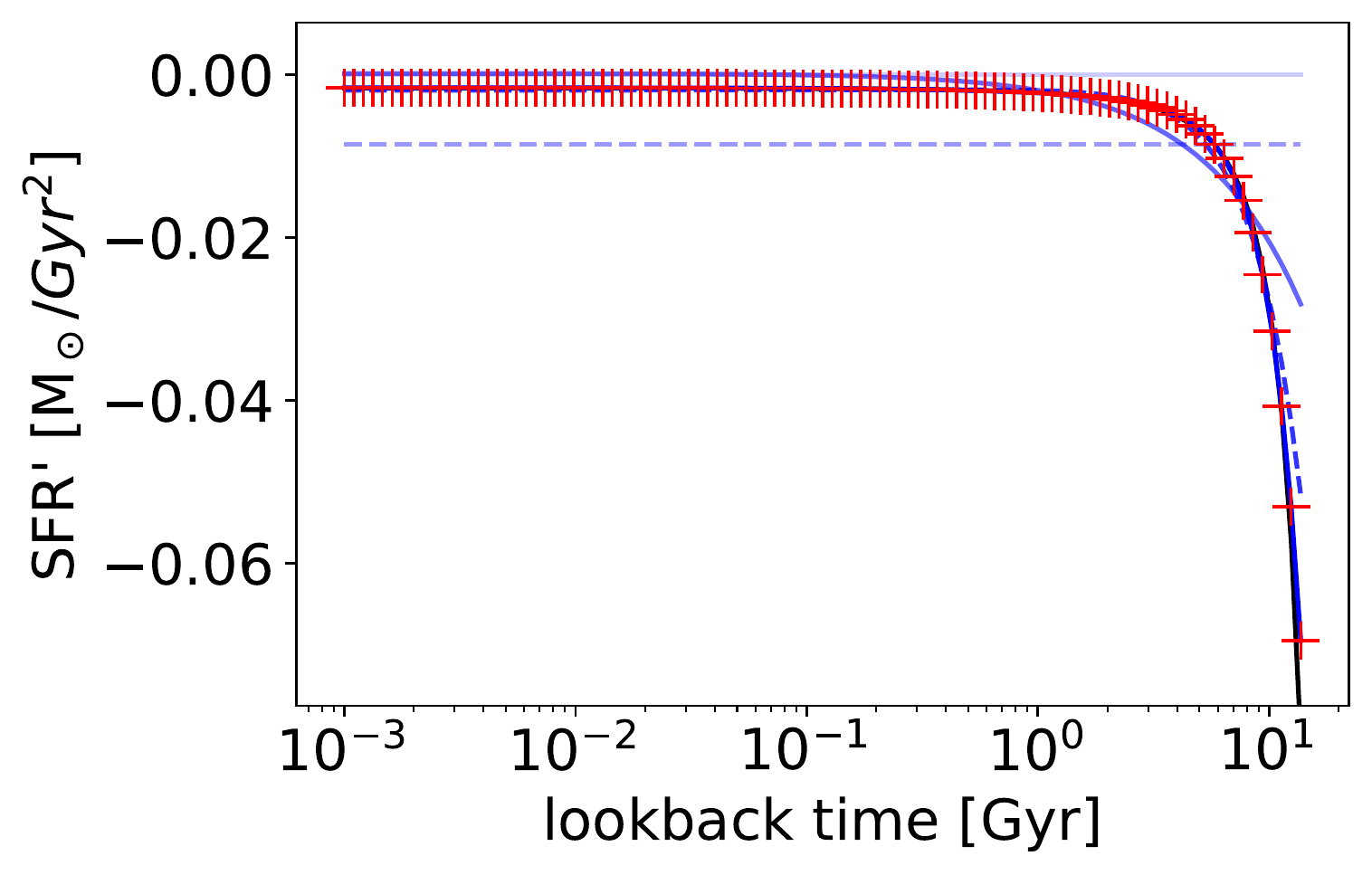} &
    \includegraphics[width=.23\textwidth]{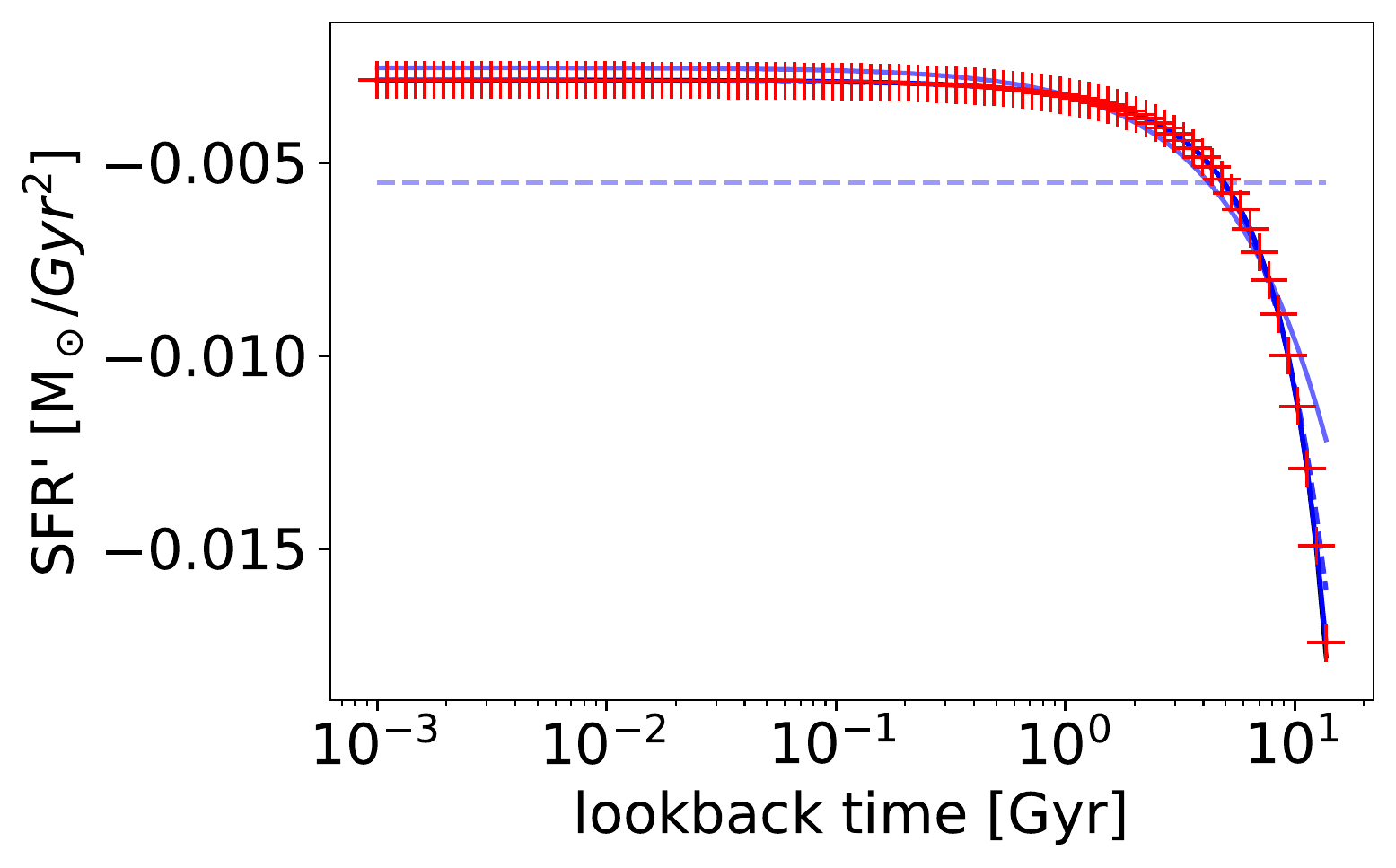} &
    \includegraphics[width=.23\textwidth]{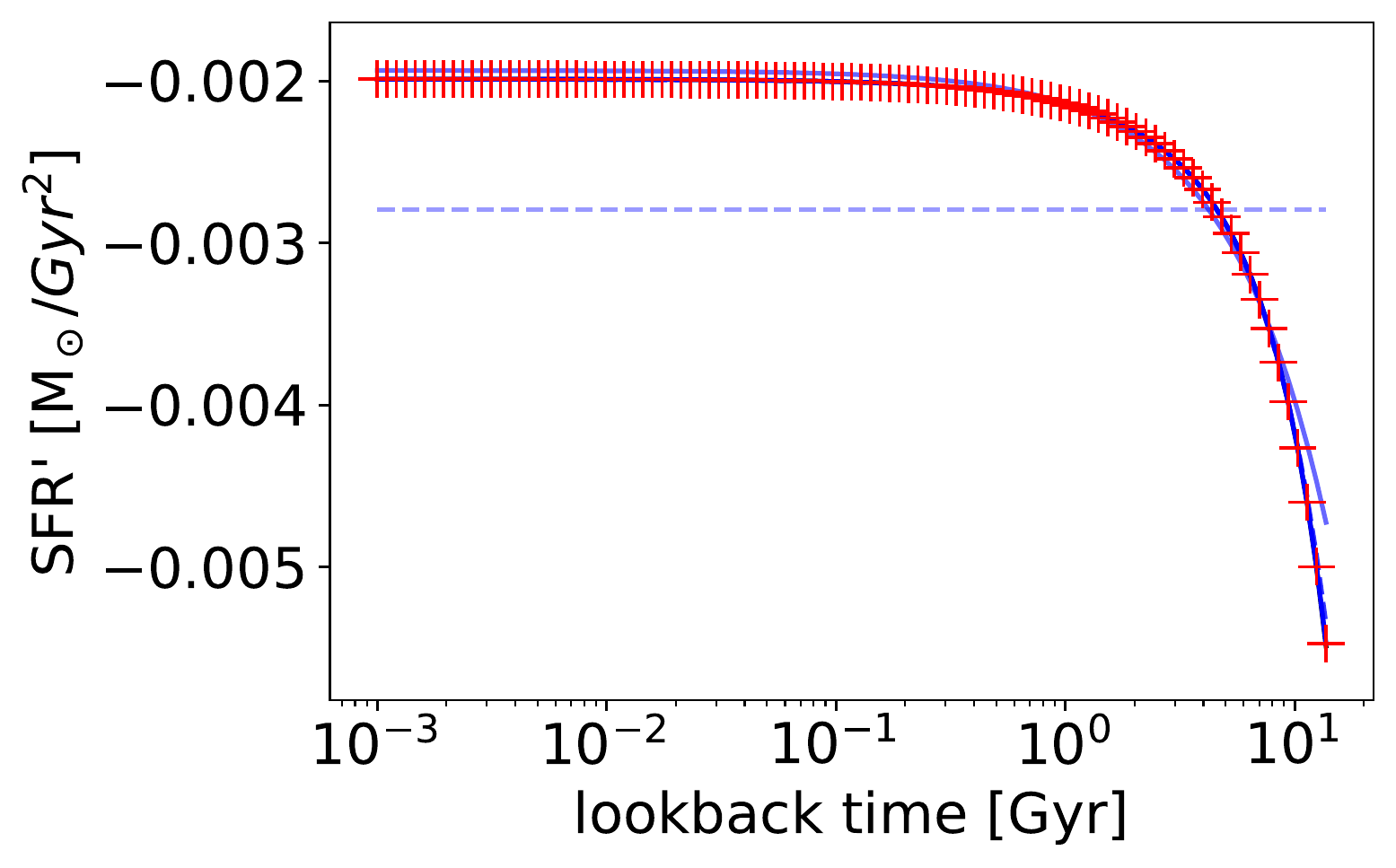}   \\
    \includegraphics[width=.23\textwidth]{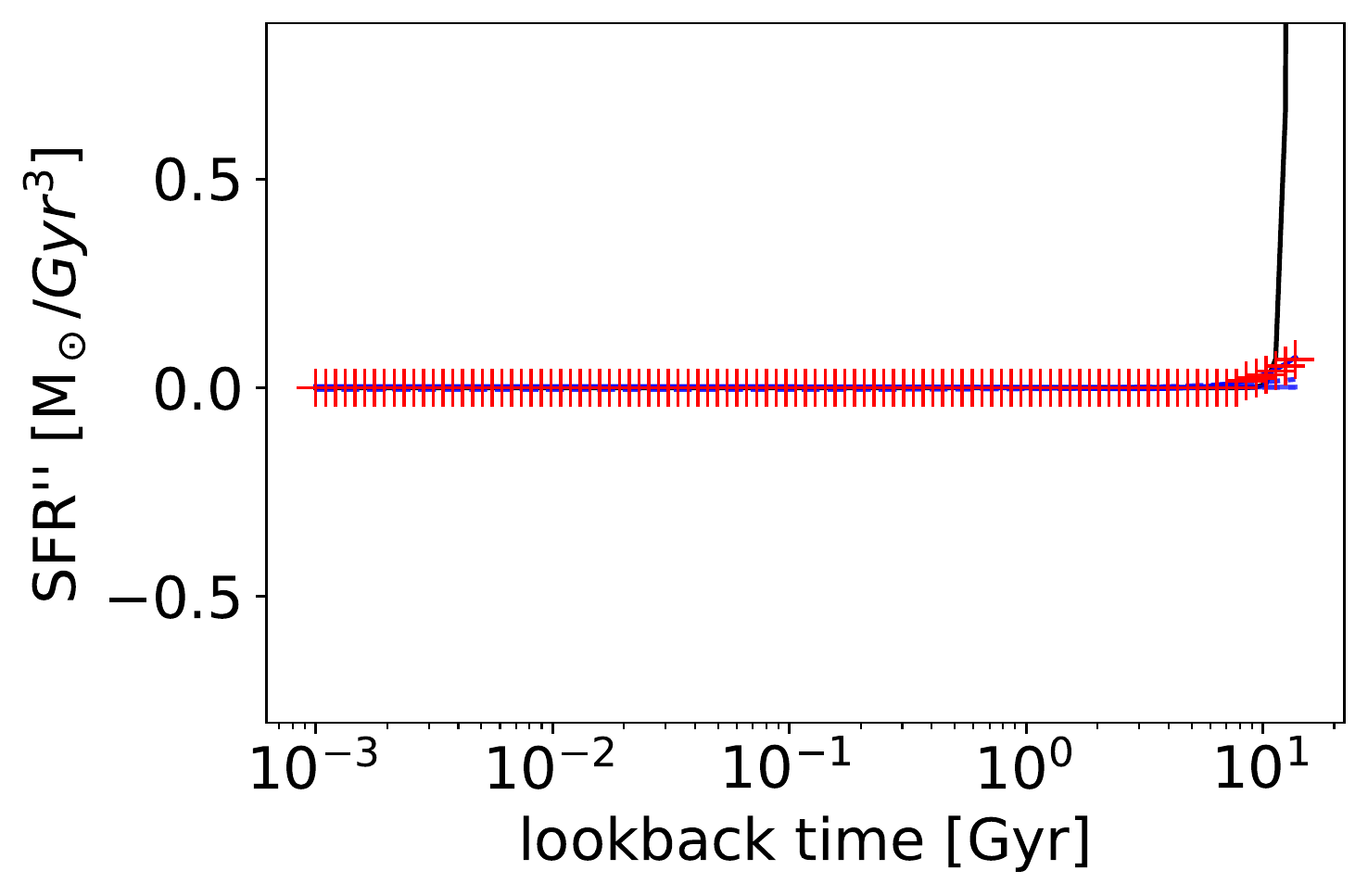} &
    \includegraphics[width=.23\textwidth]{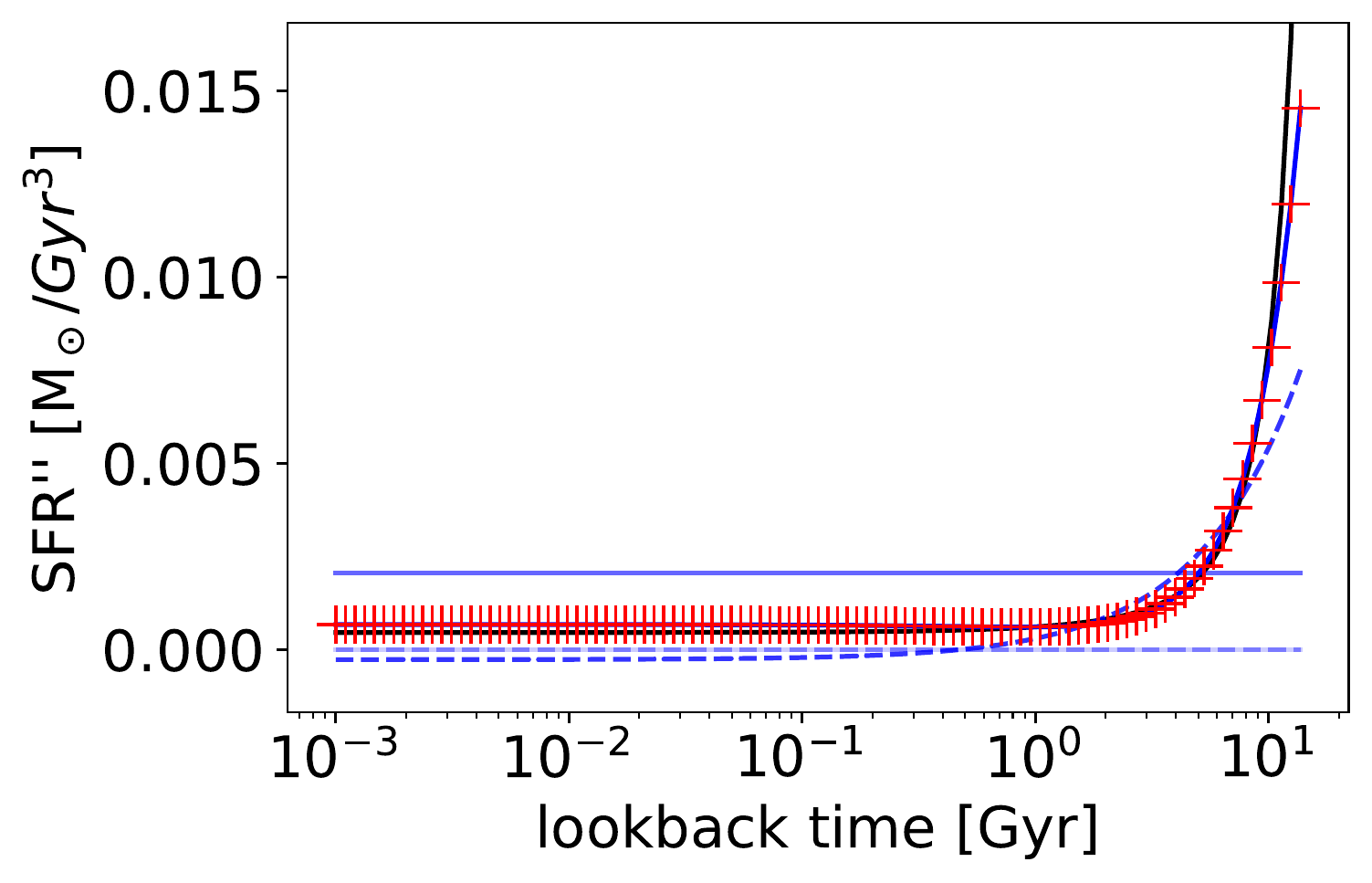} &
    \includegraphics[width=.23\textwidth]{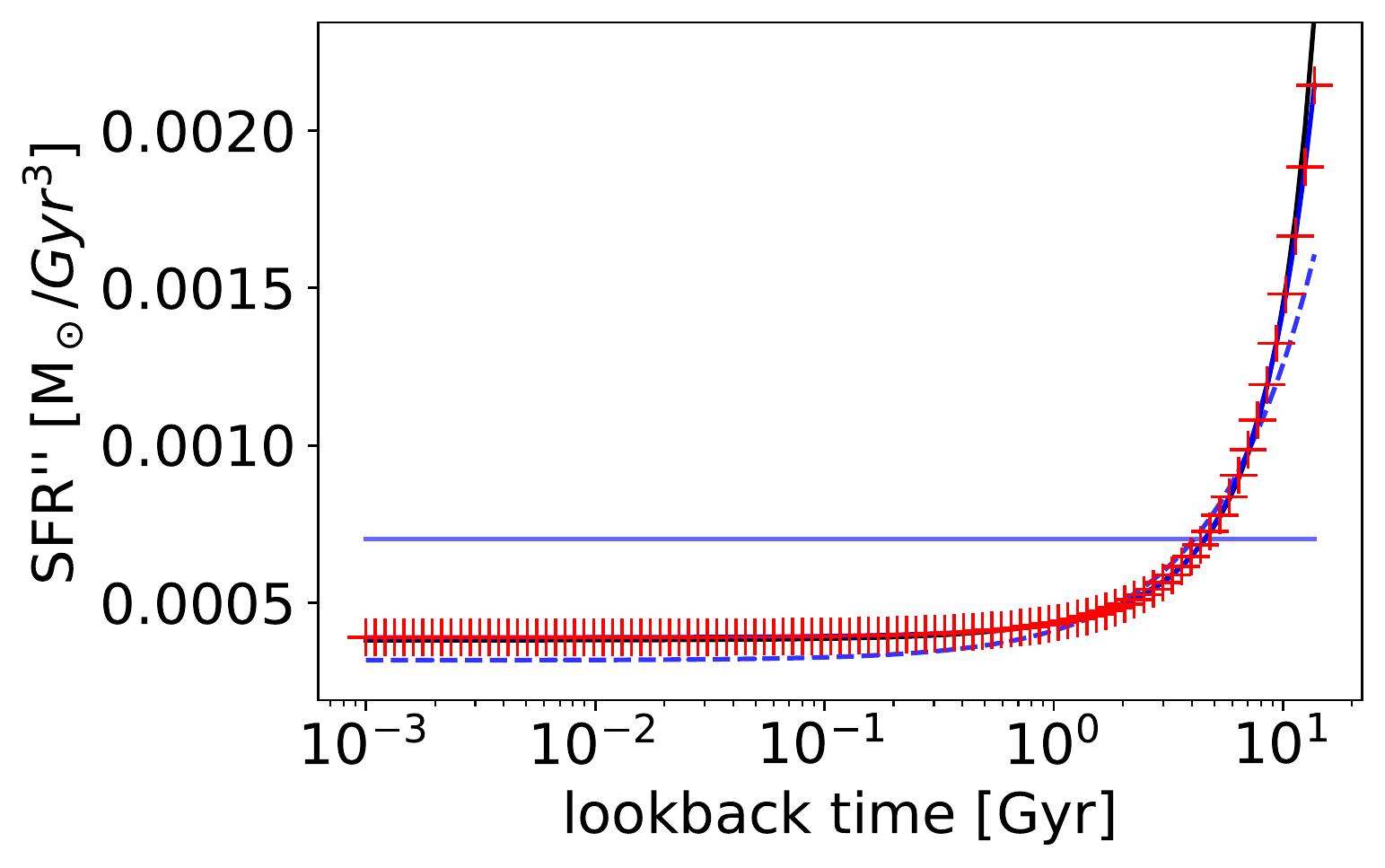} &
    \includegraphics[width=.23\textwidth]{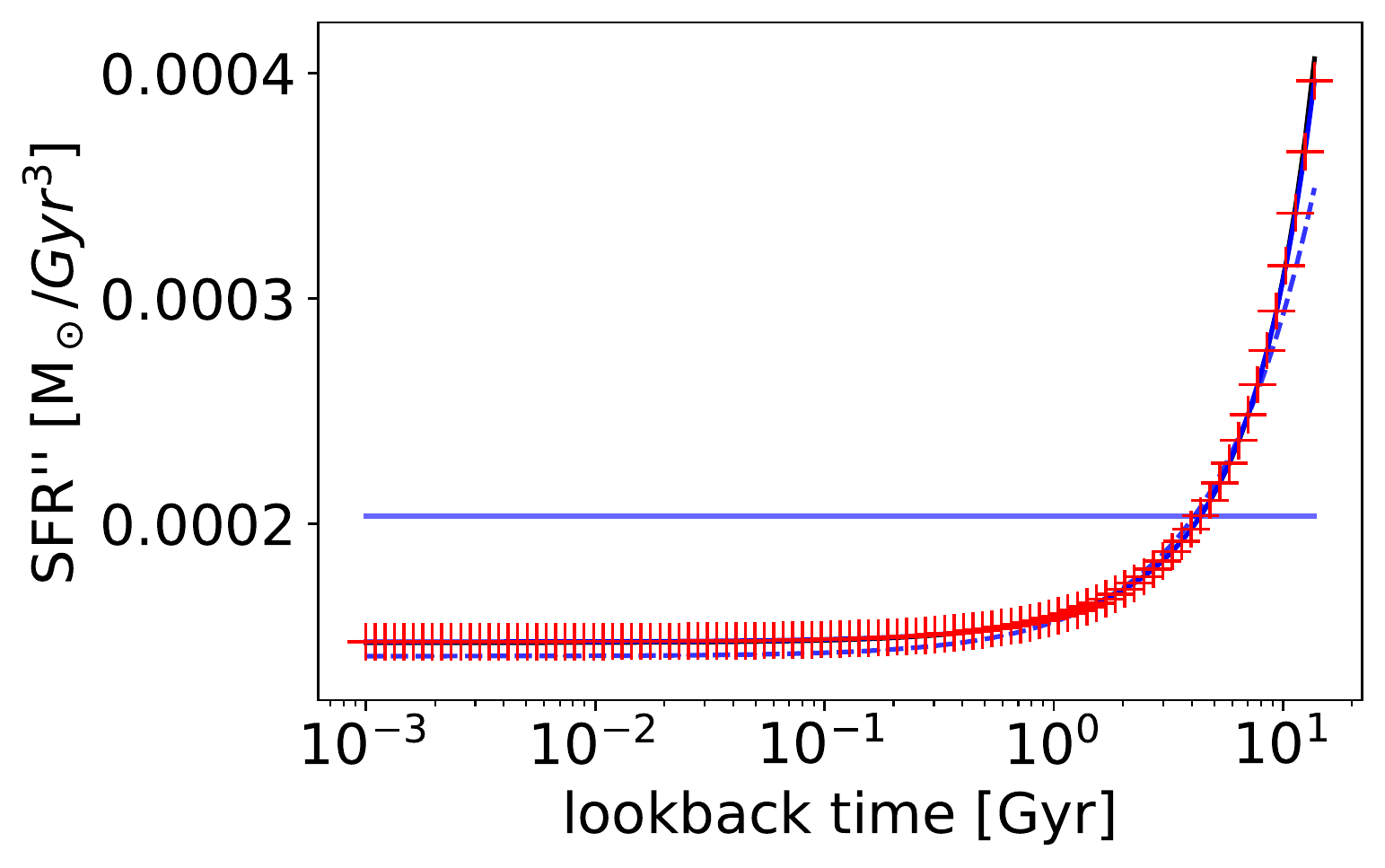}   
  \end{tabular}
  \caption{Sample of a MFHs reconstructed with our parametric model from the luminosities of a synthetic exponential SFR with different timescales, $\tau$. With blue lines we plot the stellar mass formed (first row), the SFR (second row), the SFR first derivative (third row), and the SFR second derivative (fourth row), compared to the input model (solid black line). Colour gradients and line styles indicate the degree of the polynomial reconstruction from $N=5$ (solid) to lower degrees (more diffuse lines; solid and dashed for odd and even $N$, respectively). Red crosses correspond to the best positive-SFR fit.}
\label{fig:rec_exp}
\end{figure*}

\begin{figure*}
\centering
  \begin{tabular}{@{}cccc@{}}
  $\tau=0.5$~Gyr & $\tau=3.5$~Gyr & $\tau=7.5$~Gyr & $\tau=13.5$~Gyr \\
    \includegraphics[width=.23\textwidth]{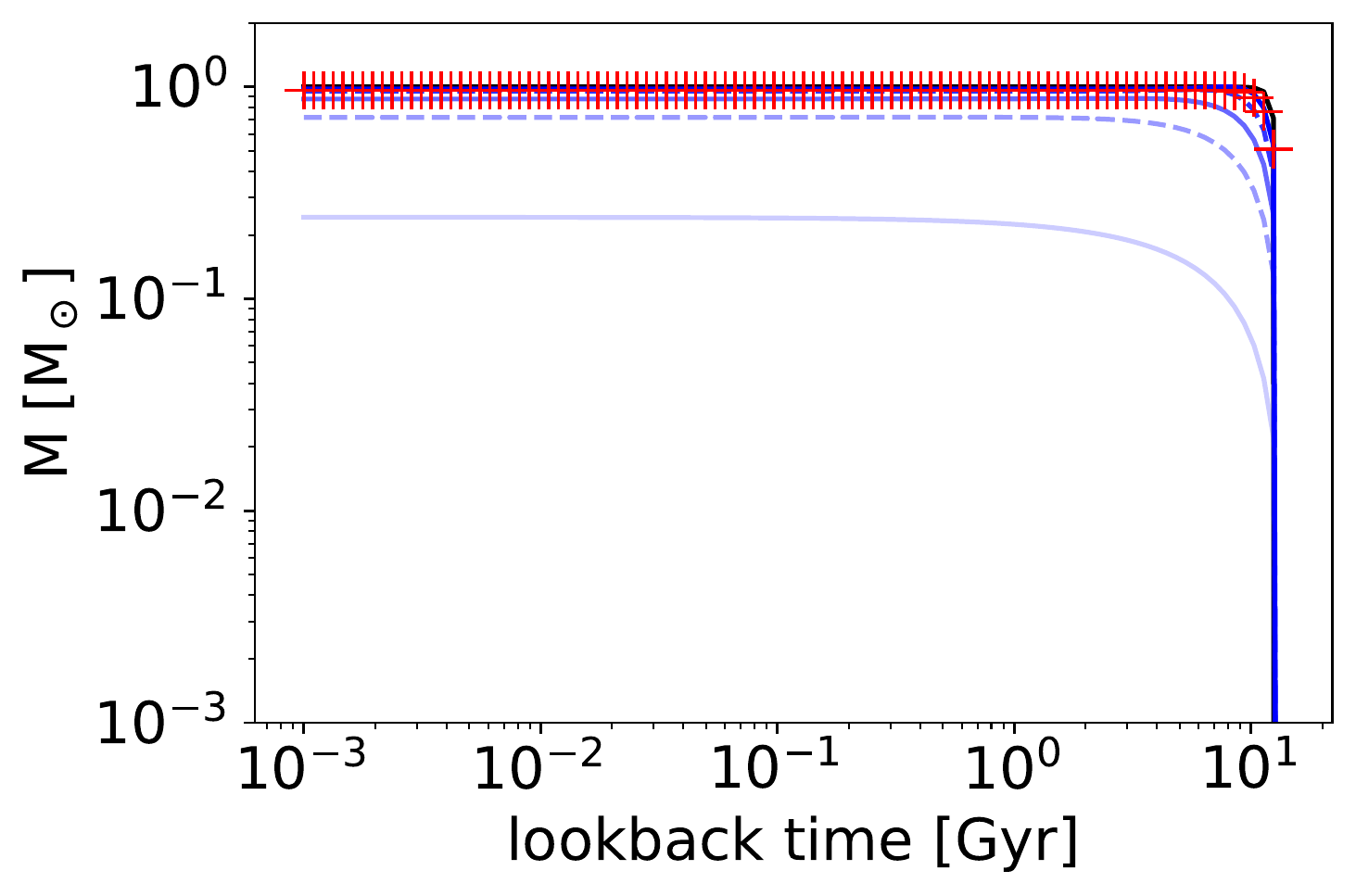} &
    \includegraphics[width=.23\textwidth]{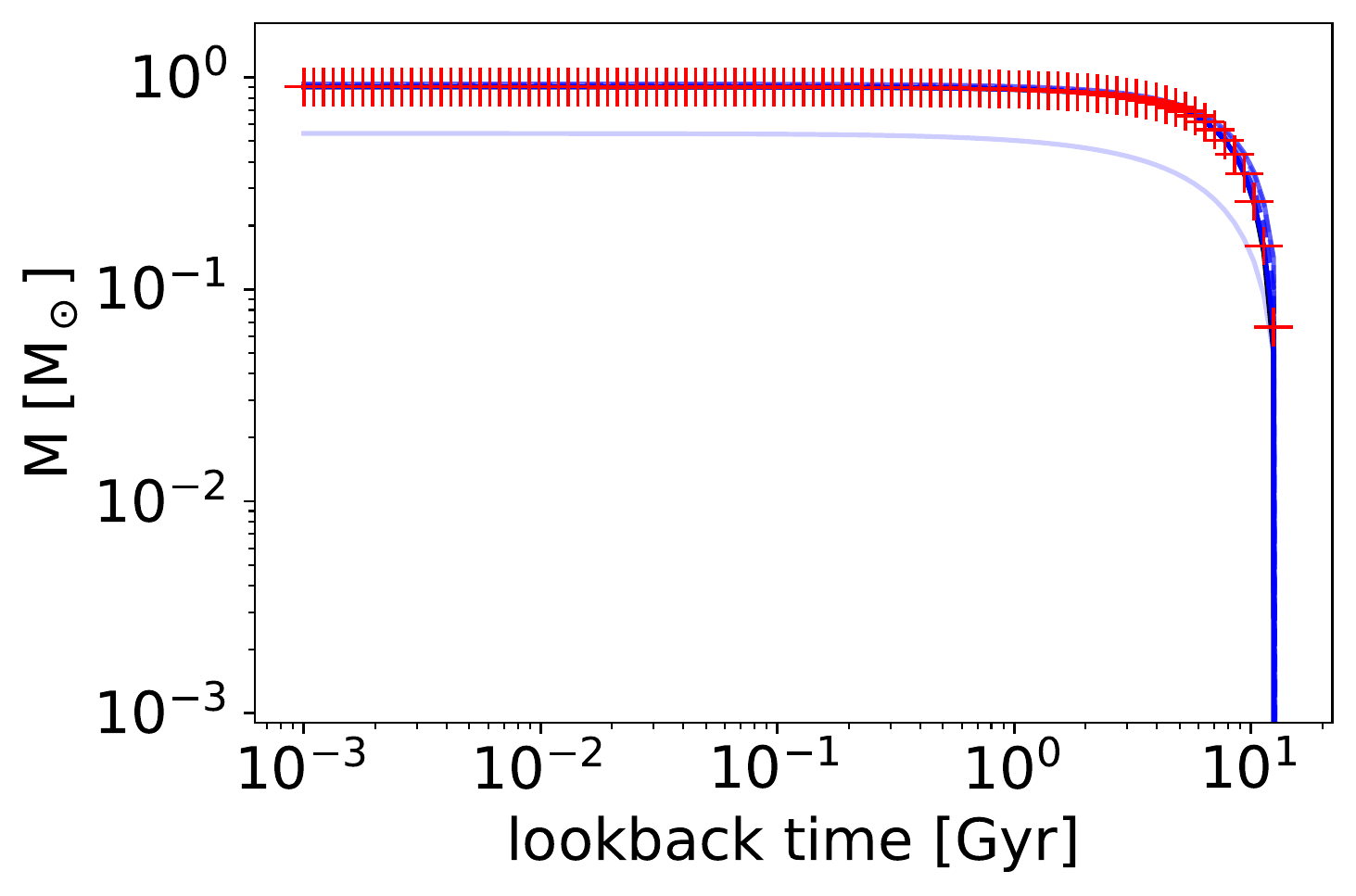} &
    \includegraphics[width=.23\textwidth]{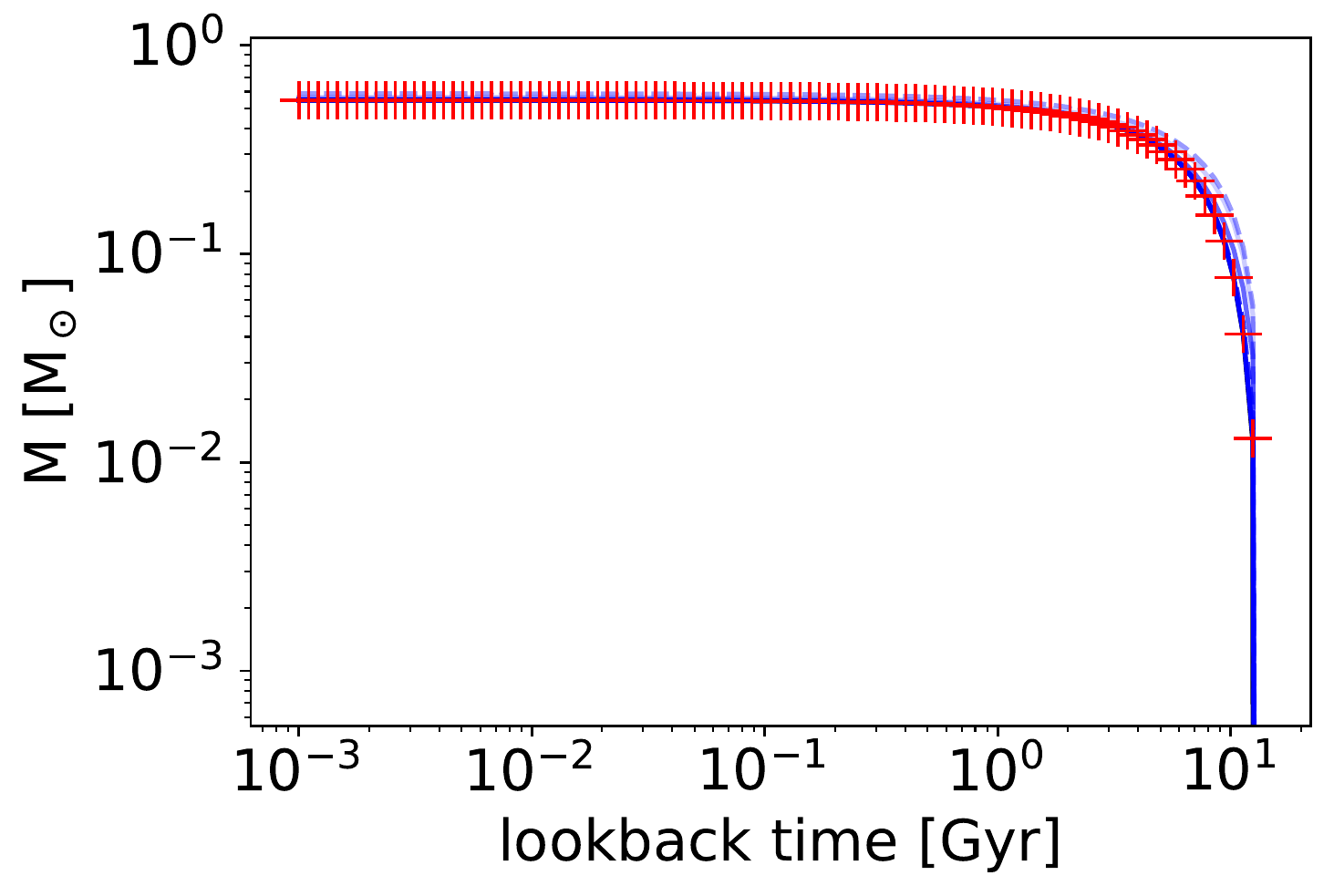} &
    \includegraphics[width=.23\textwidth]{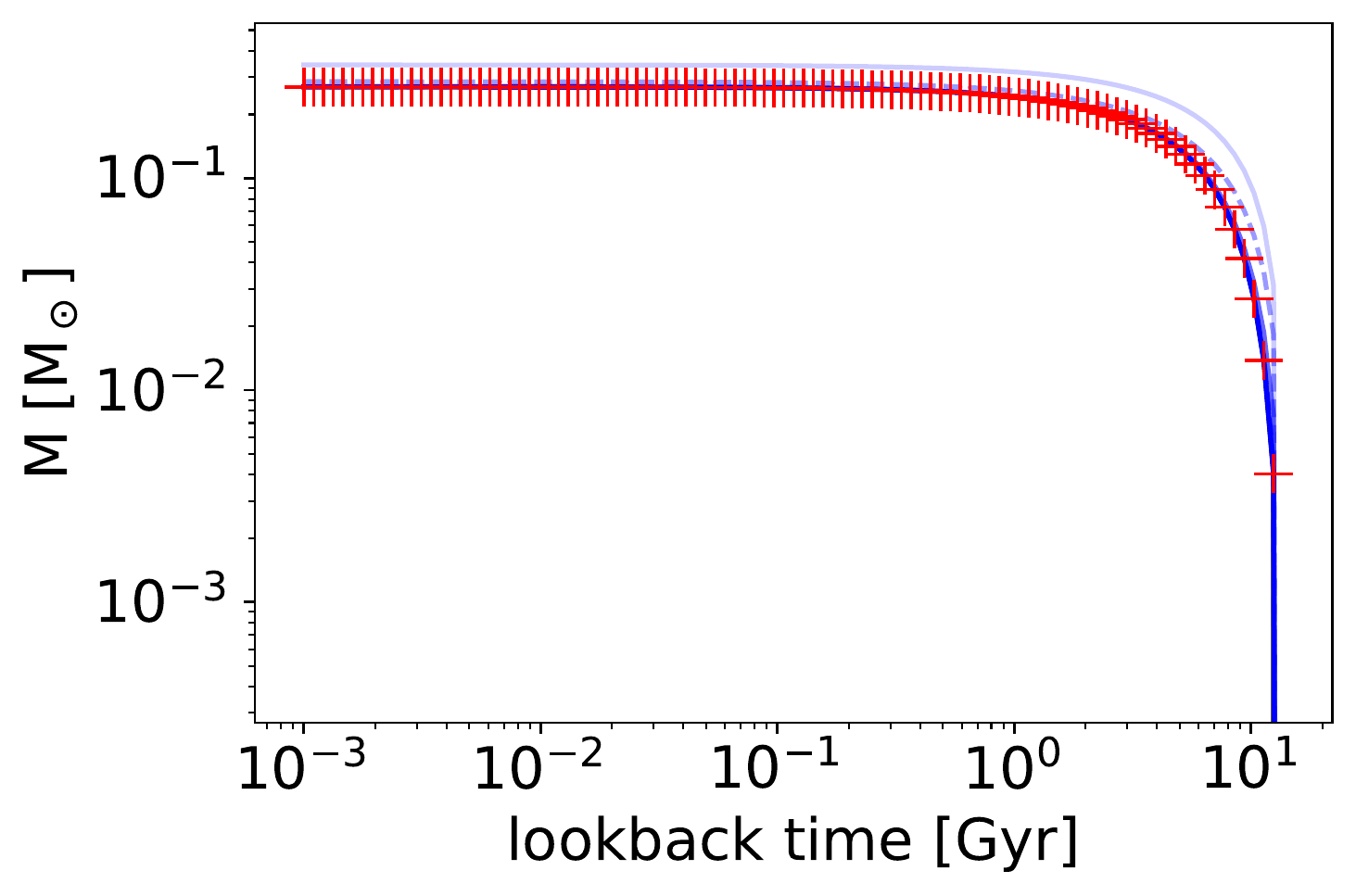}   \\
    \includegraphics[width=.23\textwidth]{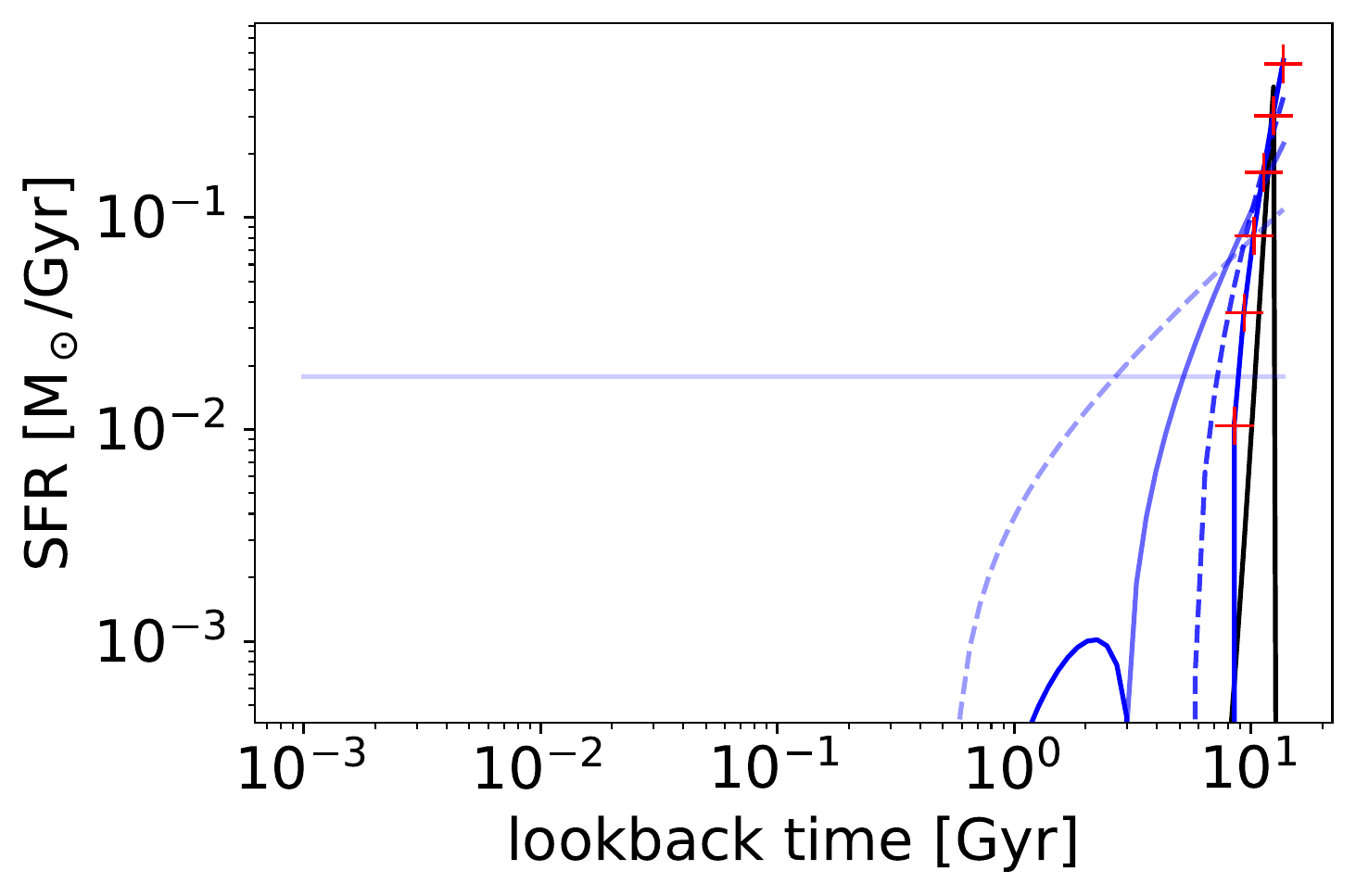} &
    \includegraphics[width=.23\textwidth]{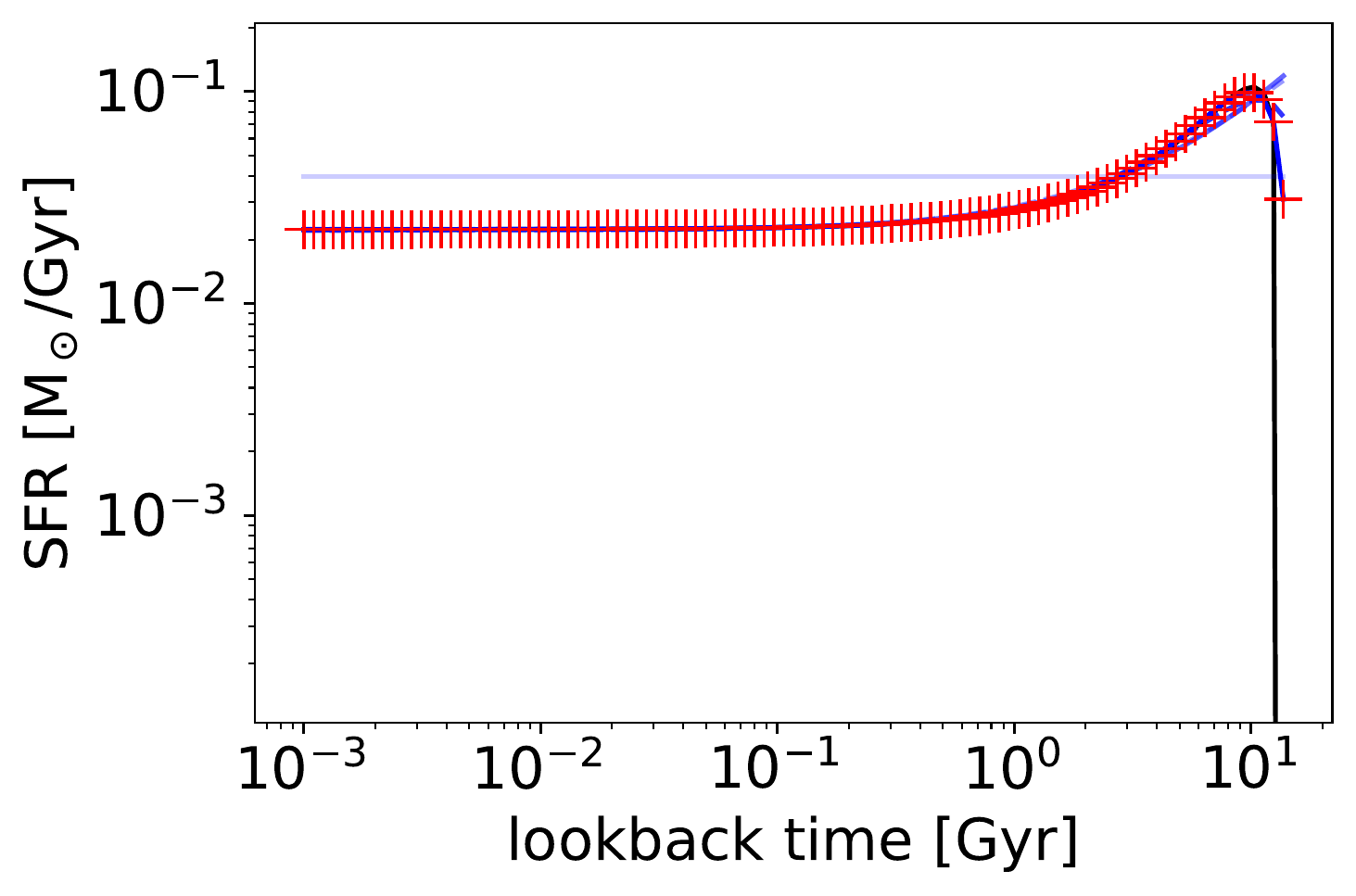} &
    \includegraphics[width=.23\textwidth]{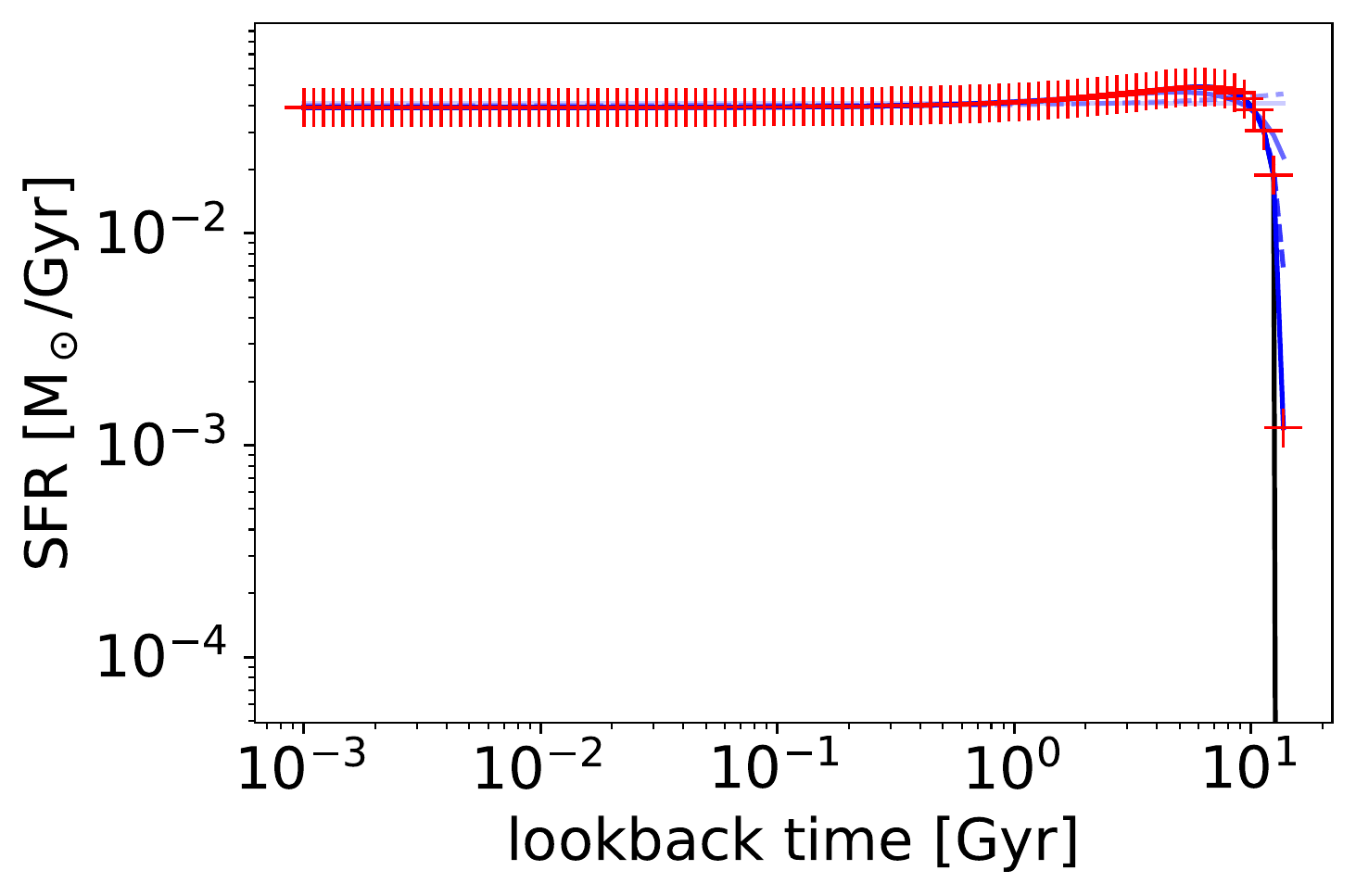} &
    \includegraphics[width=.23\textwidth]{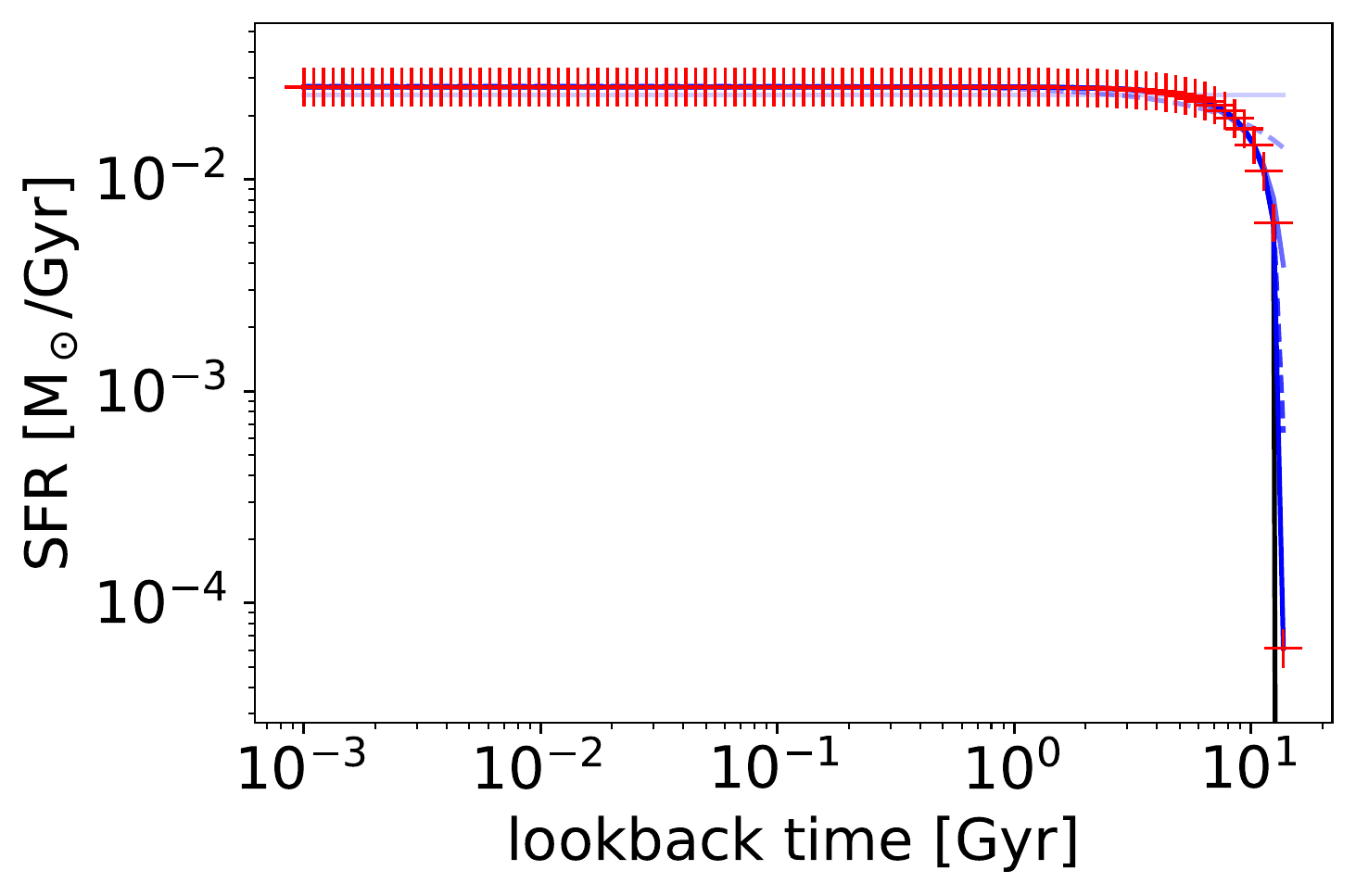}   \\
    \includegraphics[width=.23\textwidth]{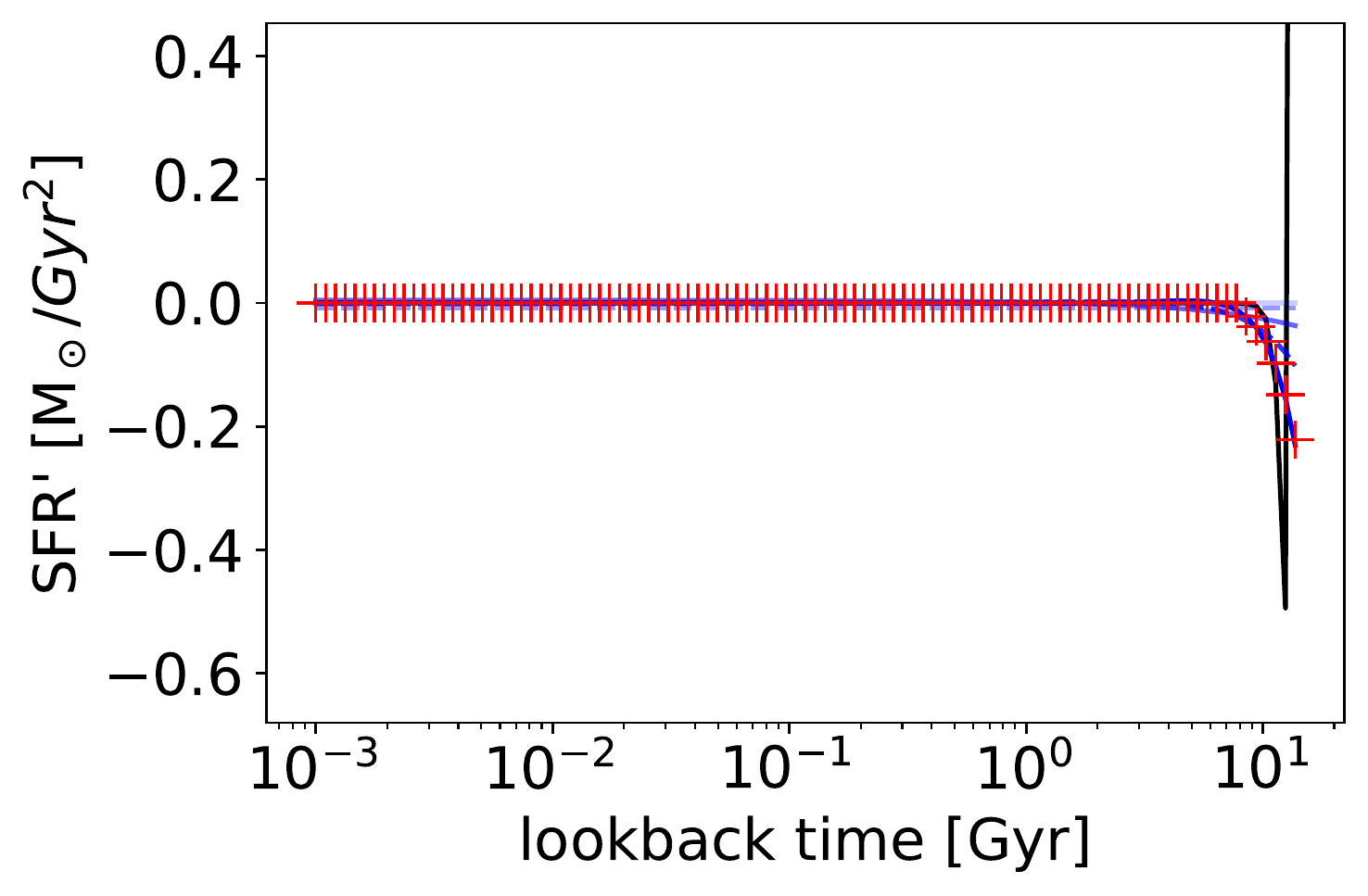} &
    \includegraphics[width=.23\textwidth]{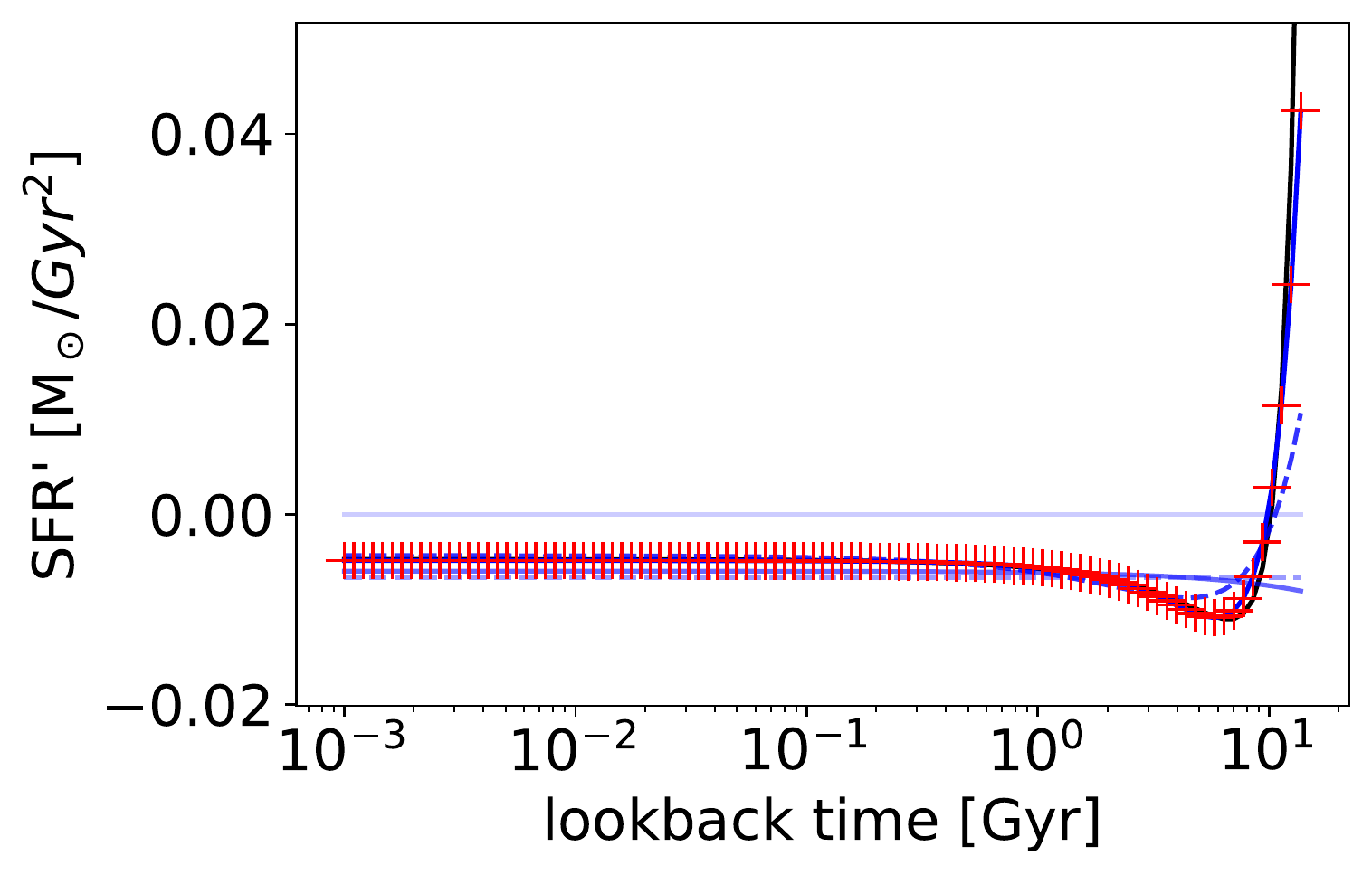} &
    \includegraphics[width=.23\textwidth]{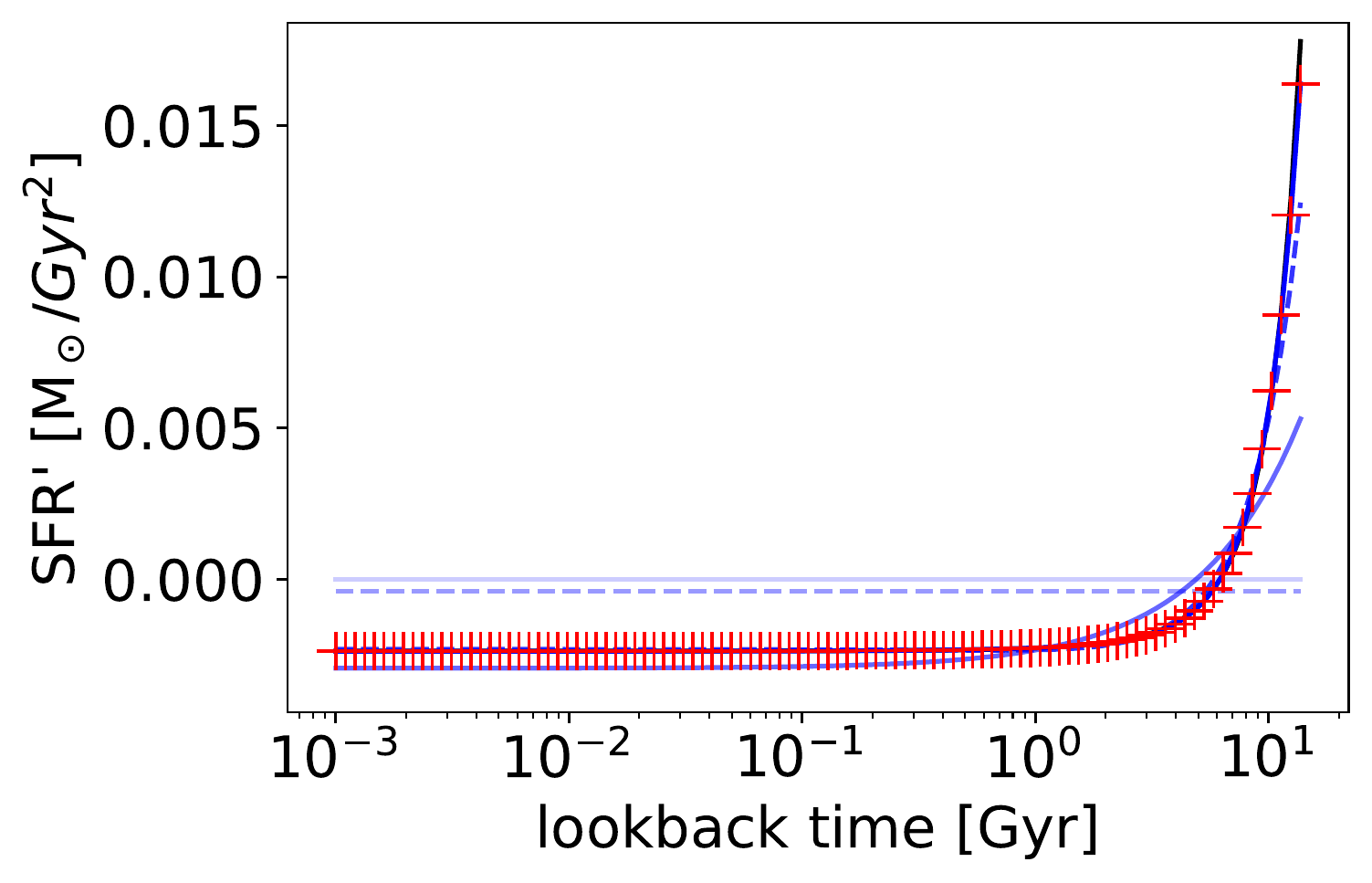} &
    \includegraphics[width=.23\textwidth]{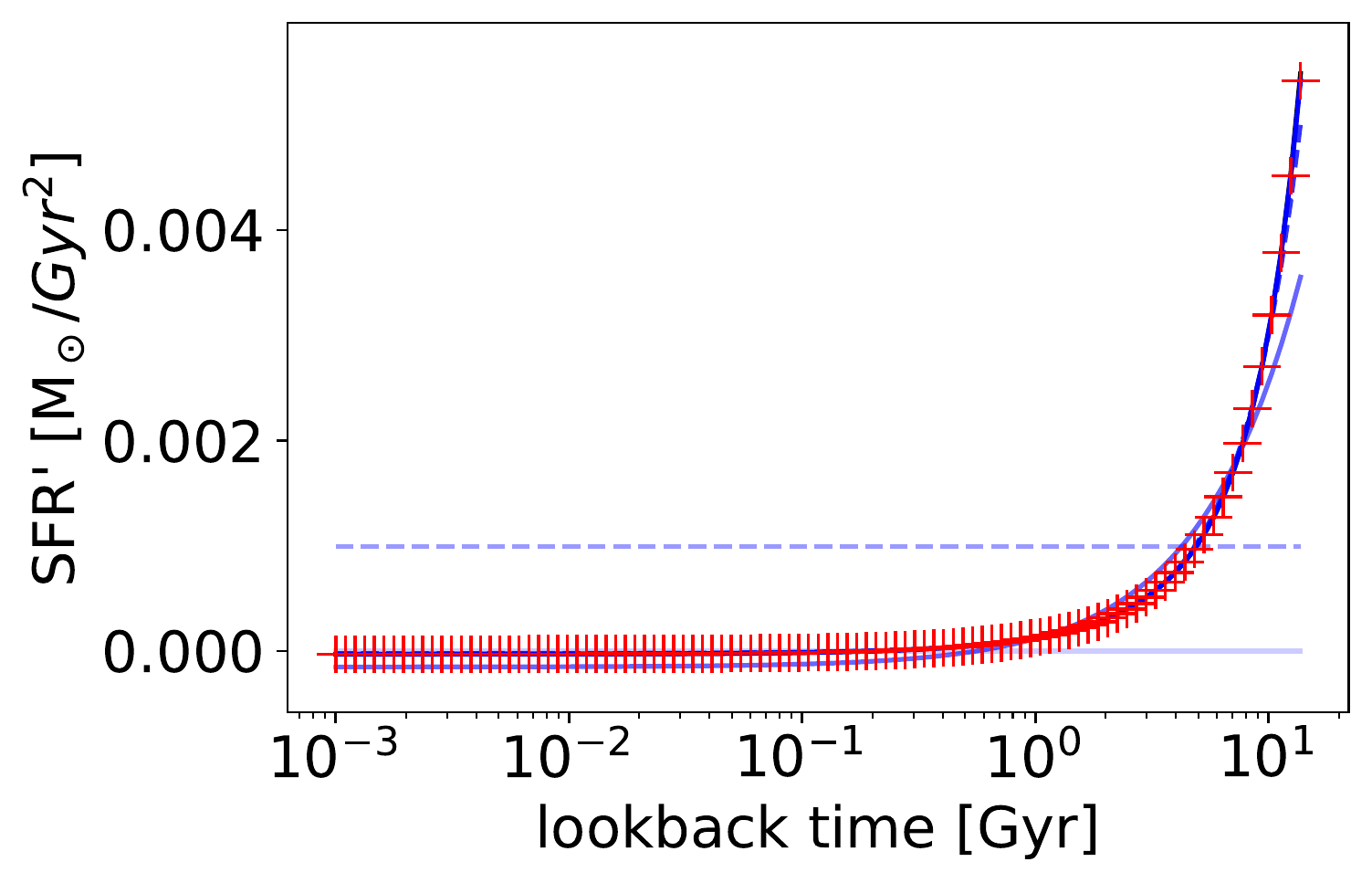}   \\
    \includegraphics[width=.23\textwidth]{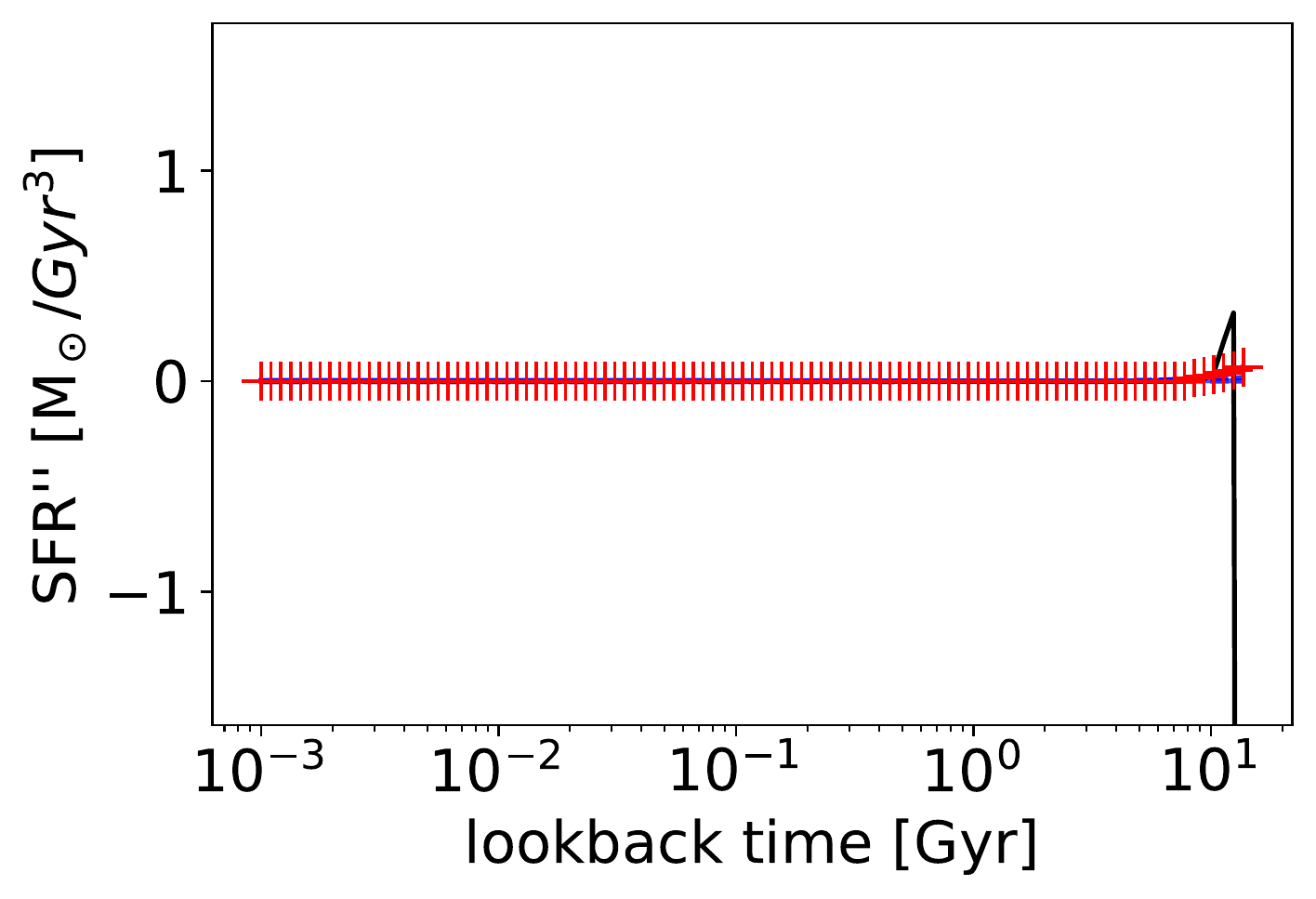} &
    \includegraphics[width=.23\textwidth]{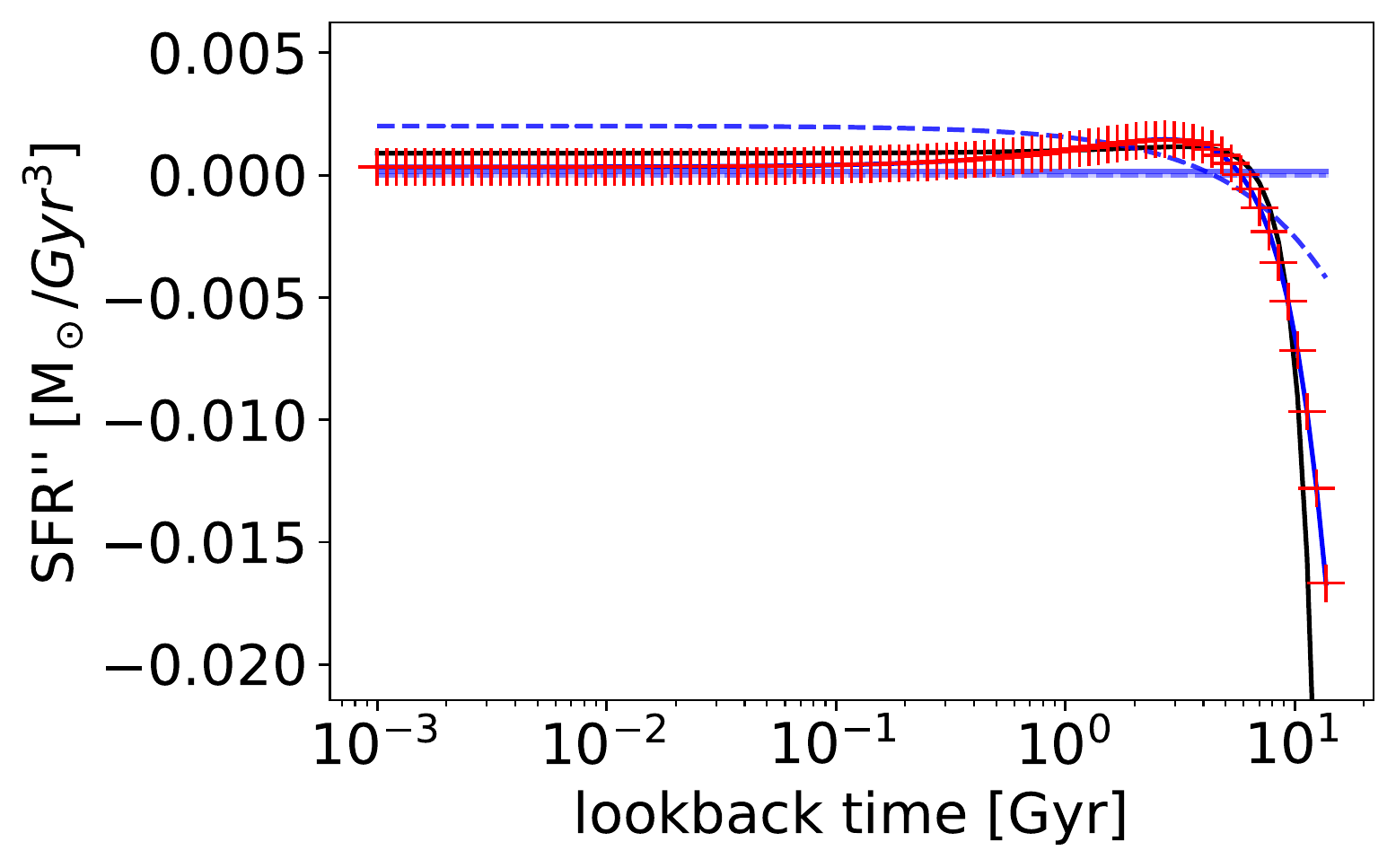} &
    \includegraphics[width=.23\textwidth]{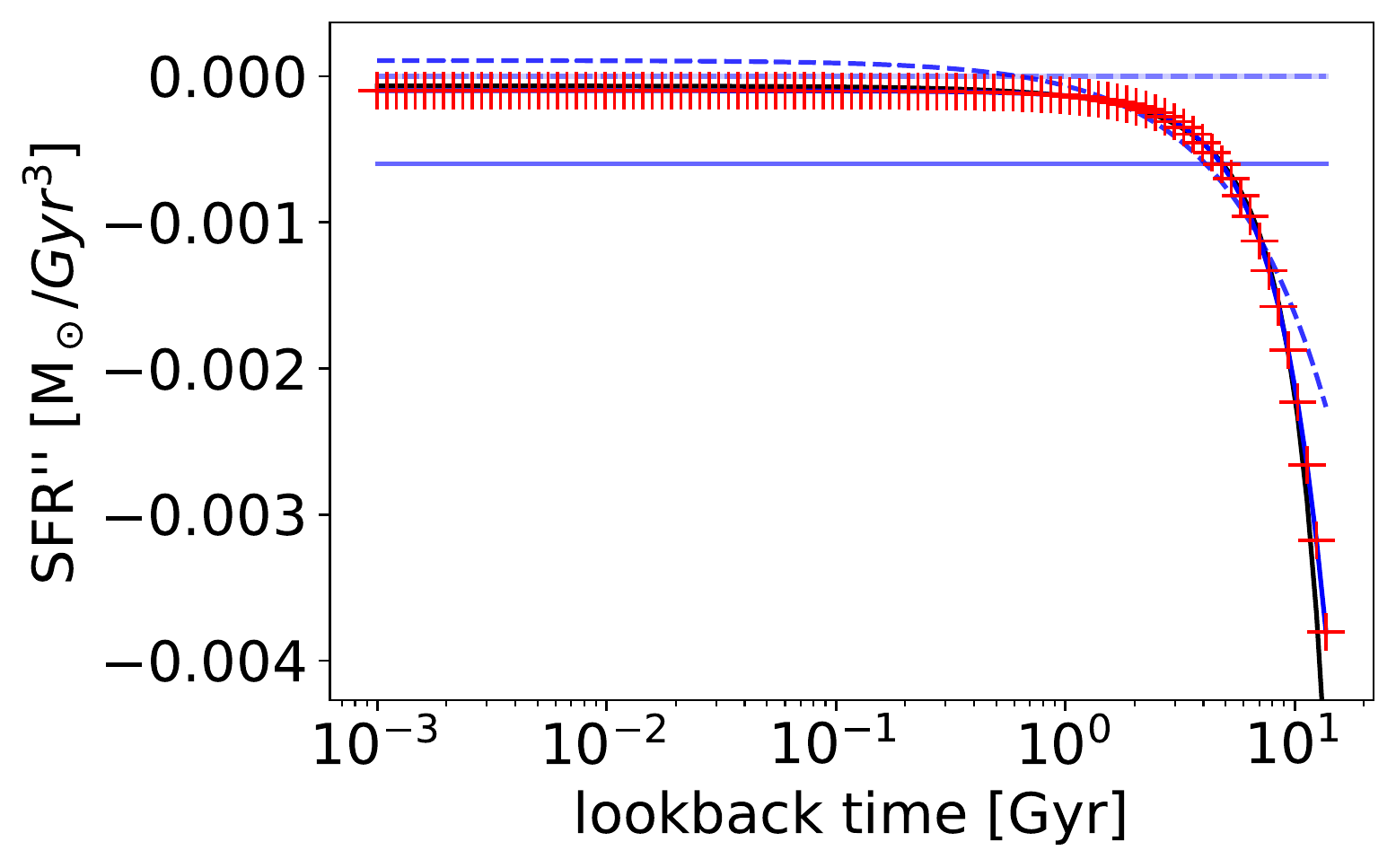} &
    \includegraphics[width=.23\textwidth]{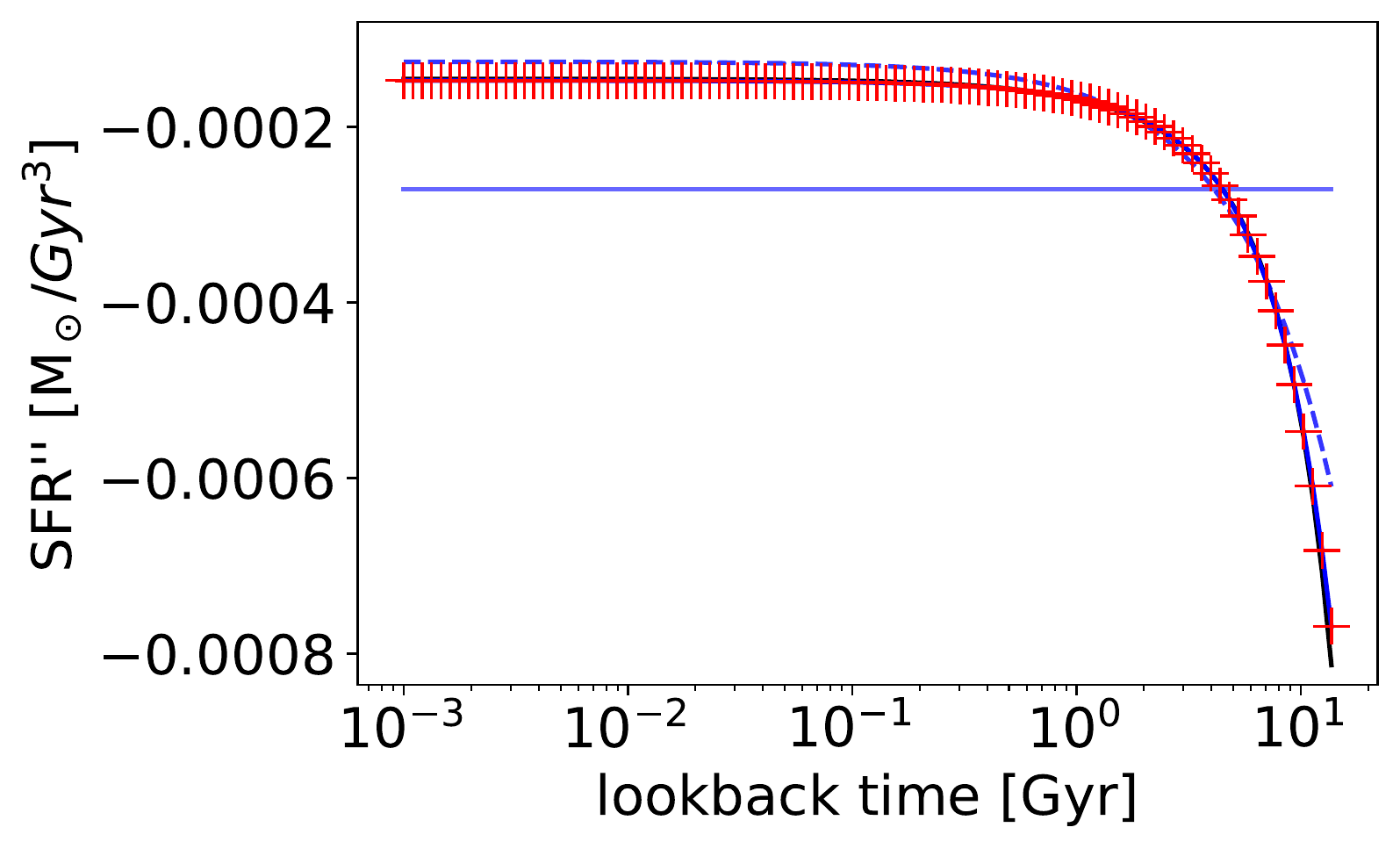}     
  \end{tabular}
  \caption{Sample of MFHs reconstructed with our parametric model from the luminosities of a synthetic exponential-delayed SFR with different timescales, $\tau$. Using blue lines, we plot the stellar mass formed (first row), the SFR (second row), the SFR first derivative (third row), and the SFR second derivative (fourth row), compared to the input model (solid black line). Colour gradients and line styles indicate the degree of the polynomial reconstruction from $N=5$ (solid) to lower degrees (more diffuse lines; solid and dashed for odd and even $N$, respectively). Red crosses correspond to the best positive-SFR fit.}
  \label{fig:rec_dt}
\end{figure*}

\begin{figure*}
  \centering
  \begin{tabular}{@{}ccccc@{}}
  FWHM=0.5~Gyr & FWHM=1.0~Gyr & FWHM=3.0~Gyr & FWHM=10.0~Gyr \\
    \turnbox{90}{\hspace{5 mm} Age = 0.5~Gyr}
    \includegraphics[width=.23\textwidth]{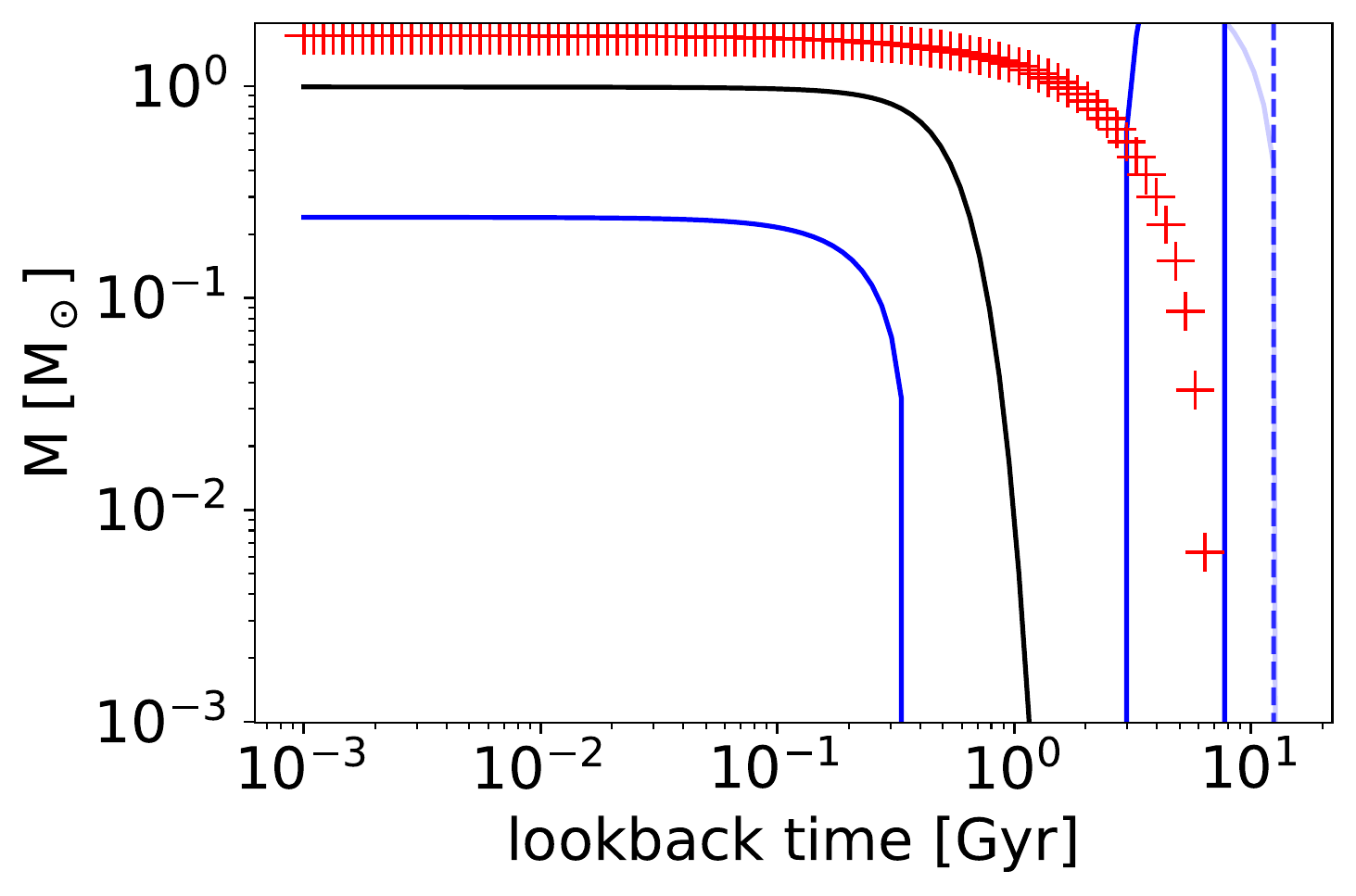} &
    \includegraphics[width=.23\textwidth]{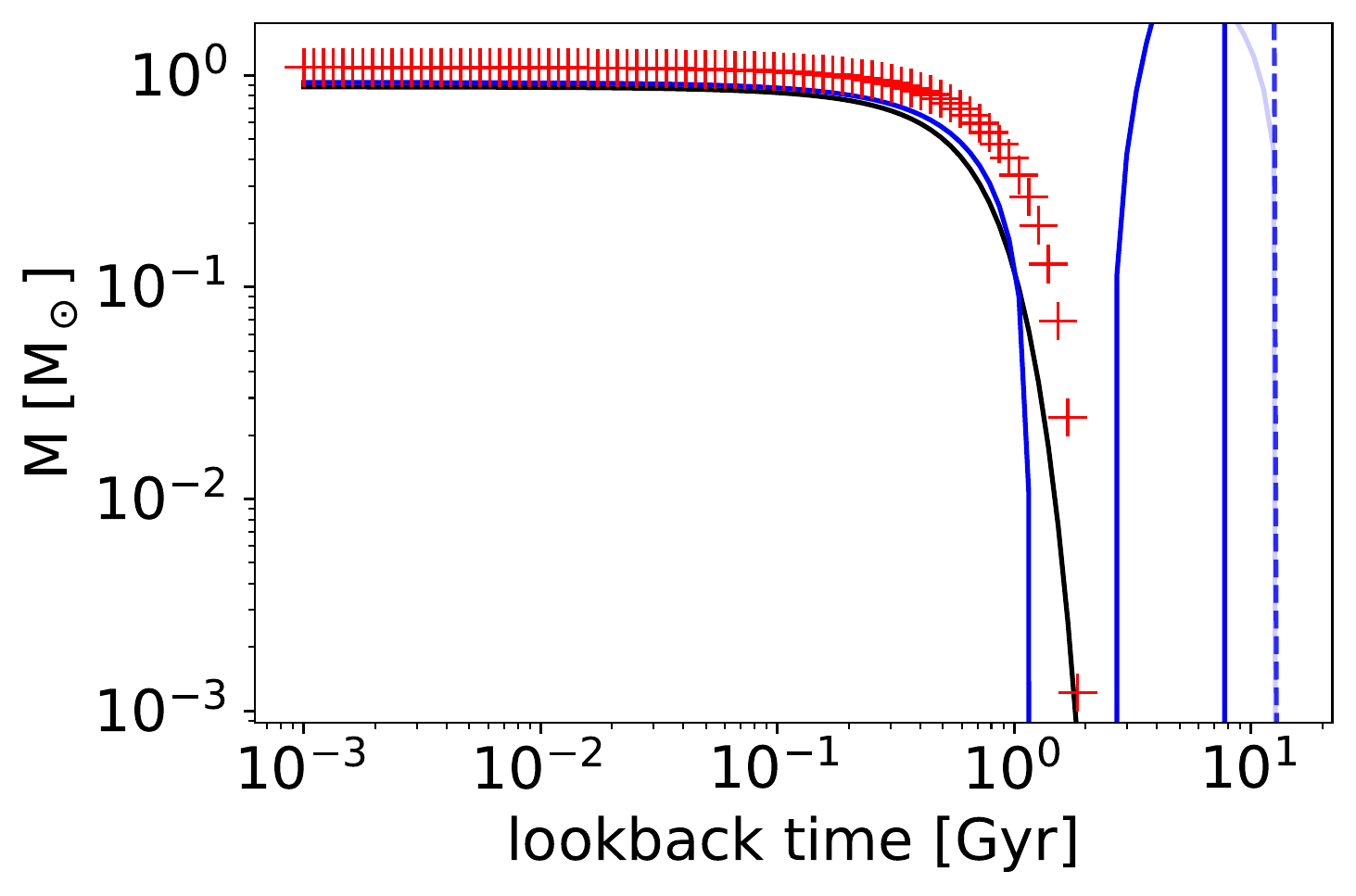} &
    \includegraphics[width=.23\textwidth]{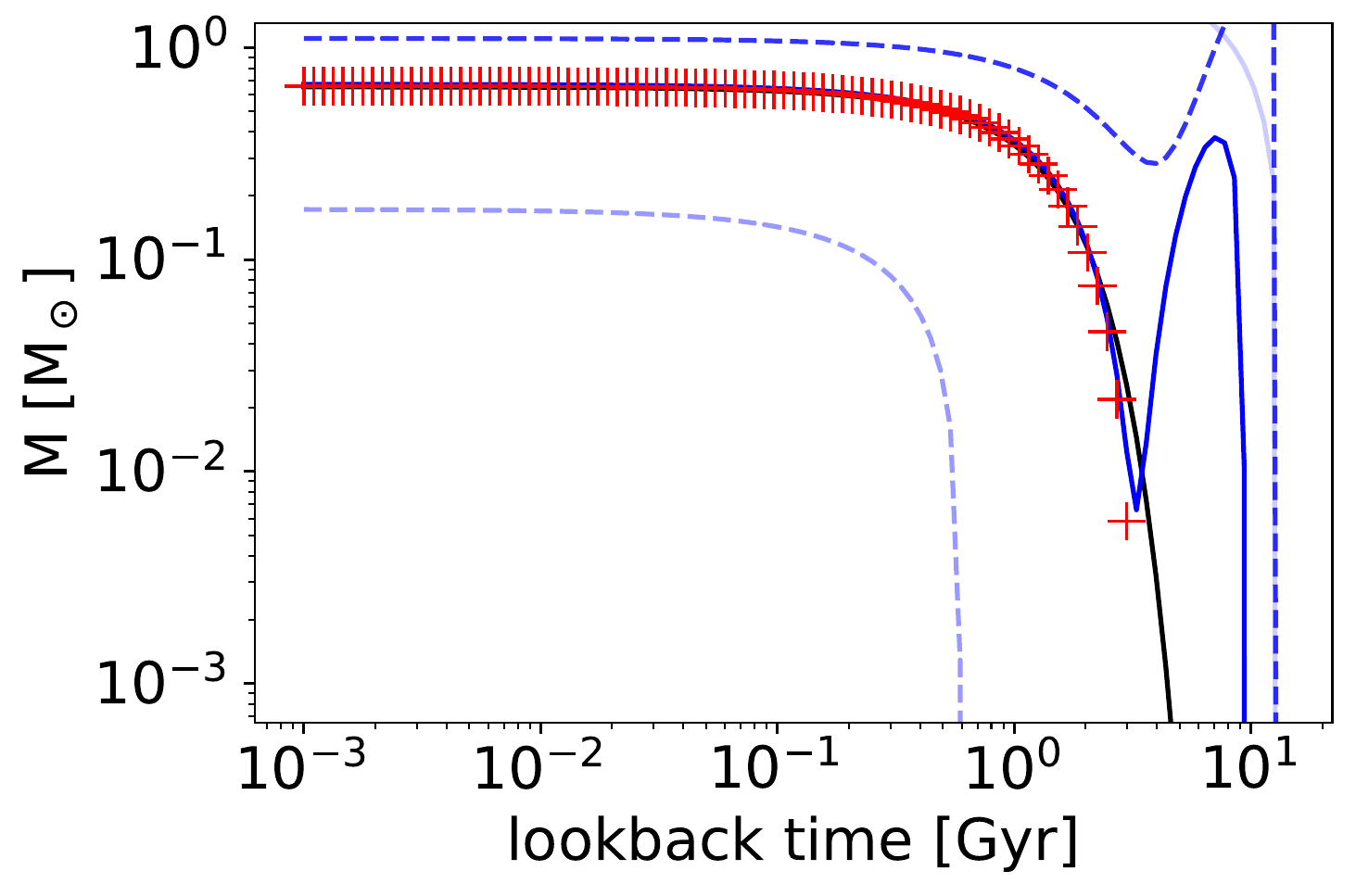} &
    \includegraphics[width=.23\textwidth]{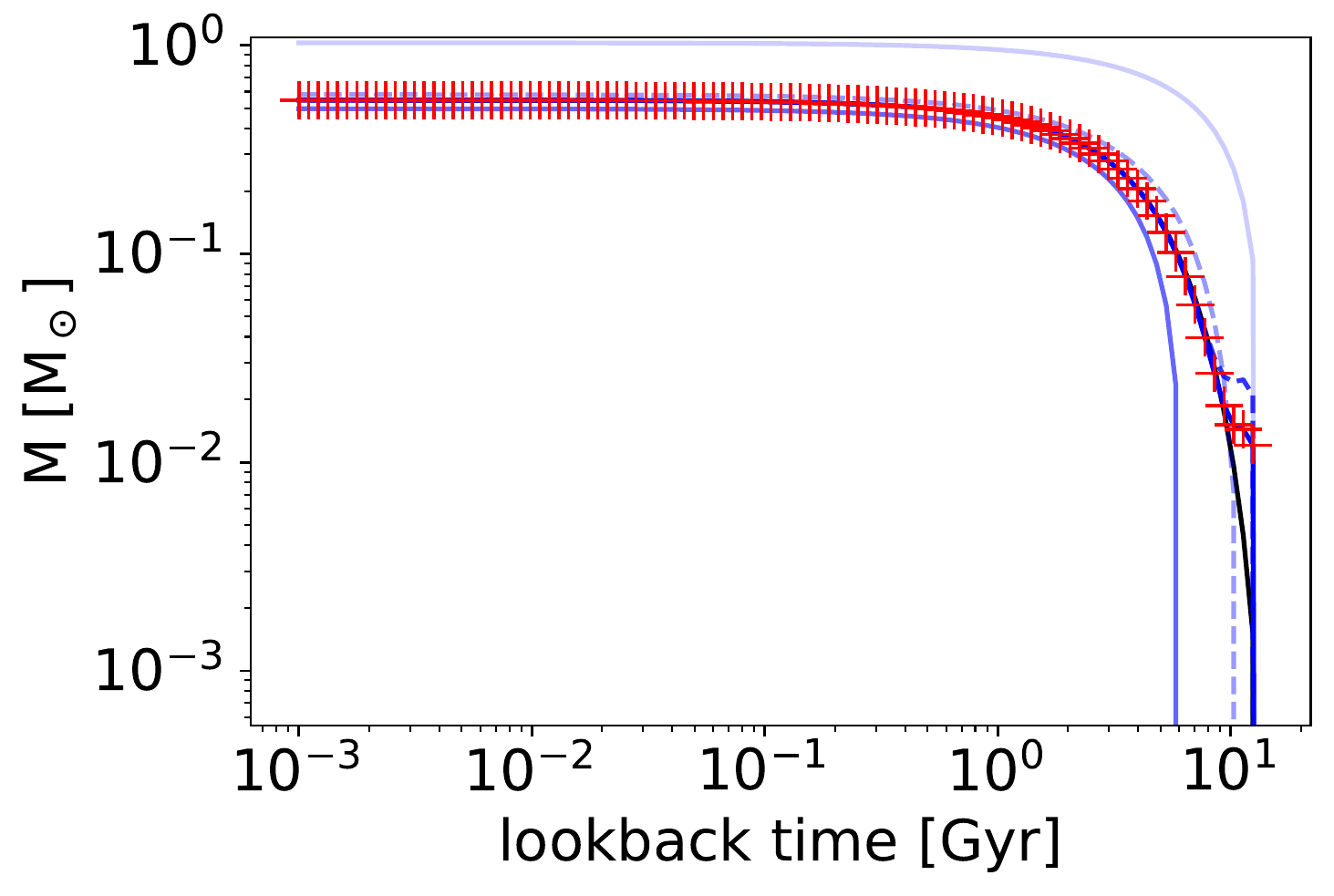} \\
    \turnbox{90}{\hspace{5 mm} Age = 1.0~Gyr}
    \includegraphics[width=.23\textwidth]{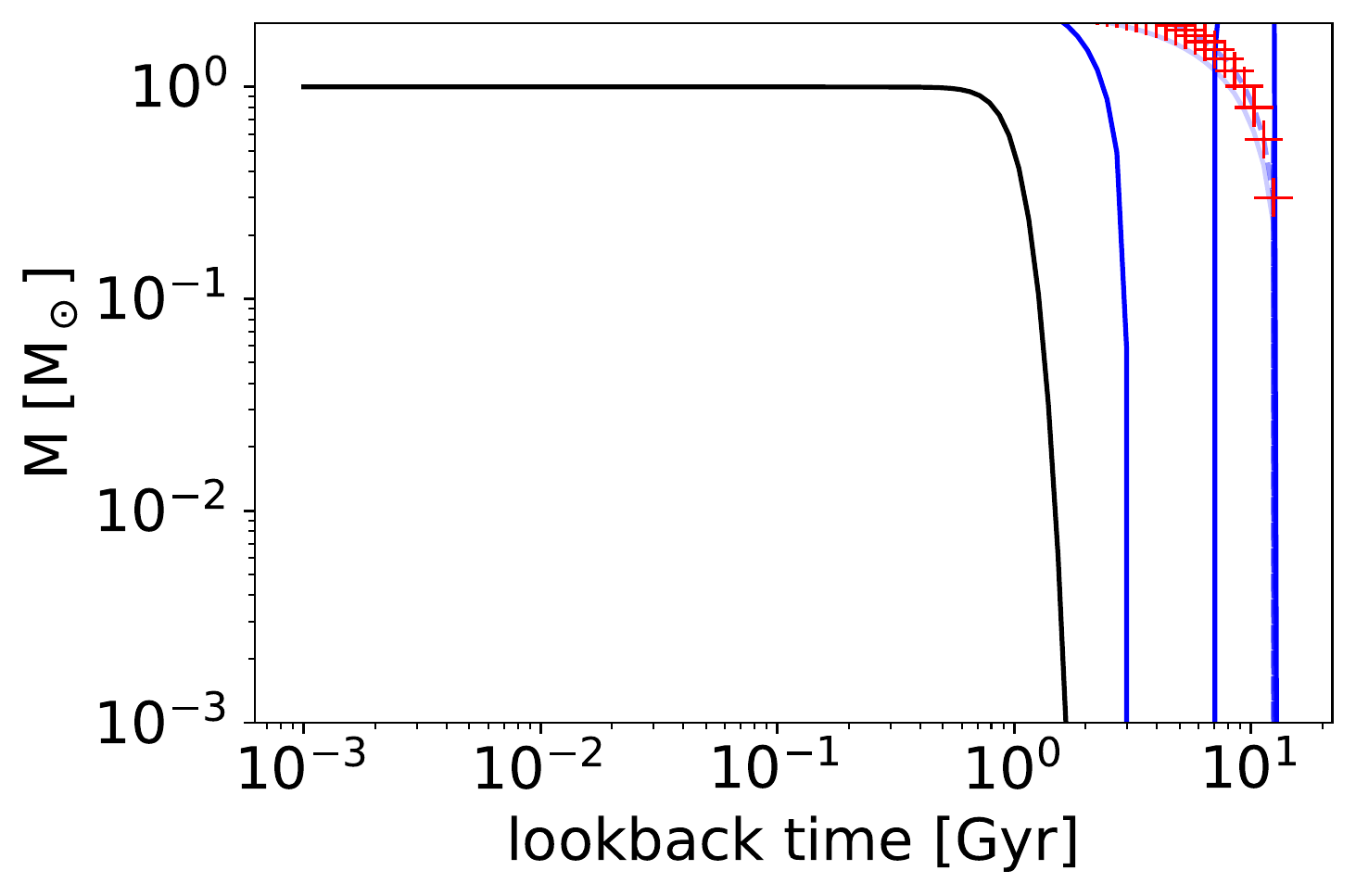} &
    \includegraphics[width=.23\textwidth]{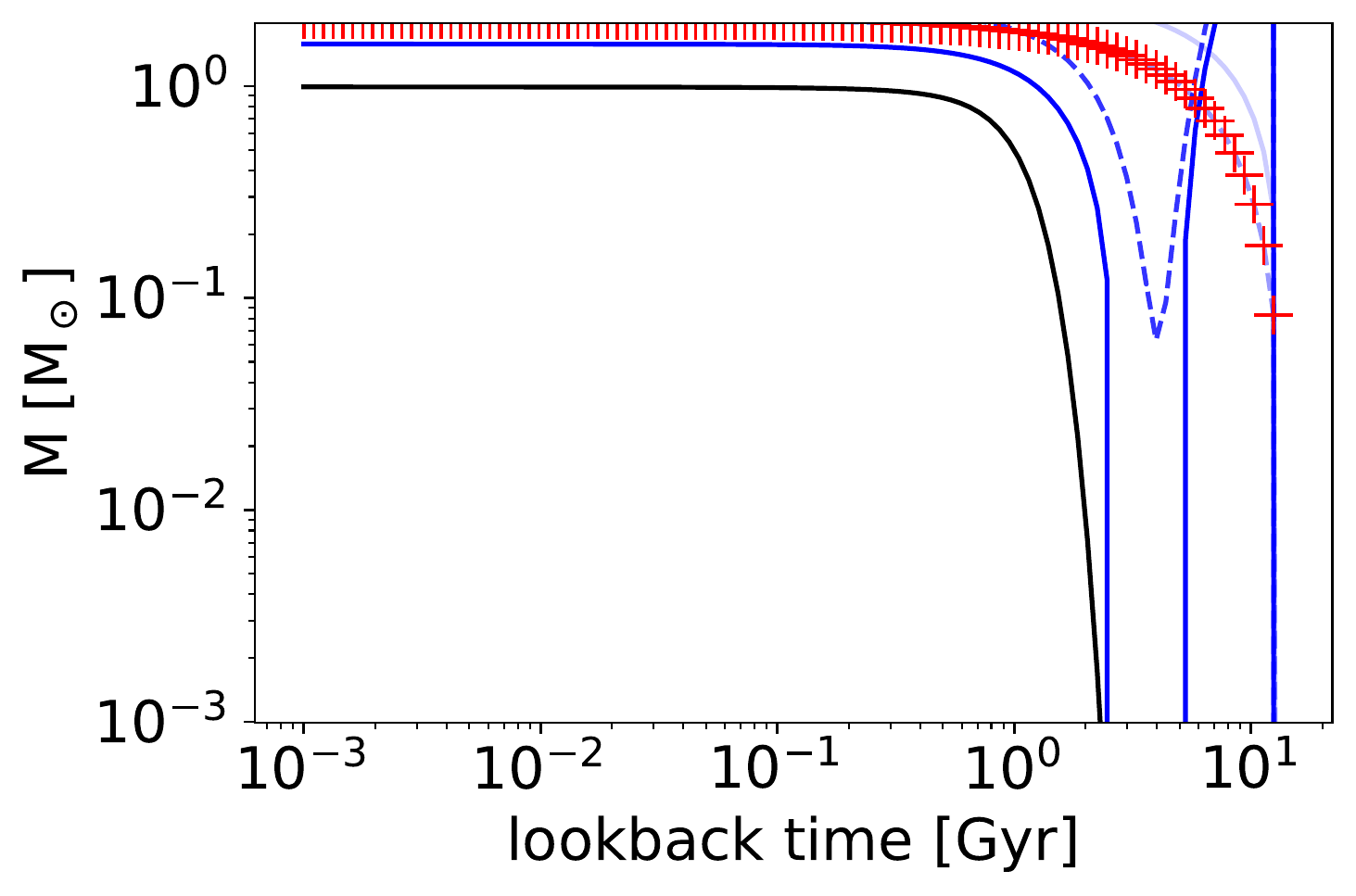} &
    \includegraphics[width=.23\textwidth]{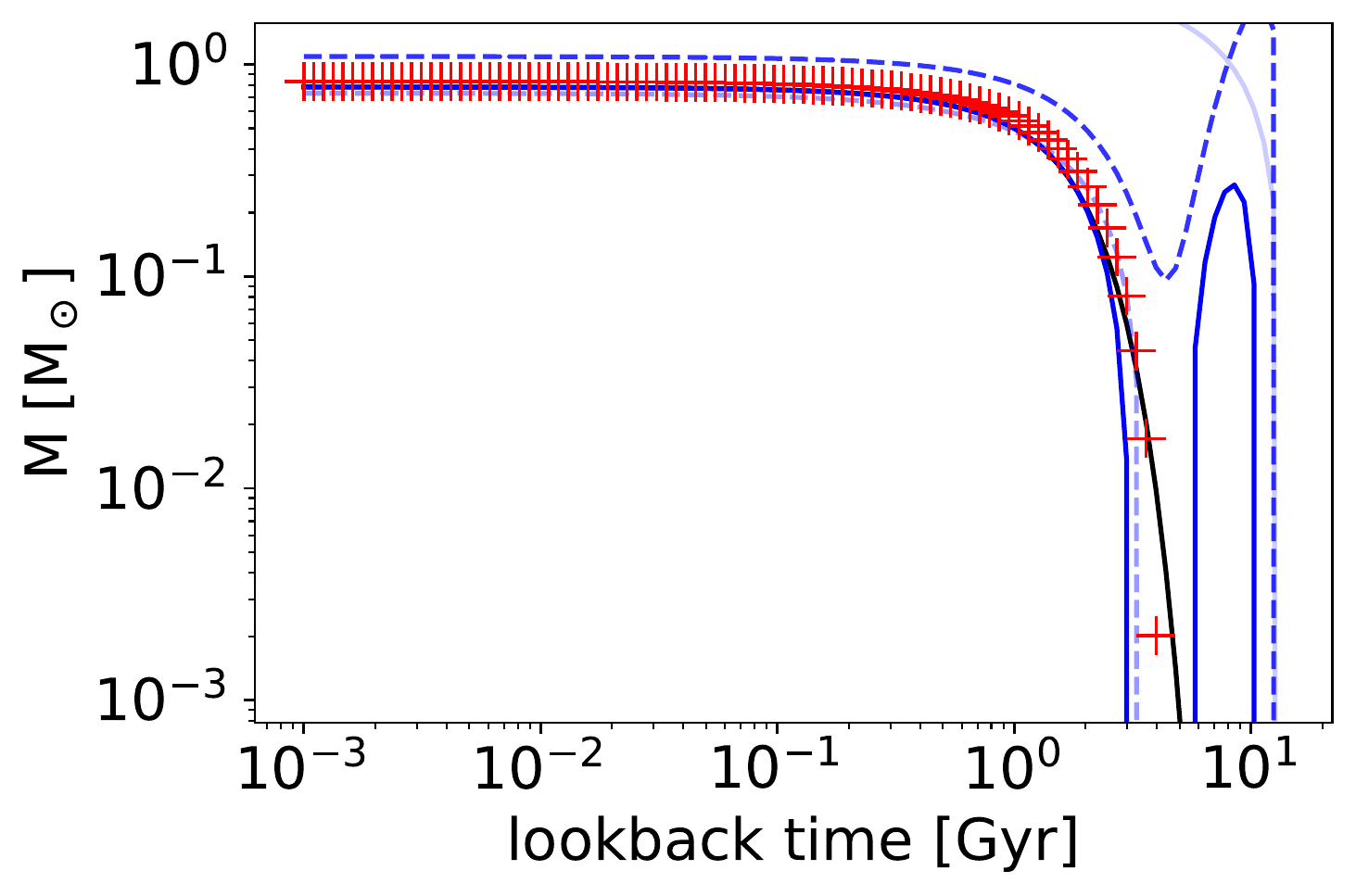} &
    \includegraphics[width=.23\textwidth]{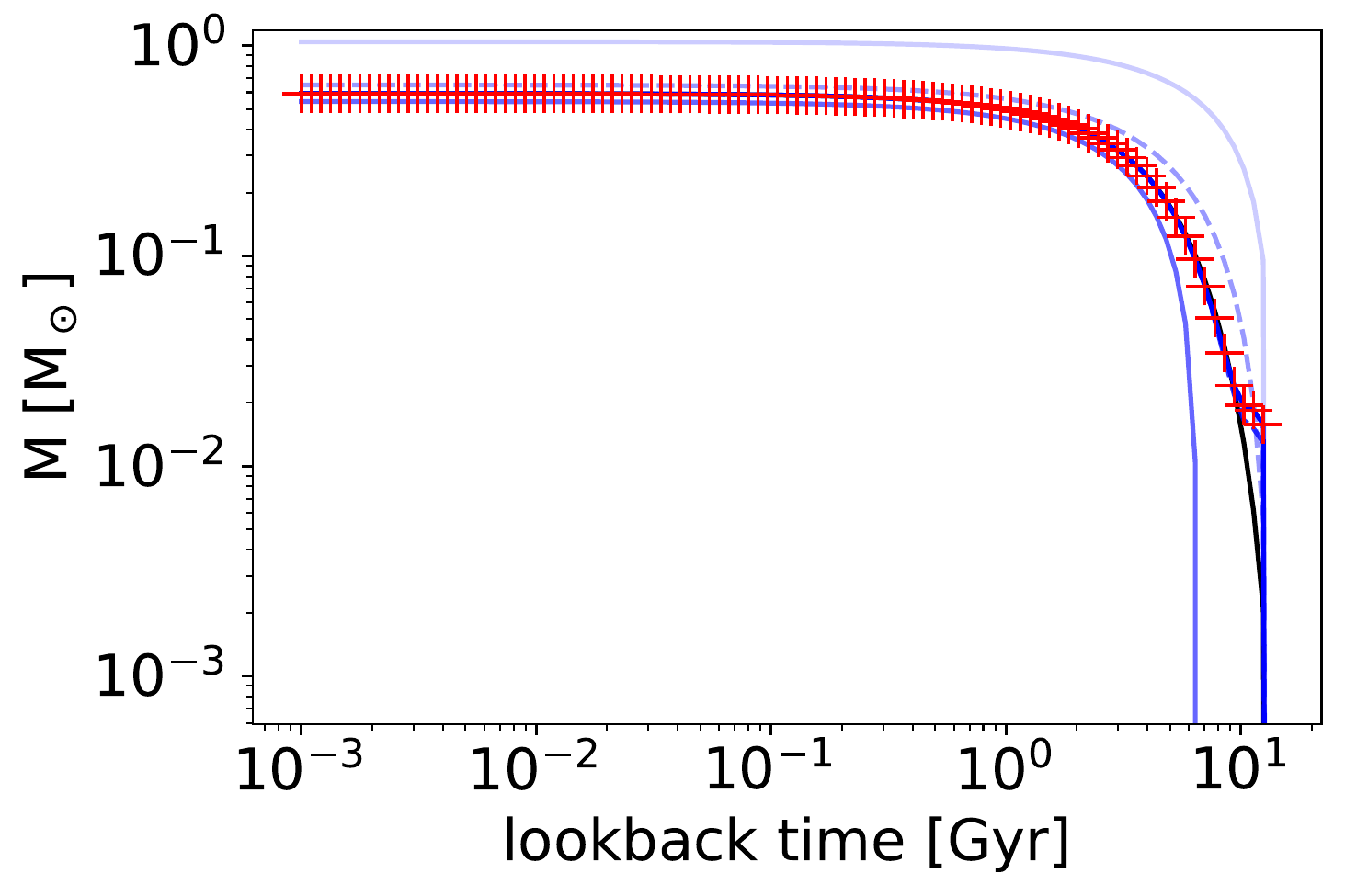} \\
    \turnbox{90}{\hspace{5 mm} Age = 3.0~Gyr}
    \includegraphics[width=.23\textwidth]{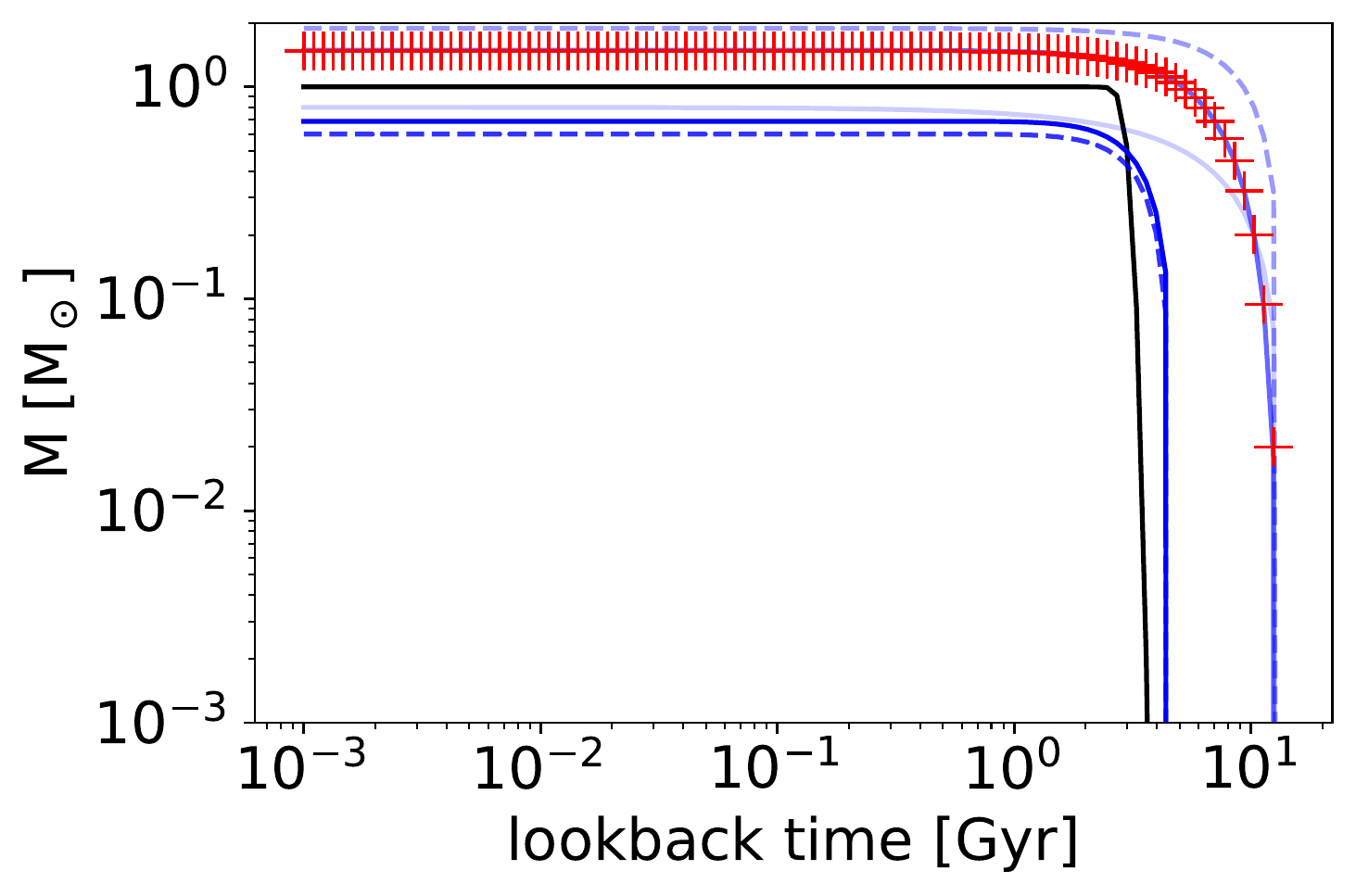} &
    \includegraphics[width=.23\textwidth]{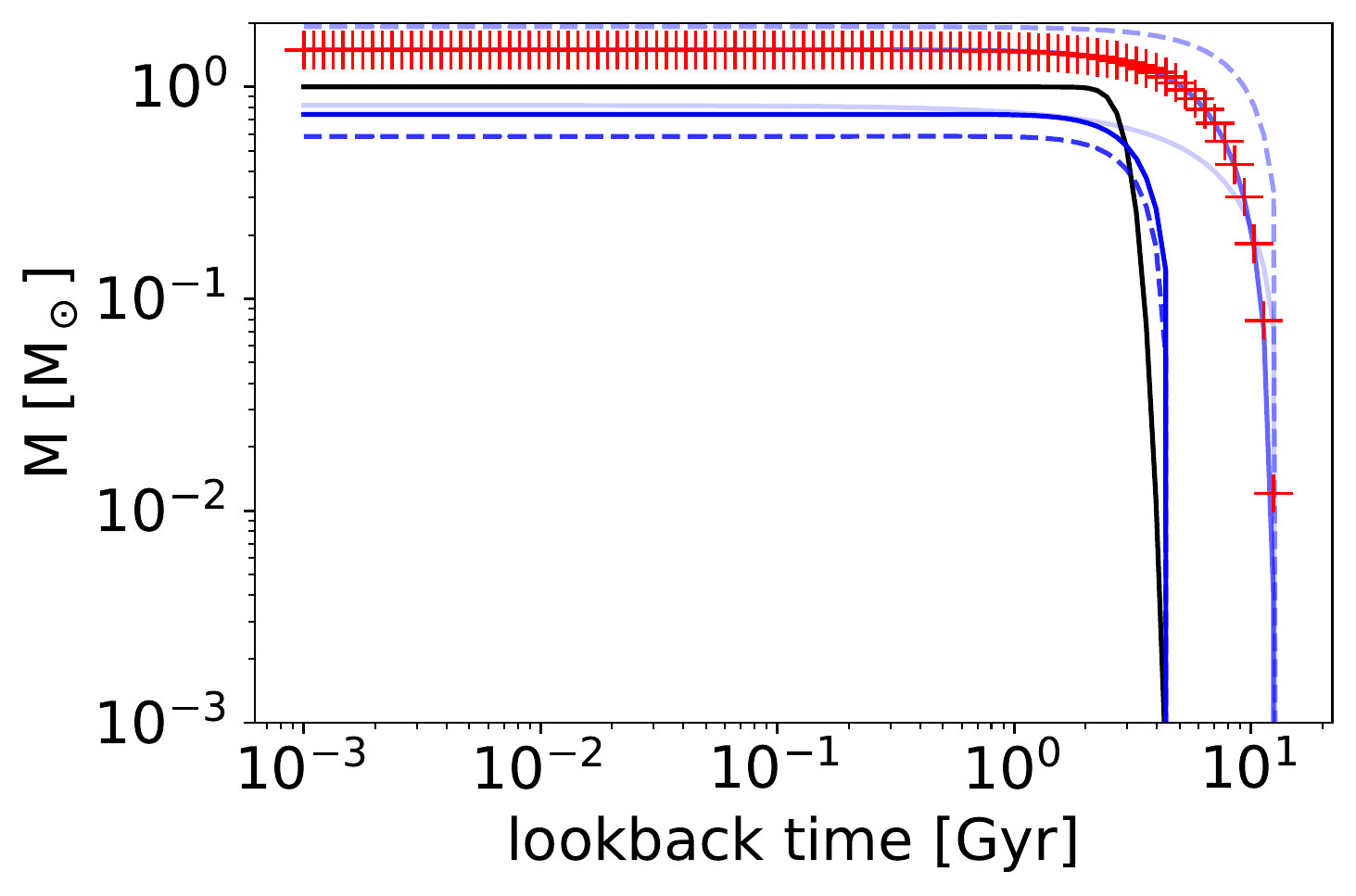} &
    \includegraphics[width=.23\textwidth]{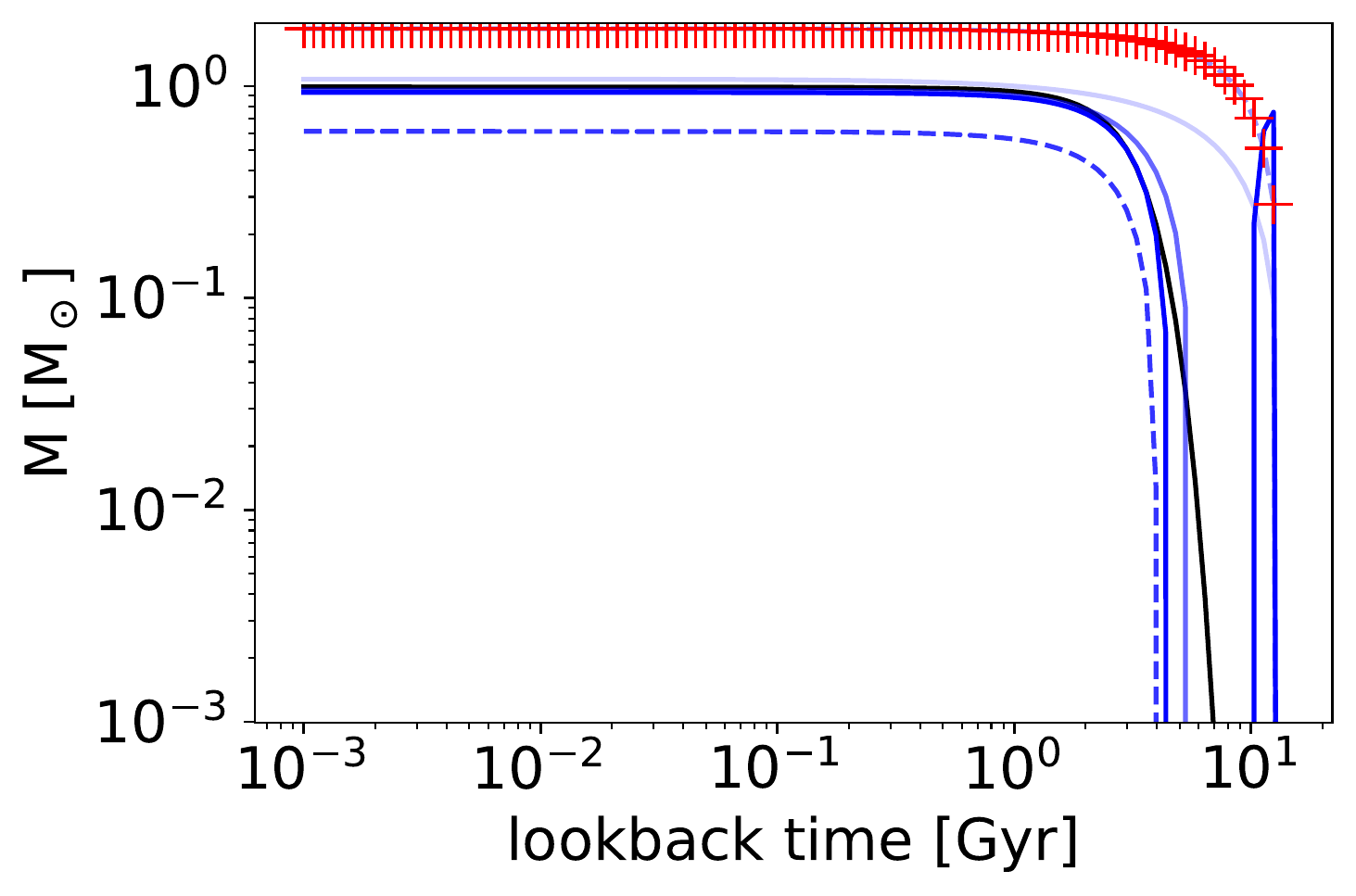} &
    \includegraphics[width=.23\textwidth]{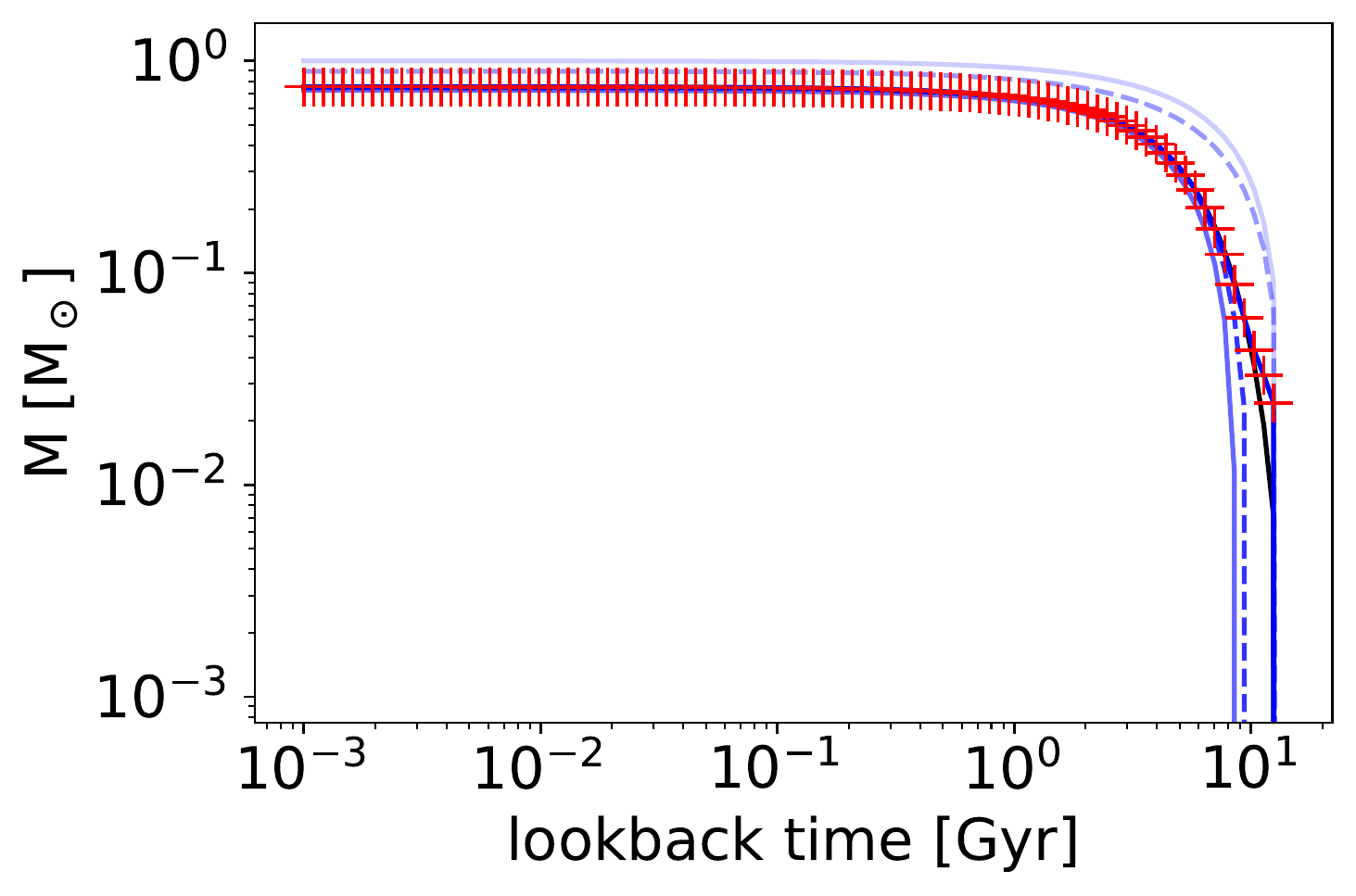} \\
    \turnbox{90}{\hspace{5 mm} Age = 10.0~Gyr}
    \includegraphics[width=.23\textwidth]{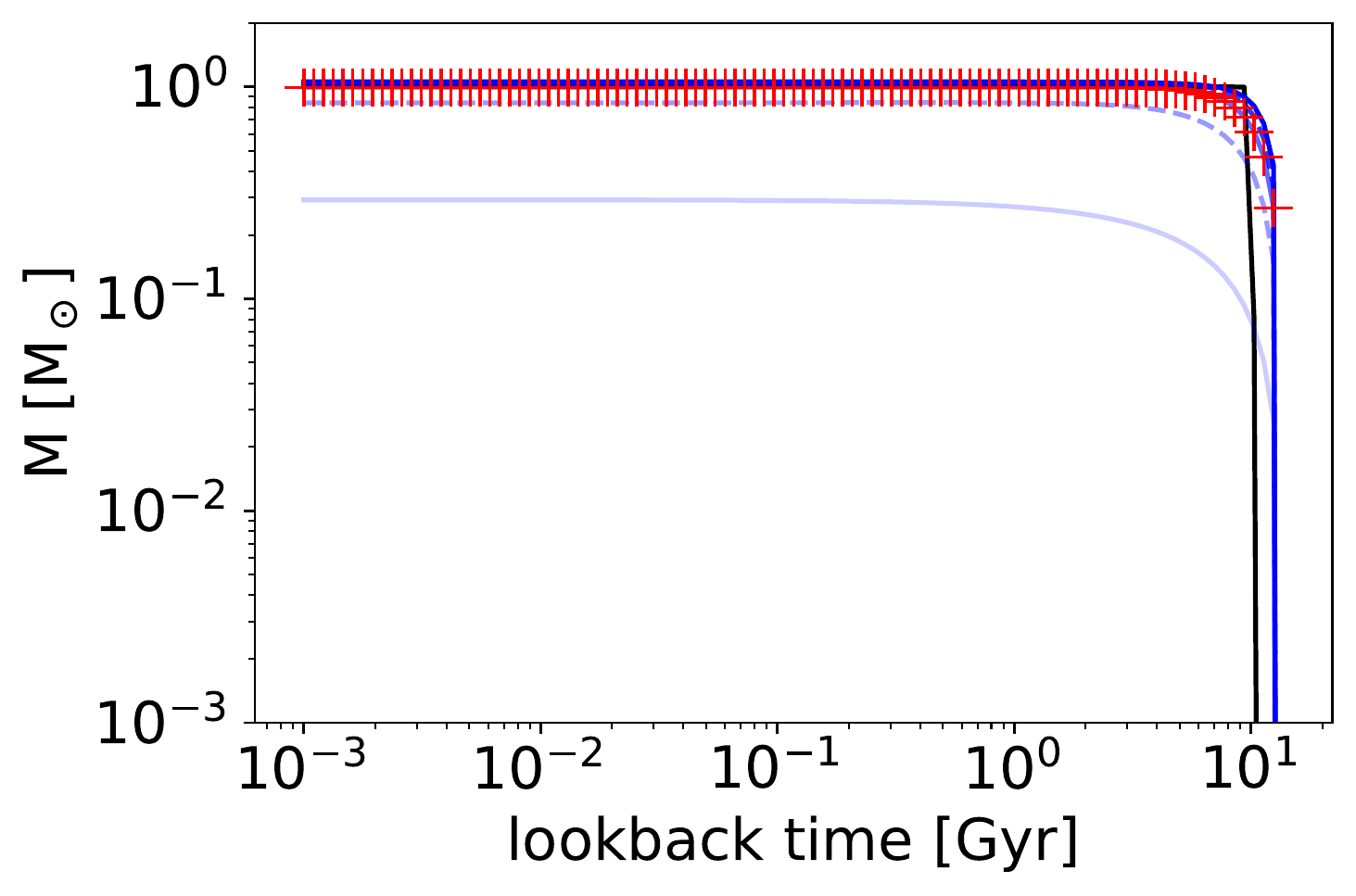} &
    \includegraphics[width=.23\textwidth]{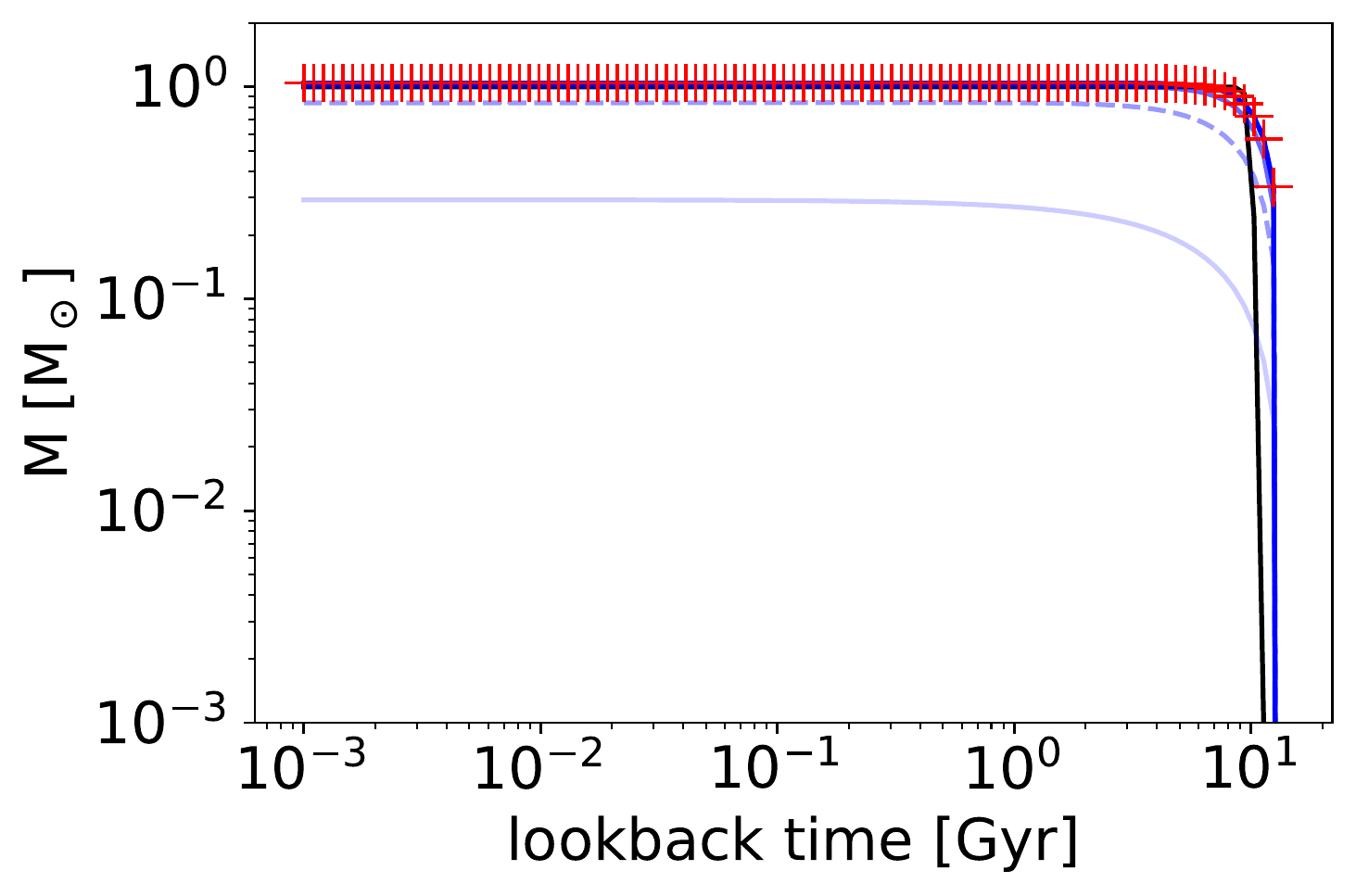} &
    \includegraphics[width=.23\textwidth]{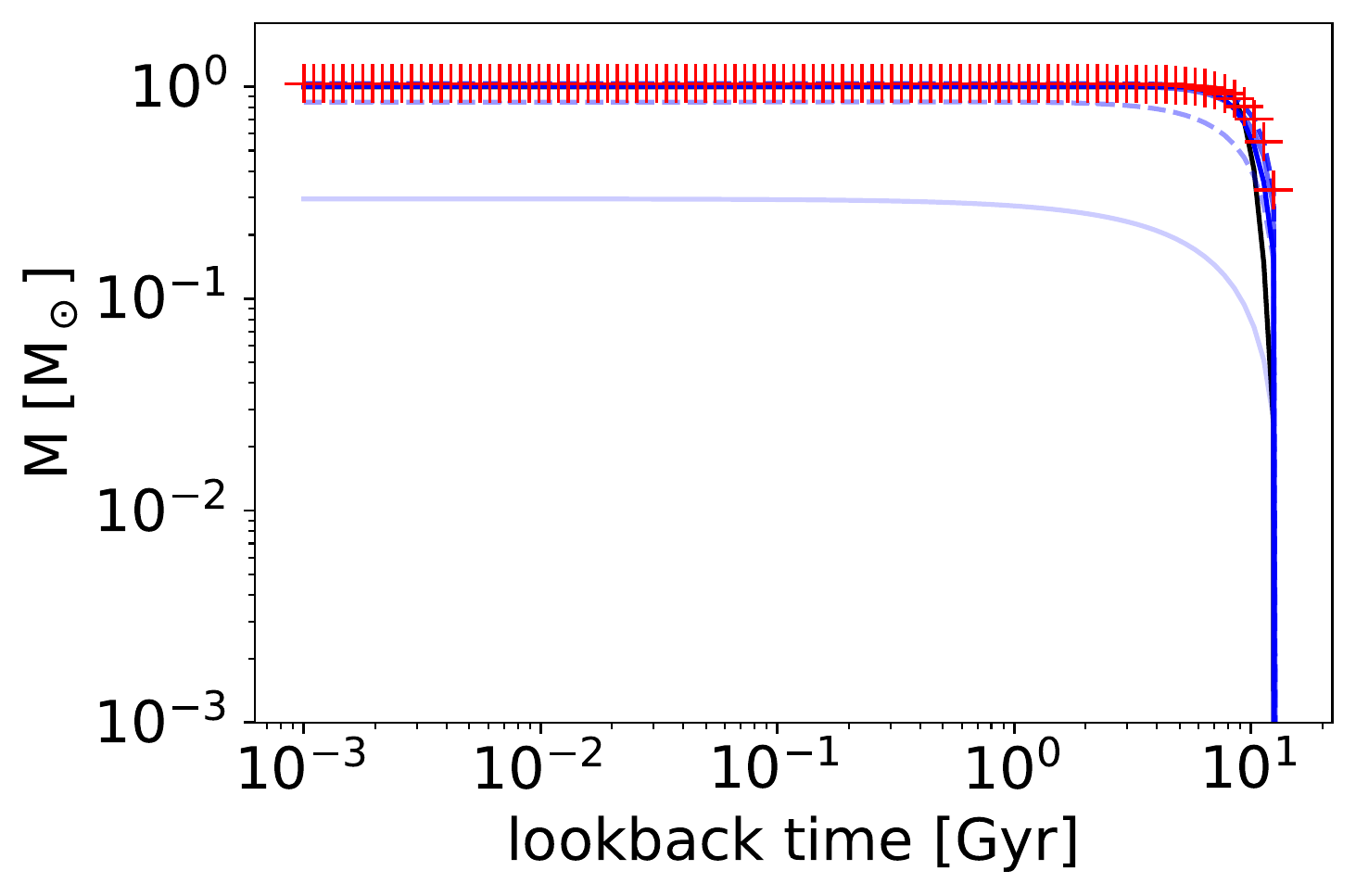} &
    \includegraphics[width=.23\textwidth]{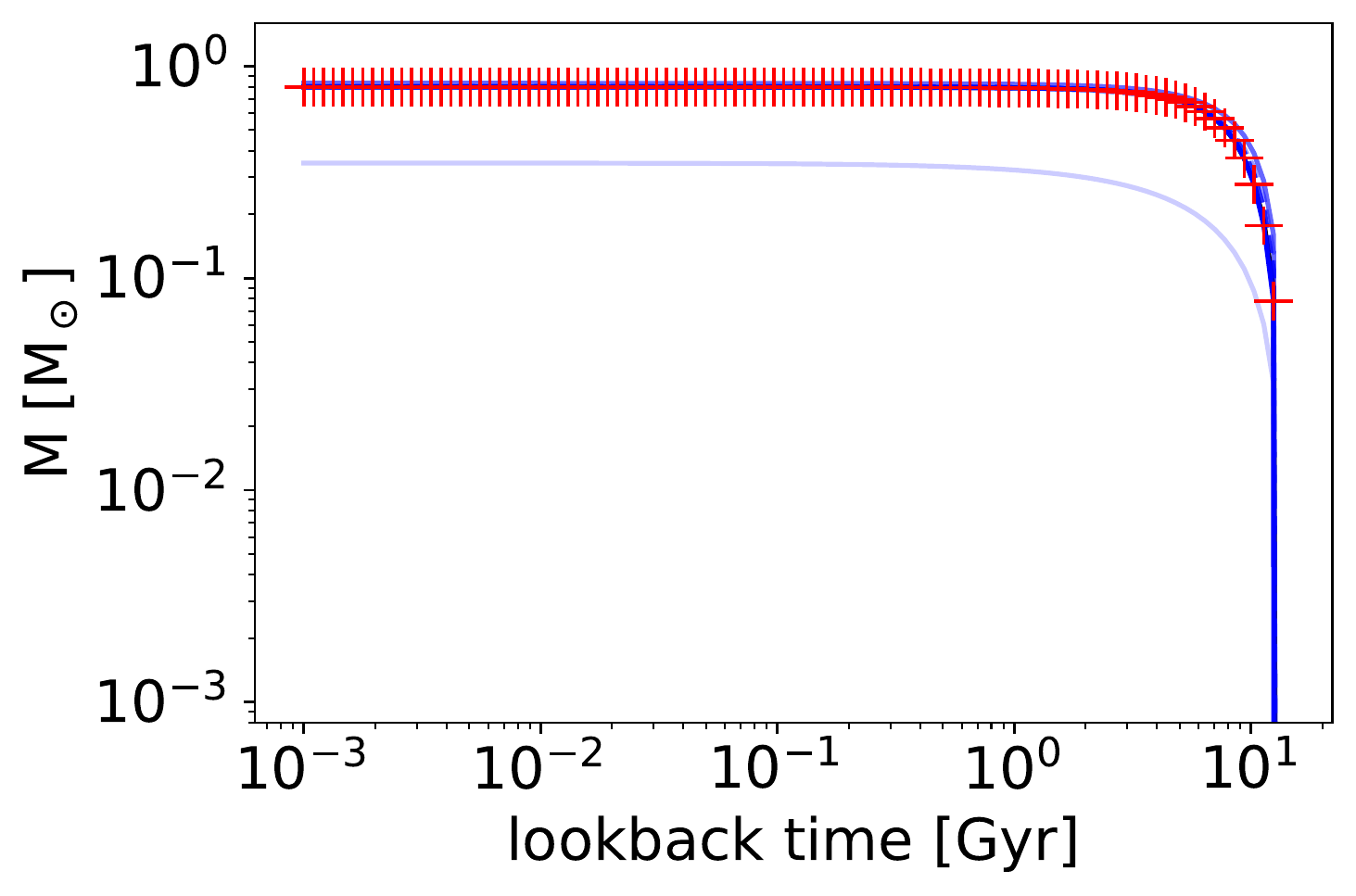} 
  \end{tabular}
  \caption{Mass formation history reconstructed with our parametric model from the luminosities of a synthetic Gaussian SFR with different peak ages (increasing from top to bottom) and FWHMs (increasing from left to right), compared to the input model (solid black line). Colour gradients and line styles indicate the degree of the polynomial reconstruction from $N=5$ (solid) to lower degrees (more diffuse lines; solid and dashed for odd and even $N$). Red crosses correspond to the best positive-SFR fit.}
  \label{fig:Gauss_rec_mass}
\end{figure*}

\begin{figure*}
\centering
  \begin{tabular}{@{}cccc@{}}
  FWHM=0.5~Gyr & FWHM=1.0~Gyr & FWHM=3.0~Gyr & FWHM=10.0~Gyr \\
  \turnbox{90}{\hspace{5 mm} Age = 0.5~Gyr}
    \includegraphics[width=.23\textwidth]{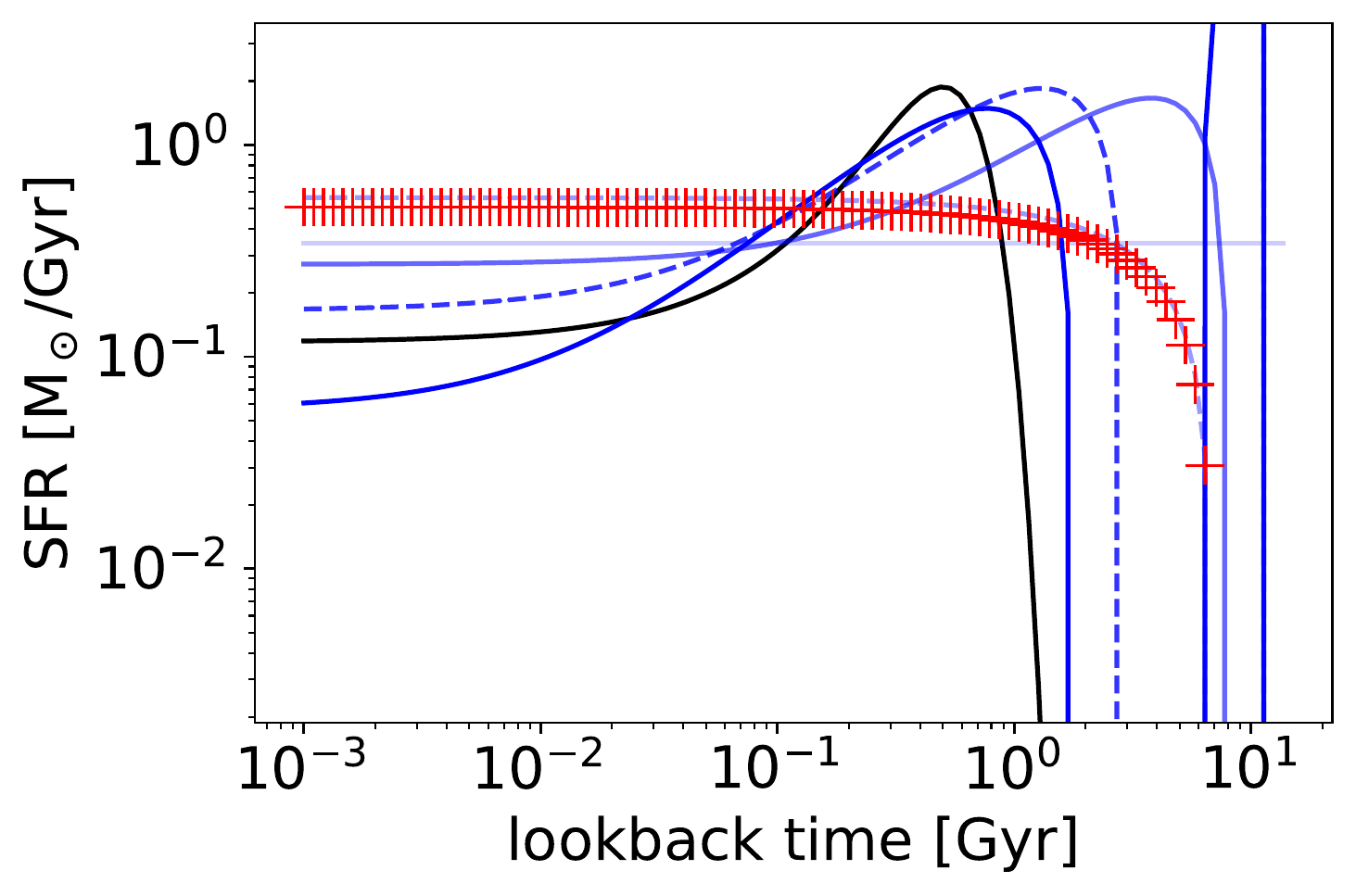} &
    \includegraphics[width=.23\textwidth]{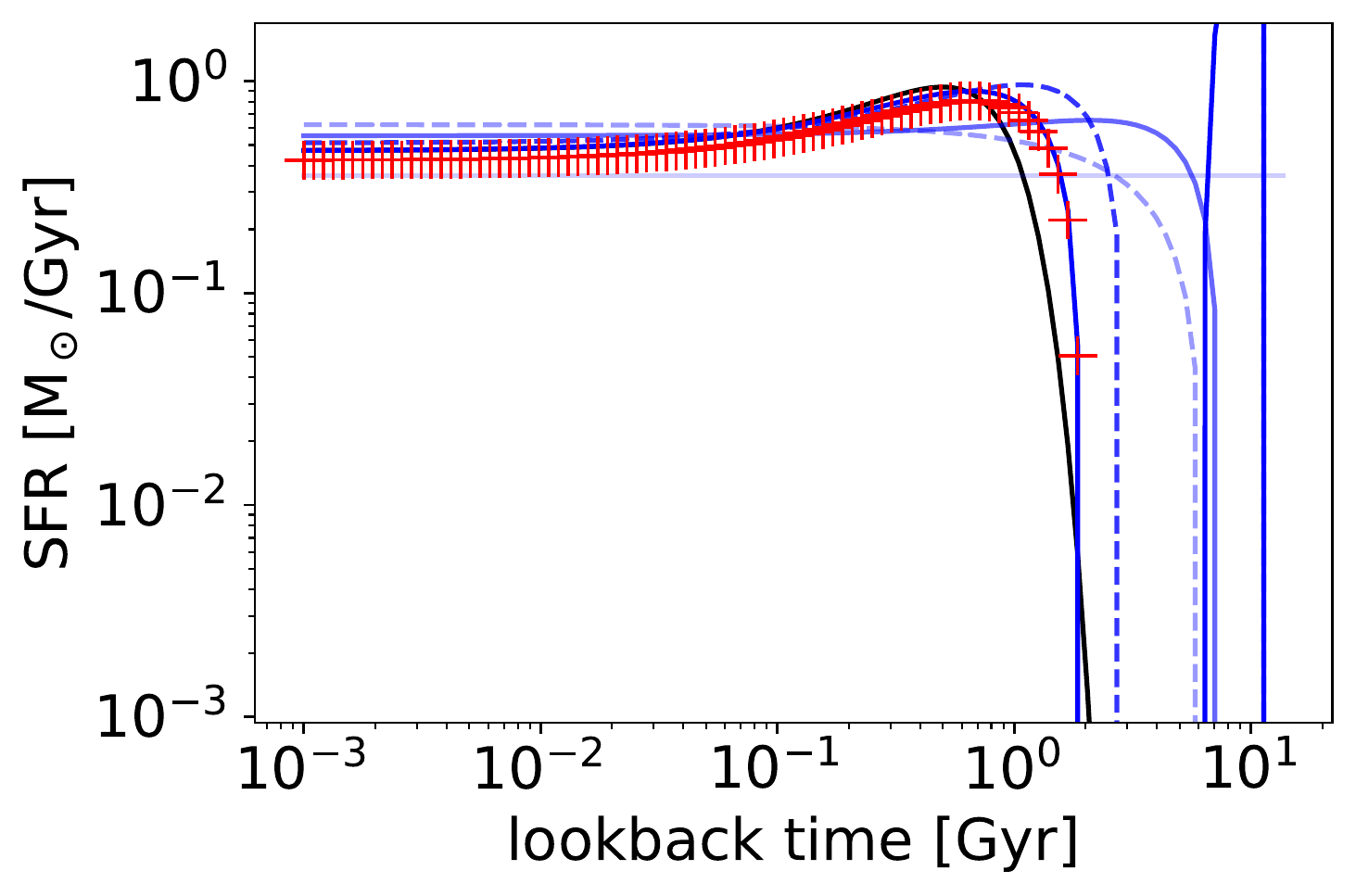} &
    \includegraphics[width=.23\textwidth]{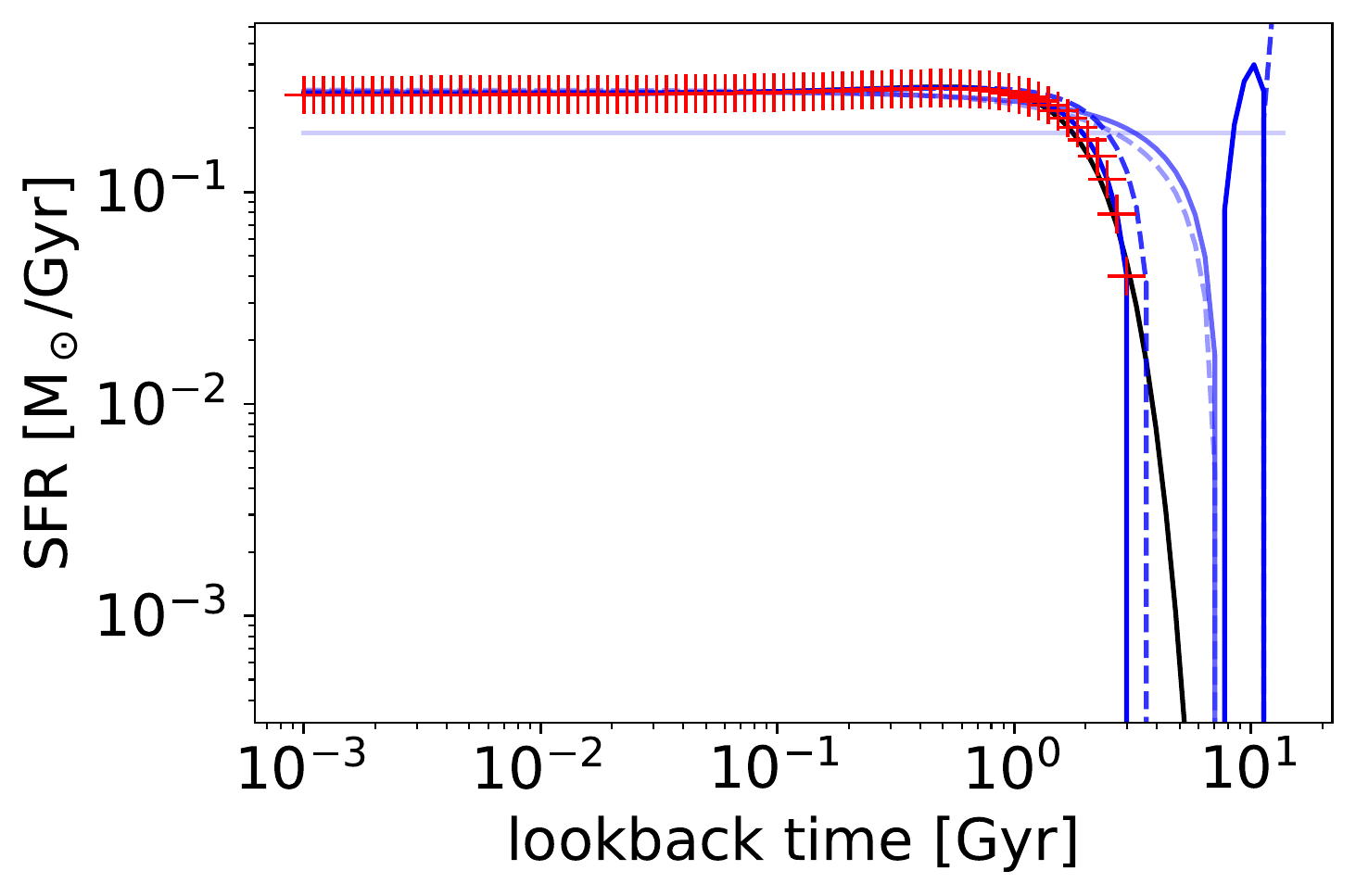} &
    \includegraphics[width=.23\textwidth]{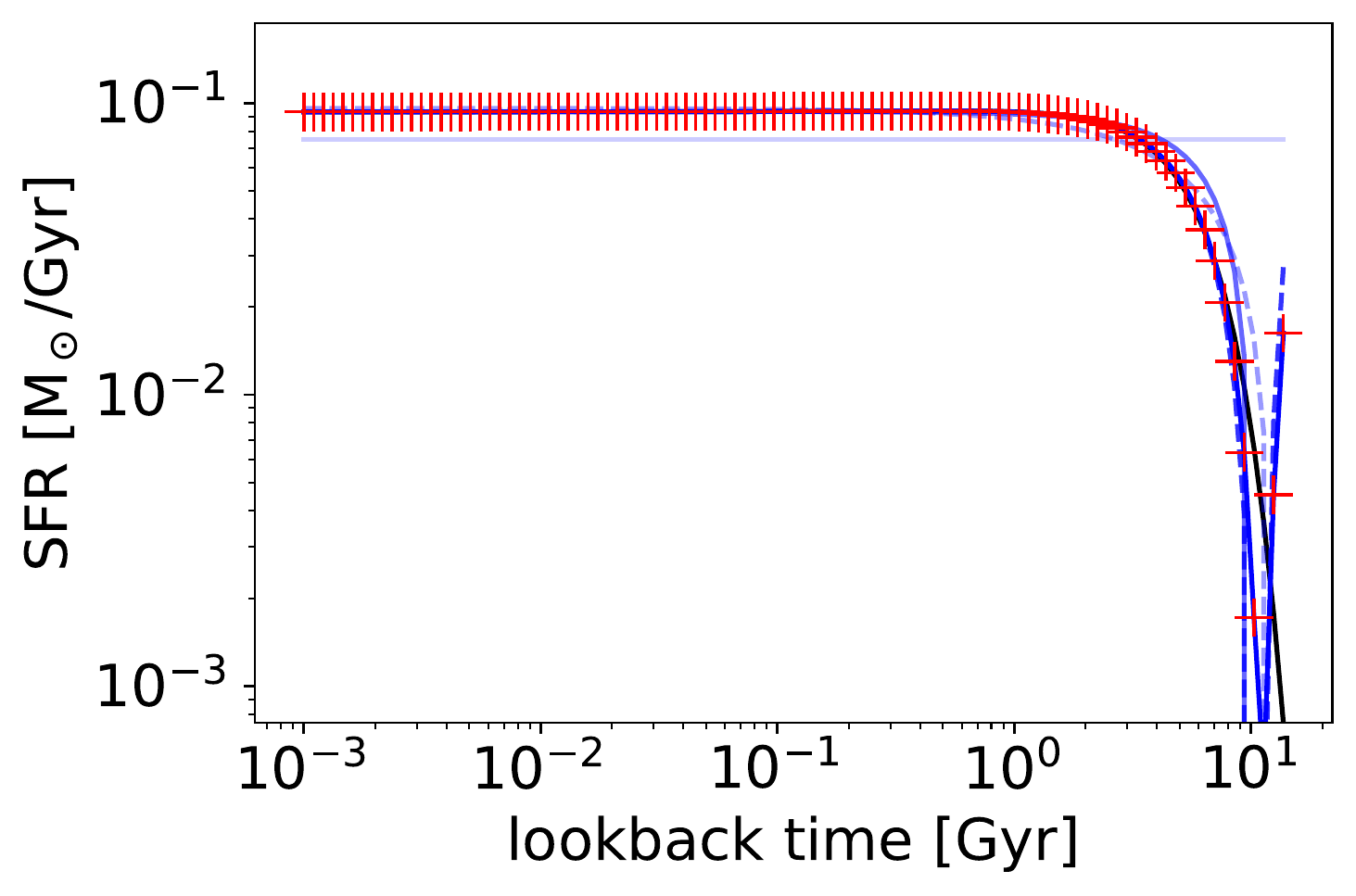} \\
    \turnbox{90}{\hspace{5 mm} Age = 1.0~Gyr}
    \includegraphics[width=.23\textwidth]{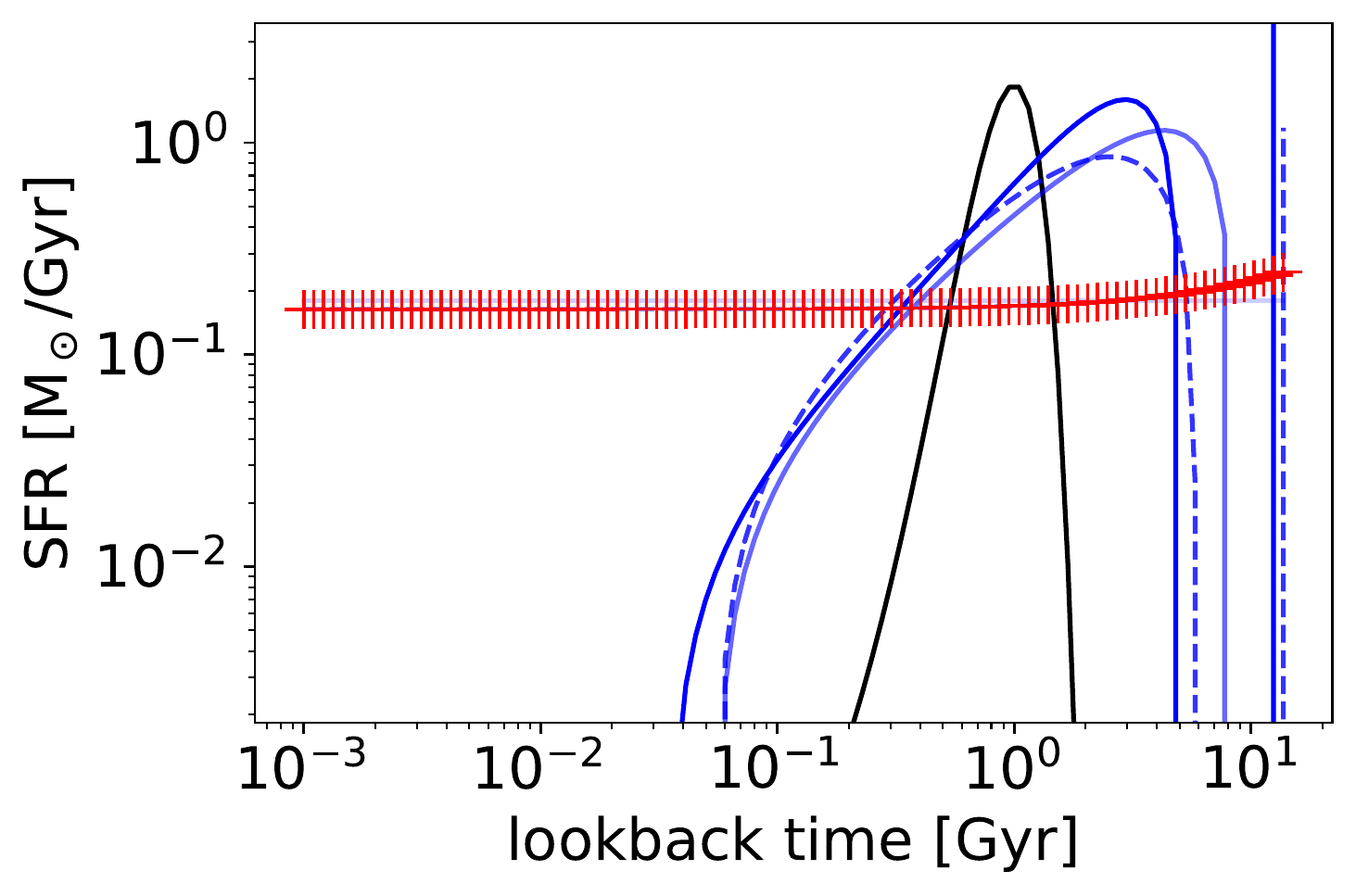} &
    \includegraphics[width=.23\textwidth]{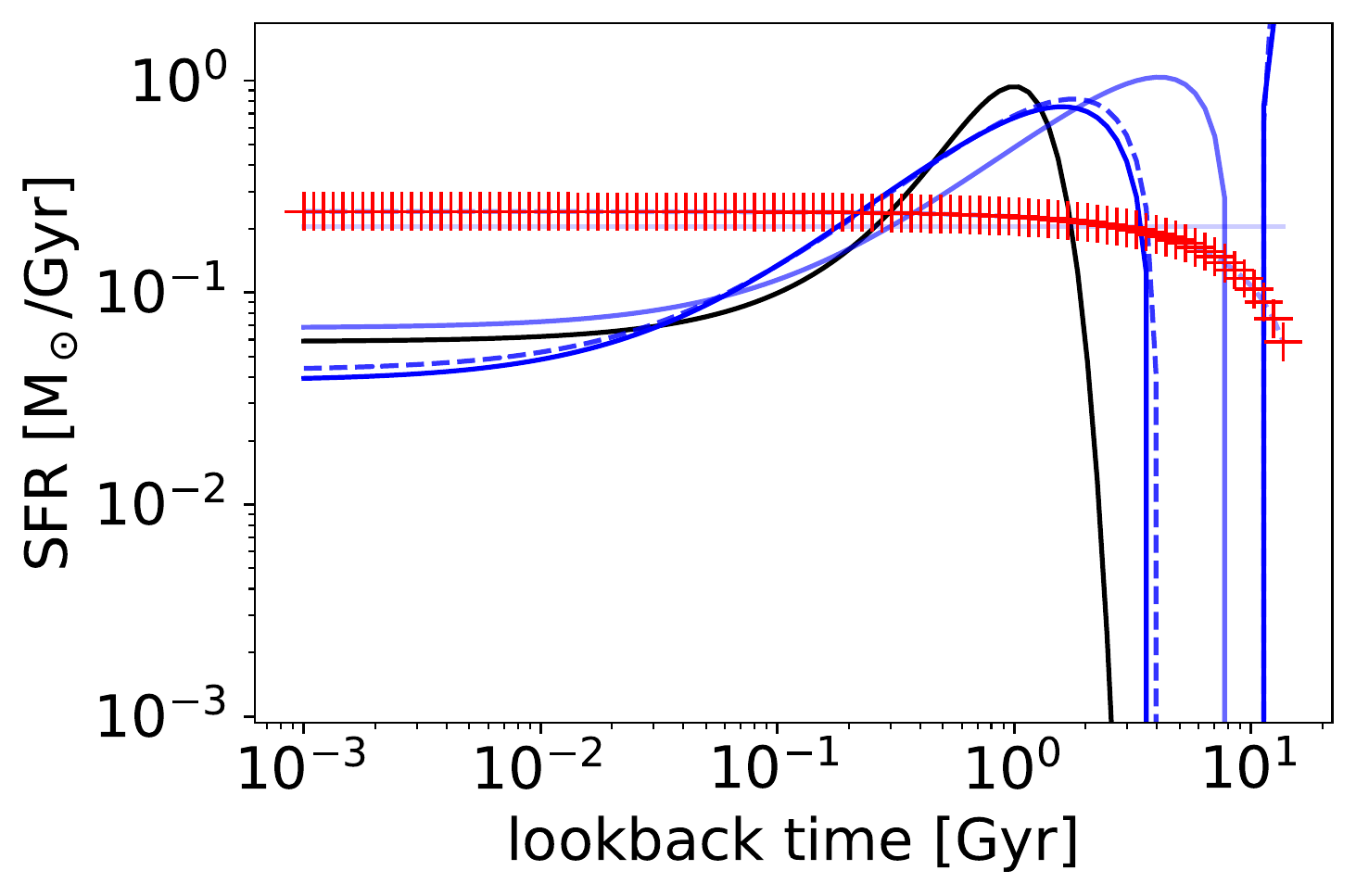} &
    \includegraphics[width=.23\textwidth]{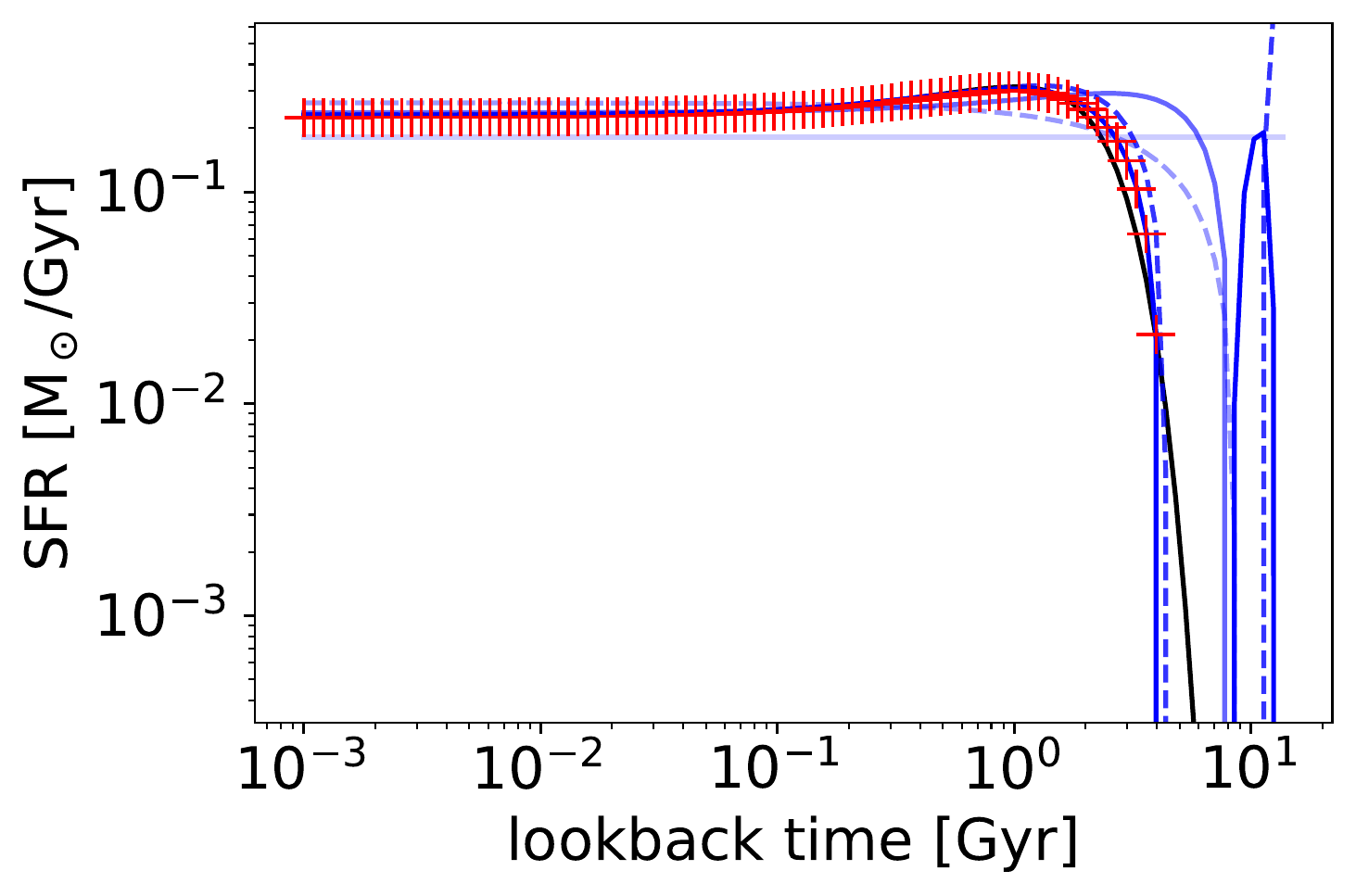} &
    \includegraphics[width=.23\textwidth]{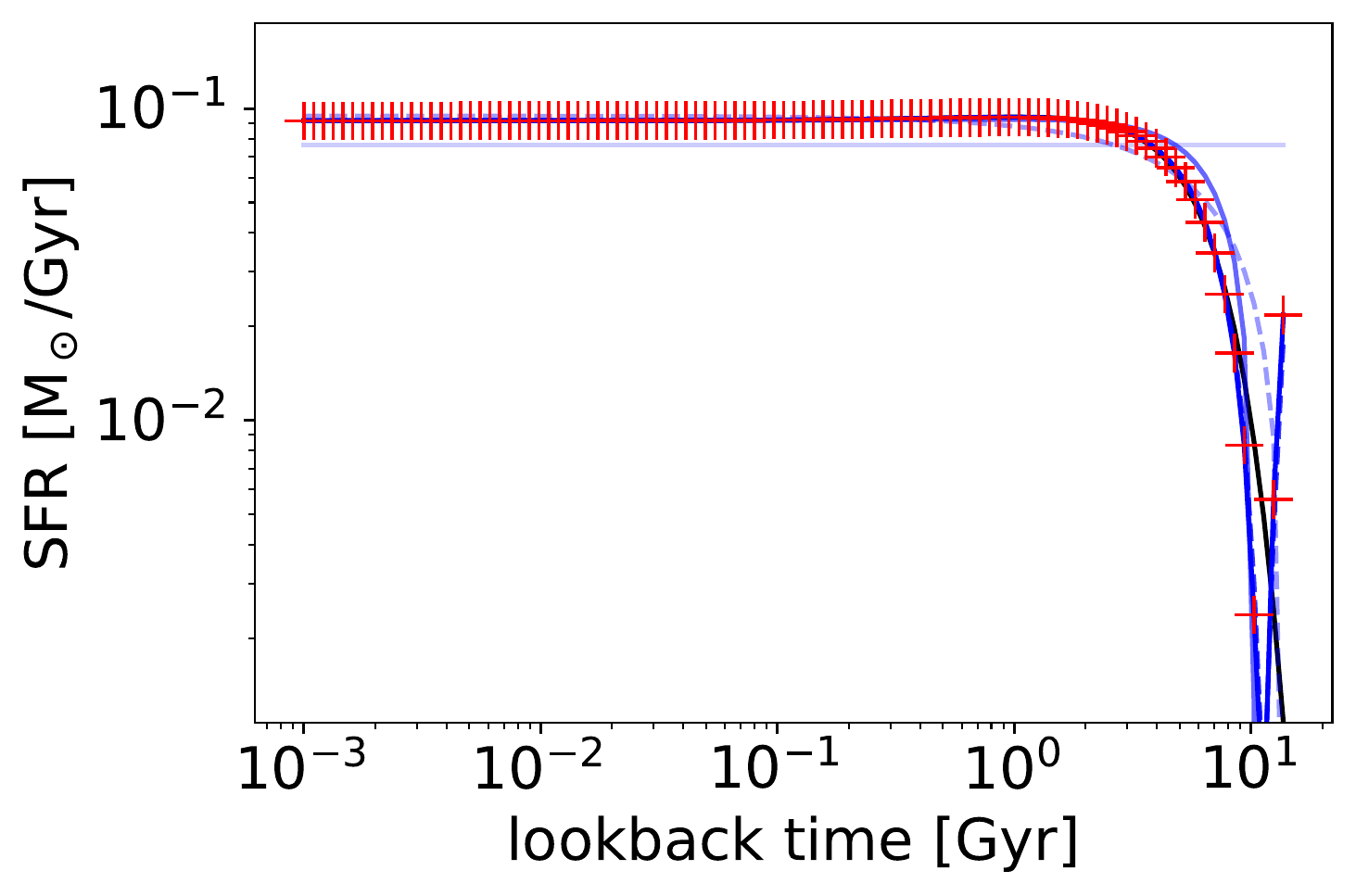} \\
    \turnbox{90}{\hspace{5 mm} Age = 3.0~Gyr}
    \includegraphics[width=.23\textwidth]{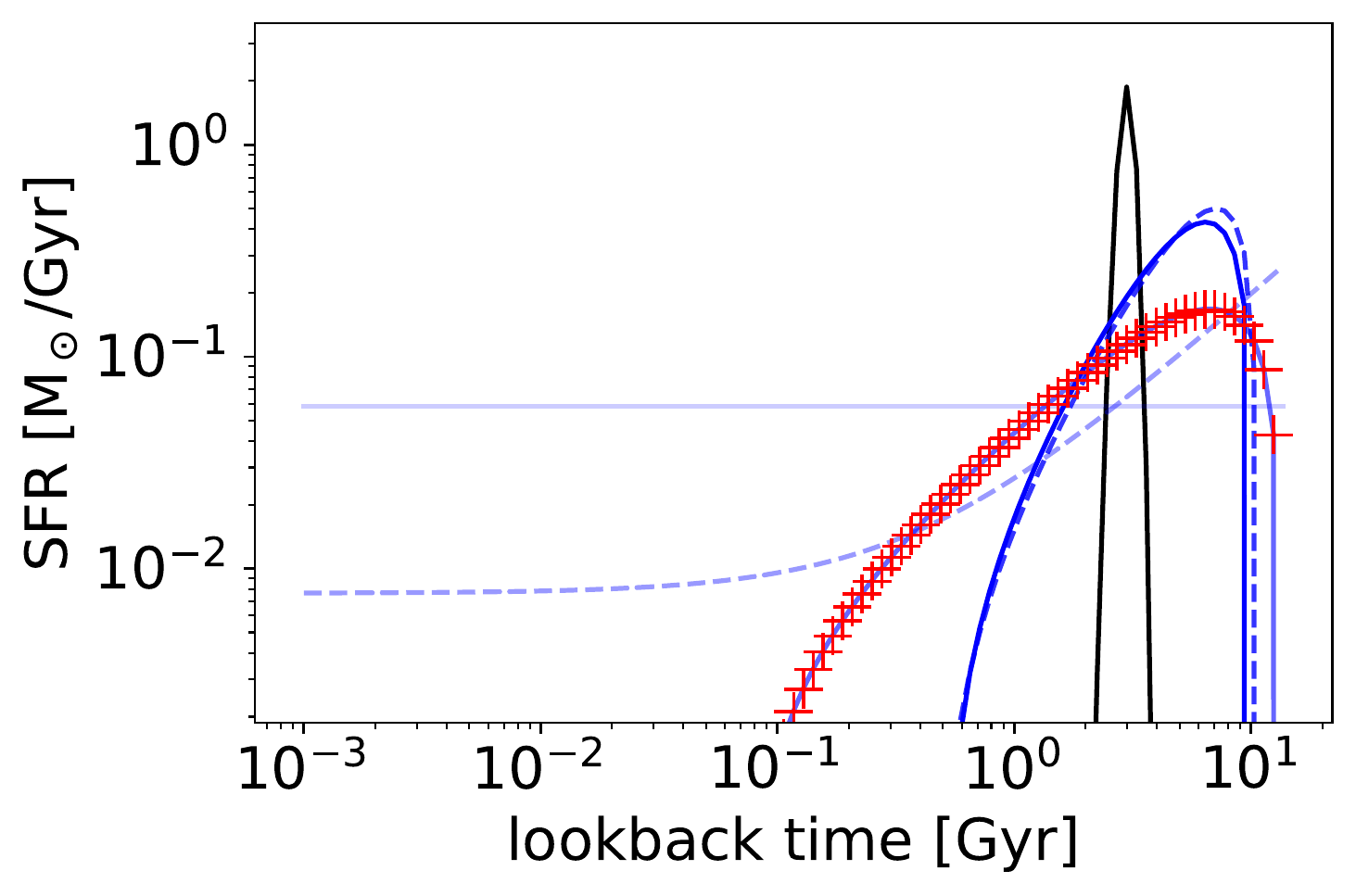} &
    \includegraphics[width=.23\textwidth]{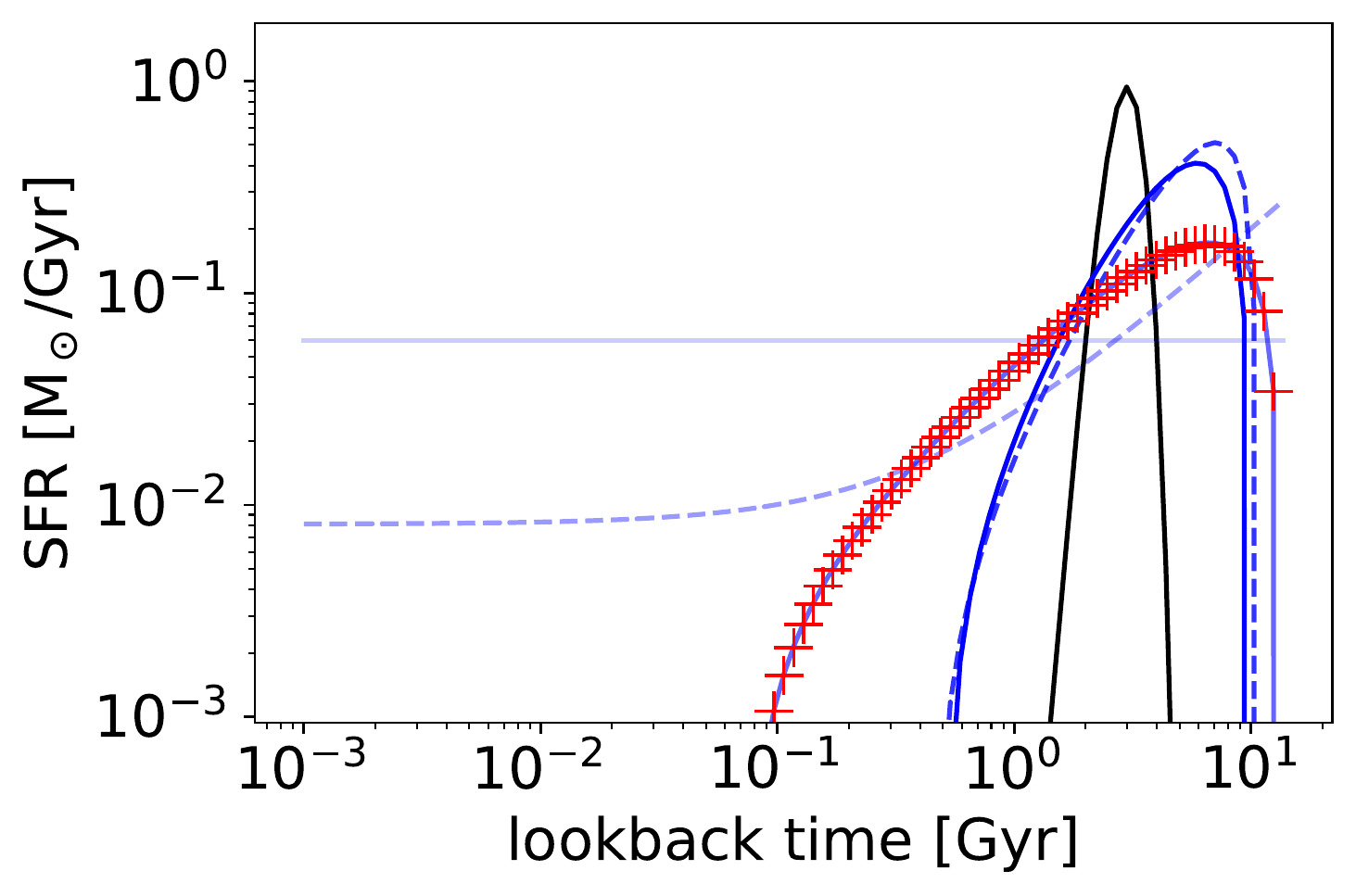} &
    \includegraphics[width=.23\textwidth]{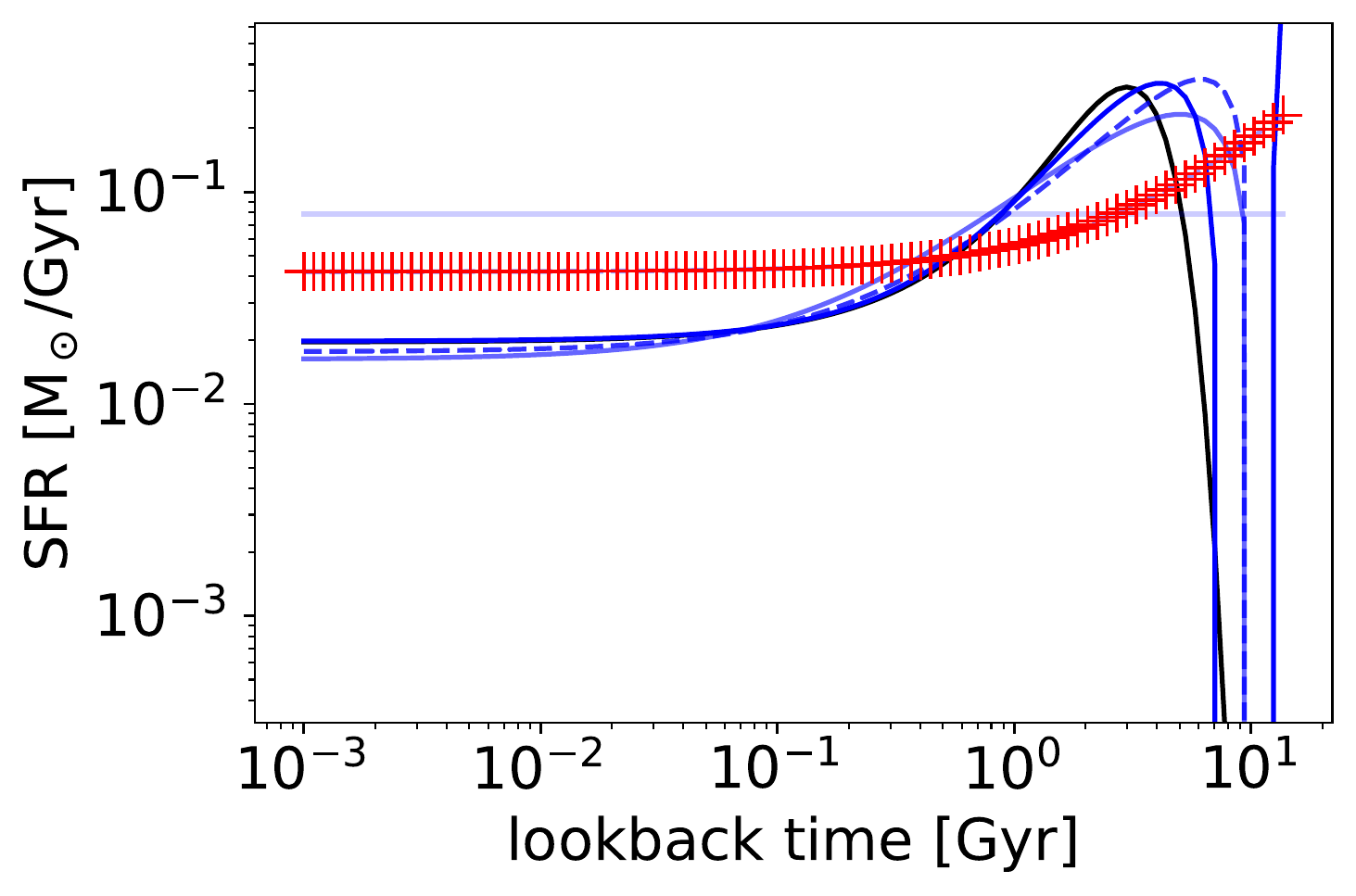} &
    \includegraphics[width=.23\textwidth]{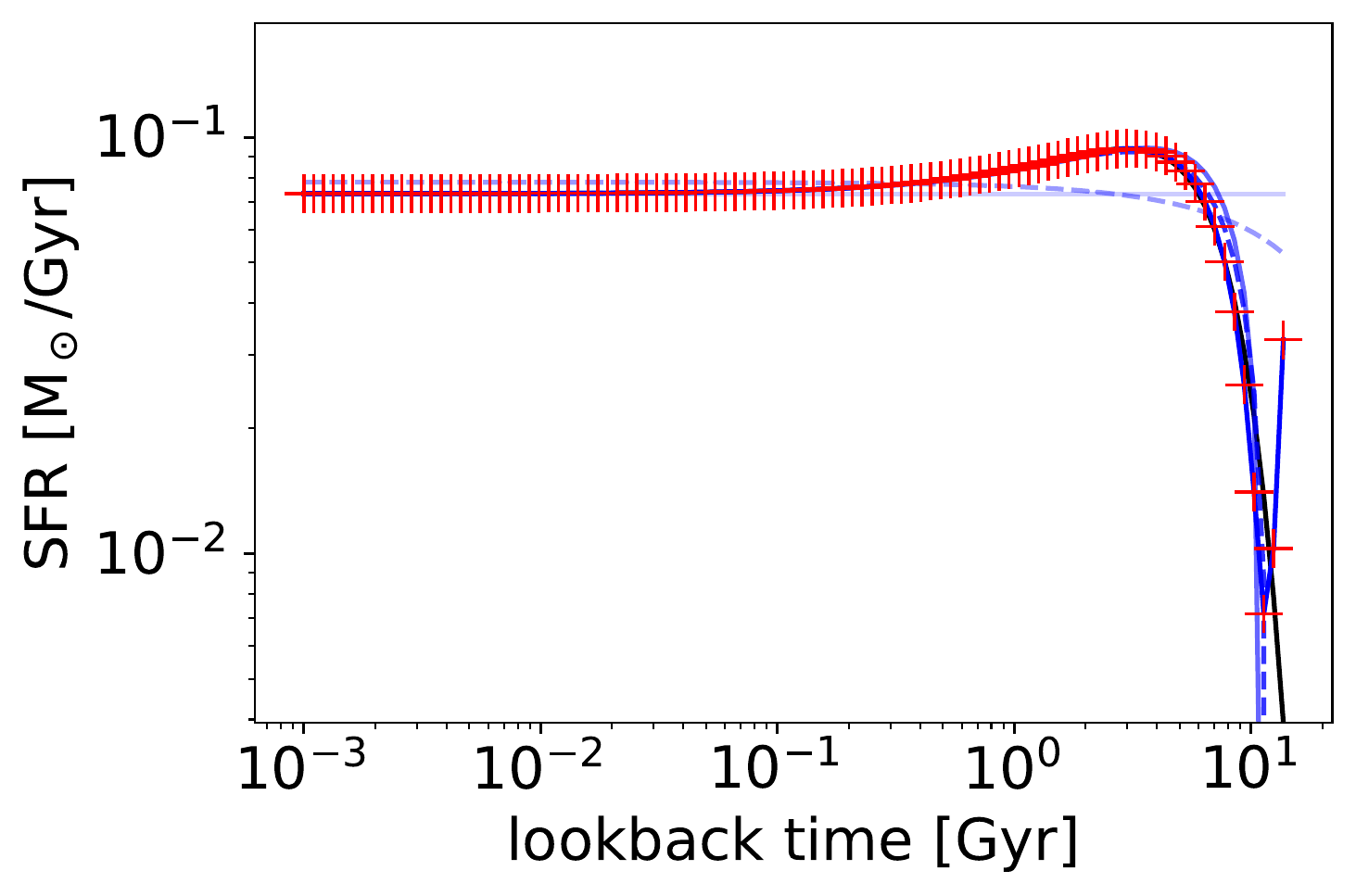} \\
    \turnbox{90}{\hspace{5 mm} Age = 10.0~Gyr}
    \includegraphics[width=.23\textwidth]{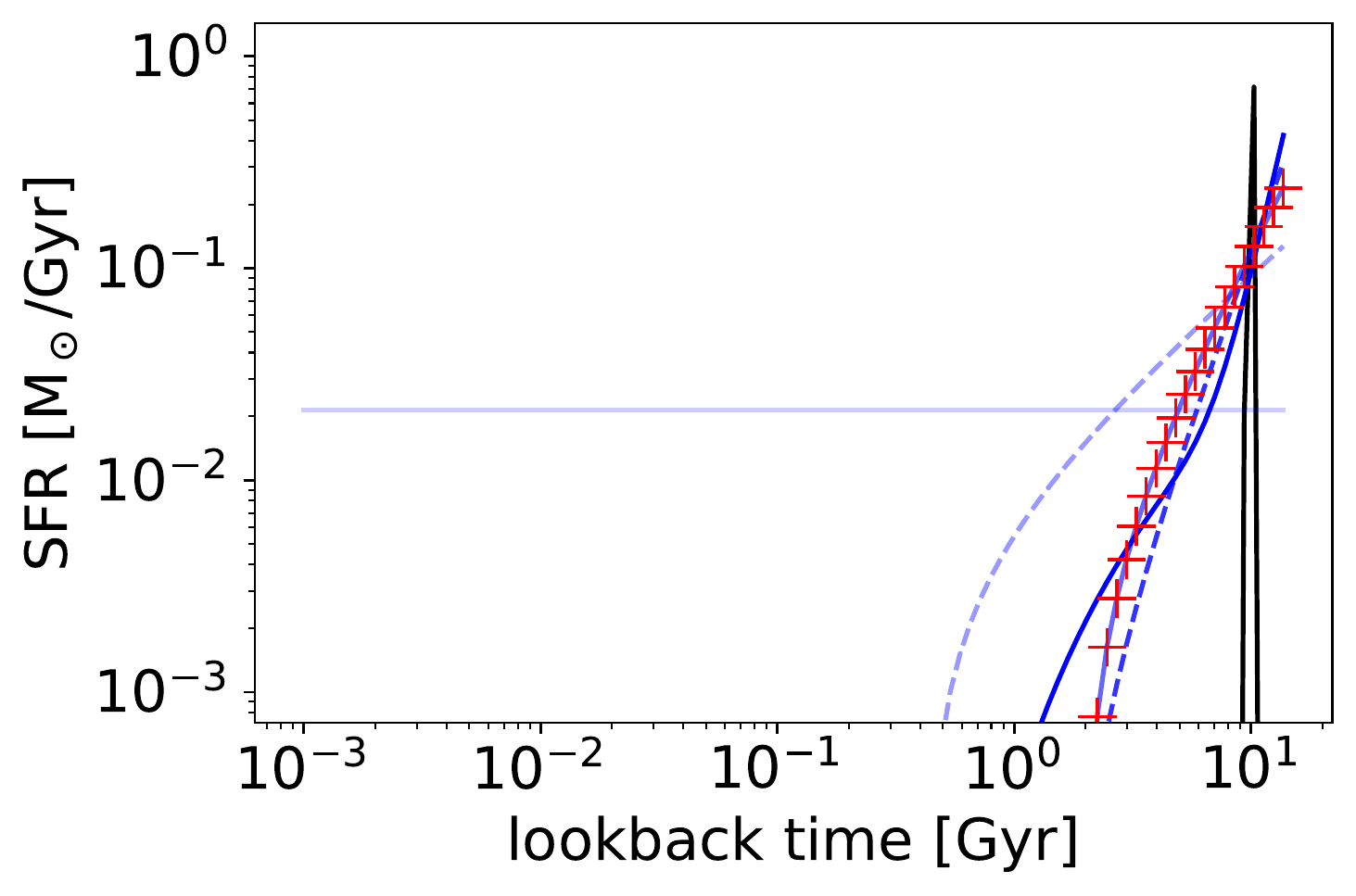} &
    \includegraphics[width=.23\textwidth]{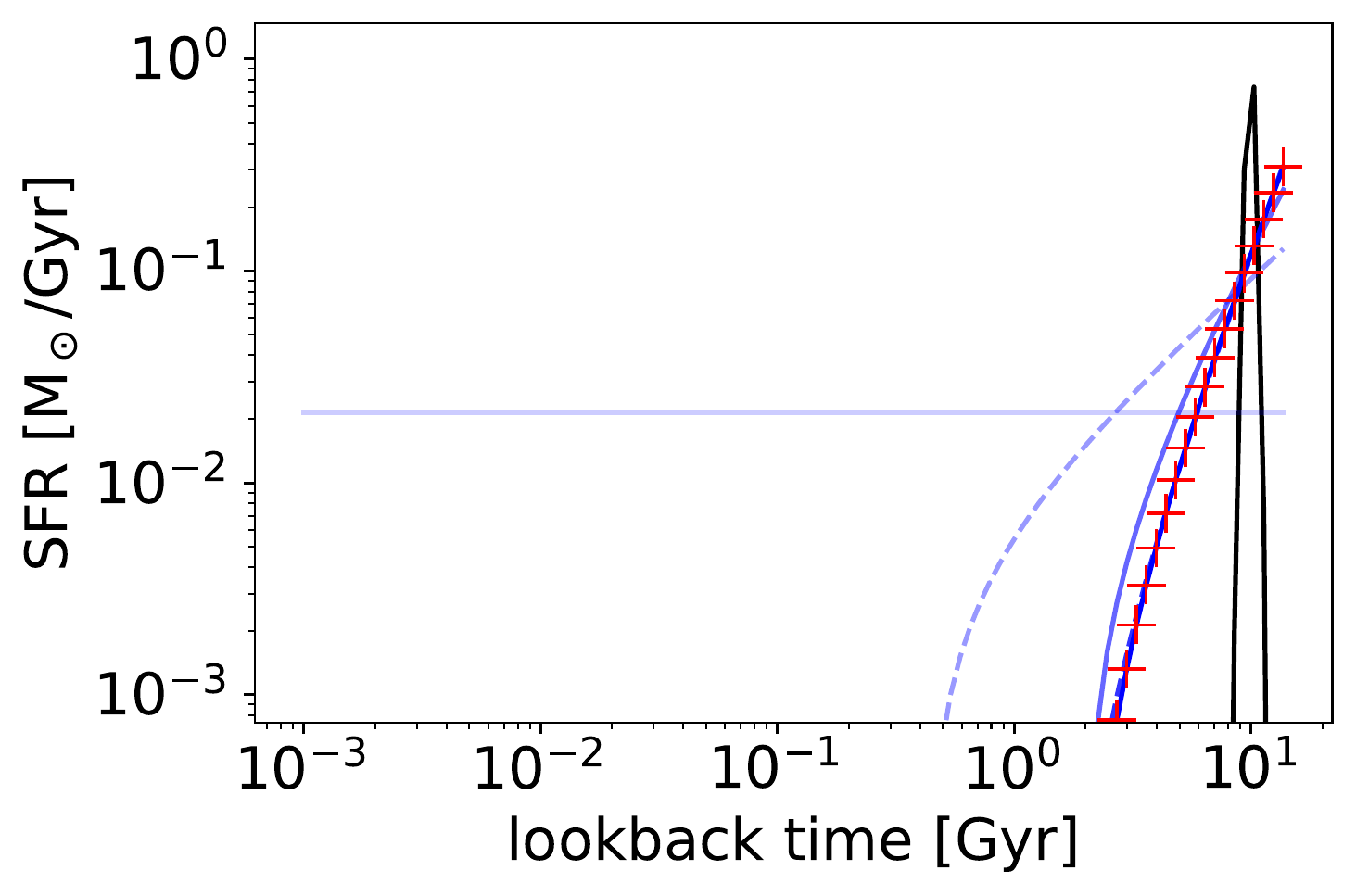} &
    \includegraphics[width=.23\textwidth]{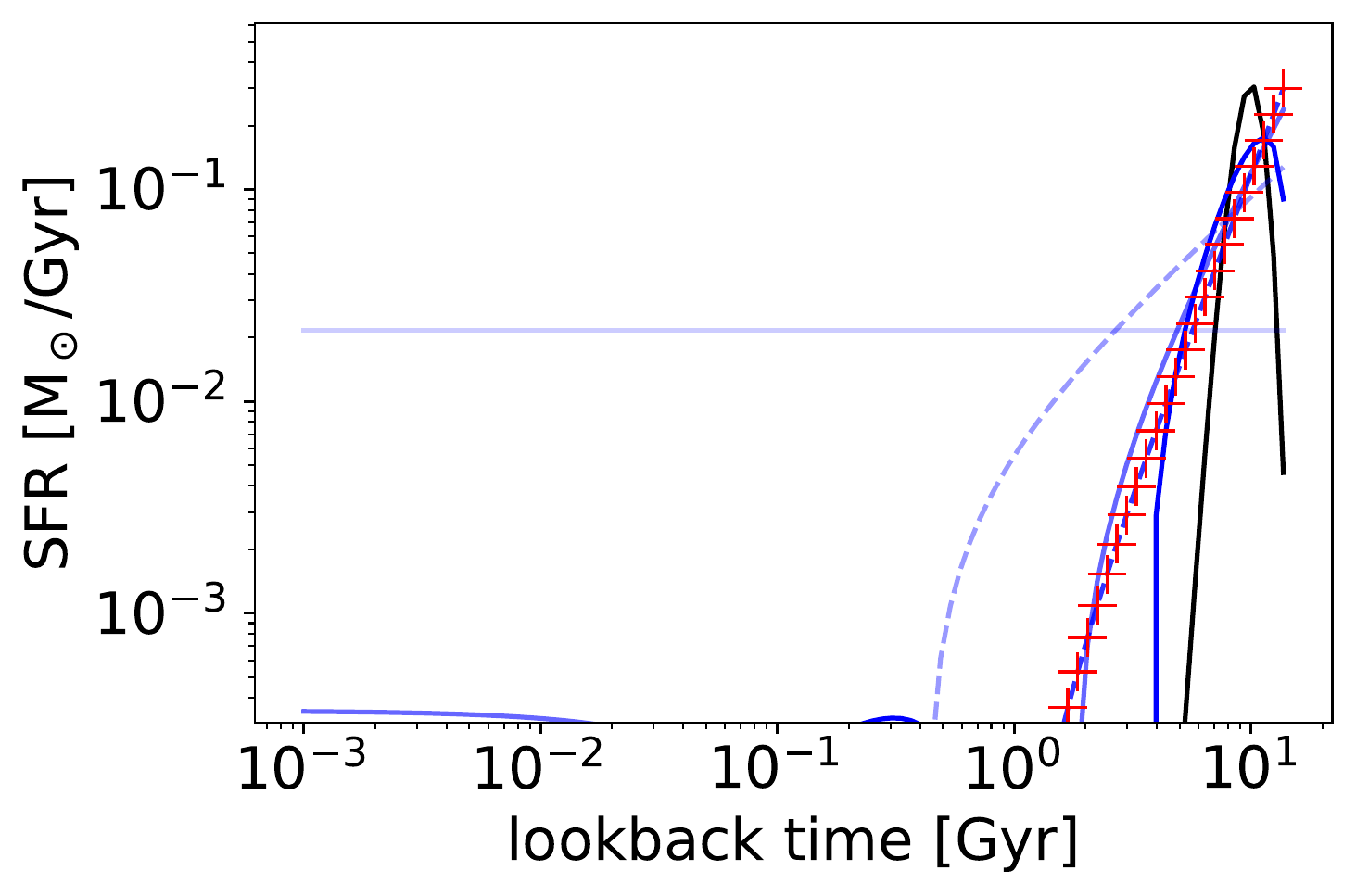} &
    \includegraphics[width=.23\textwidth]{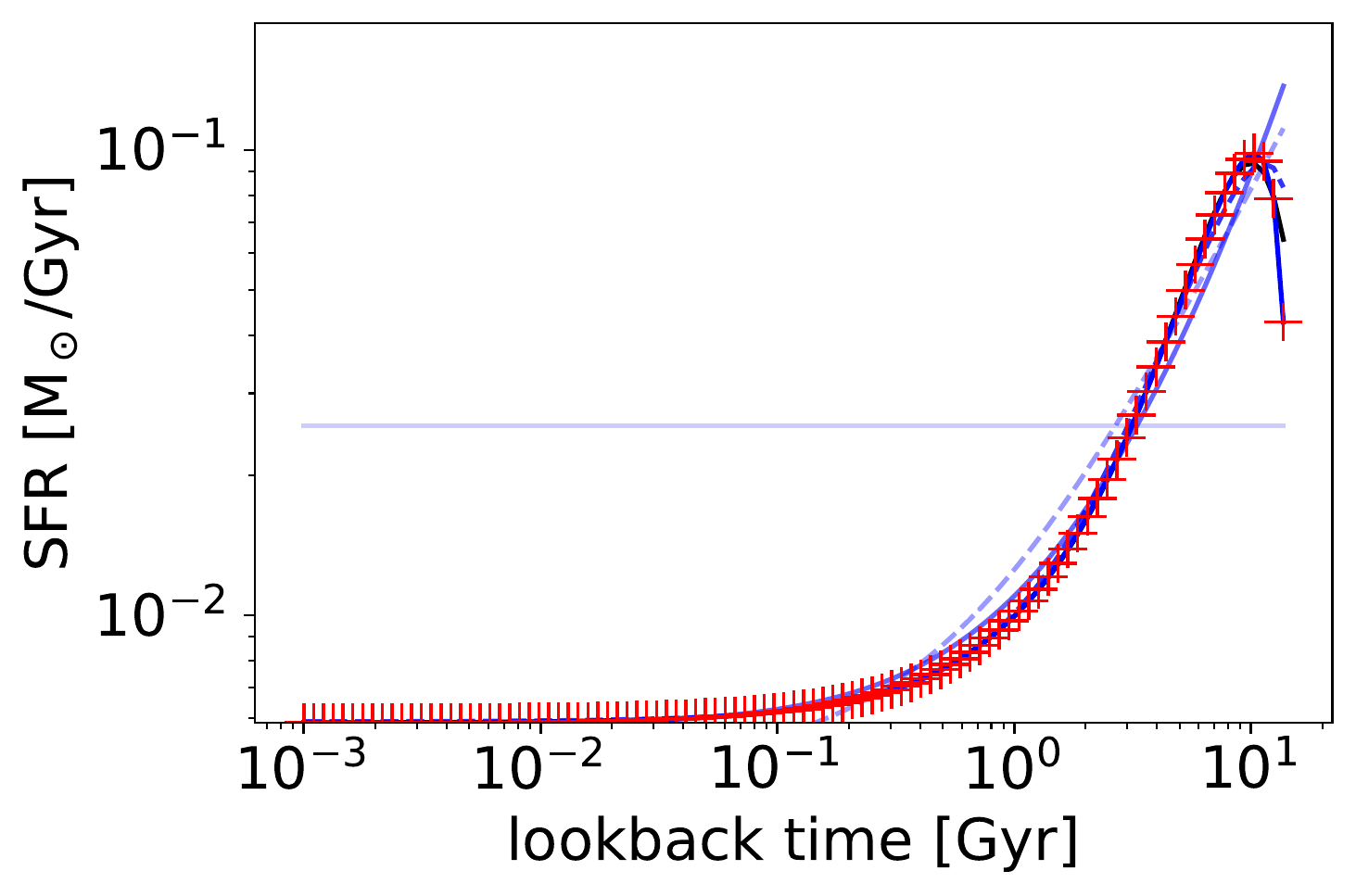} 
  \end{tabular}
  \caption{Star formation rate reconstructed with our parametric model from the luminosities of a synthetic Gaussian SFR with different peak ages (increasing from top to bottom) and FWHMs (increasing from left to right), compared to the input model (solid black line). Colour gradients and line styles indicate the degree of the polynomial reconstruction from $N=5$ (solid) to lower degrees (more diffuse lines; solid and dashed for odd and even $N$). Red crosses correspond to the best positive-SFR fit.}
\label{fig:Gauss_rec_SFR}
\end{figure*}

\begin{figure*}
\centering
  \begin{tabular}{@{}cccc@{}}
  FWHM=0.5~Gyr & FWHM=1.0~Gyr & FWHM=3.0~Gyr & FWHM=10.0~Gyr \\
  \turnbox{90}{\hspace{5 mm} Age = 0.5~Gyr}
    \includegraphics[width=.23\textwidth]{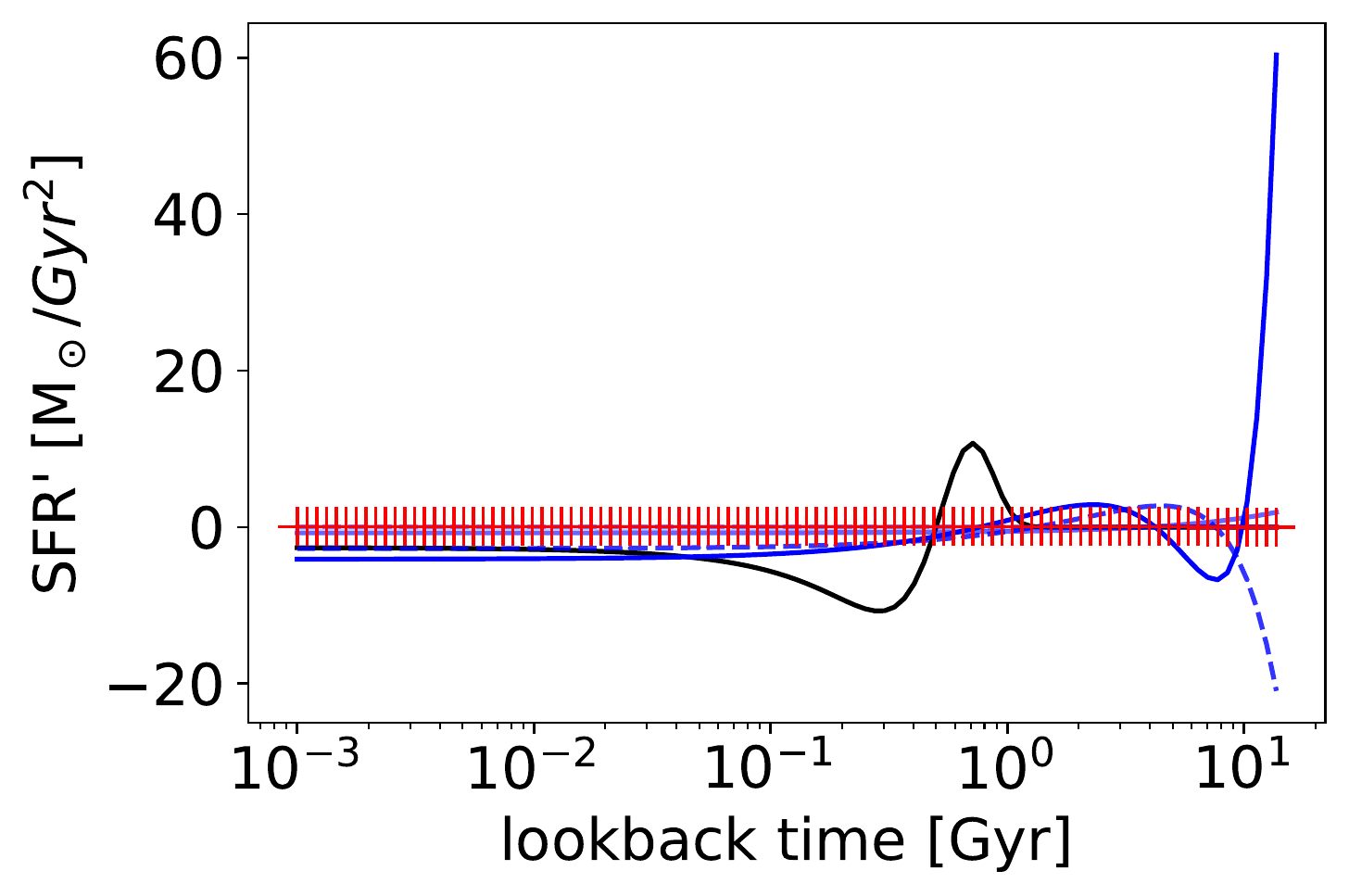} &
    \includegraphics[width=.23\textwidth]{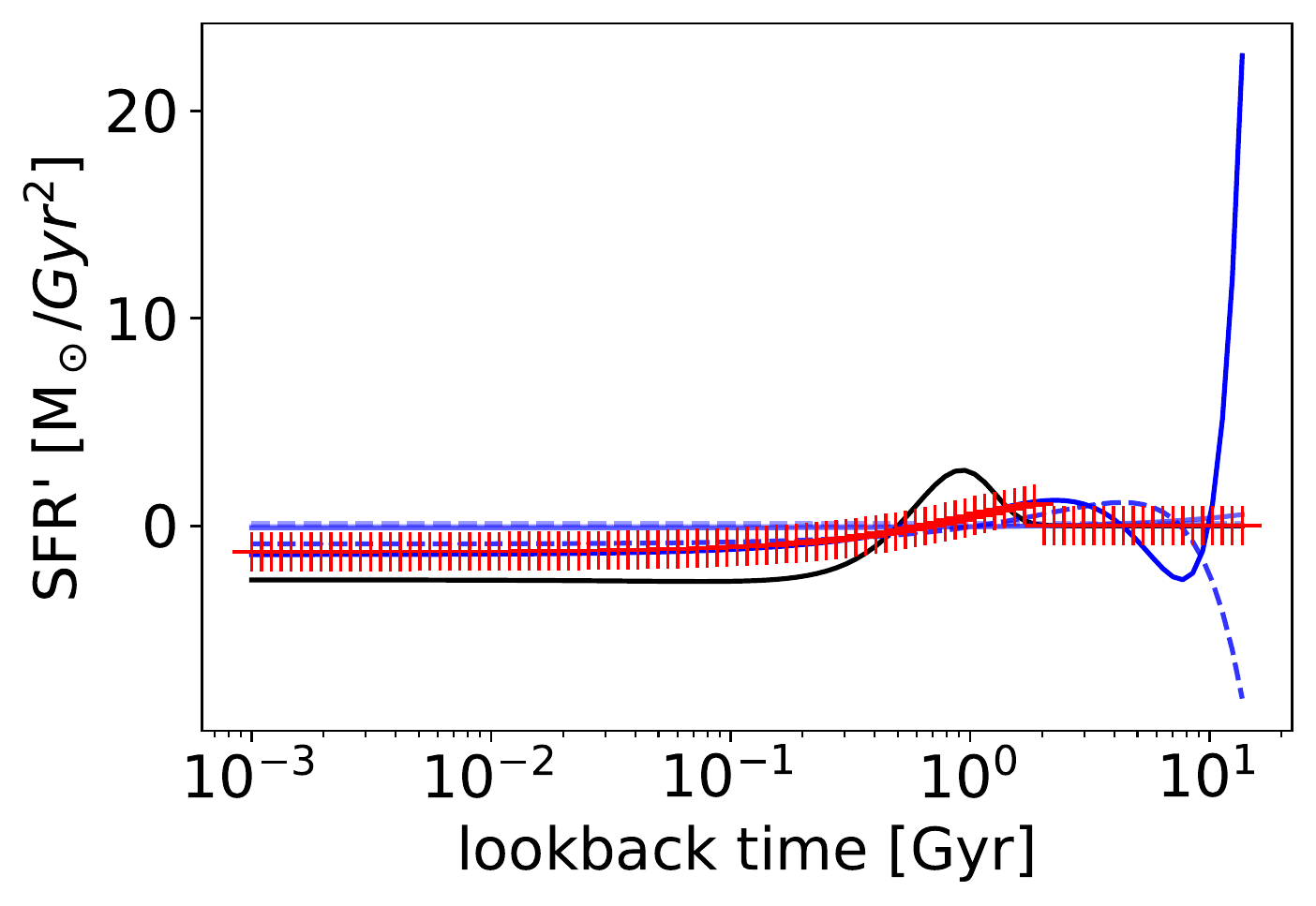} &
    \includegraphics[width=.23\textwidth]{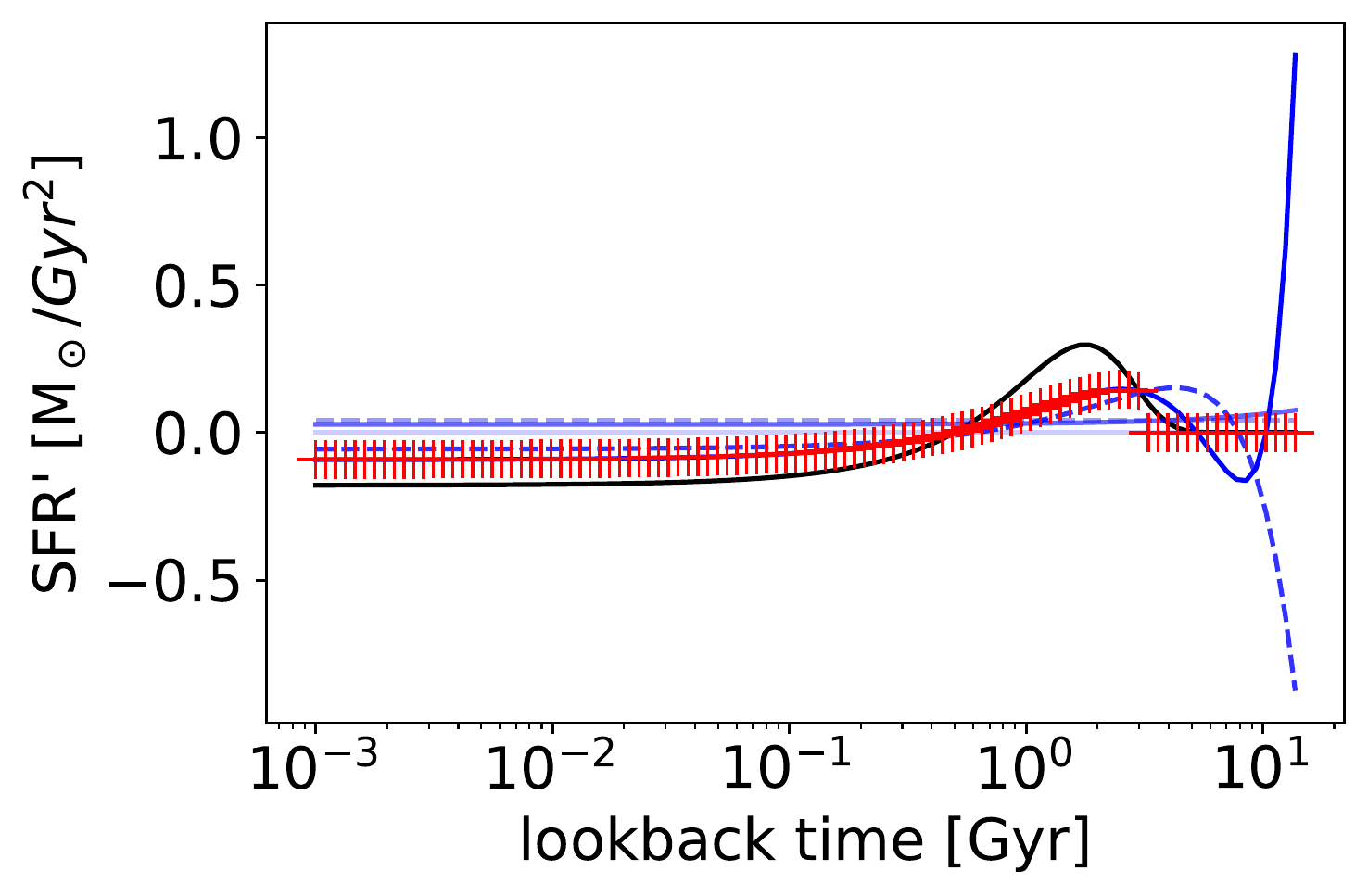} &
    \includegraphics[width=.23\textwidth]{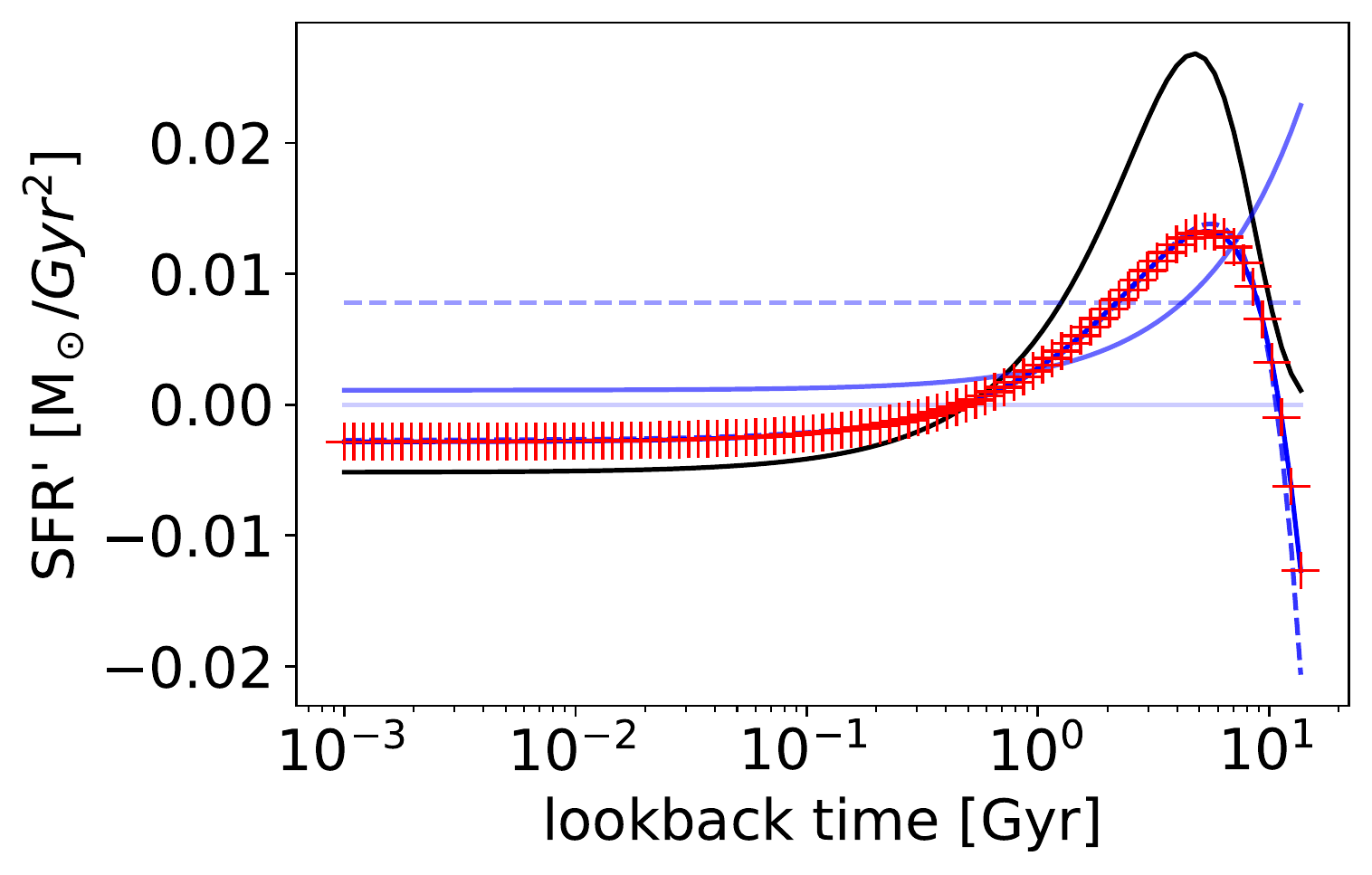} \\
    \turnbox{90}{\hspace{5 mm} Age = 1.0~Gyr}
    \includegraphics[width=.23\textwidth]{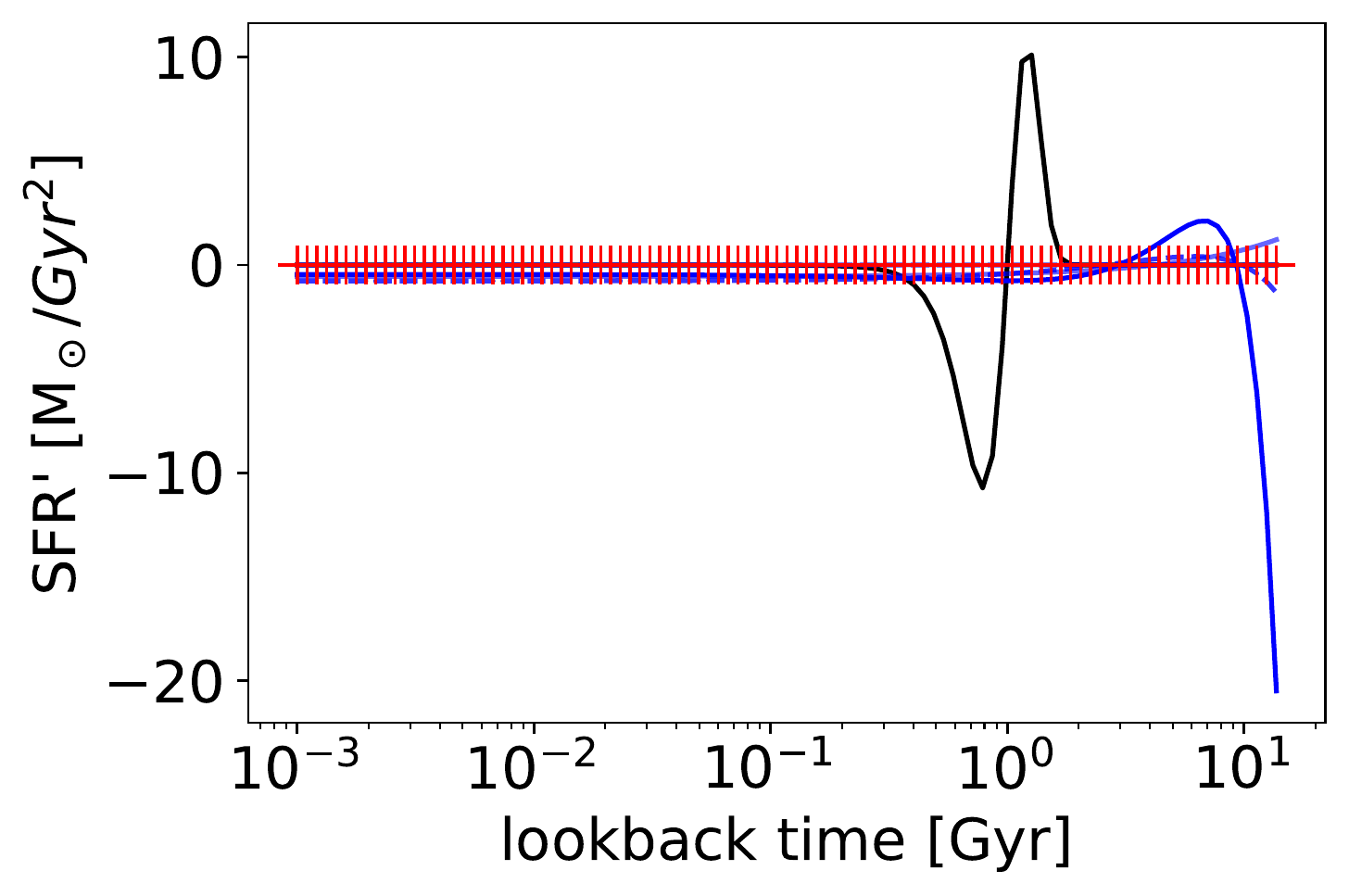} &
    \includegraphics[width=.23\textwidth]{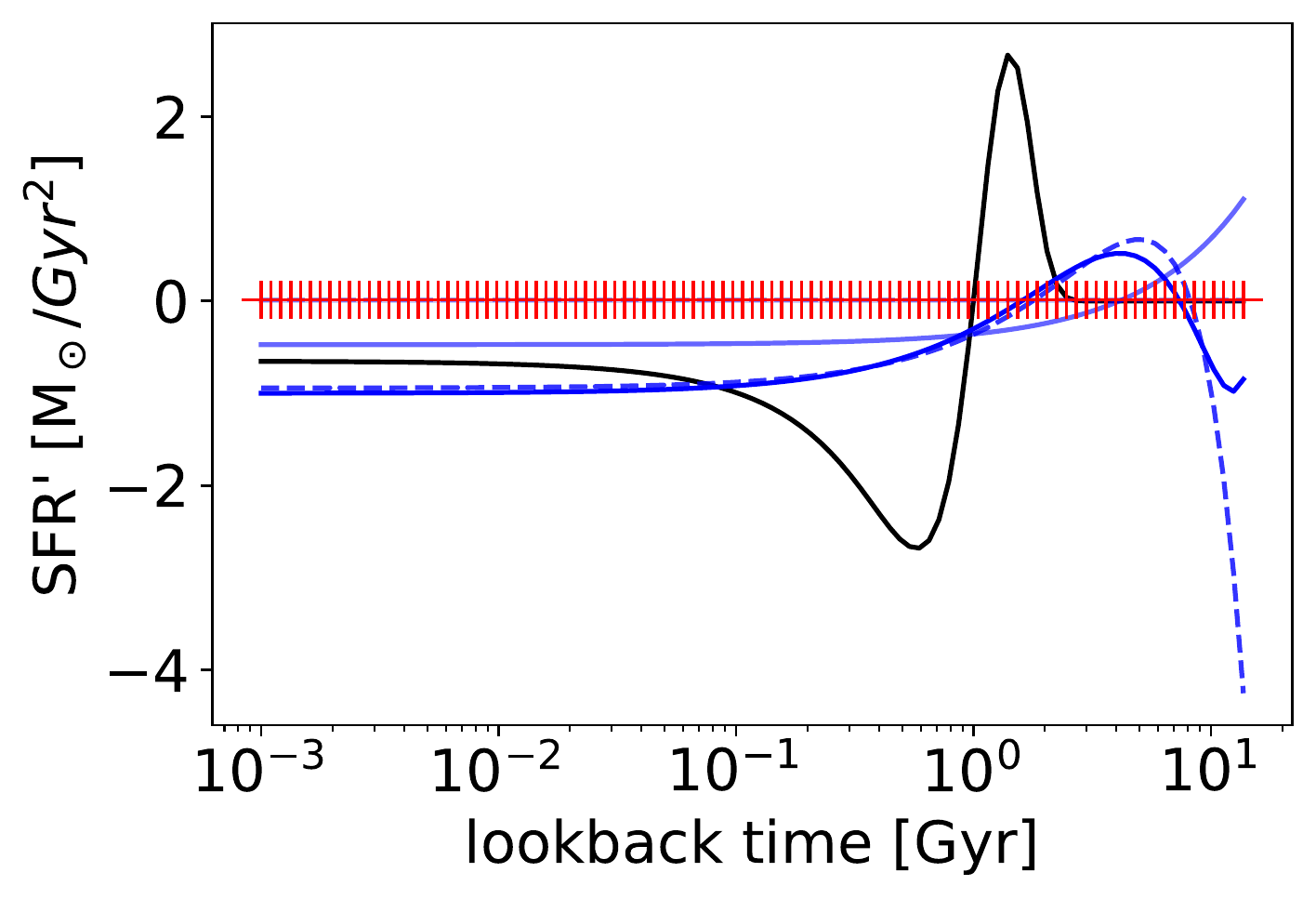} &
    \includegraphics[width=.23\textwidth]{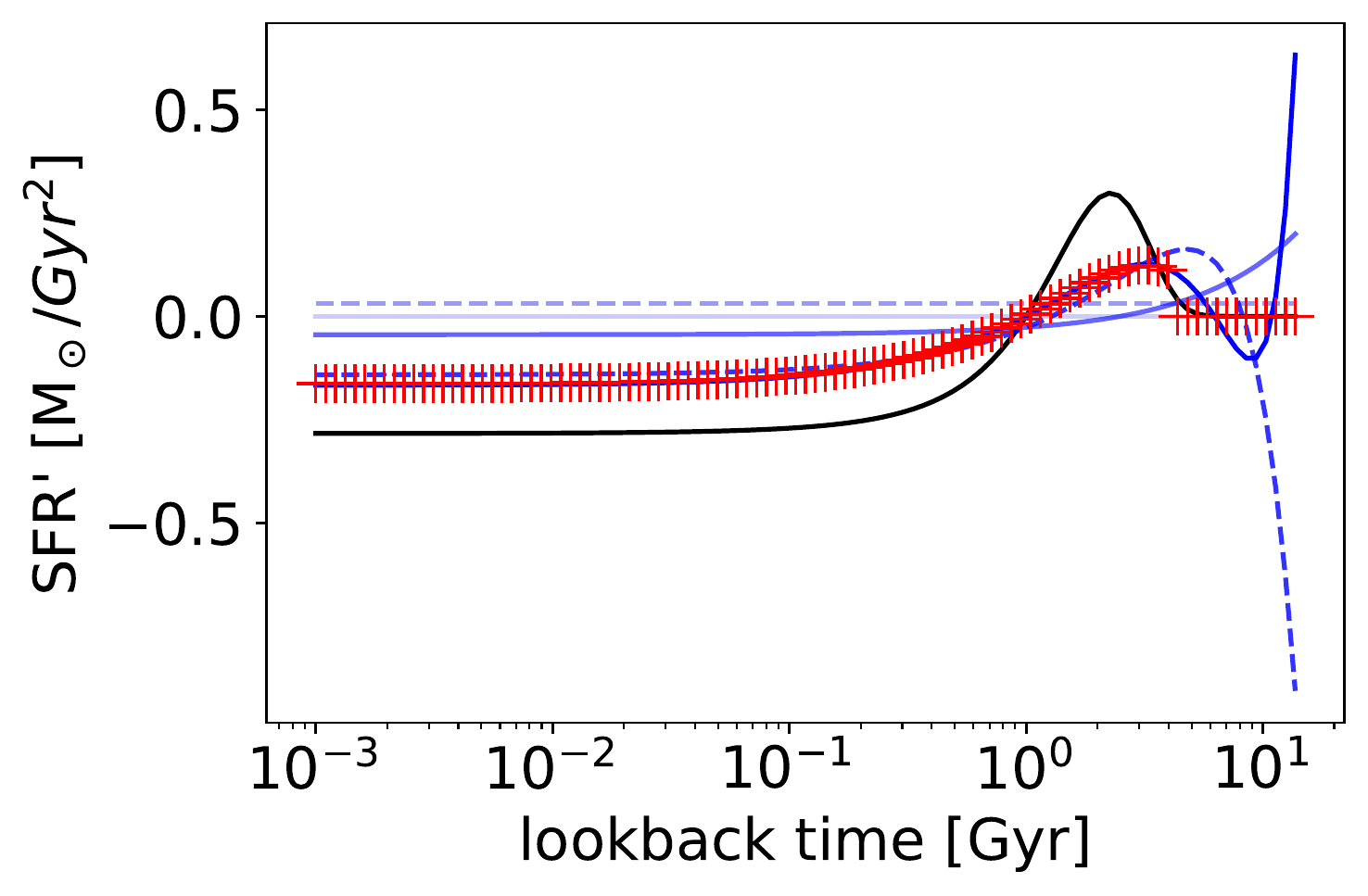} &
    \includegraphics[width=.23\textwidth]{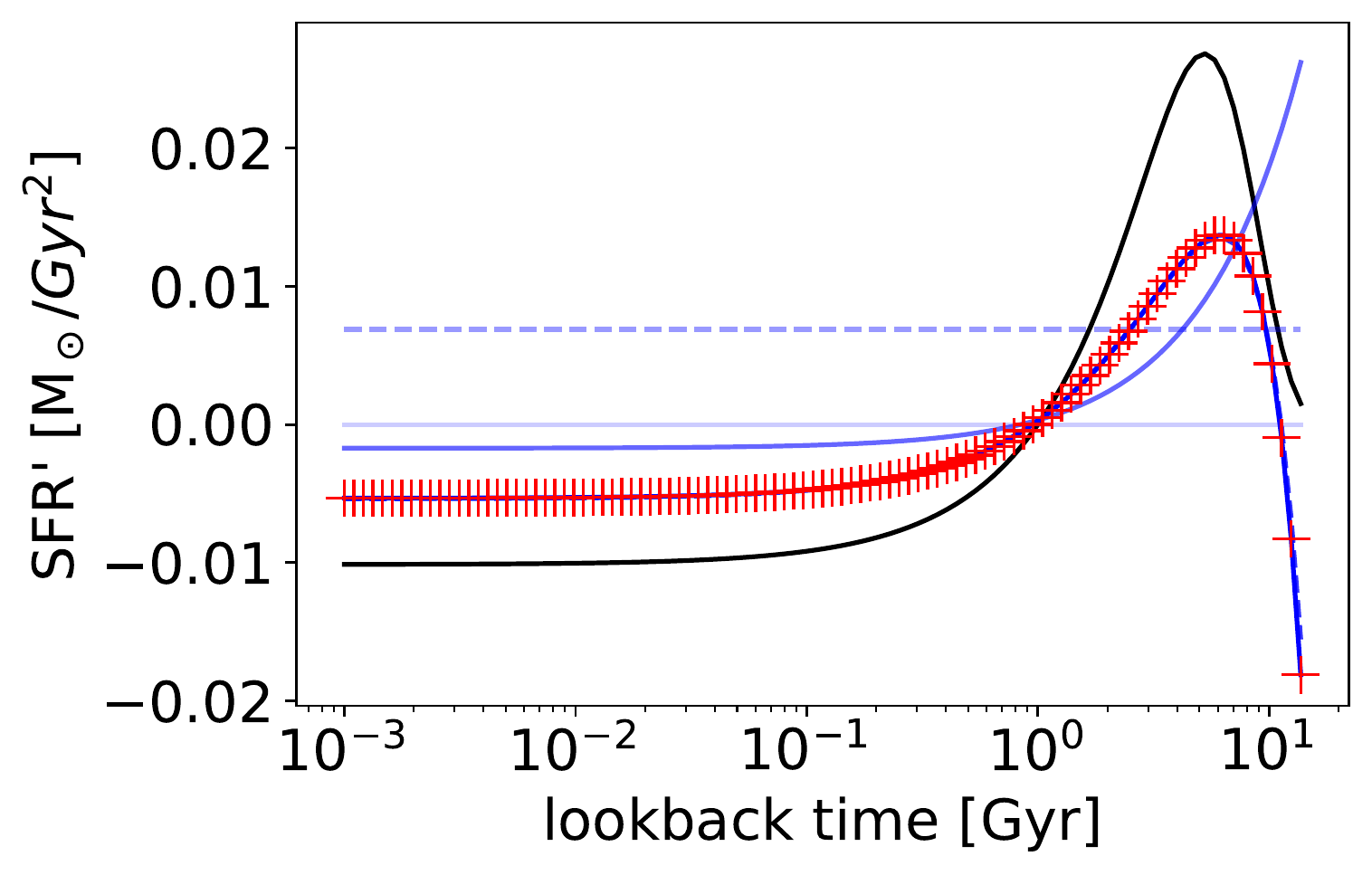} \\
    \turnbox{90}{\hspace{5 mm} Age = 3.0~Gyr}
    \includegraphics[width=.23\textwidth]{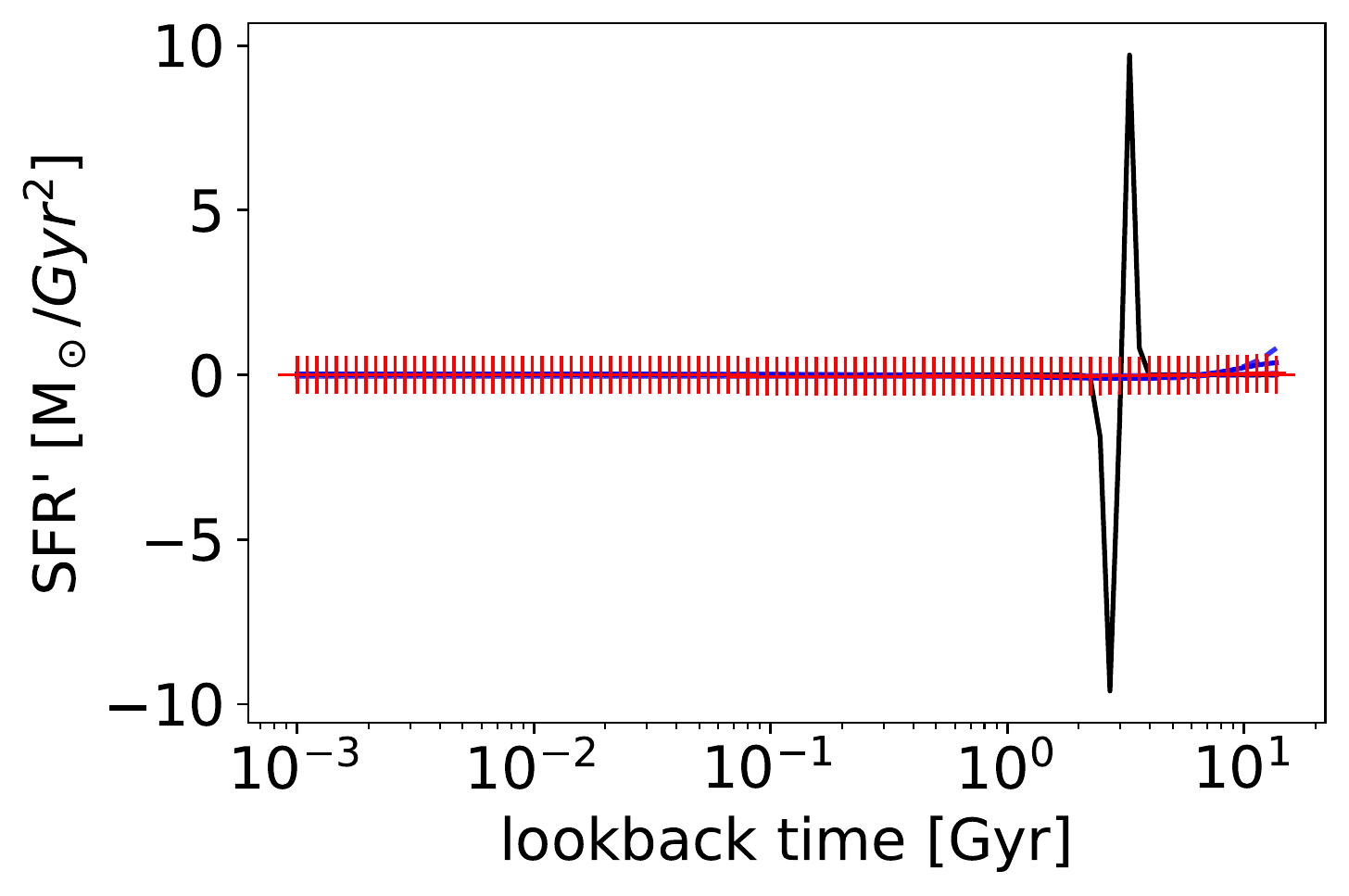} &
    \includegraphics[width=.23\textwidth]{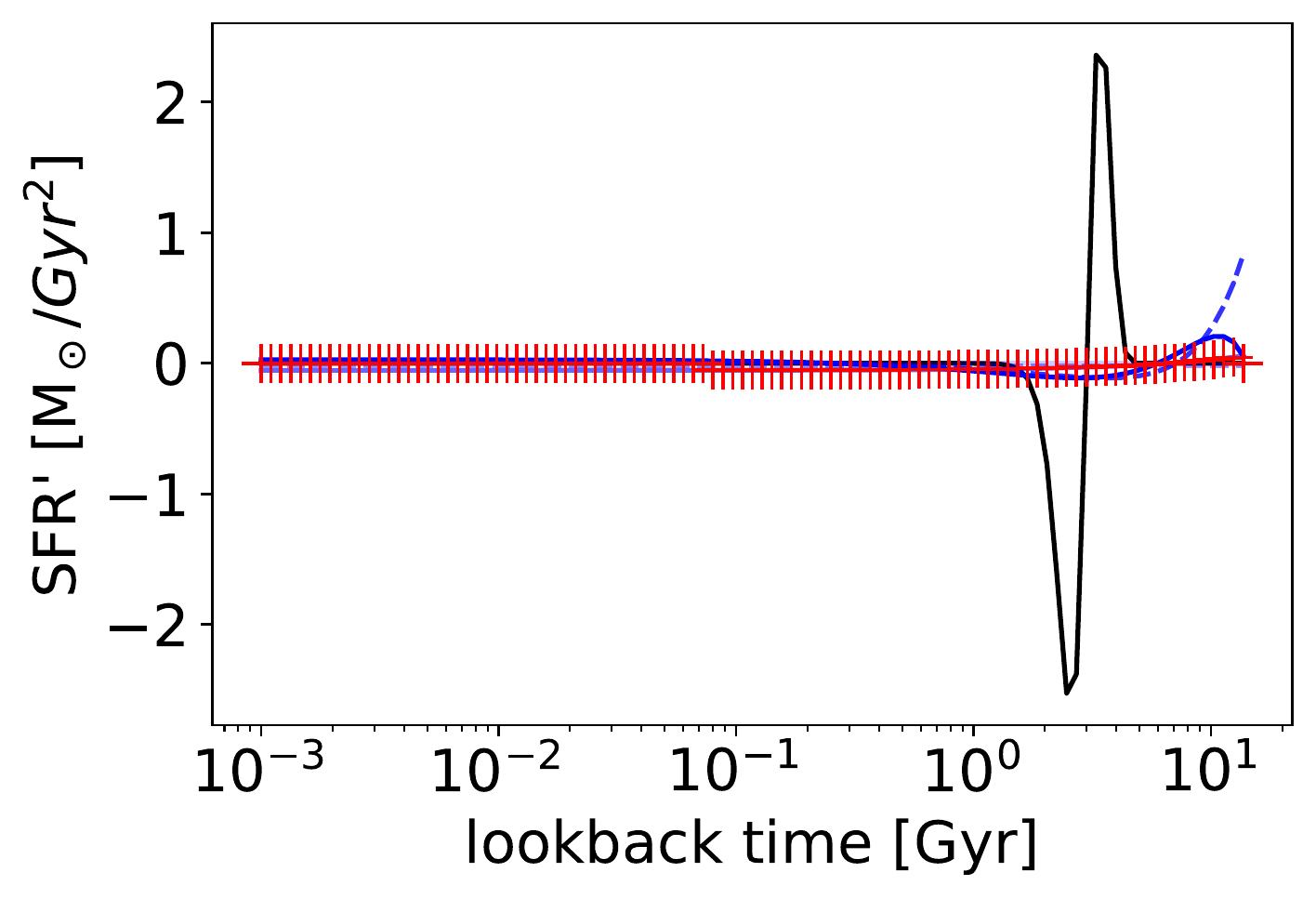} &
    \includegraphics[width=.23\textwidth]{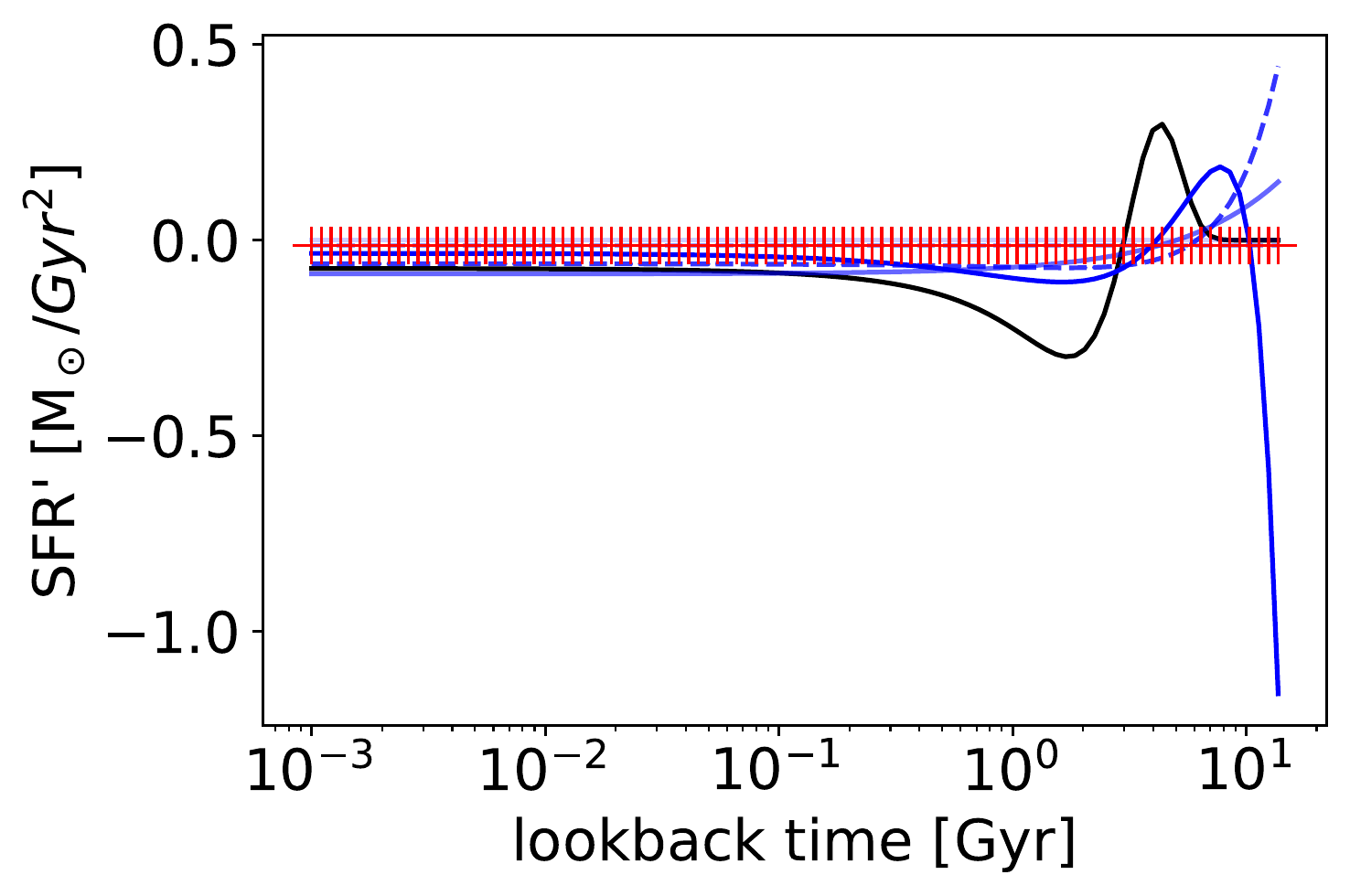} &
    \includegraphics[width=.23\textwidth]{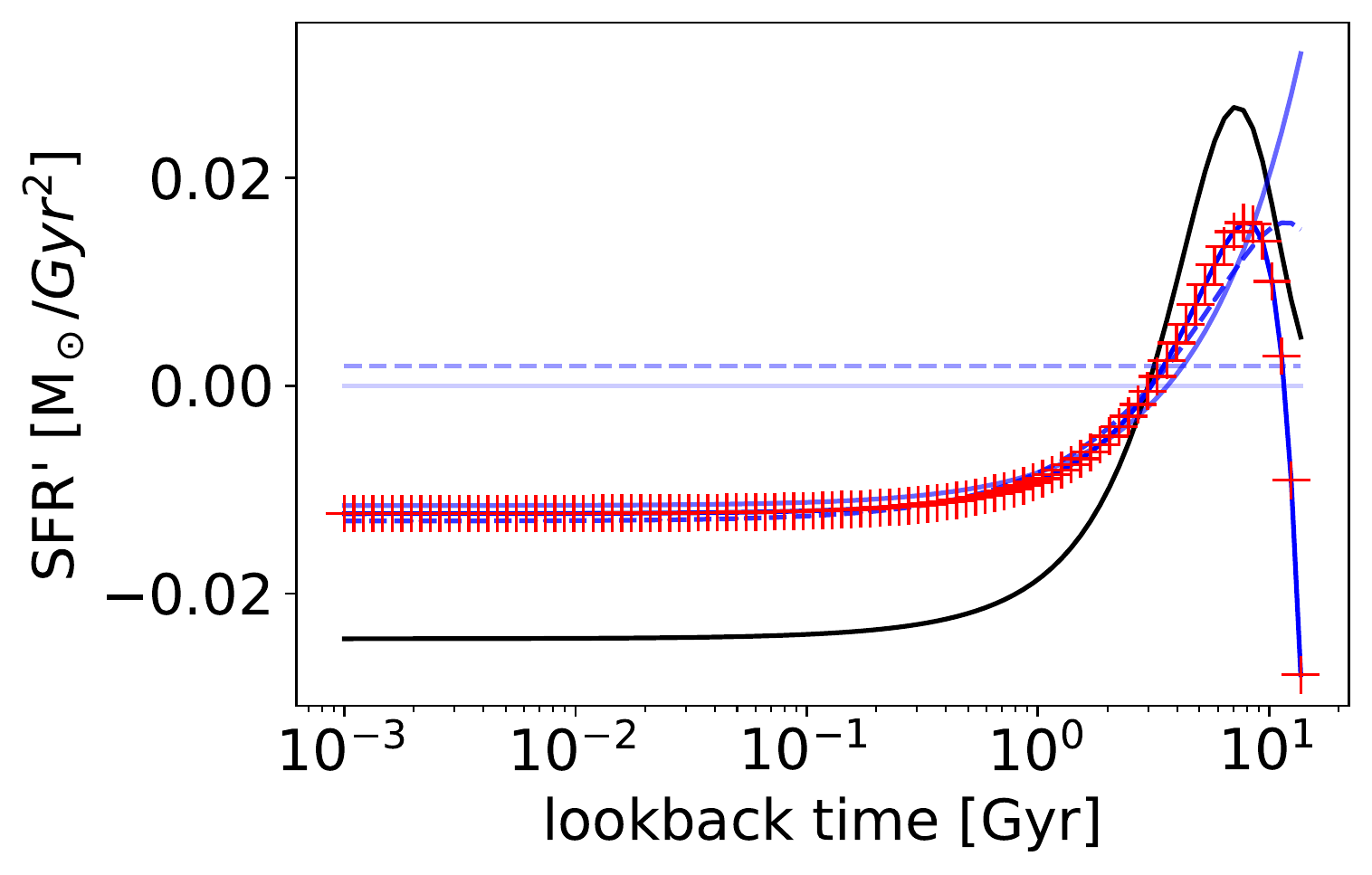} \\
    \turnbox{90}{\hspace{5 mm} Age = 10.0~Gyr}
    \includegraphics[width=.23\textwidth]{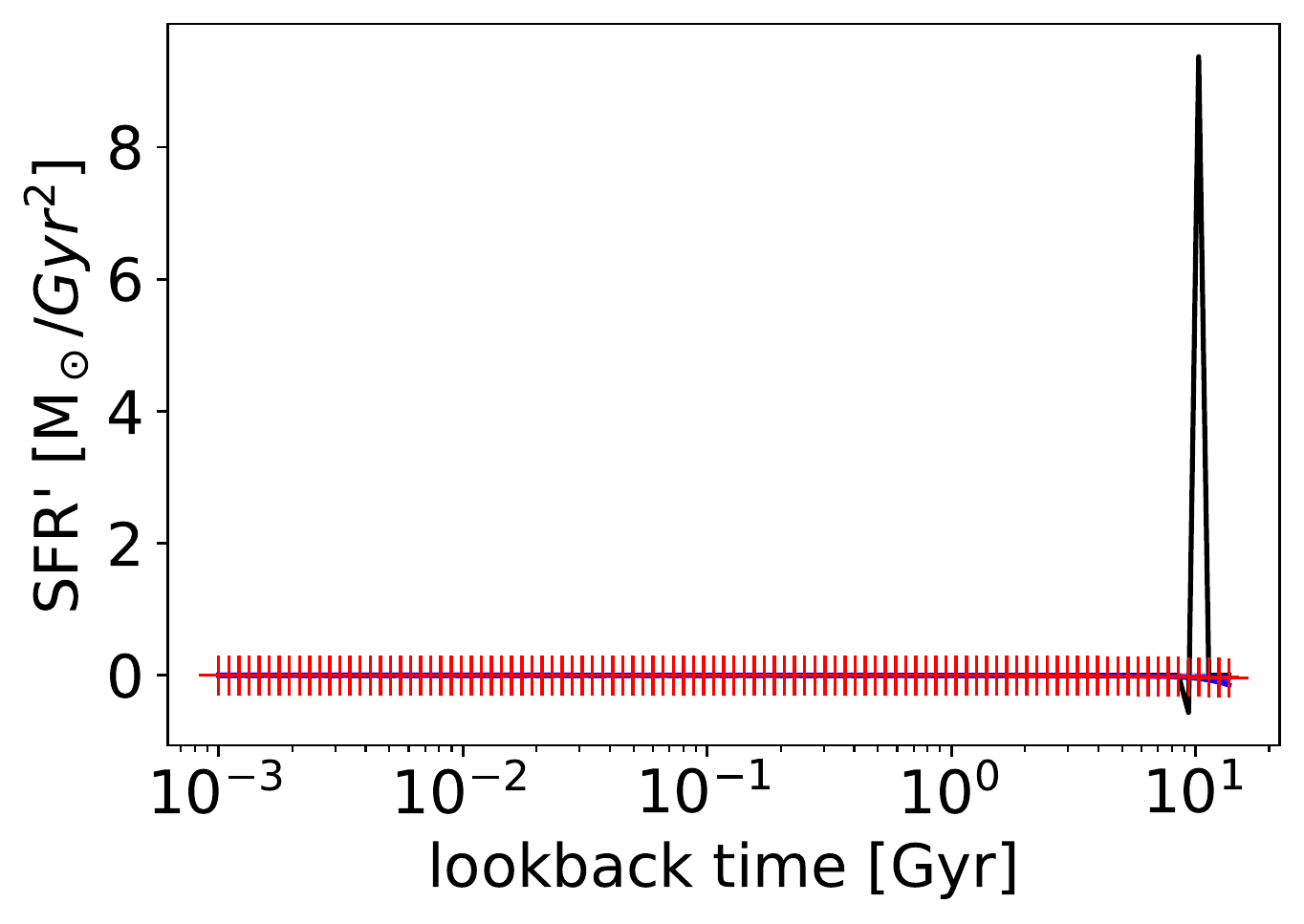} &
    \includegraphics[width=.23\textwidth]{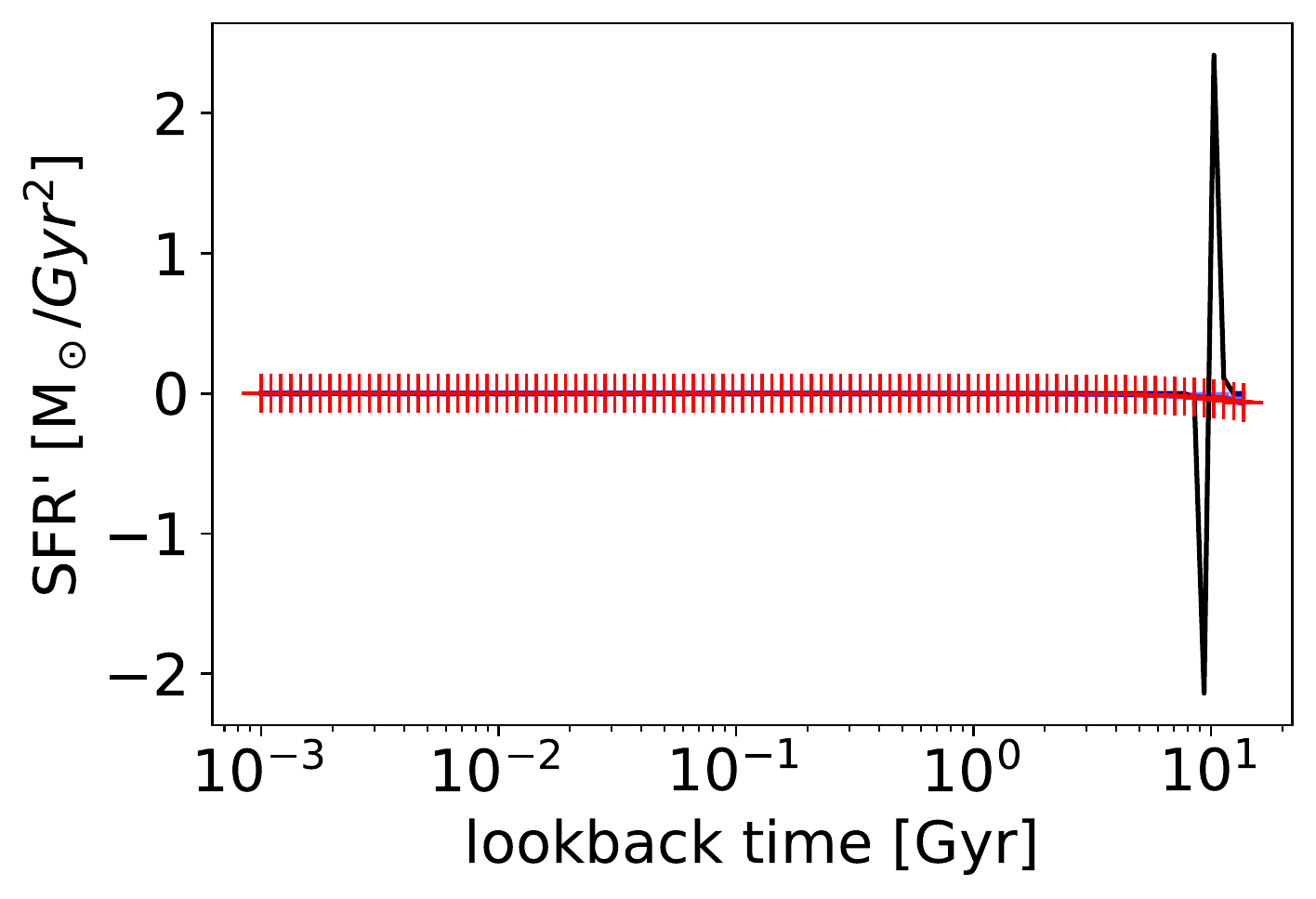} &
    \includegraphics[width=.23\textwidth]{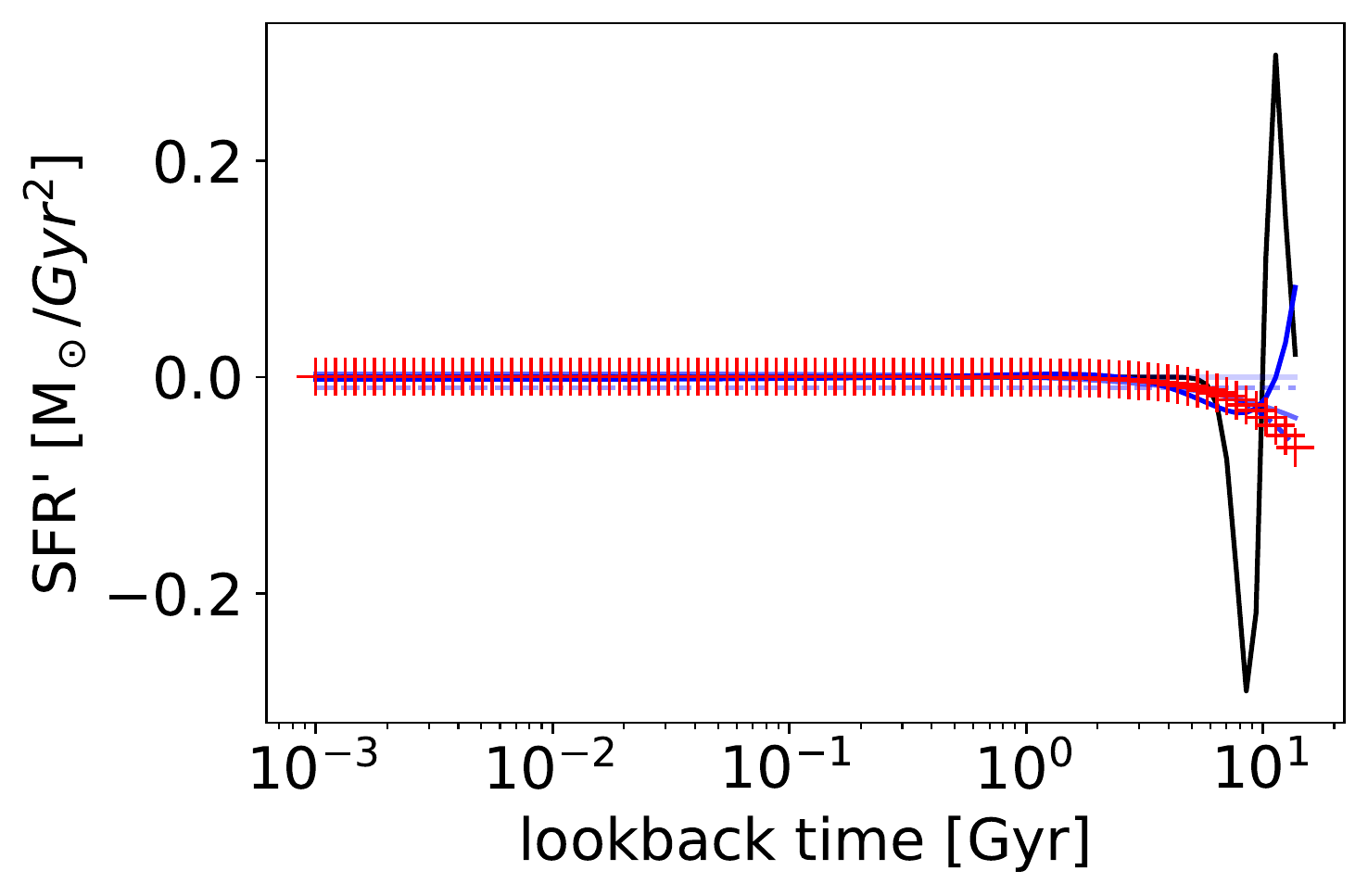} &
    \includegraphics[width=.23\textwidth]{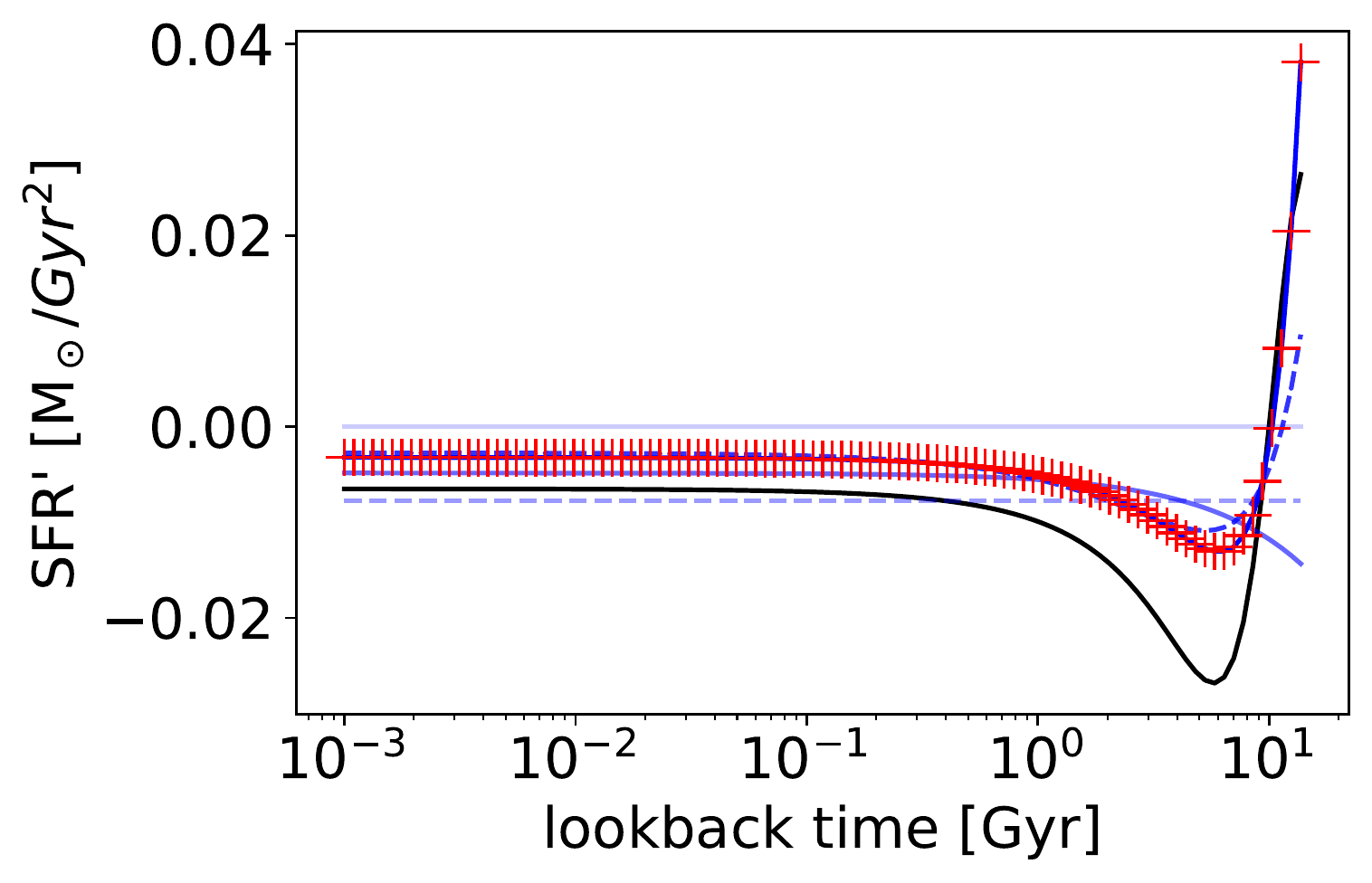} 
  \end{tabular}
  \caption{Time derivative of the SFR reconstructed with our parametric model from the luminosities of a synthetic Gaussian SFR with different peak ages (increasing from top to bottom) and FWHMs (increasing from left to right), compared to the input model (solid black line). Colour gradients and line styles indicate the degree of the polynomial reconstruction from $N=5$ (solid) to lower degrees (more diffuse lines; solid and dashed for odd and even $N$). Red crosses correspond to the best positive-SFR fit.}
  \label{fig:Gauss_rec_SFRP}
\end{figure*}

\begin{figure*}
\centering
  \begin{tabular}{@{}cccc@{}}
  
    \includegraphics[width=.33\textwidth]{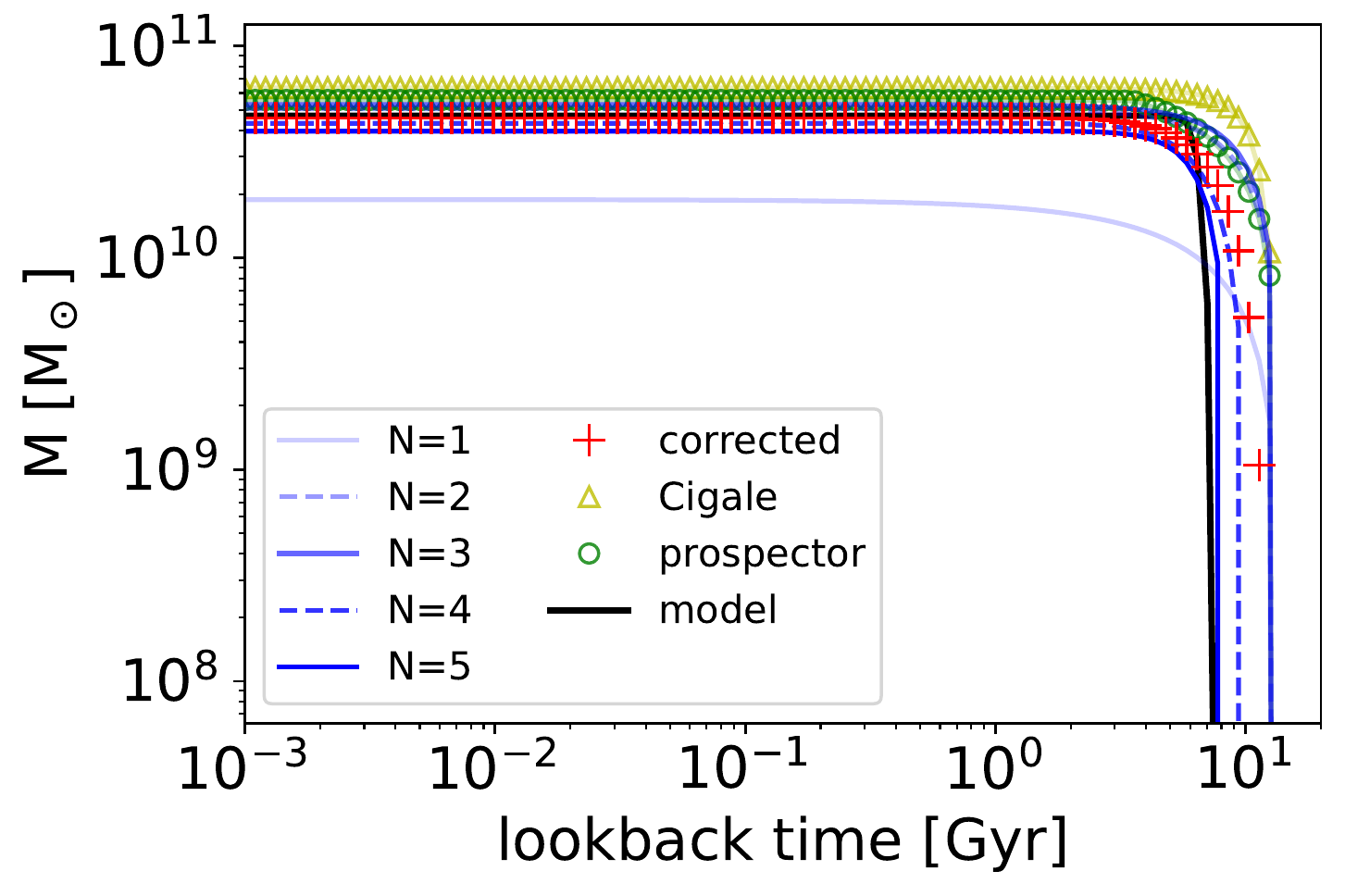} &
    \includegraphics[width=.33\textwidth]{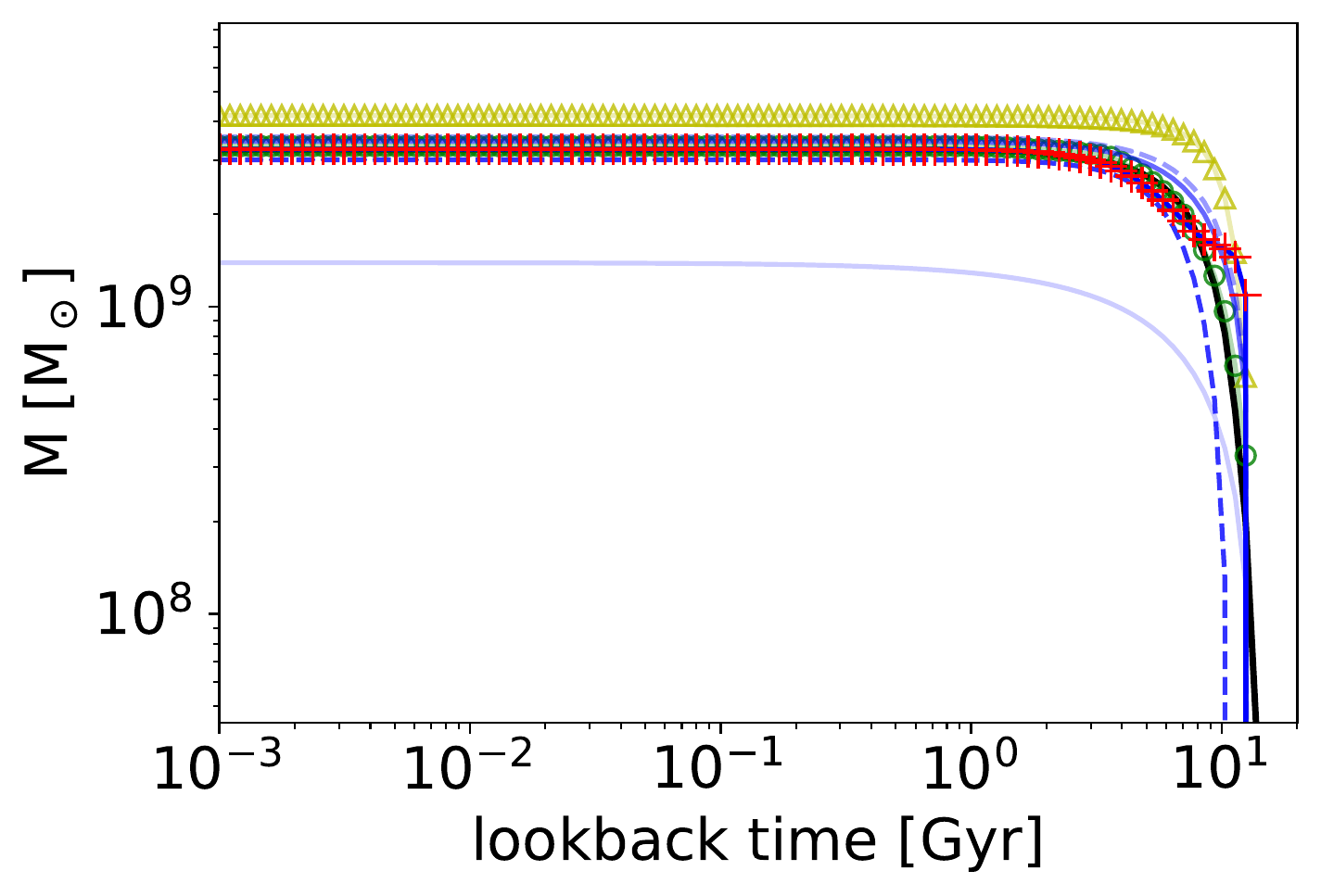} &
    \includegraphics[width=.33\textwidth]{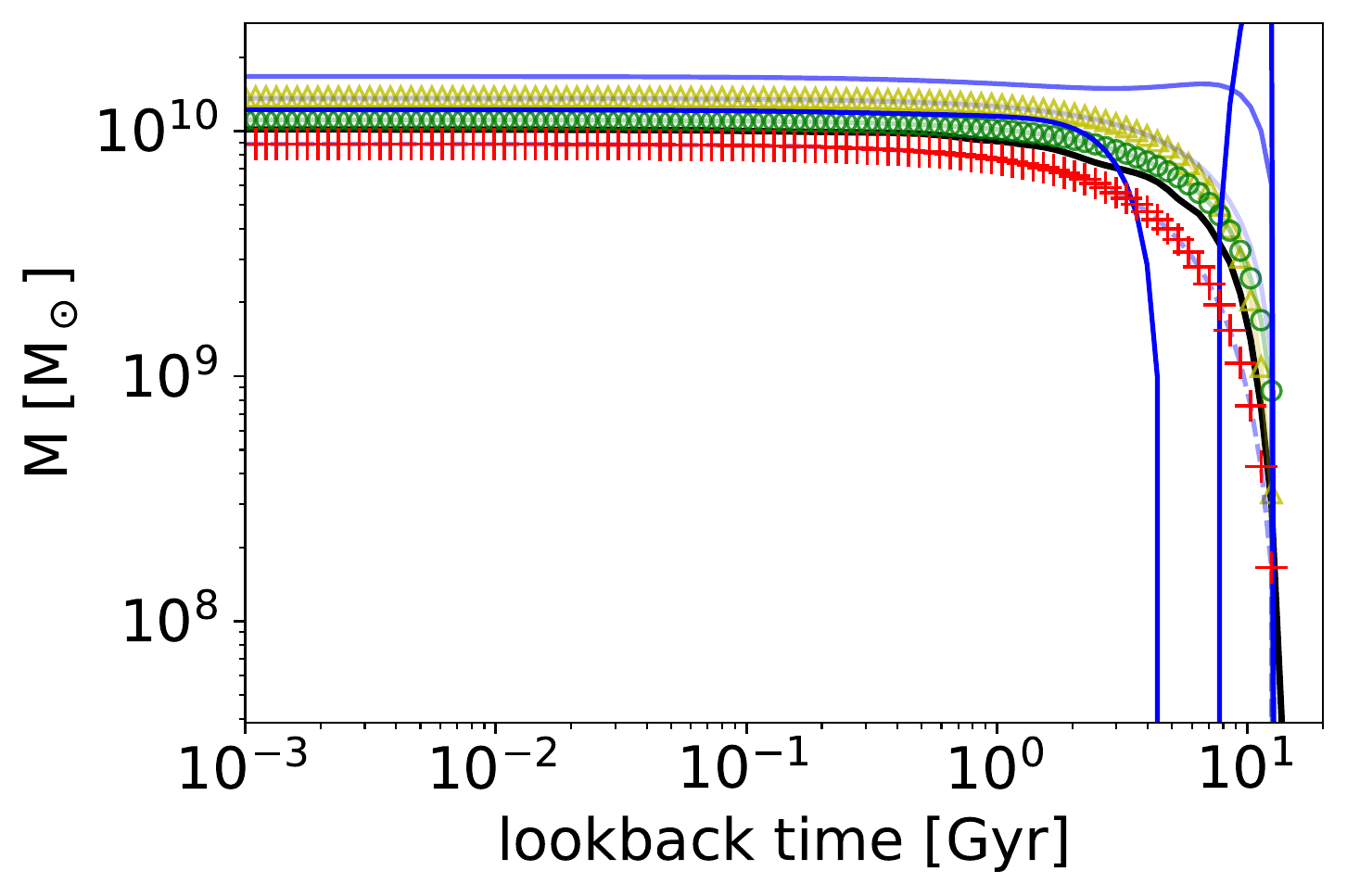} \\

    \includegraphics[width=.33\textwidth]{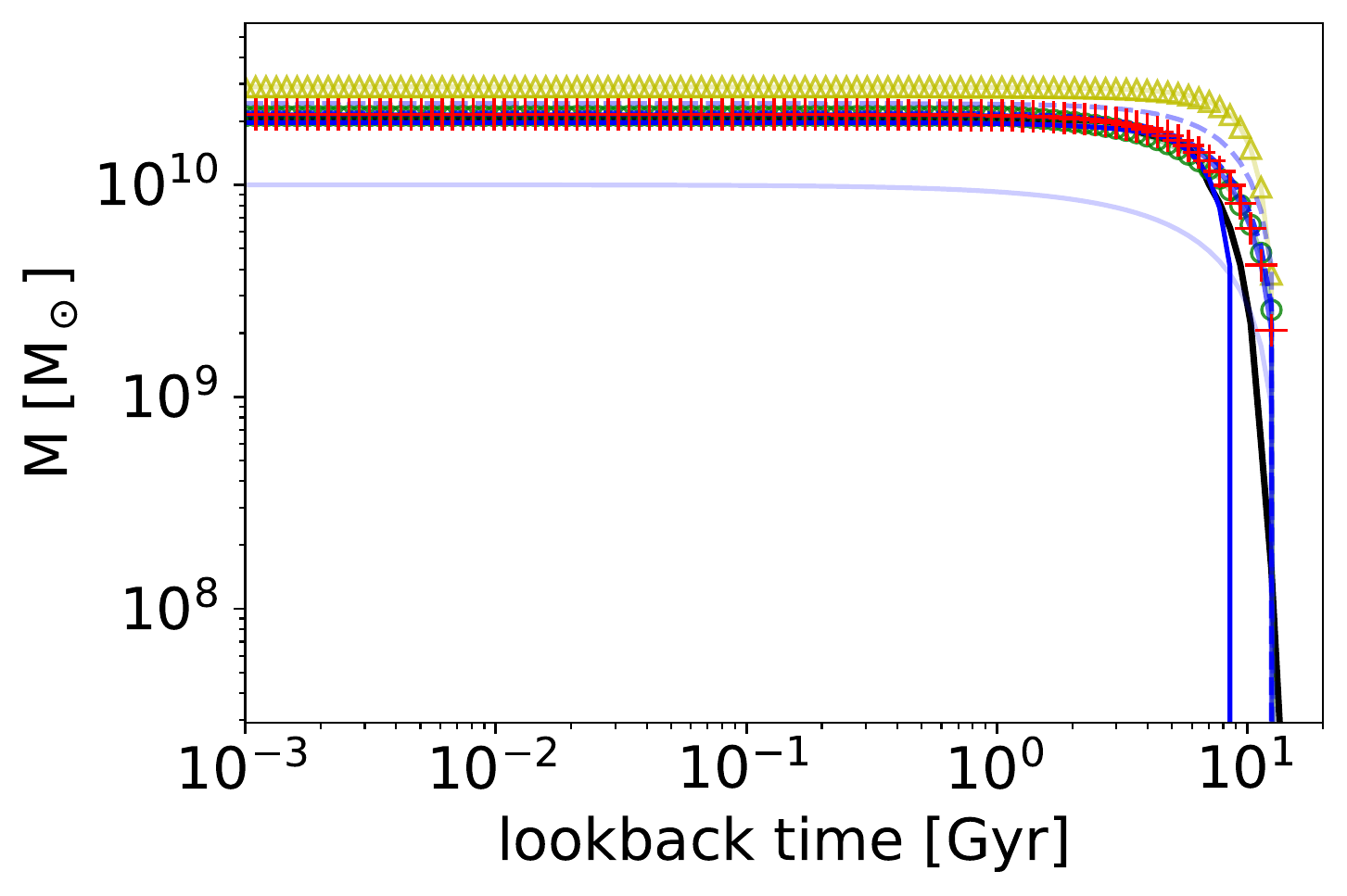} &
    \includegraphics[width=.33\textwidth]{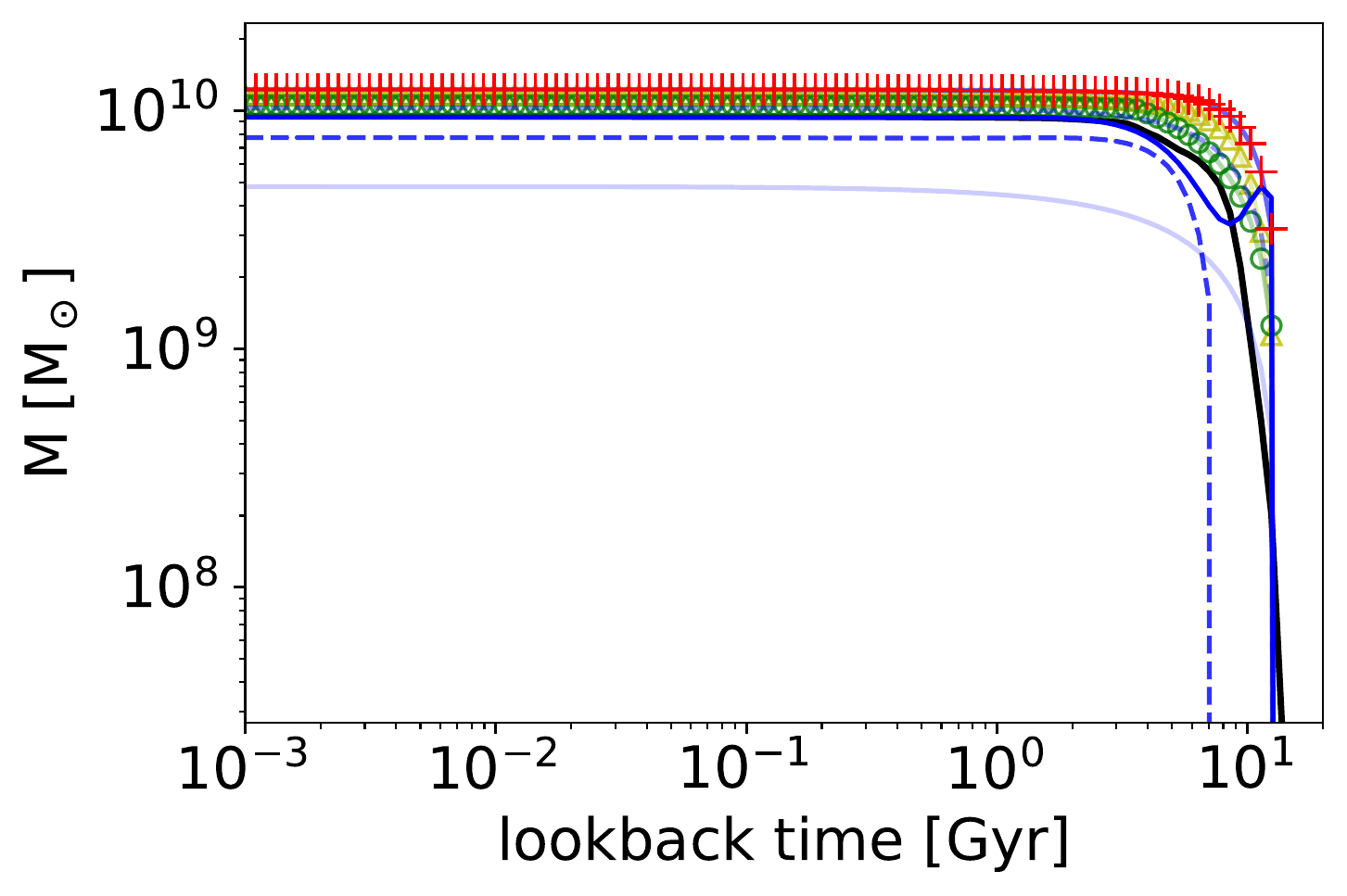} &
    \includegraphics[width=.33\textwidth]{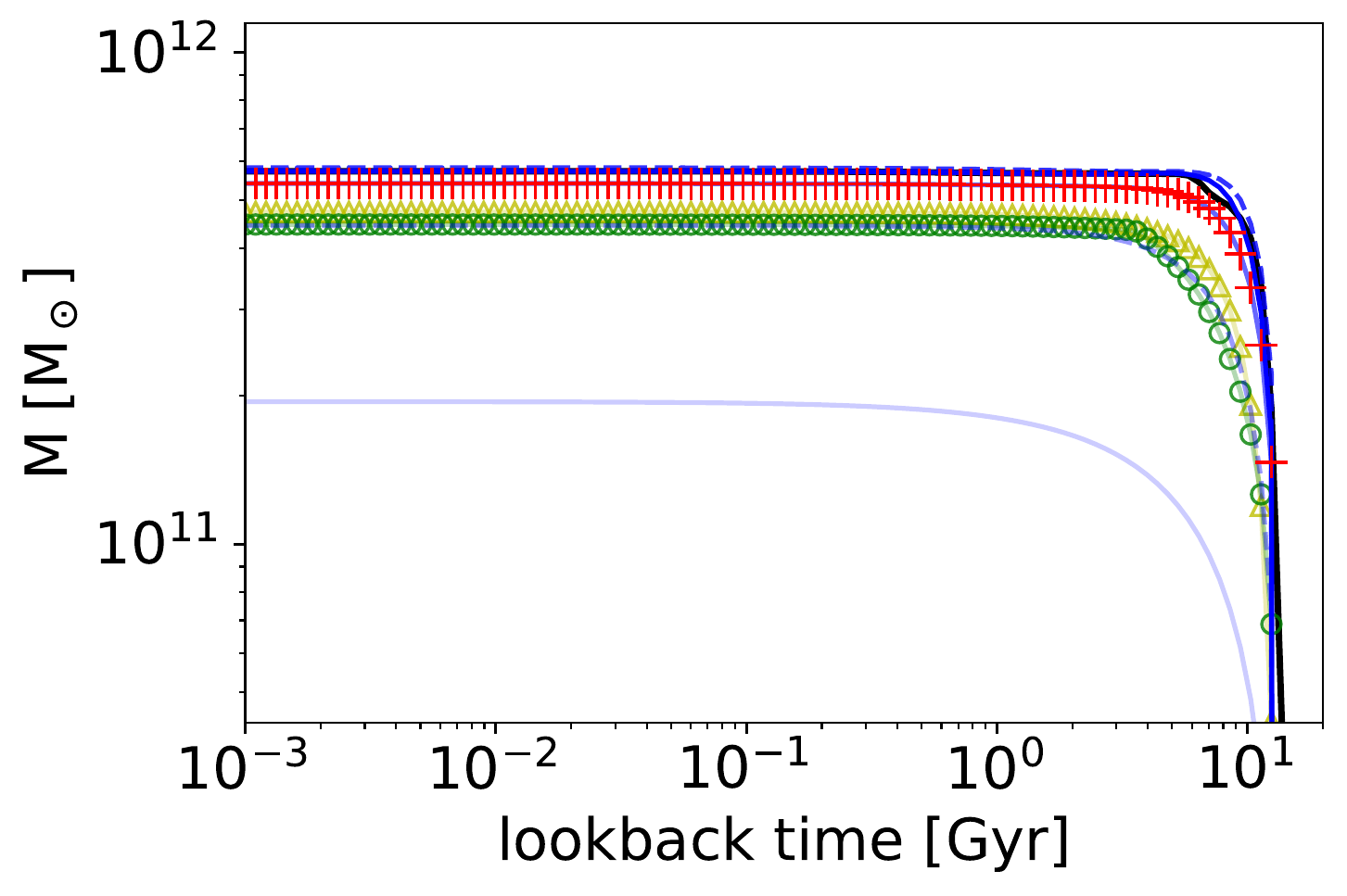} \\
    
    \includegraphics[width=.33\textwidth]{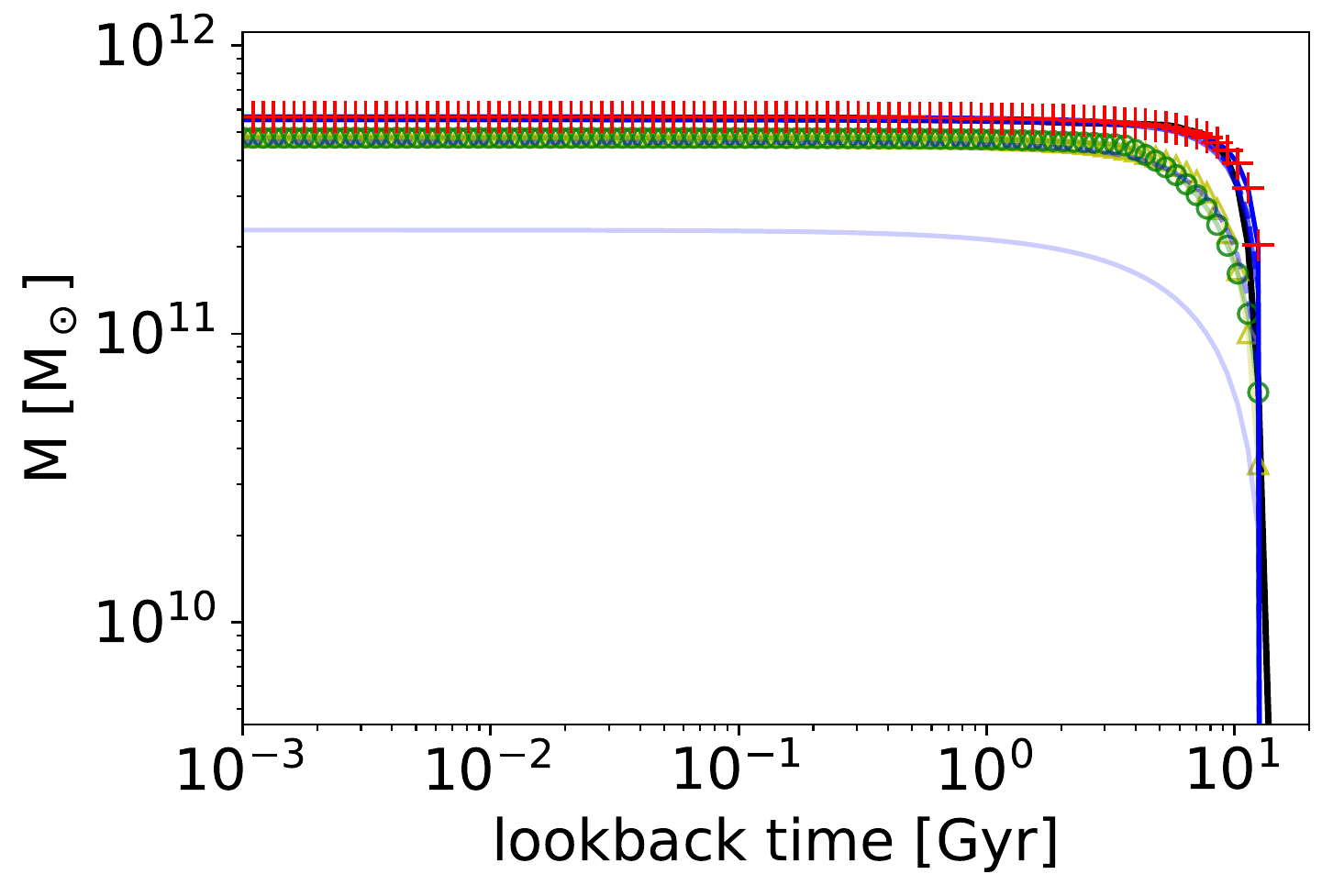} &
    \includegraphics[width=.33\textwidth]{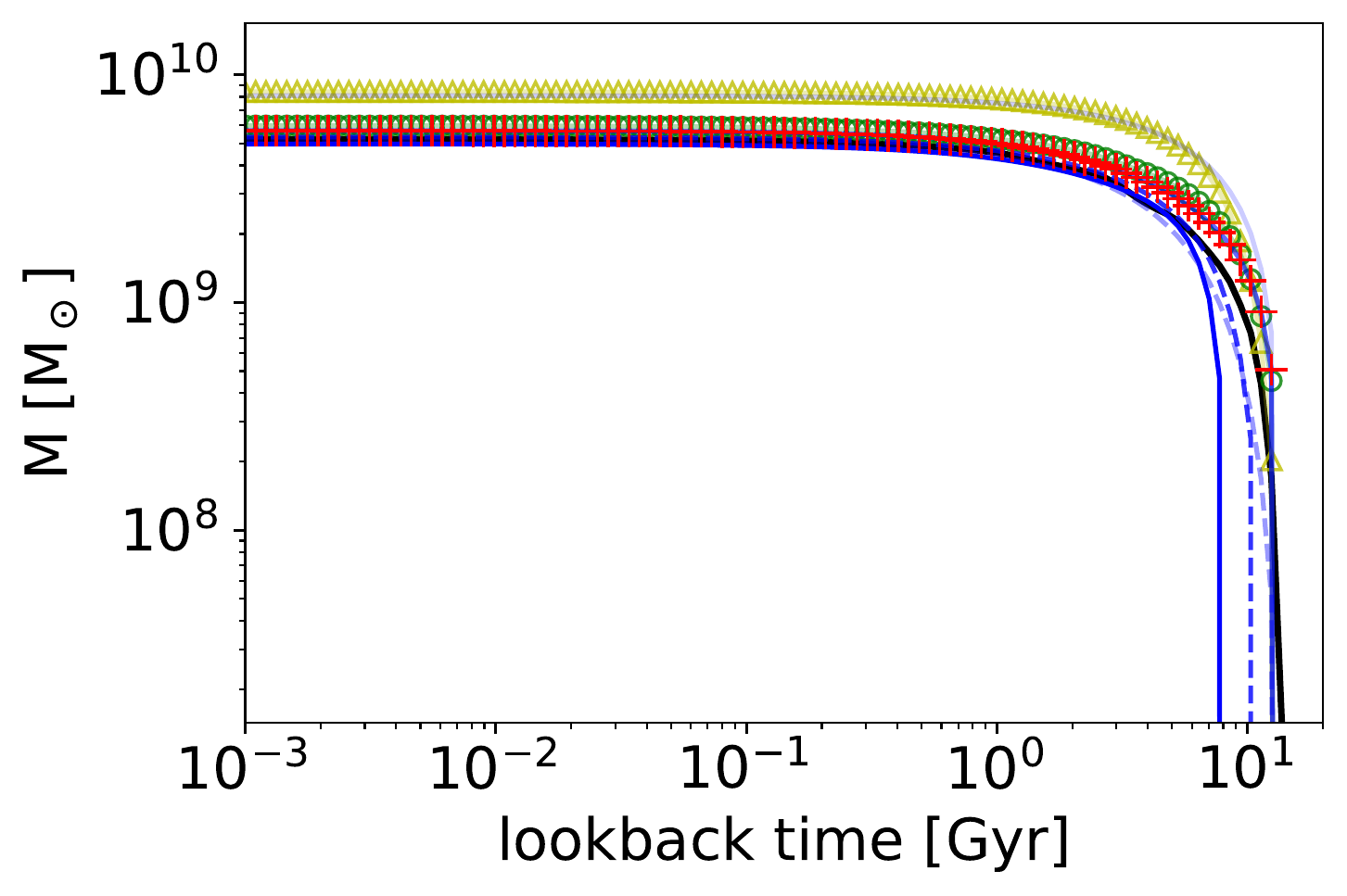} &
    \includegraphics[width=.33\textwidth]{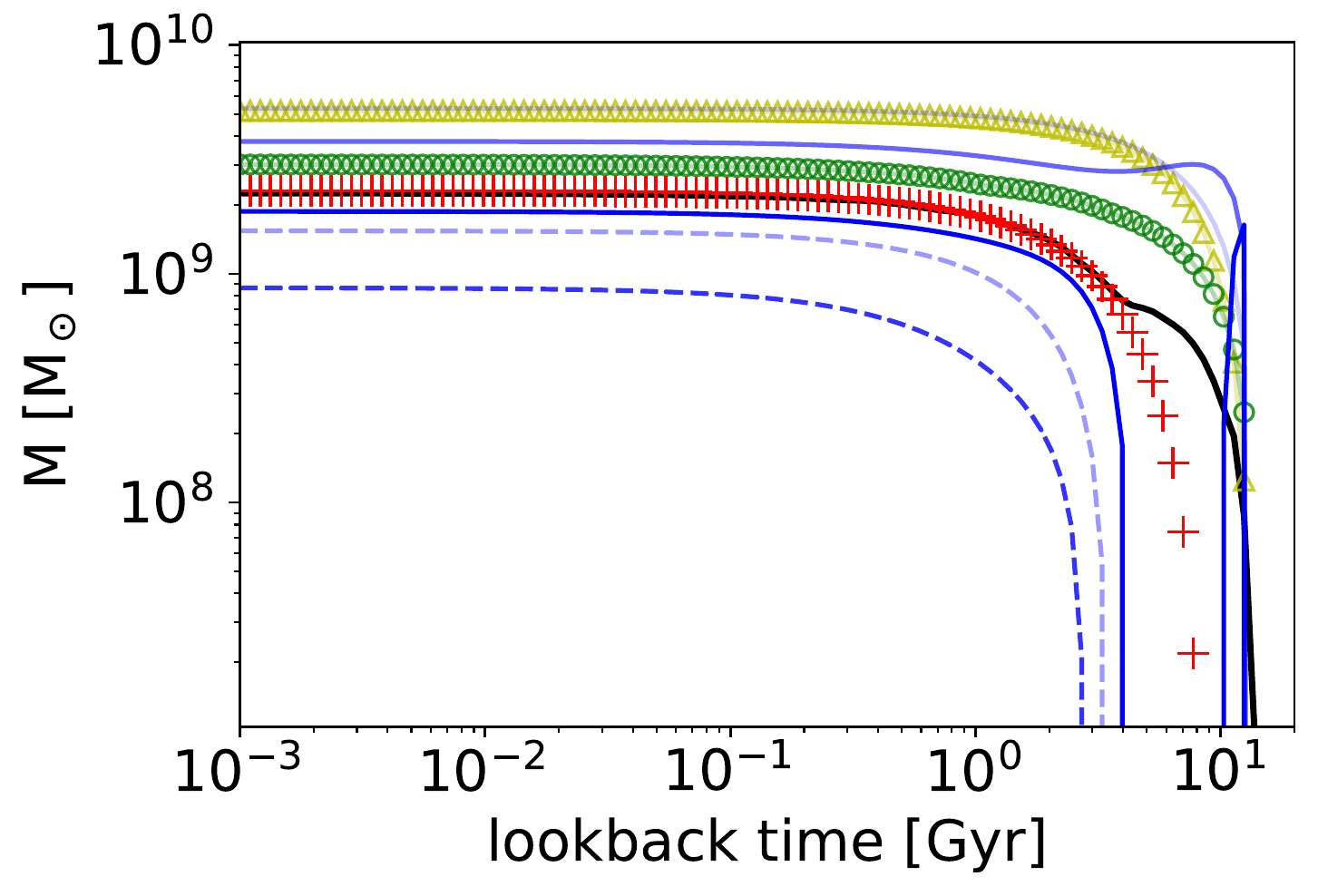} \\

  \end{tabular}
  \caption{Mass formation history for different galaxies of the Illustris sample (black line). The polynomial expansion reconstruction is shown as the different solid (odd degree) and dashed (even degree) blue lines. Results from \prosp\ are plotted as green circles, and the results from \cig\ are plotted as yellow triangles. Red crosses correspond to the best positive-SFR fit.}
  \label{fig:Illustris_reconstruction_mass}
\end{figure*}

\begin{figure*}
\centering
  \begin{tabular}{@{}cccc@{}}
  
    \includegraphics[width=.33\textwidth]{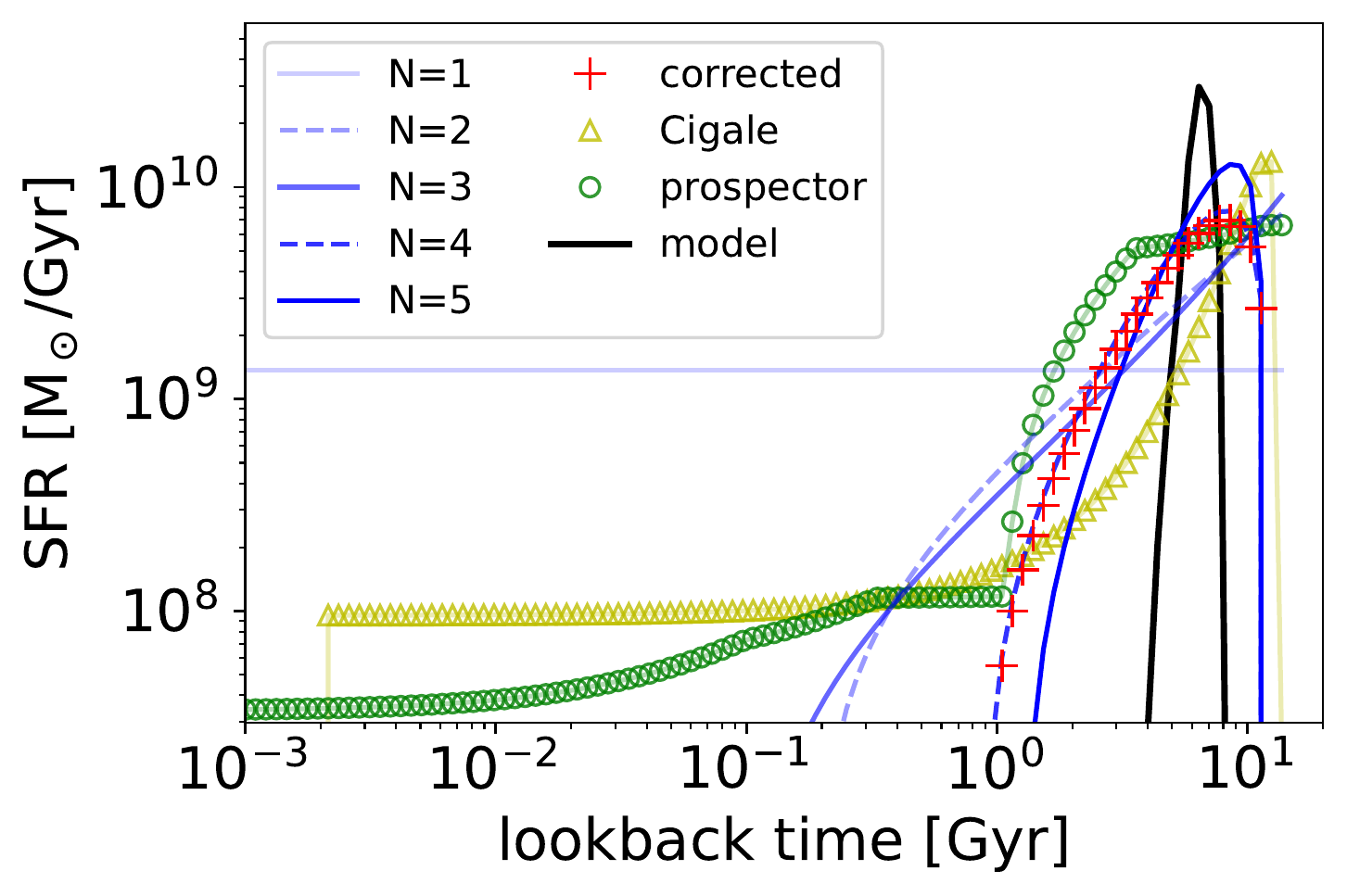} &
    \includegraphics[width=.33\textwidth]{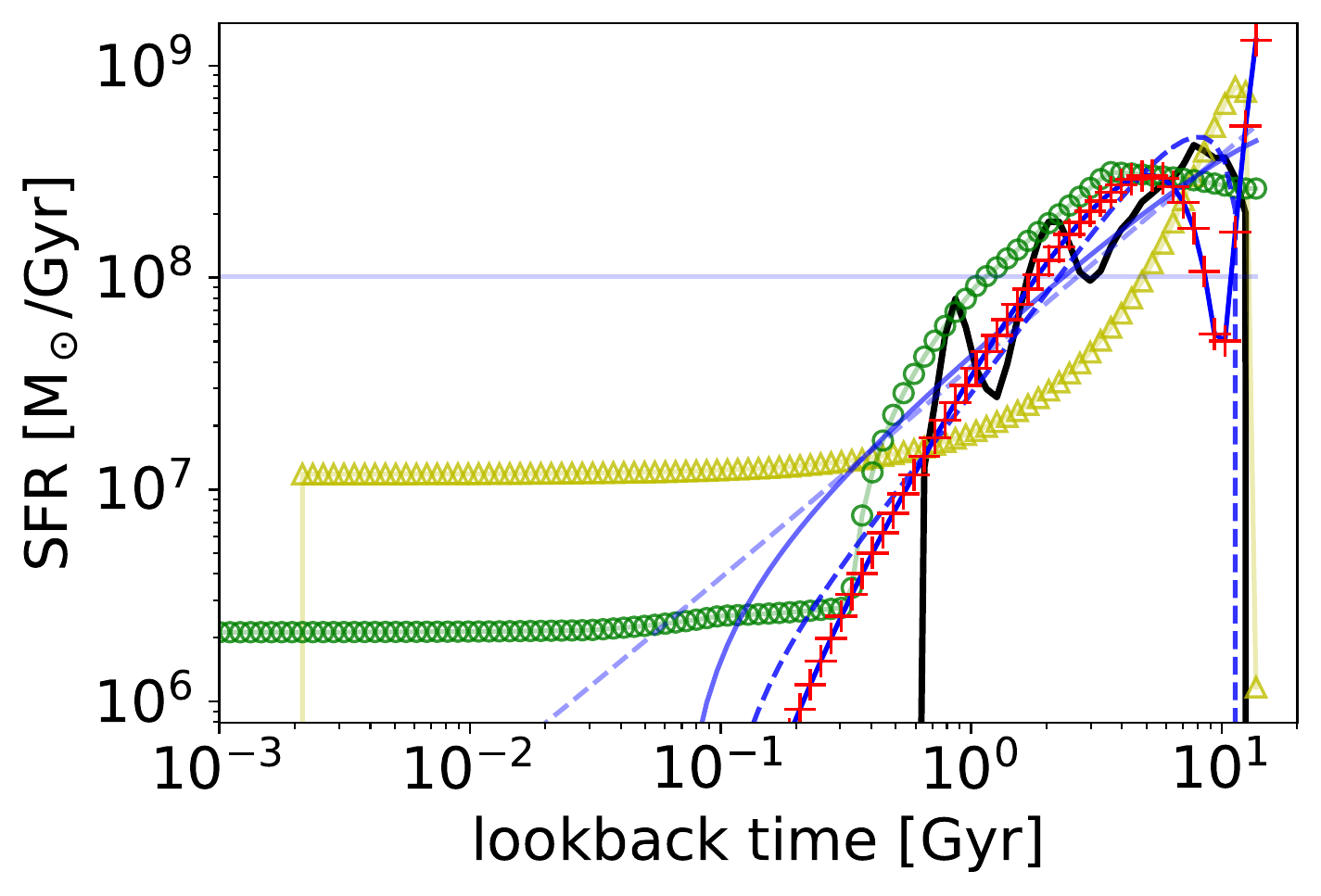} &
    \includegraphics[width=.33\textwidth]{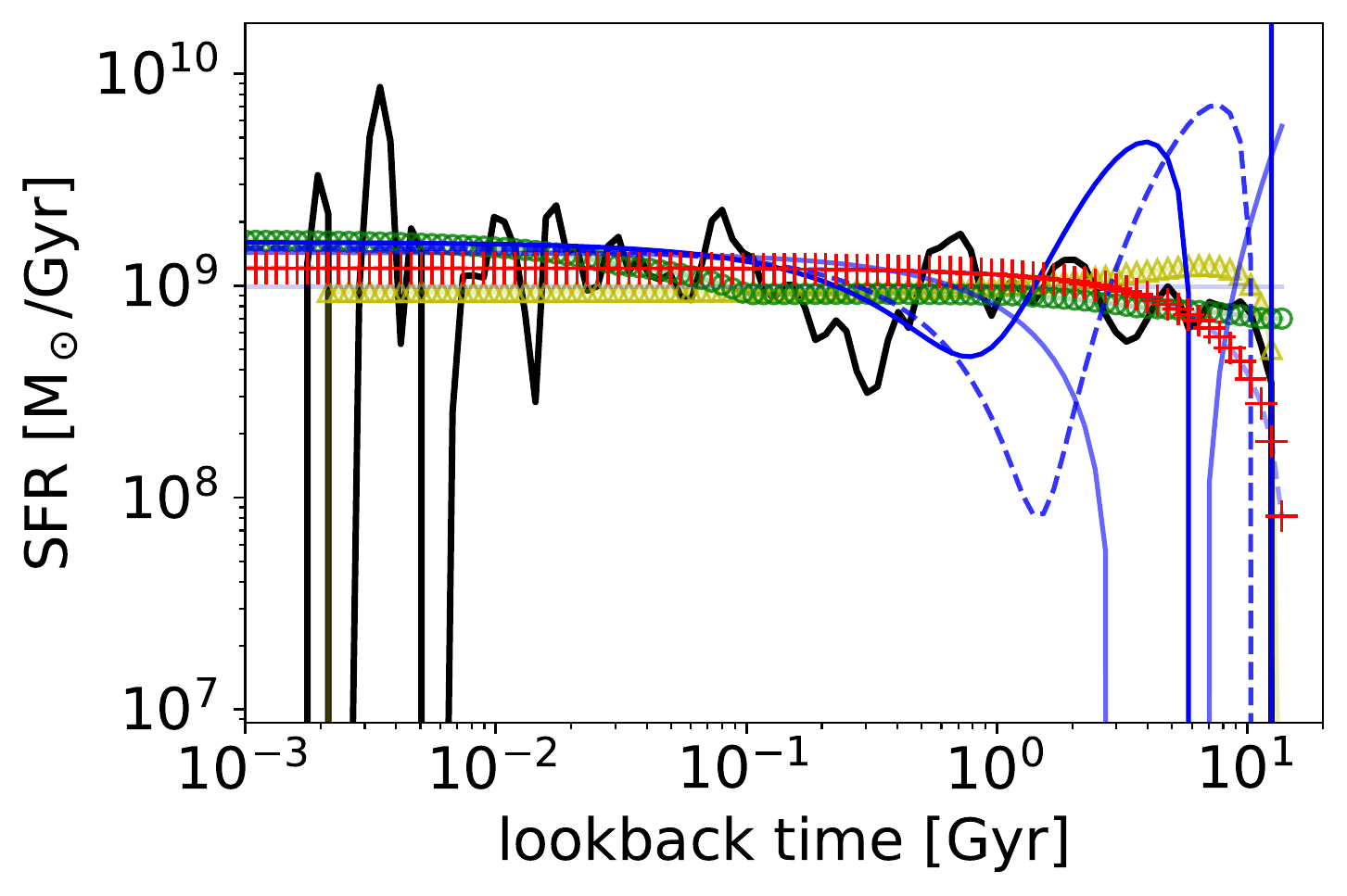} \\

    \includegraphics[width=.33\textwidth]{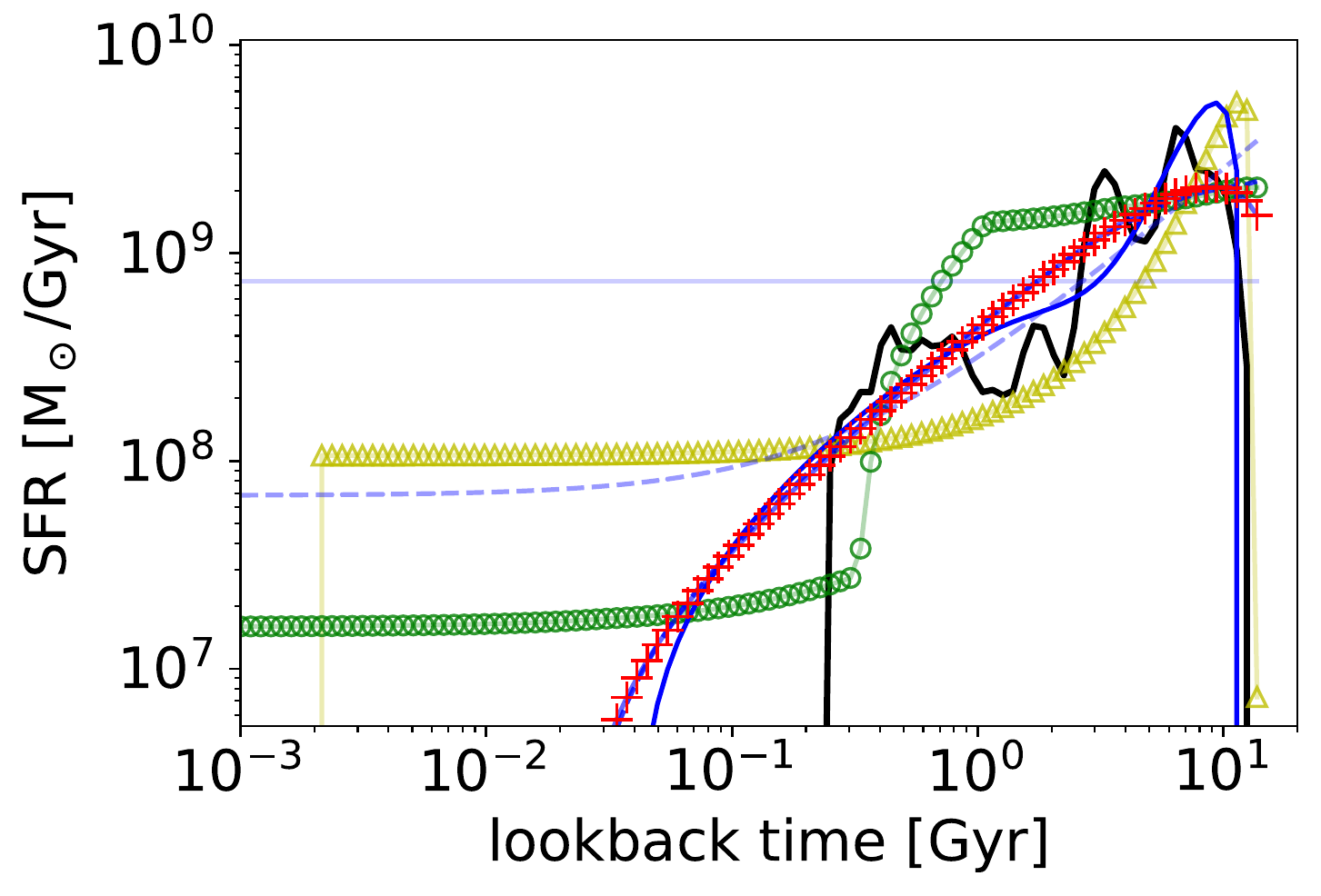} &
    \includegraphics[width=.33\textwidth]{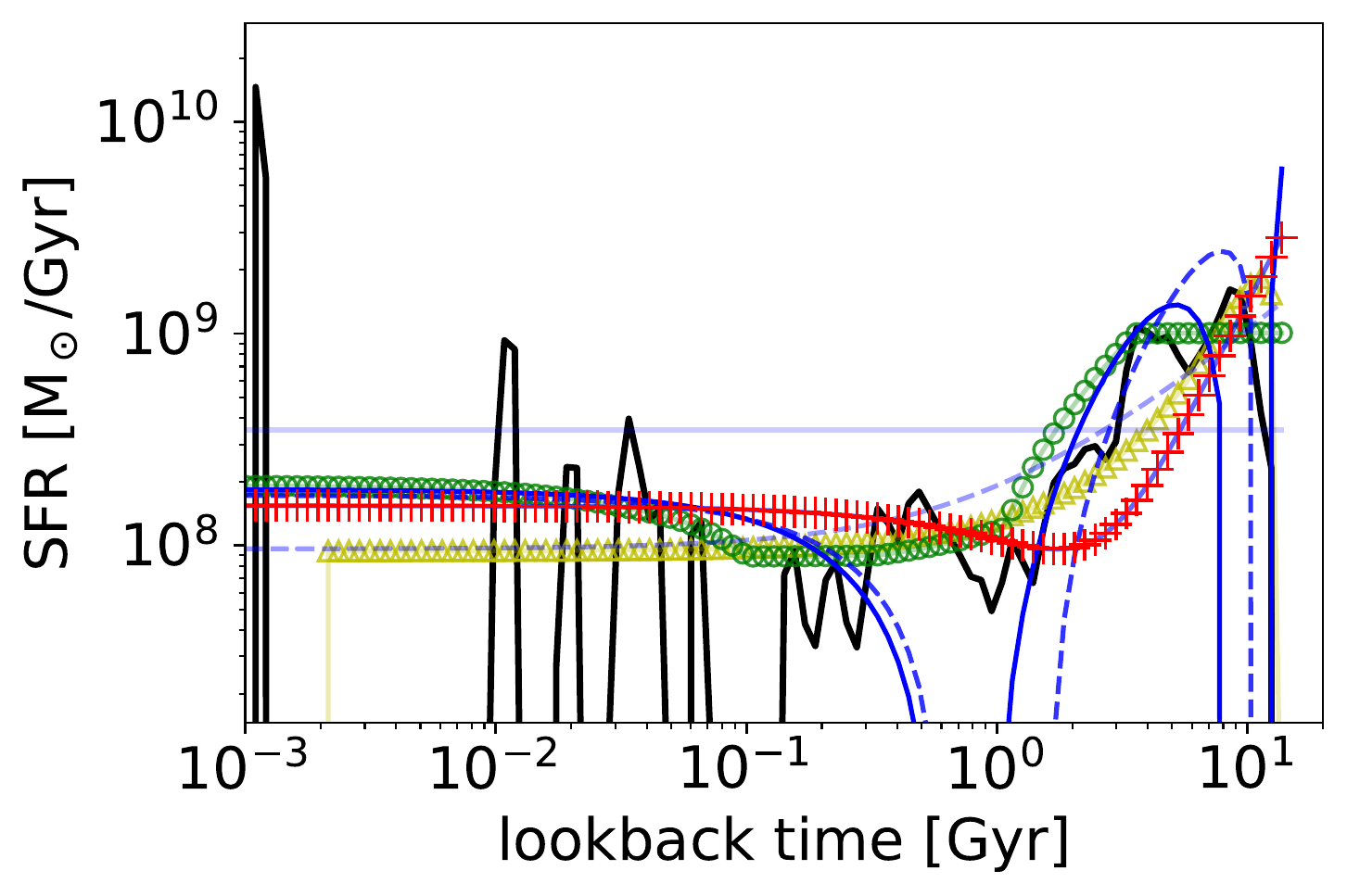} &
    \includegraphics[width=.33\textwidth]{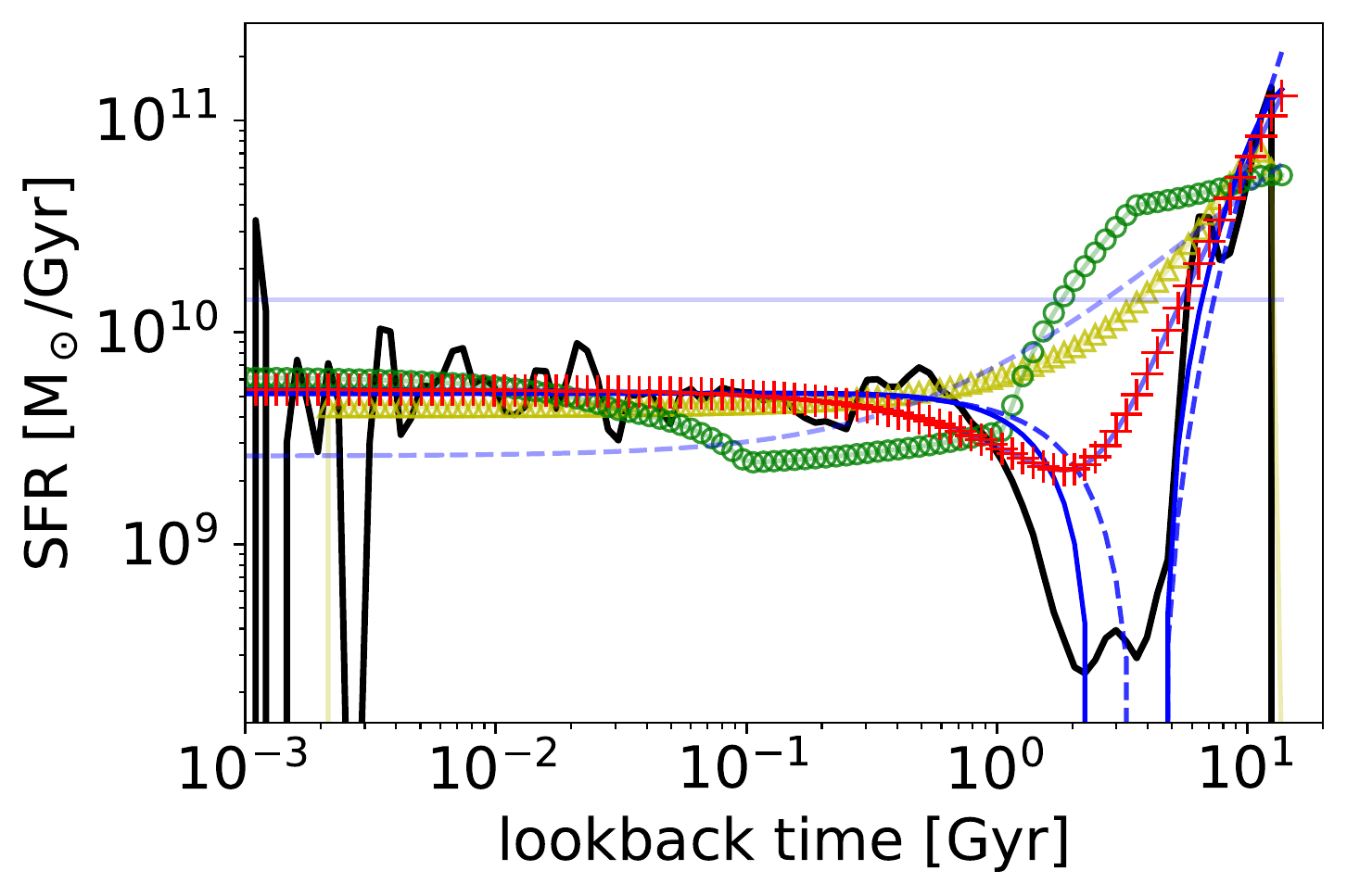} \\
    
    \includegraphics[width=.33\textwidth]{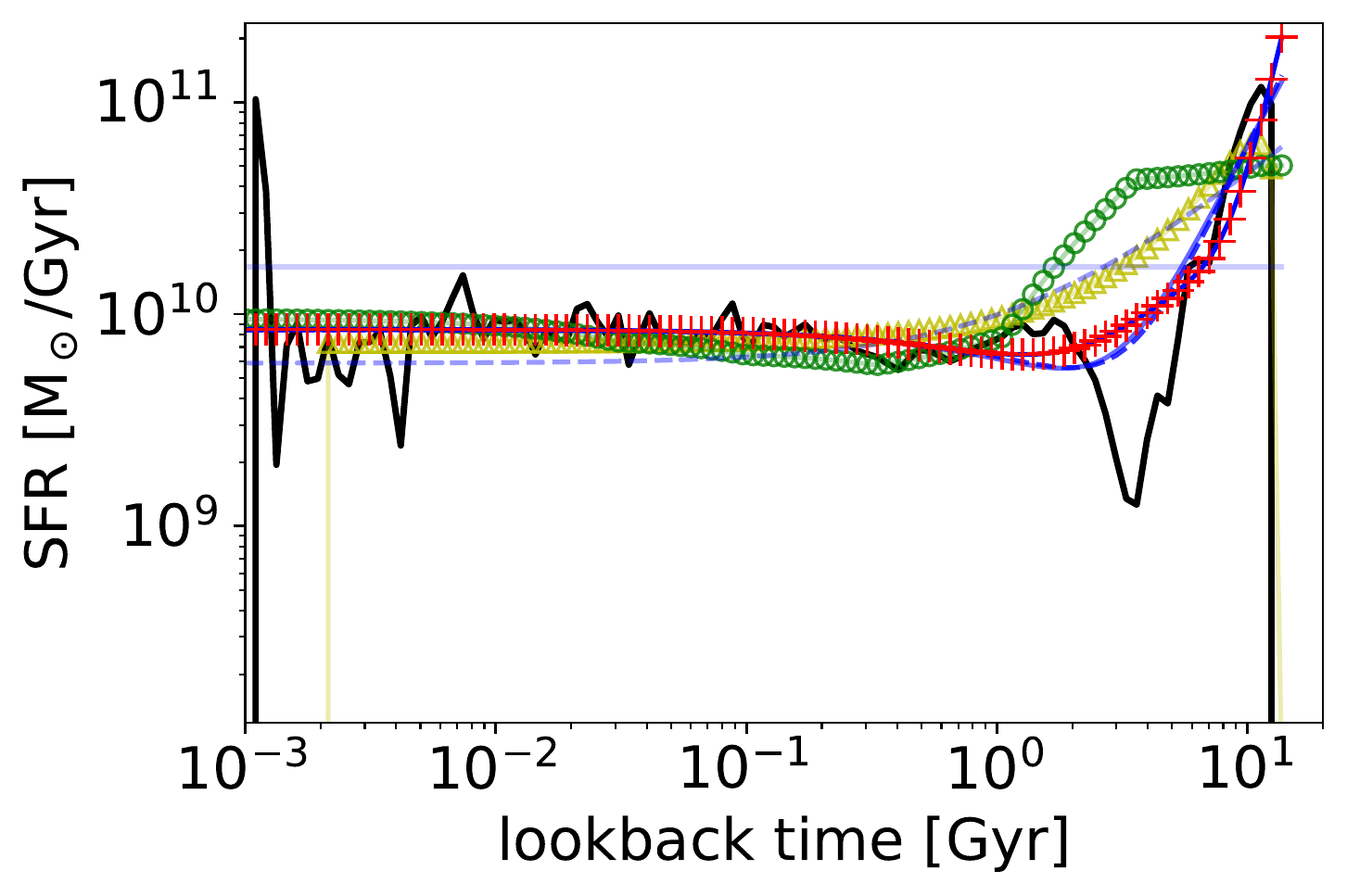} &
    \includegraphics[width=.33\textwidth]{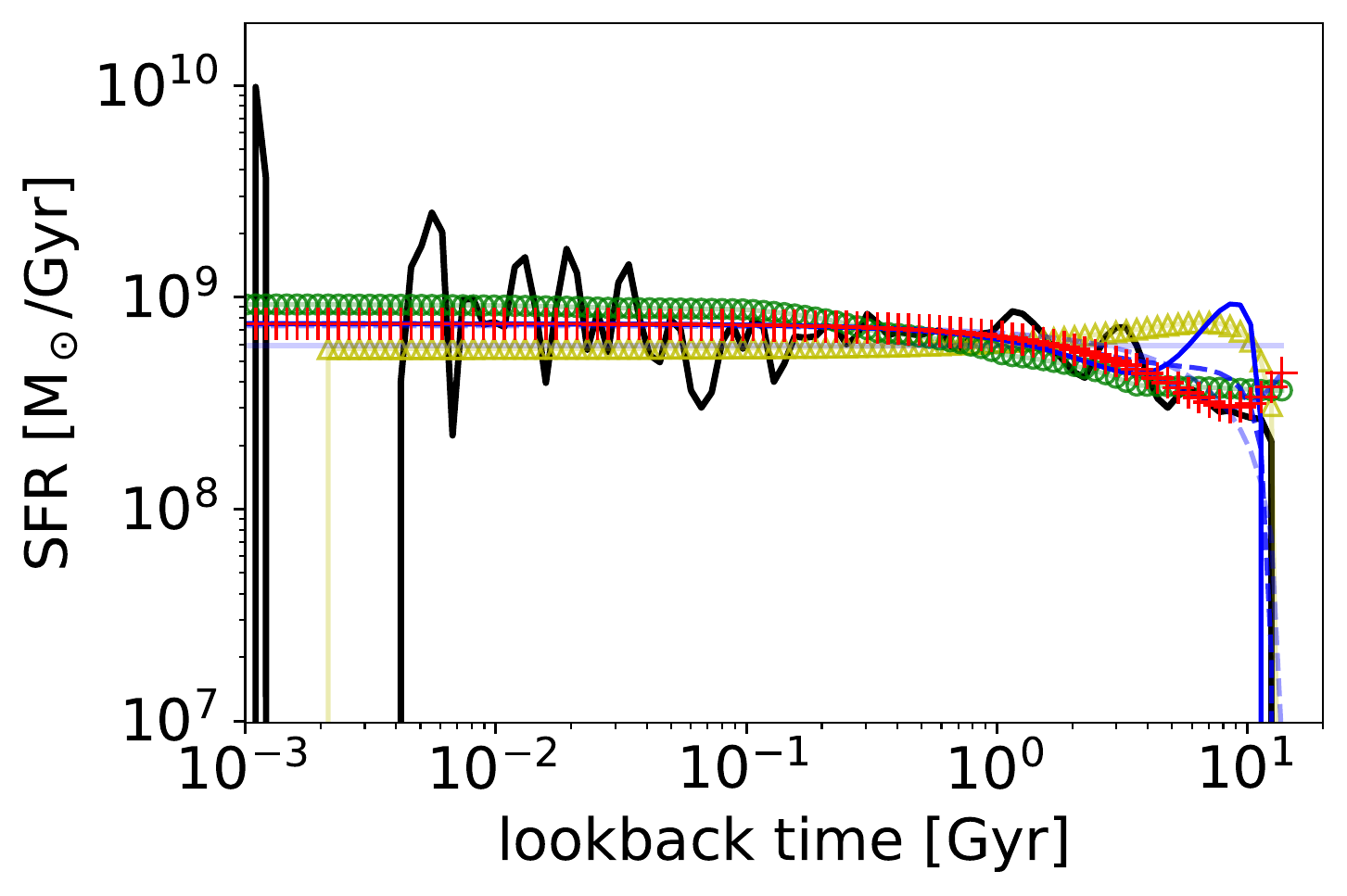} &
    \includegraphics[width=.33\textwidth]{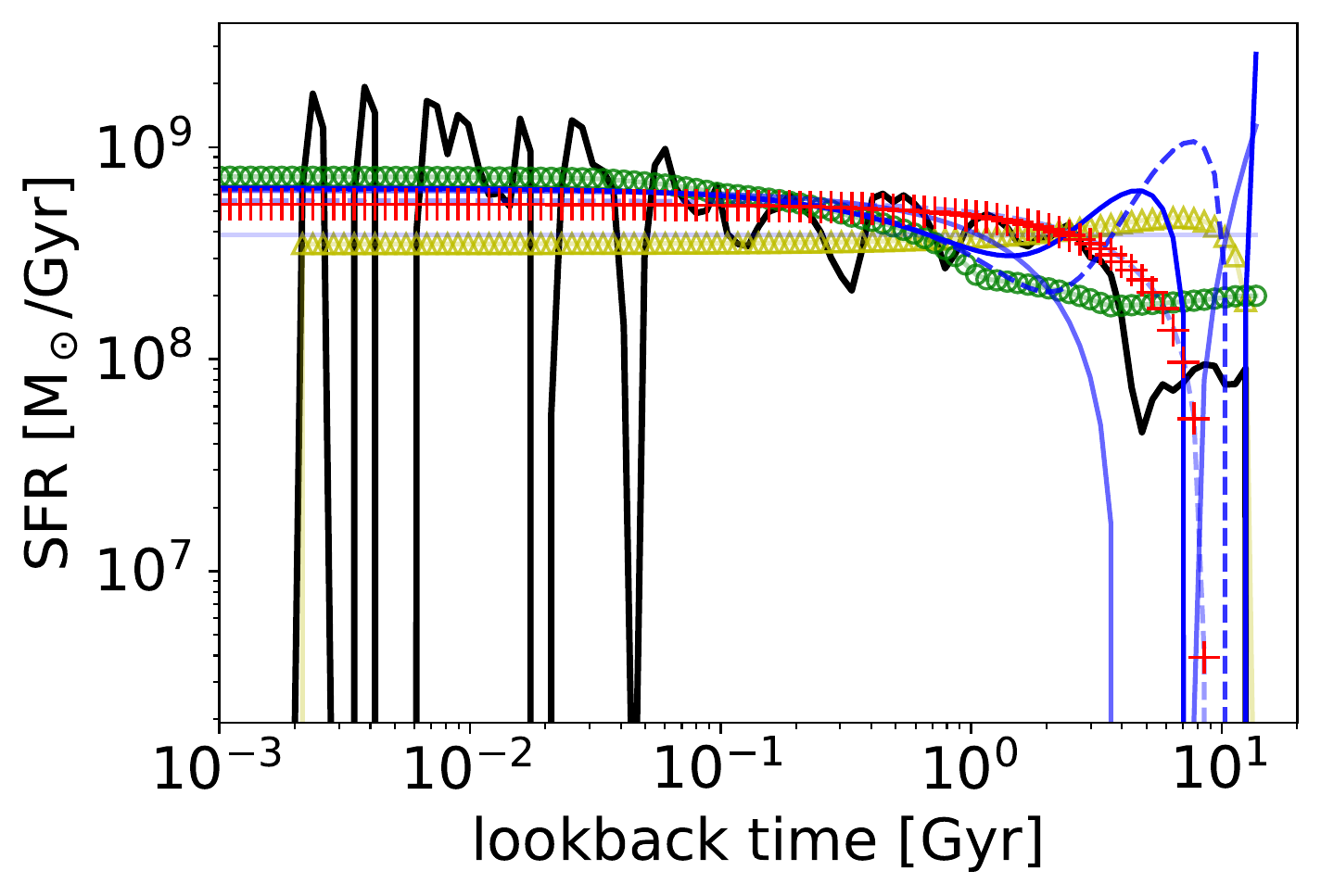} \\
    
  \end{tabular}
  \caption{Star formation rate for different galaxies of the Illustris sample (black line). The polynomial expansion reconstruction is shown as the different solid (odd degree) and dashed (even degree) blue lines. Results from \prosp\ are plotted as green circles, and the results from \cig\ are plotted as yellow triangles. Red crosses correspond to the best positive-SFR fit.}
  \label{fig:Illustris_reconstruction_sfr}
\end{figure*}

The reconstructed MFHs, $M(t)$, SFHs, $\Psi(t) \equiv \dot M(t)$, and their time derivatives, $\dot\Psi(t)$ and $\ddot\Psi(t)$, are illustrated in Figs.~\ref{fig:rec_exp} and \ref{fig:rec_dt} for the exponential and delayed-$\tau$ models, respectively, with characteristic timescales $\tau=\{0.5,\ 3.5,\ 7.5,\ 13.5\}$~Gyr.
Figures~\ref{fig:Gauss_rec_mass}, \ref{fig:Gauss_rec_SFR}, and~\ref{fig:Gauss_rec_SFRP} show $M$, $\Psi$, and $\dot\Psi$, respectively, for the Gaussian models with representative values of their peak age and FWHM.
Finally, Figs.~\ref{fig:Illustris_reconstruction_mass} and~\ref{fig:Illustris_reconstruction_sfr} show the reconstructed histories from the Illustris-TNG sample, as well as the comparison with the results obtained with \cig\ and \prosp.

\end{appendix}

\label{lastpage}
\end{document}